\documentclass[12pt]{article}
\usepackage{epsfig}
\usepackage{graphicx}
\usepackage{amsmath}

\begin{document}

 \newcommand{\beq}{\begin{equation}}
\newcommand{\eeq}{\end{equation}}
\newcommand{\bea}{\begin{eqnarray}}
\newcommand{\eea}{\end{eqnarray}}
\newcommand{\beqn}{\begin{eqnarray}}
\newcommand{\eeqn}{\end{eqnarray}}
\newcommand{\beas}{\begin{eqnarray*}}
\newcommand{\eeas}{\end{eqnarray*}}
\newcommand{\defi}{\stackrel{\rm def}{=}}
\newcommand{\non}{\nonumber}
\newcommand{\bquo}{\begin{quote}}
\newcommand{\enqu}{\end{quote}}
\newcommand{\qt}{\tilde q}


\def\de{\partial}
\def\Tr{ \hbox{\rm Tr}}
\def\const{\hbox {\rm const.}}
\def\o{\over}
\def\im{\hbox{\rm Im}}
\def\re{\hbox{\rm Re}}
\def\bra{\langle}\def\ket{\rangle}
\def\Arg{\hbox {\rm Arg}}
\def\Re{\hbox {\rm Re}}
\def\Im{\hbox {\rm Im}}
\def\diag{\hbox{\rm diag}}


\def\QATOPD#1#2#3#4{{#3 \atopwithdelims#1#2 #4}}
\def\stackunder#1#2{\mathrel{\mathop{#2}\limits_{#1}}}
\def\stackreb#1#2{\mathrel{\mathop{#2}\limits_{#1}}}
\def\Tr{{\rm Tr}}
\def\res{{\rm res}}
\def\Bf#1{\mbox{\boldmath $#1$}}
\def\balpha{{\Bf\alpha}}
\def\bbeta{{\Bf\beta}}
\def\bgamma{{\Bf\gamma}}
\def\bnu{{\Bf\nu}}
\def\bmu{{\Bf\mu}}
\def\bphi{{\Bf\phi}}
\def\bPhi{{\Bf\Phi}}
\def\bomega{{\Bf\omega}}
\def\blambda{{\Bf\lambda}}
\def\brho{{\Bf\rho}}
\def\bsigma{{\bfit\sigma}}
\def\bxi{{\Bf\xi}}
\def\bbeta{{\Bf\eta}}
\def\d{\partial}
\def\der#1#2{\frac{\d{#1}}{\d{#2}}}
\def\Im{{\rm Im}}
\def\Re{{\rm Re}}
\def\rank{{\rm rank}}
\def\diag{{\rm diag}}
\def\2{{1\over 2}}
\def\ntwo{${\mathcal N}=2\,$}
\def\nfour{${\mathcal N}=4\;$}
\def\none{${\mathcal N}=1\,$}
\def\x{\stackrel{\otimes}{,}}

\def\ba{\beq\new\begin{array}{c}}
\def\ea{\end{array}\eeq}
\def\be{\ba}
\def\ee{\ea}
\def\stackreb#1#2{\mathrel{\mathop{#2}\limits_{#1}}}

\def\Tr{{\rm Tr}}
\newcommand{\vp}{\varphi}
\newcommand{\pt}{\partial}
\newcommand{\ve}{\varepsilon}
\renewcommand{\theequation}{\thesubsection.\arabic{equation}}

\setcounter{footnote}0

\vfill

\begin{titlepage}

\begin{flushright}
FTPI-MINN-07/06, UMN-TH-2539/07\\
ITEP-TH-12/07 \\
April 16, 2007
\end{flushright}

\vspace{1.5cm}

\begin{center}
{ \Large \bf  Supersymmetric Solitons and
How They Help \\[2mm]
Us Understand Non-Abelian Gauge Theories}
\end{center}

\vspace{5mm}

\begin{center}

 {\large
 \bf    M.~Shifman$^{\,a}$ and \bf A.~Yung$^{\,\,a,b,c}$}
\end {center}

\begin{center}

\vspace{5mm}

$^a${\it  William I. Fine Theoretical Physics Institute,
University of Minnesota,
Minneapolis, MN 55455, USA}\\[1mm]
$^{b}${\it Petersburg Nuclear Physics Institute, Gatchina, St. Petersburg
188300, Russia\\[1mm]
$^c${\it Institute of Theoretical and Experimental Physics, Moscow
117259, Russia}}
\end{center}

\vspace*{.45cm}
\begin{center}
{\large\bf Abstract}
\end{center}

In the last decade it became clear that methods and techniques
based on supersymmetry provide deep insights in
quantum chromodynamics and other supersymmetric and
non-supersymmetric gauge theories
at strong coupling. In this review we summarize major advances in
the critical (Bogomol'nyi--Prasad--Sommerfield-saturated, BPS for short) solitons
in supersymmetric theories and their implications for understanding basic
dynamical regularities of non-supersymmetric  theories.
After a brief introduction in the theory of critical solitons (including a historical introduction)  we focus on three topics: (i) non-Abelian strings in \ntwo and confined
monopoles; (ii) reducing the level of supersymmetry; and
(iii) domain walls as D brane prototypes.

\vspace*{.05cm}

\end{titlepage}

\tableofcontents

\newpage

\section {Introduction}
\label{intro}
\setcounter{equation}{0}

It is well known that supersymmetric theories may have
BPS sectors in which some data can be computed
at strong coupling even when the full theory is not solvable. Historically,
this is how the first exact results on particle spectra  were obtained
\cite{A4}. Seiberg--Witten's breakthrough
results \cite{SW1,SW2} in the mid-1990's gave an additional
motivation to the studies of the BPS sectors.

BPS solitons can emerge in those supersymmetric theories
in which superalgebras are centrally extended.
In many instances the corresponding central charges are seen at
the classical level.  In some interesting models central charges
appear as quantum anomalies.

First studies of BPS solitons (sometimes referred to as critical solitons)
in supersymmetric theories at weak coupling date back to 1970s.
De Vega and Schaposnik were the first to point out \cite{dvsch}
that a model in which classical equations of motion
can be reduced to first-order Bogomol'nyi--Prasad--Sommerfield (BPS)
equations \cite{B,PS} is, in fact, a bosonic reduction of a supersymmetric
theory. Already in 1977 critical soliton solutions were
obtained in the superfield form in some two-dimensional models \cite{VecFe}.
In the same year miraculous cancellations
occurring in calculations of quantum corrections to soliton masses
were noted in \cite{dahodiv} (see also \cite{hruby}).
It was observed that for BPS solitons the boson and fermion modes
are degenerate and their number is balanced. It was believed
(incorrectly, we hasten to add) that the soliton masses receive
no quantum corrections. The modern --- correct --- version of this statement is
as follows:
if a soliton is BPS-saturated at the classical level and belongs to a shortened supermultiplet, it stays BPS-saturated after quantum corrections,
and its mass exactly coincides with the  central charge
it saturates. The latter may or may not be renormalized.
Often --- but not always ---  central charges that do not vanish at the classical level
and have quantum anomalies {\em are} renormalized. Those that
emerge as anomalies and have no classical part typically receive
no renormalizations. In many instances holomorphy
protects central charges against renormalizations.

Critical solitons play a special role in gauge field theories. Numerous parallels between such
solitonic objects and basic elements of string theory were revealed in
the recent years.
At first, the relation between string theory
and supersymmetric gauge theories
was mostly a ``one-way street" --- from strings to field theory.
Now it is becoming exceedingly more evident
that field-theoretic methods and results, in their turn,
provide insights in string theory.

String theory which emerged from dual hadronic models
in the late 1960's and 70's, elevated to the
``theory of everything" in the 1980's and 90's when it
experienced  an unprecedented expansion,  has
seemingly  entered a ``return-to-roots"
stage. The task of finding solutions to ``down-to-earth"
problems of QCD and other gauge theories by using
results and techniques
of string/D-brane theory
is currently recognized by many as one of the most
important and exciting goals of the community.
In this area the internal logic of development of string theory
is fertilized by insights and hints obtained from field theory.
In fact, this is a very healthy process of cross-fertilization.

If  supersymmetric
gauge theories  are, in a sense, dual to
string/D-brane theory ---  as is generally believed to be the case ---
they must support  domain walls (of the D-brane type) \cite{P},
and we know, they do \cite{DvSh,Witten:1997ep}.
A D brane is defined as a hypersurface on which a string may end.
In field theory both the brane and the string arise as BPS solitons,
the brane as a domain wall and the string as a flux tube. If their properties
reflect those inherent to string theory, at least to an extent,
the flux tube must end on the wall. Moreover, the wall must house  gauge fields
living on its worldvolume, under which the end of the string is charged.

The purpose of this review is to summarize developments in critical solitons
in two, three and four dimensions, with emphasis on four dimensions and on
most recent results. A large variety of BPS-saturated solitons exist in
four-dimensional field theories: domain walls, flux tubes (strings), monopoles and
dyons, and various junctions of the above objects. A list of recent discoveries
includes localization of gauge fields on domain walls,
non-Abelian strings that can end on domain walls, developed boojums,
confined monopoles attached to strings,
and other remarkable findings. The BPS nature of these objects allows one
to obtain a number of exact results. In many instances
nontrivial dynamics of the bulk theories
we will consider lead to effective low-energy theories
in the world volumes of domain walls and strings (they are related to zero modes)
\ exhibiting novel dynamical features that
are interesting by themselves.

We do not  try to review the vast literature accumulated
since the mid-1990's in its entirety.
A comparison with a huge country the exploration of which is not yet
completed is  in order here. Instead, we suggest what may be called
``travel diaries'' of the participants of the exploratory expedition.
Recent publications \cite{ob1,ob2,fidels,harvey,ken}
facilitate our task since they present the current developments in  this field
from a complementary point of view.

The  ``diaries'' are organized in two parts.
The first part (entitled ``Short  Excursion'')
is a bird's eye view of the territory.
It gives a brief and  largely  nontechnical
 introduction to basic ideas lying behind
supersymmetric solitons  and particular applications.
It is designed in such a way as to present
a general perspective that would be understandable to anyone
with an elementary  knowledge
in classical and quantum fields,  and supersymmetry.

Here we present some historic remarks, catalog relevant centrally
extended superalgebras and review basic building blocks
we consistently deal with -- domain walls, flux tubes, and  monopoles --
in their classic form. The word ``classic" is used here not in the meaning
``before quantization" but, rather, in the meaning
``recognized and cherished in the community for years."

The second  part (entitled ``Long Journey'')
is built on other principles.
It is intended for those who would like to delve in this subject thoroughly,
with its specific methods and technical devices.
We put special emphasis on recent
developments having direct relevance to QCD and gauge theories at
large, such as non-Abelian flux tubes (strings),
non-Abelian monopoles confined on these strings, gauge field localization
on domain walls, etc. We start from  presenting our benchmark model,
which has extended ${\mathcal N}=2$ supersymmetry.
Here we go well beyond conceptual foundations,
investing efforts in detailed discussions
of particular problems and aspects of our choosing.
Naturally, we choose those problems and  aspects which
are instrumental in the novel phenomena mentioned above.
In addition to walls, strings and monopoles,
we also dwell on the string-wall junctions
which play a special role in the context of dualization.

Our subsequent logic is from ${\mathcal N}=2$ to ${\mathcal N}=1$ and further on.
Indeed, in certain instances we are able do descend to non-supersymmetric
gauge theories which are very close relatives of QCD.
In particular, we present a fully controllable weakly coupled
model of the Meissner effect which exhibits quite nontrivial
(strongly coupled) dynamics on the string world sheet.
One can draw direct parallels between this consideration
and the issue of $k$-strings in QCD.

\newpage

\vspace{7mm}

\begin{center}

{\Huge PART I: Short
Excursion}

\end{center}
\addcontentsline{toc}{section}{\large PART I: Short Excursion}

\vspace{5cm}

\vspace{10mm}

\centerline{\includegraphics[width=1.8in]{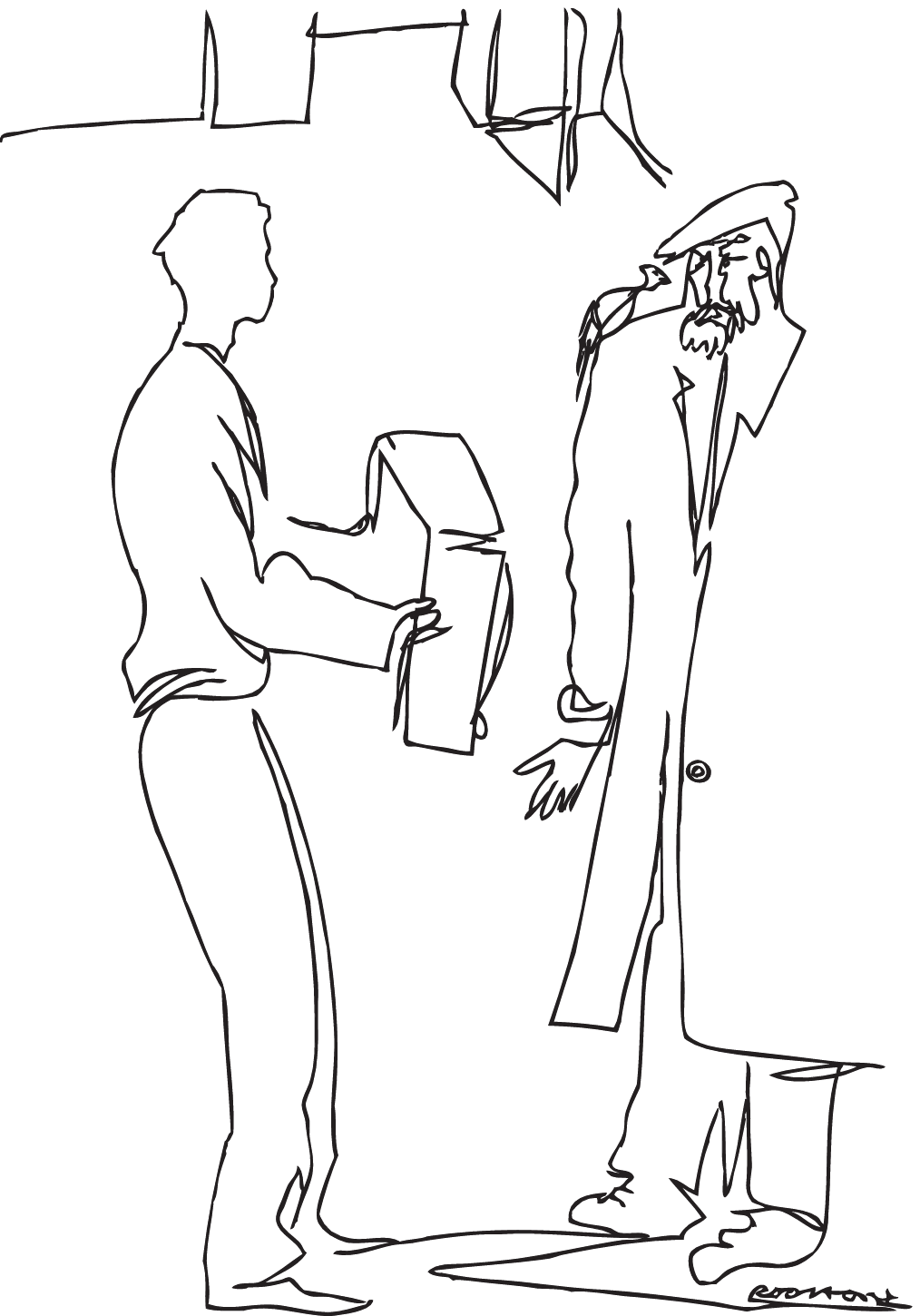}}

\newpage

\section{Central charges in superalgebras}
\label{central}
\setcounter{equation}{0}

In this Section we will briefly review
general issues related to central charges (CC) in superalgebras.

\subsection{History}
\setcounter{equation}{0}

The first superalgebra in four-dimensional field theory was derived by Golfand and
Likhtman \cite{A1} in the form
\beq
\{\bar Q_{\dot\alpha} Q_\beta\} = 2 P_\mu \left( \sigma^\mu \right)_{\alpha\beta}\,,\qquad
\{\bar Q_\alpha \bar Q_\beta\}= \{  Q_\alpha Q_\beta\}=0\,,
\label{glsa}
\eeq
i.e. with no central charges. Possible occurrence of CC (elements of
superalgebra commuting with all other operators) was first mentioned in an
unpublished paper of Lopuszanski and Sohnius \cite{A2}
where the last two anticommutators were modified as
\beq
\{  Q_\alpha^I  Q_\beta^G \} = Z_{\alpha \beta}^{IG}\,.
\eeq
The superscripts $I,G$ mark extended supersymmetry.
A more complete description of superalgebras with CC in
quantum field theory was worked out in \cite{A3}.
The only central charges analyzed in this paper were Lorentz scalars (in four dimensions),
$Z_{\alpha\beta} \sim \varepsilon_{\alpha\beta}$.
Thus, by construction, they could be relevant only to
extended supersymmetries.

A few years later, Witten and Olive  \cite{A4} showed
that in supersymmetric theories with solitons,  central extension of
superalgebras is typical; topological quantum numbers
play the role of central charges.

It was generally understood that superalgebras with (Lorentz-scalar)
central charges can be obtained from superalgebras without
central charges in higher-dimen\-sional space-time by interpreting some
of the extra components of the momentum as CC's (see e.g. \cite{Grisaru}).
When one compactifies extra dimensions one obtains an extended supersymmetry;
the extra components of the momentum act as scalar central charges.

Algebraic analysis extending that of \cite{A3} carried out in the early
1980s  (see e.g. \cite{vHvP}) indicated that the super-Poincar\'e algebra admits
CC's of a more general form, but the dynamical role
of   additional tensorial charges was not recognized
until much later.
Now it is common knowledge that central charges that originate
from operators other than the energy-momentum
operator  in higher dimensions can play a crucial role.
These tensorial central charges   take non-vanishing
values on extended objects such as strings and membranes.

Central charges that are antisymmetric tensors in various dimensions
were introduced (in the supergravity context, in the presence of $p$-branes) in
Ref.~\cite{agit} (see also \cite{at,atp}).  These
CC's are relevant to  extended objects of  the domain wall type (membranes).
Their occurrence in four-dimensional super-Yang--Mills theory
(as a quantum anomaly) was first observed in \cite{DvSh}.
A general theory of central extensions of superalgebras in three and
four dimensions
was discussed in Ref.~\cite{A7}. It is worth noting that those
central charges that have the Lorentz structure of Lorentz vectors
were not considered in \cite{A7}. The gap was closed in \cite{gorskys}.

\subsection{Minimal supersymmetry}
\setcounter{equation}{0}

The minimal number of supercharges $\nu_Q$ in various dimensions
is given in Table 1. Two-dimensional theories with a single
supercharge, although algebraically possible, are quite exotic.
In ``conventional" models  in $D=2$ with local interactions
the minimal number of supercharges is two.

\begin{table}
\begin{center}
\begin{tabular}{|c|c|c|c|c|c|c|c|c|c|}
\hline
$D$ &2 & 3 &4&5&6&7&8&9&10
\\[3mm]
\hline\hline
$\nu_Q$  & $(1^*$) 2 &2 & 4
 & 8 & 8 & 8 & 16 & 16 & 16
 \\[2mm]
\hline
Dim$(\psi )_C$ &2 &2& 4 & 4& 8 & 8 & 16 & 16 & 32
 \\[2mm]
\hline
\# cond. &2& 1 & 1 &0& 1&1 &1& 1 &2 \\
\hline
\end{tabular}
\end{center}
\caption{\small The minimal number of supercharges, the complex dimension of the spinorial
representation and the number of additional conditions (i.e. the Majorana
 and/or Weyl conditions).}
\label{table1}
\end{table}

The minimal number of supercharges in Table 1 is given for a real representation. Then, it is clear that, generally speaking,
the maximal possible number of CC's is
determined by the dimension
of the symmetric matrix $\{Q_i Q_j    \}$ of the size $\nu_Q\times \nu_Q$, namely,
\beq
\nu_{\rm CC} = \frac{\nu_Q (\nu_Q +1)}{2}\,.
\eeq
In fact, $D$ anticommutators have the Lorentz structure of the
energy-momen\-tum operator $P_\mu$. Therefore, up to $D$ central charges
could be absorbed in $P_\mu$, generally speaking. In particular situations
this number can be smaller, since although algebraically
the corresponding CC's have the same structure as $P_\mu$, they are dynamically distinguishable.
The point is that $P_\mu$ is uniquely
defined through the conserved and symmetric energy-momentum tensor of the theory.

Additional dynamical and symmetry constraints  can
further diminish the number of independent central charges, see e.g. Sect.~\ref{sssd2}.

The total set of CC's can be arranged by classifying CC's with respect to their Lorentz structure.
Below we will present
this classification for $D=2,3$ and 4, with special emphasis on the four-dimensional case.
In Sect.~\ref{extsus} we will deal with ${\mathcal N}=2$ superalgebras.

\subsubsection{$D=2$}
\label{sssd2}

Consider two-dimensional theories with two supercharges.
From the discussion above, on purely algebraic grounds,
three CC's are possible: one Lorentz-scalar and a two-component
vector,
\beq
\{Q_\alpha,\, Q_\beta\} = 2(\gamma^\mu\gamma^0)_{\alpha\beta}(P_\mu+Z_\mu)+ i (\gamma^5\gamma_0)_{\alpha\beta}Z\, .
\label{2dcc}
\eeq
We refer to Appendix A for our conventions regarding gamma matrices.
$Z^\mu\neq 0$ would require existence of a
vector  order parameter taking distinct values in different vacua.
 Indeed, if this central charge existed, its current
would have the form
$$
\zeta_{\nu}^{\,\mu} =
\varepsilon_{\nu\rho}\,\partial^\rho A^\mu\,,\qquad
Z^\mu =\int \zeta_{0}^{\,\mu}\, dz\,,
$$
where $A^\mu$ is the above-mentioned order parameter. However,
$\langle A^\mu\rangle\neq 0$ will break  Lorentz invariance
and supersymmetry of the vacuum state. This option will not be considered.
Limiting ourselves to supersymmetric vacua we conclude that a single
(real) Lorentz-scalar central charge $Z$ is possible in ${\mathcal N}=1$ theories.
This central charge is saturated by kinks.

\subsubsection{$D=3$}
\label{sssd3}

The central charge allowed in this case is a Lorentz-vector $Z_{\mu}$,
i.e.
\beq
 \{Q_\alpha,Q_\beta\} = 2(\gamma^\mu\gamma^0)_{\alpha\beta}(P_\mu+Z_\mu).
 \label{gcc3}
\eeq
One should arrange $Z_\mu$ to be orthogonal to $P_\mu$.
In fact, this is
the scalar central charge of Sect.~\ref{sssd2} elevated by one dimension.
Its topological current can be written as
\beq
\zeta_{\mu\nu} =
\varepsilon_{\mu \nu \rho}\,\partial^\rho A\,,\qquad Z_{\mu}=\int d^2x\,
\zeta_{\mu 0}\,.
\eeq
By an appropriate choice of the reference frame
$Z_{\mu}$ can always be reduced to a real number times  $(0,0,1)$.
This central charge is associated with a domain line oriented along the
second axis.

Although from the general relation (\ref{gcc3})
it is pretty clear why BPS vortices cannot appear in theories
with two supercharges, it is instructive to discuss this
question from a slightly different standpoint. Vortices in
three-dimensional theories are localized objects, particles
(BPS vortices in 2+1 dimensions were previously considered in \cite{3Dvor};
see also references therein).
The number
of broken translational generators is $d$,
where $d$ is soliton's co-dimension, $d=2$ in the case at hand.
Then {\it at least}
$d$ supercharges are  broken. Since we have only two supercharges
in the problem at hand, both must be broken.
This simple argument tells us that for
a 1/2-BPS vortex the
minimal matching between bosonic and fermionic zero modes in the
(super) translational sector is one-to-one.

Consider now a putative BPS vortex in a theory with minimal
${\mathcal N}=1$
supersymmetry (SUSY) in 2+1D. Such a configuration would require a world volume
description with two bosonic zero modes, but only one fermionic mode.
This is not permitted by the argument above, and indeed no
configurations of this type are known. Vortices always exhibit
at least two fermionic zero modes and can be
BPS-saturated only in ${\mathcal N}=2$ theories.

\subsubsection{$D=4$}
\label{ms14}

Maximally one can have 10 CC's which are decomposed
into Lorentz representations as (0,1) + (1,0) + (1/2, 1/2):
\begin{eqnarray}
 \{Q_{\alpha},\bar{Q}_{\dot \alpha}\} &=& 2(\gamma^\mu)_{\alpha\dot\alpha}(P_\mu+Z_\mu),
 \label{zee}
 \\[2mm]
 \{Q_{\alpha},Q_{\beta}\} &=& (\Sigma^{\mu\nu})_{\alpha\beta}Z_{[\mu\nu]},
 \label{zeee}
 \\[2mm]
 \{\bar Q_{\dot\alpha},\bar Q_{\dot \beta}\} &=& (\bar \Sigma^{\mu\nu})_{\dot \alpha \dot \beta}\bar{Z}_{[\mu\nu]}\,,
  \label{zeeee}
\end{eqnarray}
where $(\Sigma^{\mu\nu})_{\alpha\beta} = (\sigma^\mu)_{\alpha\dot\alpha}(\bar{\sigma}^\nu)^{\dot\alpha}_\beta$ is a chiral
version of $\sigma^{\mu\nu}$ (see e.g. \cite{SVinst}).
The antisymmetric tensors $Z_{[\mu\nu]}$ and $\bar{Z}_{[\mu\nu]}$ are
associated with domain walls, and reduce to a complex number and a
spatial vector orthogonal to the domain wall.
The (1/2, 1/2) CC $Z_\mu$ is a Lorentz vector orthogonal to
$P_\mu$. It is associated with strings (flux tubes), and
reduces to one real number and a three-dimensional unit
spatial vector parallel to the string.

\subsection{Extended SUSY}
\label{extsus}
\setcounter{equation}{0}

In four dimensions one can extend superalgebra up to ${\mathcal N}=4$,
which corresponds to sixteen supercharges. Reducing this to lower dimensions
we get a rich variety of extended superalgebras in $D=3$ and 2.
In fact, in two dimensions the
Lorentz invariance provides a much weaker constraint than in higher dimensions, and one can consider a wider set of
$(p,q)$ superalgebras comprising $p+q=4$, 8, or 16 supercharges.
We will not pursue a general solution; instead,
we will limit our task  to (i) analysis of central charges in ${\mathcal N}=2$
in four dimensions; (ii)
reduction of the minimal SUSY algebra
in $D=4$ to $D=2$ and 3, namely the ${\mathcal N}=2$ SUSY algebra in those dimensions. Thus, in two dimensions we will
consider only the non chiral ${\mathcal N}=(2,2)$ case.
As should be clear from the discussion above, in the dimensional reduction
the maximal number of CC's stays intact.
What changes is the decomposition in
Lorentz and $R$-symmetry irreducible representations.

\subsubsection{${\mathcal N}=2$ in $D=2$}
\label{sssd22}

Let us
focus on the non chiral ${\mathcal N}=(2,2)$ case corresponding to dimensional reduction of the
${\mathcal N}=1$, $D=4$ algebra. The tensorial decomposition is as follows:
\beqn
\{Q_\alpha^I,Q_\beta^J\}& =& 2(\gamma^\mu\gamma^0)_{\alpha\beta}
\left[(P_\mu+Z_\mu)\delta^{IJ} + Z_{\mu}^{(IJ)}\right]+
 2i \, (\gamma^5\gamma^0)_{\alpha\beta}\,  Z^{\{IJ\}}
 \nonumber\\[3mm]
 & +& 2i\, \gamma^0_{\alpha\beta}Z^{[IJ]}\,,\qquad\qquad I,J = 1,2\,.
\eeqn
Here $Z^{[IJ]}$ is antisymmetric in $I,J$; $Z^{\{IJ\}}$ is symmetric while $Z^{(IJ)}$
is symmetric and traceless.
We can discard all vectorial central charges $Z_{\mu}^{IJ}$ for the same reasons as in Sect.~\ref{sssd2}. Then we are left with two
Lorentz singlets $Z^{(IJ)}$, which represent the reduction of the domain wall charges in $D=4$  and two Lorentz singlets Tr$Z^{\{ IJ\}}$ and $Z^{[IJ]}$, arising  from $P_2$ and the vortex charge in $D=3$ (see Sect.~\ref{sssd23}).  These central charges are saturated by kinks.

Summarizing, the $(2,2)$ superalgebra in $D=2$ is
\beq
\{Q_\alpha^I,Q_\beta^J\} = 2(\gamma^\mu\gamma^0)_{\alpha\beta}\, P_\mu\,
\delta^{IJ}
+
 2i \, (\gamma^5\gamma^0)_{\alpha\beta}\, Z^{\{IJ\}}
+2i\, \gamma^0_{\alpha\beta}Z^{[IJ]}\,.
\label{2esa2}
\eeq
It is instructive to rewrite Eq.~(\ref{2esa2}) in terms of complex supercharges
$Q_{\alpha}$ and
$Q^{\dagger}_\beta$ corresponding to four-dimensional $Q_\alpha ,\,\, \bar Q_{\dot\alpha}$, see Sect.~\ref{ms14}.  Then
\beq
\begin{split}
&\big\{Q_{\alpha}, Q_{\beta}^{\dagger}\big\}(\gamma^{0})_{\beta\gamma}=2\left[ P_{\mu}\gamma^{\mu}
+ {Z}\,\frac{1-\gamma_{5}}{2} +{Z}^\dagger\,\frac{1+\gamma_{5}}{2}\right]_{\alpha\gamma},
\\[1mm]
&\big\{Q_{\alpha},Q_{\beta}\big\}(\gamma^{0})_{\beta\gamma}=-
2 {Z'}\,(\gamma_{5})_{\alpha\gamma}\,,\qquad
\big\{Q_{\alpha}^{\dagger}, Q_{\beta}^{\dagger}\big\}(\gamma^{0})_{\beta\gamma}=
2 {Z'}^{\dagger}\,(\gamma_{5})_{\alpha\gamma}\,.
\end{split}
\label{eq:supalg}
\eeq
The algebra contains two complex central charges, $Z$ and $Z'$.
In terms of  components $Q_{\alpha}=(Q_{R}, Q_{L})$ the non vanishing anticommutators are
\beq
\begin{split}
&\{ Q_L , Q_L^\dagger\} = 2(H+ P)\,,\qquad \{Q_R, Q_R^\dagger    \} = 2(H-P)\,,
\\[1mm]
&\{Q_L, Q_R^\dagger  \} =2i{Z}\,,\qquad\qquad
~~\{Q_R, Q_L^\dagger \}= -2i {Z}^{\dagger}\,,
\\[1mm]
&\{Q_L, Q_R \} =2i{Z'}\,,\qquad\qquad
~\{Q_R^{\dagger}, Q_L^\dagger \}= -2i {Z'}^{\dagger}\,.
\end{split}
\label{compalg}
\eeq
It exhibits the automorphism
$Q_{R}\leftrightarrow Q^{\dagger}_{R},\, Z\leftrightarrow Z'$  associated
\cite{Dorey} with the transition to a mirror representation \cite{HH}. The complex central charges
$Z$ and $Z^\prime$ can be readily expressed in terms
of real $Z^{\{IJ\}}$ and
$Z^{[IJ]}$,
\beq
Z= Z^{[12]}+\frac{i}{2}\, \left(Z^{\{11\}} +Z^{\{22\}}
\right),\qquad
Z^\prime =\frac{Z^{\{12\}} + Z^{\{21\}}}{2} - i\,\frac{Z^{\{11\}}-Z^{\{22\}}}{2}\,.
\eeq
Typically, in a given  model either $Z$ or $Z^\prime$
vanish. A practically important example to which we will repeatedly
turn below (e.g. Sect.~\ref{tscl}) is provided by the so-called
twisted-mass-deformed CP$(N-1)$ model
\cite{twisted}. The central charge $Z$ emerges in this model at the
classical level. At the quantum level it acquires additional anomalous terms
\cite{LosSh, SVZ06}.
Both $Z\neq 0$ and $Z^\prime \neq 0$ simultaneously
in a contrived model  \cite{LosSh} in which the
Lorentz symmetry and a part of supersymmetry are spontaneously broken.

\subsubsection{${\mathcal N}=2$ in $D=3$}
\label{sssd23}

The superalgebra can be decomposed into Lorentz and $R$-symmetry tensorial structures as follows:
\beq
 \{Q_\alpha^I,Q_\beta^J\} = 2(\gamma^\mu\gamma^0)_{\alpha\beta}[(P_\mu+Z_\mu)\delta^{IJ} + Z_{\mu}^{(IJ)}]
 +2i\, \gamma^0_{\alpha\beta}Z^{[IJ]},
 \label{tdecomp}
\eeq
where all central charges above are real. The maximal set of 10 CC's
enter as a triplet of space-time vectors $Z_{\mu}^{IJ}$ and a singlet $Z^{[IJ]}$.
The singlet CC is associated with vortices (or lumps),
and corresponds to the reduction of the (1/2,1/2) charge or the $4^{th}$ component of the
momentum vector in $D=4$. The triplet $Z_{\mu}^{IJ}$ is decomposed into an
$R$-symmetry
singlet $Z_\mu$, algebraically indistinguishable from the momentum,
and a traceless symmetric combination $Z_\mu^{(IJ)}$. The former is equivalent to
the vectorial charge in the ${\mathcal N}=1$ algebra, while $Z_\mu^{(IJ)}$ can be reduced to
a complex number and vectors specifying the orientation.  We see that these are the direct  reduction of the (0,1) and (1,0) wall charges in $D=4$.
They are saturated by domain lines.

\subsubsection{On extended supersymmetry (eight supercharges)
in $D=4$}
\label{extsc}

Complete algebraic analysis of all tensorial  central charges
in this problem is analogous to the previous cases and is rather
straightforward. With eight supercharges the maximal number of CC's is 36.
Dynamical aspect is less developed -- only a modest fraction of the above 36
CC's are known to be non-trivially realized in models studied in the literature.
We will limit ourselves  to a few remarks regarding the
well-established CC's. We will use a complex (holomorphic) representation
of the supercharges. Then the supercharges are labeled as follows
\beq
Q_\alpha^F\,,\quad \bar Q_{\dot\alpha\,G}\;,\qquad \alpha ,\dot\alpha =1,2\,,\qquad
F,G =1,2\,.
\eeq
On general grounds one can write
\begin{eqnarray}
\{Q_\alpha^F,\,\bar Q_{\dot\alpha\,G}\} &=& 2\delta^F_G\, P_{\alpha \dot\alpha}
+ 2 (Z^F_G)_{\alpha \dot\alpha}\,,\nonumber\\[3mm]
\{Q_\alpha^F,\, Q_\beta^G\} &=& 2\, Z_{\{\alpha \beta\}}^{\{FG\}}
+2\,  \varepsilon_{ \alpha \beta }\, \varepsilon^{ FG }\,  Z
\,,\nonumber\\[3mm]
\{\bar Q_{\dot\alpha\, F},\, \bar Q_{\dot\beta\, G}\} &=&
2\left( \bar Z_{\{FG\}}\right)_{\{\dot\alpha \dot\beta\}}
+ 2\, \varepsilon_{ \dot\alpha\dot \beta }\, \varepsilon_{ FG }\, \bar Z\,.
\label{grounds}
\end{eqnarray}
Here $(Z^F_G)_{\alpha \dot\alpha}$ are four vectorial central charges (1/2, 1/2),
(16 components altogether) while $Z_{\{\alpha \beta\}}^{\{FG\}}$ and the complex conjugate are (1,0) and (0,1) central charges. Since the matrix
$Z_{\{\alpha \beta\}}^{\{FG\}}$ is symmetric with respect to $F,G$,
there are three flavor
components, while the total number of components
residing in (1,0) and (0,1) central charges is 18. Finally, there are two scalar
central charges, $Z$ and $\bar Z$.

Dynamically the above central charges can be described as follows.
The scalar CC's $Z$ and $\bar Z$ are saturated by monopoles/dyons.
One vectorial central charge $Z_\mu$ (with the additional condition
$P^\mu Z_\mu =0$) is saturated \cite{VY}
by Abrikosov--Nielsen--Olesen
string (ANO for short) \cite{ANO}.  A (1,0) central charge with $F=G$
is saturated by domain walls \cite{SYnawall}.

Let us briefly  discuss the Lorentz-scalar central charges
in Eq.~(\ref{grounds}) that are saturated by monopoles/dyons.
They will be referred to as monopole central charges.
A rather dramatic story is associated with them.
Historically they were the first to be introduced within the framework
of  an extended
4D superalgebra \cite{A2,A3}. On the dynamical side, they
appeared as the first example of the ``topological charge $\leftrightarrow$ central charge" relation revealed by Witten and Olive in their pioneering paper \cite{A4}.
Twenty years later, the ${\mathcal N}=2$ model where these central
charges first appeared, was solved by Seiberg and Witten \cite{SW1,SW2},
and the exact masses of the BPS-saturated monopoles/dyons found.
No direct comparison with the operator expression for the central charges
was carried out, however.  In Ref.~\cite{Rebhan:2004vn}
it was noted that for the Seiberg--Witten formula to be valid,
a boson-term anomaly should exist in the monopole central charges.
Even before \cite{Rebhan:2004vn} a fermion-term anomaly
was identified \cite{SYnawall},  which plays a crucial
role \cite{Shifman:2004dr} for the monopoles in the Higgs regime (confined monopoles).

\vspace{2cm}

\centerline{\includegraphics[width=1.3in]{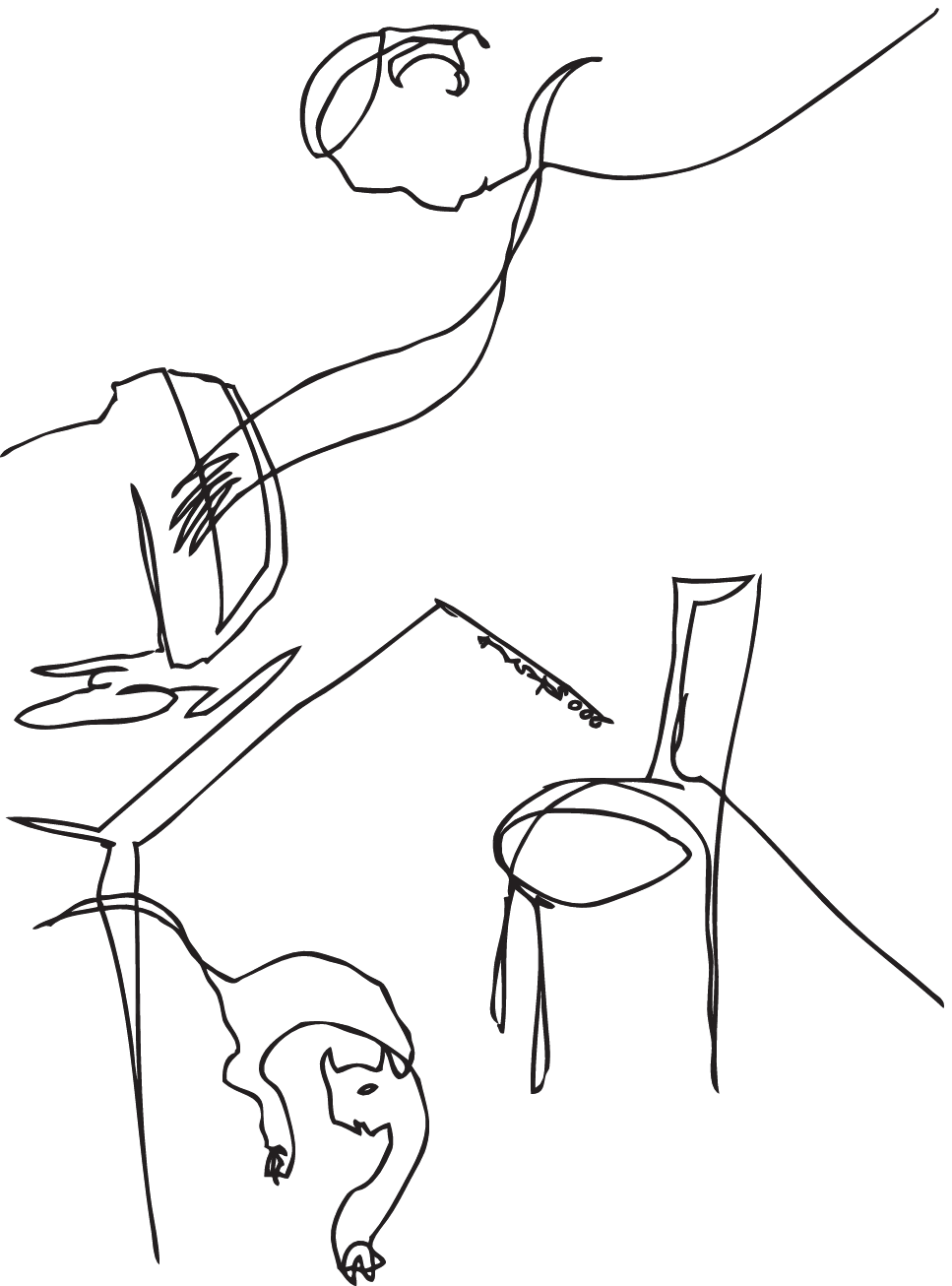}}

\newpage

\section{The main building blocks}
\label{main}
\setcounter{equation}{0}

\subsection{Domain walls}
\setcounter{equation}{0}
\subsubsection{Preliminaries}

In four dimensions domain walls are two-dimensional extended objects.
In three dimensions they become domain lines, while in two dimensions
they reduce to kinks which can be considered as particles since they are localized.
Embeddings of bosonic models supporting kinks in \none supersymmetric models in two dimensions were first discussed in \cite{A4,VecFe}. Occasional remarks on kinks in models
with four supercharges of the type of the Wess--Zumino models \cite{BW} can be found in the literature in the 1980s but they went unnoticed. The only issue which caused much interest and debate  in the 1980s
was the issue of quantum corrections to the BPS kink mass in 2D
models with ${\mathcal N} =1$ supersymmetry.

The mass of the BPS saturated
kinks in two dimensions
must be  equal to the central charge $Z$ in Eq.~(\ref{2dcc}). The simplest
two-dimensional model with the minimal superalgebra, admitting
solitons, was considered in \cite{DADI}.
In components the Lagrangian takes the form
\beq
{\mathcal L} = \frac{1}{2}\left( \partial_\mu\phi \,\partial^\mu\phi
+\bar\psi\, i\! \!\not\!\partial \psi +F^2 \right) + {\mathcal W}'(\phi)F
-\frac{1}{2}{\mathcal W}''(\phi)\bar\psi\psi\, .
\label{minlag}
\eeq
where ${\mathcal W}(\phi)$ is a real ``superpotential"
which in the simplest case takes the form
\beq
{\mathcal W}(\Phi) = \frac{m^2}{4\lambda}\, \Phi-\frac{\lambda}{3}\, \Phi^3\,.
\eeq
Moreover,
the auxiliary field $F$ can be eliminated by
virtue of the classical equation of motion, $ F = - {\mathcal W}^\prime$.
This is a real reduction (two supercharges) of the Wess--Zumino model  which has
one real scalar field $\phi$ and one two-component real spinor $\psi$.

The story of kinks in this model is long and dramatic.
In the very beginning it was argued
\cite{DADI} that, due to a residual supersymmetry, the mass of the
soliton calculated at the classical level remains intact at  the
one-loop level. A few years later it was noted \cite{KR} that the
non-renormalization theorem \cite{DADI} cannot possibly be correct,
since the classical soliton mass is proportional to $m^3/\lambda^2$
(where $m$ and $\lambda$ are the bare mass parameter and coupling
constant, respectively), and the physical mass of the scalar field gets
a logarithmically infinite renormalization. Since the soliton mass is an
observable physical parameter, it must stay finite in the limit $M_{\rm
uv}\to\infty$, where $M_{\rm uv}$ is the ultraviolet cut off. This
implies, in turn, that the quantum corrections cannot vanish -- they
``dress" $m$ in the classical expression, converting  the bare mass
parameter into the renormalized one.  The one-loop renormalization of
the soliton mass was first calculated in \cite{KR}. Technically the
emergence of the one-loop correction was attributed to a ``difference in
the density of states in continuum in the boson and fermion operators in
the soliton background field".  The subsequent work \cite{IM} dealt with
the renormalization of the central charge, with the conclusion that the
central charge is renormalized in just the same way as the kink mass,
so that the saturation condition is not violated.

Then many authors repeated one-loop calculations for the kink mass
and/or central charge
\cite{AHV,JFS,SR,AU,AU1,HY,CM1,CM,RN,NSNR,Jaffe}. The results reported
and the  conclusion of saturation/non-saturation oscillated with time,
with little
sign of convergence. Needless to say, all authors agreed that the
logarithmically divergent term in $Z$  matched the renormalization of
$m$. However,  the finite (non logarithmic) term varied from work to work,
sometimes even in the successive works of the same authors, e.g.
\cite{RN,NSNR} or \cite{AU,AU1}. Polemics continued unabated through the 1990s.
For instance,
according to publication
\cite{NSNR}, the BPS saturation is violated at one loop. This assertion
reversed the earlier trend \cite{KR,HY,CM1}, according to which
the kink  mass and the corresponding central charge
are renormalized in a concerted way.

The story culminated in 1998 with the discovery of a quantum  anomaly
in the central charge \cite{SVVo}. Classically, the kink central charge
${Z}$ is equal to the difference between the values of the
superpotential ${\mathcal W}$ at spatial infinities,
\begin{equation}
Z = {\mathcal  W}[\phi (z=\infty )]-{\mathcal W}[\phi (z=-\infty )]\, .
\label{zfirst}
\end{equation}
This is known from the pioneering paper \cite{A4}.   Due to the anomaly, the central charge gets
modified in the following way
\beq
{\mathcal W}\longrightarrow
{\mathcal W} +\frac{{\mathcal W}'' }{4\pi}\, ,
\label{qa}
\eeq
where the term proportional to ${\mathcal W}''$ is anomalous \cite{SVVo}.
The right-hand side of Eq.~(\ref{qa}) must be substituted  in the expression
for the central charge (\ref{zfirst}) instead of ${\mathcal W}$.
Inclusion of the additional anomalous term restores the equality
between the kink mass and its central charge.
The BPS nature is preserved, which is correlated with the fact that
the kink supermultiplet is short in the case at hand \cite{losevwe}. All subsequent investigations
confirmed this conclusion (see e.g. the review paper \cite{GRVN}).

Critical domain walls in theories with four supercharges started attracting attention
in the 1990s. The most popular model of this time supporting such domain walls
is the generalized Wess--Zumino model with the Lagrangian
\beq
{\mathcal L} =\,\int d^2\theta\,d^2\bar\theta \, K (\bar \Phi, \Phi ) +
\left( \int d^2\theta \, {\mathcal W} \,(\Phi ) +{\rm h.c.}\right)
\eeq
where $K$ is the K\"ahler potential and $\Phi$ stands for a set of the chiral superfields.
This model can be considered in two and four dimensions.
A popular choice was  a trivial K\"ahler potential,
$$
K =\bar \Phi \Phi \,.
$$
BPS walls in this system satisfy  the first-order differential
equations \cite{fmvw,at,cvetic,cv,CSh}
\beq
 g_{{\bar a}b}\, \partial_z \Phi^b = e^{i\gamma}\, \partial_{\bar a}\bar{{\mathcal W}},
   \label{bpseqn}
\eeq
where the K\"ahler metric is given by
\beq
g_{\bar{a}b}= \frac{\partial^2 \, K}{\partial \bar\Phi^{\bar a}\,\partial \Phi^{b} }
\equiv\partial_{\bar{a}}\partial_{b} {K}\,,
\eeq
and  $\gamma $ is the the phase of the (1,0)
central charge ${Z}$ as defined in (\ref{zeee}).
The phase $\gamma$ depends on the choice of the vacua
between which the given domain wall interpolates,
\beq
Z = 2\left( {\mathcal W}_{\rm vac_{\,f}} -{\mathcal W}_{\rm vac_{\,i}}\,\right)\,.
\label{wzcc}
\eeq
A useful consequence of the BPS equations is that
\beq
\partial_z {\mathcal W} = e^{i\gamma}\,\|\partial_a\,{\mathcal W}\|^2\,,
\label{Wline}
\eeq
and thus the domain wall describes a straight line in the ${\mathcal W}$-plane
connecting the two vacua.

Construction and analysis of BPS saturated domain walls in four dimensions
crucially depends on the realization of the fact that the  central charges
relevant to critical domain walls are not Lorentz scalars; rather they transform as
(1,0)+(0,1) under the Lorenz transformations. It was a textbook
statement ascending to the pioneering paper \cite{A3}
that \none superalgebras in four dimensions leave place to no central charges.
This statement is correct only with respect to Lorenz-scalar central charges.
Townsend was the first to note  \cite{tow} that ``supersymmetric branes,"
being BPS saturated, require the existence of tensorial central charges
antisymmetric in the Lorenz indices. That the anticommutator
$\{Q_\alpha\,,Q_\beta\}$ in four-dimensional Wess--Zumino model
contains the (1,0) central charge
is obvious. This anticommutator vanishes, however, in super-Yang--Mills theory at the classical level.

\subsubsection{ $D$ Branes in Gauge Field Theory}

In 1996 Dvali and Shifman found in supersymmetric gluodynamics \cite{DvSh}
an anomalous $(1,0)$ central charge in  superalgebra, not seen at the classical level.
They argued that this central charge is saturated by domain walls interpolating
between vacua with distinct values of the order parameter,
the gluino condensate $\langle \lambda\lambda\rangle$,
labeling $N$ distinct  vacua of super-Yang--Mills theory
with the gauge group SU($N$).

Supersymmetric gluodynamics (it is often referred to as pure super-Yang--Mills
theory) is defined by the Lagrangian
\beq
{\mathcal L} =  \frac{1}{g^2}  \int\!{\rm d}^2\theta \,\mbox{Tr}\, W^2 + \,
\mbox{H.c.}
 =\frac{1}{g^2} \left\{ -\frac 14
F_{\mu\nu}^a\, F^{a\mu\nu} +
i \lambda ^{a \alpha}
{\mathcal D}_{\alpha\dot\beta}\bar\lambda^{a\dot\beta}
\right\}   ,
\eeq
where $\lambda ^{a \alpha}$ is the Weyl spinor in the adjoint
representation of SU($N$).

\begin{figure}[h]
\centerline{\includegraphics[width=2.5in]{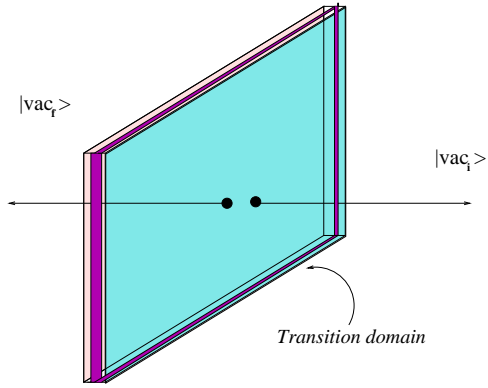}}
\caption{\small A field configuration interpolating between two distinct degenerate vacua }
\label{dwa}
\end{figure}

What is the domain wall?
It is a field configuration interpolating between vacuum i and vacuum f
with some transition domain in the middle. Say, to the left you have vacuum i,
to the right  you have vacuum f, in the middle you have a transition
domain which, for obvious reasons, is referred to
as the wall (Fig.~\ref{dwa}).

\begin{figure}[h]
\centerline{\includegraphics[width=2.5in]{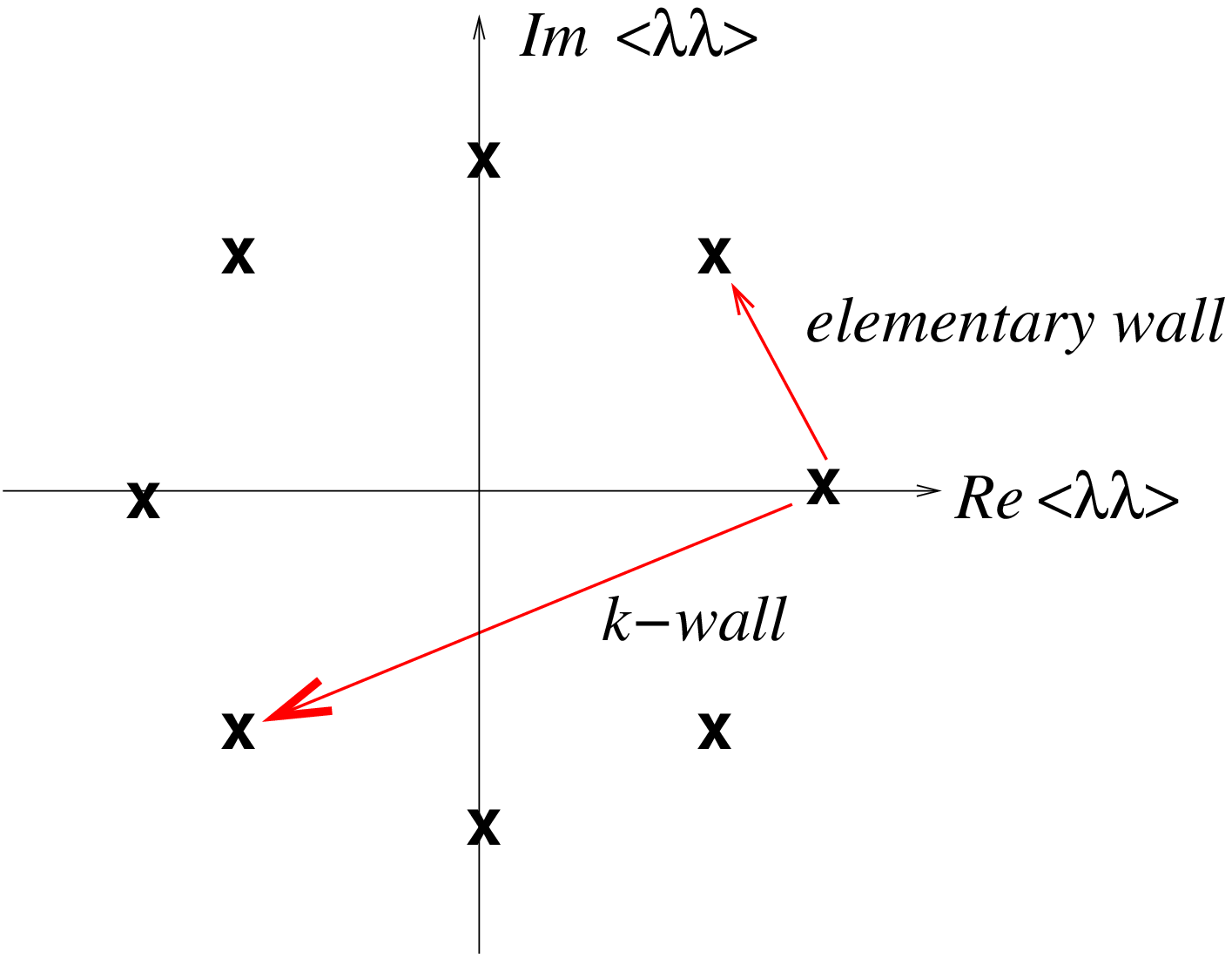}}
\caption{\small $N$ vacua for SU($N$). The vacua are labeled by the vacuum expectation value
$\langle\lambda\lambda\rangle = -6\,N\,\Lambda^3\, \exp(2\pi\,i\,k/N)$ where
$k=0,1,...,N-1$.
Elementary walls interpolate between two neighboring vacua }
\label{susym}
\end{figure}

There is a large variety of walls in supersymmetric gluodynamics.
 Minimal, or elementary,  walls interpolate between
vacua $n$ and $n+1$, while $k$-walls interpolate
between $n$ and $n+k$, see Fig.~\ref{susym}.
In \cite{DvSh} a mechanism was suggested for
localizing gauge fields on the wall through bulk confinement.
Later this mechanism was implemented
in  models at weak coupling, as we will see below.

Shortly after
Witten interpreted the   BPS walls in
supersymmetric gluodynamics as analogs of D-branes \cite{Witten:1997ep}.
This is because their tension scales as $N\sim 1/g_s$ rather than
$1/g_s^2$ typical of solitonic objects (here $g_s$ is the string constant).
Many promising consequences ensued. One of them was the Acharya--Vafa
derivation of the wall worldvolume theory \cite{Achar}. Using a wrapped $D$-brane
picture and certain dualities they identified the $k$-wall worldvolume theory
as 1+2 dimensional U($k$) gauge theory with the field content of
${\mathcal N}=2$ and the Chern-Simons term at level $N$ breaking ${\mathcal N}=2$ down to
${\mathcal N}=1$.

In \none gauge theories with arbitrary matter content and superpotential
the general relation (\ref{zee}) takes the form

\begin{equation}
\left\{Q_\alpha\,,Q_\beta\right\}=-4\,\Sigma_{\alpha\beta}\,\bar {
Z}\,,
\label{cccdw}
\end{equation}
where
\begin{equation}
\Sigma_{\alpha\beta}=-\frac 12\int {\rm d} x_{[\mu} {\rm d} x_{\nu ]}\,
(\sigma^\mu)_{\alpha\dot\alpha} (\bar \sigma^\nu)_{\beta}^{\dot\alpha}
\end{equation}
is the wall area tensor, and
\begin{eqnarray}
{\!\!\!\!}  {\!\!\!\!}
Z &\!\!= &\!\!\frac{2}{3}
\Delta\left\{
\left[ 3{\mathcal W } - \sum_f Q_f\, \frac{\partial{\mathcal W }}{\partial Q_f}
\right] \right.
 \nonumber\\[0.2cm]
{\!\!\!\!}{\!\!\!\!}
&\!\!- &\!\!
\left.
\left[ \frac{3N- \sum_f T(R_f)}{16\pi^2}\,{\rm Tr}\,W^2 +
\frac{1}{8}\sum_f\gamma_f
\bar{D}^2 (\bar{Q}_{\!f}\, e^{V} Q_f) \right]
\right\}_{\theta=0}
\label{achgt}
\end{eqnarray}
In this expression $\Delta$ implies taking the difference at two spatial infinities
in the direction perpendicular to the surface of the wall.
The first term in the second line  presents
the gauge anomaly in the central charge. The second term in the second line is
a total superderivative. Therefore, it vanishes after averaging over any
supersymmetric vacuum state. Hence, it can be safely omitted.
The first line presents the classical result, cf. Eq.~(\ref{wzcc}). At the classical
level
$ Q_f(\partial{\mathcal W }/\partial Q_f)$ is a total superderivative too
which can be seen from the Konishi anomaly~\cite{Konishian},
\begin{equation}
  \bar{D}^2\, (\bar{Q}_f e^{V} Q_f) =
4 \,Q_f \frac{\partial{\mathcal W }}{\partial Q_f} +
\frac{T(R_f)}{2\pi^2}\,{\rm Tr}\, W^2\, .
\label{ka1}
\end{equation}
If we discard this total superderivative  for a short while
(forgetting about quantum effects), we return to ${ Z} = 2\Delta ({\mathcal
W})$, the formula obtained in the Wess--Zumino model. At the quantum level
$ Q_f(\partial{\mathcal W }/\partial Q_f)$ ceases to be a total superderivative
because of the Konishi anomaly.
It is still  convenient to eliminate $ Q_f(\partial{\mathcal W }/\partial Q_f)$
in favor of Tr$W^2$ by virtue of the Konishi relation (\ref{ka1}). In this
way one arrives at
\begin{equation}
Z = 2
\Delta\left\{ {\mathcal W} - \frac{N- \sum_f T(R_f)}{16\pi^2}\,{\rm Tr}\,W^2
\right\}_{\theta=0}\, .
\label{achgf}
\end{equation}
We see that the superpotential ${\mathcal W}$ is amended by the anomaly;
in the operator form
\begin{equation}
 {\mathcal W} \longrightarrow  {\mathcal W} - \frac{N- \sum_f T(R_f)}{16\pi^2}\,{\rm
Tr}\,W^2\,.
\label{ofasp}
\end{equation}
Of course, in pure Yang--Mills theory only the anomaly term survives.

Beginning from 2002 we developed a benchmark
${\mathcal N}=2$ model, weakly coupled in the bulk (and, thus, fully controllable),
which supports both BPS walls and BPS flux tubes.
We demonstrated that a gauge field is indeed localized on the
wall; for the minimal wall this is a U(1) field while for non minimal walls
the localized gauge field is non-Abelian. We also found a BPS wall-string junction
related to the gauge field localization, see Sect.~\ref{walls}. The field-theory string
does end on the BPS wall, after all!
The end-point of the string on the wall, after Polyakov's dualization,
becomes a source of the electric field localized on the wall.
In 2005 Norisuke Sakai and David Tong analyzed generic wall-string configurations.
Following condensed matter physicists they called them
{\em boojums}.\footnote{``Boojum" comes from L.~Carroll's children's  book
{\em Hunting of the Snark.} Apparently,  it is fun to hunt a snark, but if the snark turns out to be a boojum, you are in trouble! Condensed matter physicists adopted the name to describe solitonic objects of the wall-string junction type in helium-3. Also:
The boojum  tree (Mexico) is the strangest plant
imaginable. For most  of the year it is leafless and looks like a giant upturned turnip. G.~Sykes, found it in 1922 and said, referring to Carrol  ``It must be a boojum!"   The Spanish  common name for this tree is Cirio, referring to its candle-like appearance.}

Equation (\ref{achgt}) implies that in pure gluodynamics (super-Yang--Mills
theory without matter) the domain wall tension is
\beq
T = \frac{N}{8\pi^2} \left| \langle {\rm Tr} \lambda^2\rangle_{\rm vac \,\,f}
-\langle {\rm Tr} \lambda^2\rangle_{\rm vac \,\,i}
\right|
\label{eqte}
\eeq
where vac$_{\rm i,f}$ stands for the initial (final) vacuum between which the given
wall interpolates.  Furthermore, the gluino condensate
$ \langle {\rm Tr} \lambda^2\rangle_{\rm vac }$ was calculated -- {\em exactly} --
long ago \cite{miar}, using the very same methods which were later advanced and perfected by Seiberg and Seiberg and Witten in their quest for dualities
 in ${\mathcal N}=1 $ super-Yang--Mills theories \cite{seiobz} and the
dual Meissner effect in ${\mathcal N}=2$ (see \cite{SW1,SW2}). Namely,
\beq
2\,\langle {\rm Tr} \lambda^2\rangle =
\langle
\lambda^{a}_{\alpha}\lambda^{a\,,\alpha}
\rangle = -6 N\Lambda^3 \exp \left({\frac{2\pi i k}{N}}\right)\,,
\,\,\, k = 0,1,..., N-1\,.
\label{gluco}
\eeq
Here $k$ labels the $N$ distinct vacua of the theory, see Fig. \ref{susym},
and $\Lambda$ is a dynamical scale, defined in the standard manner
(i.e. in accordance with Ref.~\cite{Hinchliffe}) in terms of the ultraviolet
parameters, $M_{\rm UV}$ (the ultraviolet  regulator mass), and $g_0^2$
(the bare coupling constant),
\beq
\Lambda^3 = \frac{2}{3}\, M_{\rm UV}^3\,  \left(
\frac{8\pi^2}{Ng_0^2}\right) \exp\left(-\frac{8\pi^2}{Ng_0^2}\right) .
\label{dynscgl}
\eeq

In each given vacuum the gluino condensate scales with the number of colors as
$N$. However, the difference of the values of the gluino condensates
in two vacua which lie not too far away from each other scales as $N^0$.
Taking into account Eq.~(\ref{eqte}) we conclude that the wall tension
in supersymmetric gluodynamics $$T\sim N\,.$$ (This statement just rephrases
Witten's argument why the above walls should be considered
as analogs of D branes.)

The volume energy density in both vacua, to the left and to the right of the wall, vanish due to supersymmetry. Inside the transition domain, where the order parameter
changes its value gradually,
the volume energy density is expected to be
proportional to  $N^2$, just because there are $N^2$ excited
degrees of freedom. Therefore,
$T\sim N $ implies that the wall thickness in supersymmetric gluodynamics
must scale as  $N^{-1}$. This is very unusual, because normally we would
say: the glueball mass is $O(N^0)$, hence,
everything built of regular glueballs should have thickness of order $O(N^0)$.

If the wall thickness is indeed $O(N^{-1})$ the
question ``what consequences ensue?" immediately comes to one's mind.
This issue is far from complete understanding, for relevant discussions see
\cite{gism,gis,ashi}.

As was mentioned, there is a large variety of walls in supersymmetric gluodynamics
as they can interpolate between vacua with arbitrary values of
$k$. Even if  $k_f=k_i+1$, i.e. the wall is elementary,
in fact we deal with several walls, all having one and the same tension
--- let us call them degenerate walls. The first indication on the wall degeneracy
was obtained in Ref.~\cite{Kovner}, where two degenerate walls were observed in
SU(2) theory. Later, Acharya and Vafa calculated the $k$-wall
multiplicity \cite{Achar} within the framework of D-brane/string formalism,
\beq
\nu_k = C_N^k=\frac{N!}{k! (N-k)!}\,.
\label{wmult}
\eeq
For $N=2$ only elementary walls exist, and $\nu=2$. In the field-theoretic setting
Eq.~(\ref{wmult}) was derived in \cite{svritz}. The derivation is based on the fact
that the index $\nu$ is topologically stable --- continuous deformations of the theory
do not change $\nu$. Thus, one can add an appropriate set of
matter fields sufficient for complete Higgsing of supersymmetric gluodynamics.
The domain wall multiplicity in the effective low-energy theory obtained in this way
is the same as in supersymmetric gluodynamics albeit the effective low-energy theory,
a Wess--Zumino type model,
is much simpler.

\subsubsection{ Domain wall junctions}

\begin{figure}[h]
 \centerline{\includegraphics[width=2in]{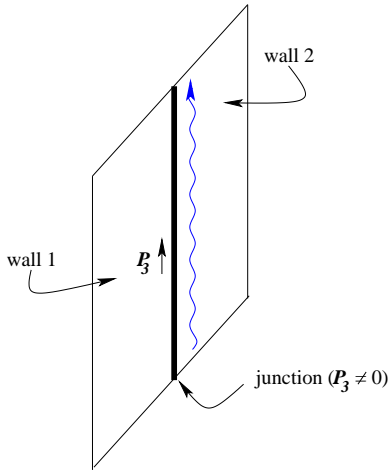}}
 \caption{\small Two distinct degenerate domain walls separated by
 the wall junction.}
 \label{dwa3}
 \end{figure}

Two degenerate domain walls can coexist in one plane --- a new phenomenon which, to the best of our knowledge, was first discussed in \cite{svritz2}. It is illustrated in
Fig.~\ref{dwa3}. Two distinct degenerate domain walls lie on the plane;
the transition domain between wall 1 and wall 2 is the domain wall junction
(domain line).

Each individual domain wall is 1/2 BPS-saturated. The wall configuration with the
junction line (Fig.~\ref{dwa3}) is 1/4 BPS-saturated. We start from ${\mathcal N}=1$
four-dimensional bulk theory (four supercharges).
Naively, the effective theory on the plane must preserve two supercharges,
while the domain line must preserve one supercharge.
In fact, they have four and two conserved supercharges, respectively. This is another
new phenomenon --- {\em supersymmetry enhancement} --- discovered in \cite{svritz2}.
. One can excite the junction line
endowing it with momentum in the direction of the line, without altering its  BPS
status.
 A domain line with a plane wave propagating on it
(Fig.~\ref{dwa3}) preserves the property of the BPS saturation, see \cite{svritz2}.

\begin{figure}[h]
 \centerline{\includegraphics[width=2in]{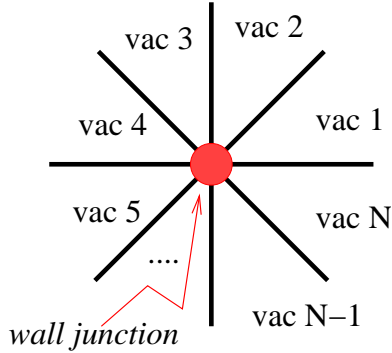}}
 \caption{\small The cross section of the wall junction.}
 \label{waj}
 \end{figure}

Let us pass now to more conventional wall junctions. Assume that the theory under
consideration has a spontaneously broken  $Z_N$
symmetry, with $N\geq 3$, and, correspondingly, $N$
vacua. Then one can have
$N$ distinct walls connected in the  asterisk-like pattern, see Fig.~\ref{waj}.
This field configuration possesses an obvious axial symmetry: the vacua
are located cyclically.

This configuration is absolutely topologically stable, as stable as
the wall itself. Moreover, it can be BPS-saturated for any value of $N$.
It was noted \cite{at} that theories with either
a U(1) or $Z_N$ global symmetry may contain 1/4-BPS objects with axial geometry.
The corresponding Bogomol'nyi  equations were derived in
\cite{CSh} and shortly after rediscovered in \cite{GT}.
Further advances in the issue of the
domain wall junctions of the hub-and-spoke type  were presented in
\cite{CHT,oda,BT,ST}, see also later works  \cite{GTTi,KakimotoSak,Jweb,Jnaweb,Jtrian}
We would like to single out Ref.~\cite{oda} where
 the first analytic solution for a BPS wall junction was found
in a specific generalized Wess--Zumino model. Among
stimulating findings in this work
is the fact that the junction tension turned out to be
negative in this model.
The model has $Z_3$ symmetry. It is derived from a
${\rm SU}(2)$ Yang--Mills theory with
extended supersymmetry (${\mathcal{N}} = 2$) and one matter
flavor perturbed by an adjoint scalar mass.
The original model contains three pairs of chiral superfields
and, in addition, one extra chiral superfield.
In fact,  the model  of \cite{oda} can be  simplified and adjusted to cover the case of arbitrary $N$, which was done in \cite{ST}. The latter work demonstrates that
the tension of the wall junctions is generically negative although exceptional models
with the positive tension are possible too. Note that the negative sign of the wall junction tension does not lead to instability since the wall junctions do not exist in isolation. They are always attached to walls which stabilize this field configuration.

Returning to  SU($N$) supersymmetric gluodynamics  ($N\geq 3$) one expects
to get in this theory the 1/4 BPS junctions of the
type depicted in Fig.~\ref{waj}. Of course, this theory is strongly coupled;
therefore, the classical Bogomol'nyi equations are irrelevant.
However, assuming that such wall junctions do exist, one
can find their tension at large $N$  even without
solving the theory. To this end one uses \cite{gis,ST}
the expression for the (1/2,1/2) central charge\,\footnote{
There is a subtle point here which must be noted. For the
wall type of the hub-and-spokes type the overall tension
is the sum of two tensions: the tension of the walls and the tension of the
hub. The first is determined by the (1,0) central charge, the second by
(1/2,1/2). Each separately is somewhat ambiguous in the case at hand. The ambiguity cancels in the sum \cite{gorskys}.
}
in terms of the contour integral over the axial current \cite{gorskys}.
At large $N$ the latter integral is  determined by two things:
the absolute value of the gluino condensate
and the overall change of the phase of the condensate when one makes the
$2\pi$ rotation around the hub. In this way one arrive at the prediction
\beq
T_{\rm wall \,\, junction} \sim N^2\,.
\label{twj}
\eeq
The coefficient in front of the $ N^2$ factor is
model dependent.

Can one interpret this $N^2$ dependence of the hub of the junction?
 Assume that
each wall has thickness $1/N$  and there are $N$ of them.
Then it is natural to expect the radius of the
intermediate domain where  all walls join together to  be of the order
$(1/N)\times N\sim N^0$.  This implies, in turn, that  the area of the hub is $O(N^0)$. If the volume  energy density inside the junction
is $N^2$ (i.e. the same as inside the walls), one immediately  gets Eq.~(\ref{twj}).

\vspace{2cm}

\centerline{\includegraphics[width=1.3in]{extra3.eps}}

\newpage

\subsection{Vortices in D=3 and
flux tubes in D=4}
\label{vortandflt}
\setcounter{equation}{0}

Vortices were among the first examples of topological defects
treated in the Bogomol'nyi limit \cite{B,dvsch,A4} (see also \cite{Taubes}).
Explicit embedding of the bosonic sector in supersymmetric models dates back to the 1980s. In  \cite{Mello} a three-dimensional Abelian Higgs model was considered.
That model had ${\mathcal N}=1$ supersymmetry (two supercharges)
and thus, according to Sect.~\ref{sssd3},
contained no central charge that could be saturated by vortices.
Hence, the vortices discussed in \cite{Mello} were not critical. BPS saturated vortices
can and do occur in ${\mathcal N}=2$ three-dimensional models (four supercharges)
with a non-vanishing Fayet--Iliopoulos term \cite{Schmidt,Edelstein}.
Such models can be obtained by dimensional reduction from
four-dimensional ${\mathcal N}=1$ models. We will start from
a brief excursion in SQED.

\subsubsection{SQED in 3D}
\label{sqed3d}

The starting point is SQED with the Fayet--Iliopoulos term $\xi$ in
four dimensions.
The SQED  Lagrangian is
\begin{eqnarray}
{\mathcal L } &=& \left\{ \frac{1}{4\, e^2}\int\!{\rm d}^2\theta \, W^2 + {\rm
H.c.}\right\} +
\int \!{\rm d}^4\theta \,\bar{Q}\, e^{n_e\,V}\, Q
\nonumber\\[3mm]
&+&
\int \!{\rm d}^4\theta \,\bar{\tilde{Q}}\, e^{-n_e\,V}\, \tilde{Q}
- n_e \, \xi  \int\! {\rm d}^2\theta {\rm d}^2\bar \theta
\,V(\! x,\theta , \bar\theta ) \, ,
\label{sqed}
\end{eqnarray}
where $e$ is the electric coupling constant, $Q$ and $\tilde{Q}$ are  chiral matter
superfields
(with charge $n_e$ and $-n_e$ respectively), and
$W_\alpha$ is the supergeneralization of the photon field strength
tensor,
\begin{equation}
{W}_{\alpha} = \frac{1}{8}\;\bar{D}^2\, D_{\alpha } V =
  i\left( \lambda_{\alpha} + i\theta_{\alpha}D - \theta^{\beta}\,
F_{\alpha\beta} -
i\theta^2{\partial}_{\alpha\dot\alpha}\bar{\lambda}^{\dot\alpha}
\right)\, .
\label{sgpfst}
\end{equation}

In  four dimensions the absence of the chiral anomaly in  SQED requires
the matter superfields enter in pairs of the opposite charge. Otherwise the theory is anomalous, the chiral anomaly renders it non-invariant under gauge transformations.
Thus, the minimal matter sector includes two chiral superfields
$Q$ and $\tilde Q$, with charges $n_e$ and $-n_e$, respectively.
(In the literature a popular choice is $n_e=1$.
In Part II we will use a different normalization,
$n_e=1/2$, which is more convenient in some problems that we address  in Part~II.)

In three dimensions there is no chirality. Therefore, one can consider
3D SQED with a single matter superfield $Q$, with charge  $n_e$.
Classically  it is perfectly fine just to discard
the superfield  $\tilde Q$ from the Lagrangian (\ref{sqed}).
However, such ``crudely truncated" theory may be inconsistent
at the quantum level  \cite{Redlich,AlvgaumeWit,AHISS}. Gauge invariance in loops
requires, as we will see shortly, simultaneous introduction
of a Chern--Simons term in the
one matter superfield model \cite{Redlich,AlvgaumeWit,AHISS}. The Chern--Simons
term breaks parity.
That's the reason why this phenomenon is sometimes referred to as
{\em parity anomaly}.

A perfectly safe way to get rid of $\tilde Q$ is as follows. Let us start from
the two-superfield model (\ref{sqed}), which is certainly self-consistent both
at the classical and quantum levels.
The one-superfield model can be obtained from that with two superfields by making
$\tilde Q$ heavy and integrating it out. If one manages to introduce a mass $\tilde m$
for $\tilde Q$ without breaking ${\mathcal N}=2$ supersymmetry,
the large $\tilde m$ limit can be viewed as an excellent regularization procedure.

Such mass terms are well-known, for a review see \cite{3Dzero,BHO,AHISS}. They go
under the name of ``real masses,"  are specific to theories with U(1) symmetries
dimensionally reduced from $D=4$ to $D=3$, and present a direct generalization
of {\em twisted masses} in two dimensions \cite{twisted}. To introduce a
``real mass" one couples matter fields to a background vector field with
a non-vanishing component along the reduced direction. For instance, in the case at
hand we introduce a background field $V_{\rm b}$ as
 \beq
 \Delta {\mathcal L}_m = \int \!{\rm d}^4\theta \,\bar{\tilde Q}\, e^{\,V_{\rm b}}\,
\tilde Q\,,\qquad V_{\rm b} =\tilde m\, (2\,i)\left(\theta^1\,\bar\theta^{\dot 2}
 -\theta^2\,\bar\theta^{\dot 1}\right).
 \eeq
The reduced spatial direction is that along the $y$ axis. We couple $V_{\rm b}$
to the U(1) current of $\tilde Q$ ascribing to $\tilde Q$ charge one with respect
to the background
field. At the same time $ Q$ is assumed to have $V_{\rm b}$ charge zero and, thus, has no coupling to $V_{\rm b}$.
Then, the background field generates a mass term only  for
$\tilde Q$, without breaking ${\mathcal N}=2$.

After reduction to three dimensions and passing to components (in the Wess--Zumino gauge) we arrive at the action in the following form (in the three-dimensional notation):
\begin{eqnarray}
S &=&\int d^3 x\,
\left\{-\frac{1}{4e^2}\, F_{\mu\nu}\,F^{\mu\nu} +\frac{1}{2e^2}\,
\left(\partial_\mu\,a\right)^2+\frac{1}{e^2}\,\bar\lambda
\,i\,\!{\not\!\partial}\,\lambda \right.
\nonumber\\[2mm]
&+&\left.
\frac{1}{2e^2}\,D^2 -n_e\, \xi\,D + n_{e}\,D\left(  \bar{q}\,q -  \bar{\tilde q}\,\tilde q\right)  \right.
\nonumber\\[2mm]
&+& \left[
{\mathcal D}^\mu\bar{q}\, {\mathcal D}_\mu q
+\bar\psi \,i\,\!{\not\!\!{\mathcal D}}\,\psi \right]
+
 \left[
{\mathcal D}^\mu\bar{\tilde q}\, {\mathcal D}_\mu \tilde q
+\bar{\tilde\psi }\,i\,\!{\not\!\!{\mathcal D}}\,\tilde\psi \right]
\nonumber\\[3mm]
&-&
 a^2 \bar{q} \,q
-(\tilde m+a)^2 \,\bar{\tilde{q}}\,\tilde{q} +
 a\, \bar\psi\,\psi -(\tilde m+a)\, \bar{\tilde\psi}\,\tilde\psi
 \nonumber\\[3mm]
&+&
 \left.
 n_{e}\left[\sqrt{2}\left(\bar\lambda\,\psi \right)\bar{q}+
{\rm h.c.}\right]-
n_{e}\left[\sqrt{2}\left(\bar\lambda\,\tilde \psi \right)\bar{\tilde q}+
{\rm h.c.}\right]
\right\}.
\label{n1bt}
\end{eqnarray}
Here $a$ is a real scalar field,
$$
a=-n_e \,A_2\,,\qquad i{\mathcal D}_\mu =i \partial_\mu + n_{e} A_\mu\,,
$$
$\lambda$ is the photino field, and
$q,\,\tilde q $ and $\psi ,\,\tilde\psi $ are matter fields belonging to
$Q$ and $\tilde Q$, respectively. Finally, $D$ is
an auxiliary field, the last component of the superfield $V$. Eliminating $D$ via the equation of motion we get the scalar potential
\beq
V=\frac{e^2}{2}\, n_e^2 \big[ \xi -
\left( \bar{q}\,q -  \bar{\tilde q}\,\tilde q\right)
\big]^2+ a^2 \bar{q} \,q
+(\tilde m+a)^2 \,\bar{\tilde{q}}\,\tilde{q}
 \,,
 \label{odynone}
\eeq
which implies a potentially rather rich vacuum structure. For our purposes --- the BPS-saturated vortices --- only the Higgs phase is of importance. We will assume that
\beq
\xi > 0\,,\qquad \tilde m\geq 0\,.
\eeq

If $\tilde \psi$ and $\tilde{q}$ are viewed as regulators (i.e. $\tilde m\to \infty$),
they can be integrated out leaving us with the one matter superfield model.
It is obvious that integrating them out we get a Chern--Simons term at one
loop,\footnote{In passing from two to one matter superfield,
in order to justify integrating out $\tilde Q$, one must consider $\tilde m\gg
e\sqrt \xi$. Given that $e^2/\xi \ll 1$, the condition
$\tilde m\gg
e\sqrt \xi$ does not necessarily imply that $\tilde m\gg \xi$.}
with a well-defined coefficient that does not vanish in the limit $\tilde m= \infty$.
We prefer to keep $\tilde m$ as a free parameter, assuming that
$\tilde m\neq 0$.

From the standpoint of vortex studies, the model (\ref{sqed}) {\em per se}
is not quite satisfactory due to the existence of the flat direction
(correspondingly, there is a gapless mode which renders the theory ill-defined
in the infrared, see Sect.~\ref{nonep}). The flat direction is eliminated
at $\tilde m \neq 0$.
Thus, there are free relevant parameters of dimension of mass,
$$e^2\,,\,\,\, \xi\,,\,\,\,\mbox{ and}\,\,\,\tilde m\,.$$
The weak coupling regime implies that
$e^2/\xi \ll 1$.

If $\tilde m \neq 0$ the field $\tilde{q}$ can (and must) be set to zero,
and $\tilde{q}\,,\,\, \tilde\psi$ play a role only at the level of quantum corrections,
providing a well-defined regularization in loops. If $\tilde{q} =0$,
the vanishing of the $D$ term in the vacuum requires $\bar{q}\,q_{\rm vac} =\xi$.
Then the term $a^2\bar{q}q$ in (\ref{odynone}) implies that
$a=0$ in the vacuum.
Up to gauge transformations the  vacuum is unique. The Higgs phase is enforced
by our choice $\tilde m\neq 0$ and $\xi \neq 0$.

\vspace{3mm}

{\em Central charge}

\vspace{2mm}

The general form of the centrally extended ${\mathcal N}=2$ superalgebra in $D=3$
was discussed in Sect.~\ref{sssd23}. The central charge relevant in the
problem at hand --- vortices ---  is
presented by the last term  in Eq.~(\ref{tdecomp}). It can be conveniently
derived using the complex representation for supercharges and reducing from
$D=4$ to $D=3$.
In four dimensions \cite{gorskys}
\begin{eqnarray}
\{Q_\alpha\, , \bar Q_{\dot\alpha}\}
 =  2 P_{\alpha\dot\alpha}
+2 Z_{\alpha\dot\alpha}
\equiv
2\left( P_\mu + Z_\mu \right)
\left(\sigma^\mu \right)_{\alpha\dot\alpha}\, ,
\label{bsa}
\end{eqnarray}
where $P_\mu$ is the momentum operator, and
\beq
Z_\mu = \xi\,  \int {\rm d}^3 x\, \epsilon_{0\mu\nu\rho}
\left( \partial^\nu A^\rho \right) + ...
\eeq
Here ellipses denote full spatial derivatives of currents\,\footnote{Moreover, these currents
are not unambiguously defined, see \cite{gorskys}.} that fall off
exponentially fast at infinity. Such terms are clearly inessential.

In three dimensions the central charge of interest reduces to
$P_2 + Z_2$. Thus, in terms of complex supercharges
the appropriate centrally extended algebra takes the form
\begin{eqnarray}
\left\{ Q,\,\left( Q^\dagger\right)\gamma^0\right\} &=& 2\left(P_0\, \gamma^0 + P_1\,\gamma^x
+P_3\,\gamma^z\right)
\nonumber\\[3mm]
&+&2\left\{ \frac{1}{e^2}\,\int  d^2x \vec\nabla\left( \vec E\,a\right) \,  + \frac{\tilde m}{2}\, q\, -n_e\, \xi\,\int d^2x B\right\},
\nonumber\\
\label{topandm}
\end{eqnarray}
where $\vec E$ is the electric field,
$B$ is the magnetic field,
\beq
B=
\frac{\partial A_z}{\partial x} - \frac{\partial A_x}{\partial z}\,,
\eeq
 and $q $ is a conserved Noether charge,
\begin{eqnarray}
q= \int d^2x\,  j^0 \,,\qquad j^\mu \equiv
-n_e\, \bar{\tilde\psi}\,\gamma^\mu\,\tilde\psi
+n_e\, \bar{\tilde{q}}\, i\, \stackrel{\leftrightarrow}{{\mathcal D}}_\mu\,\tilde{q}\,.
\end{eqnarray}
The second line  in Eq.~(\ref{topandm}) presents the vortex-related central charge.\footnote{The emergence of the U(1) Noether charge $\tilde m q/2$ in the central charge
is in one-to-one correspondence with a similar phenomenon
taking place in the two-dimensional CP($N-1$) models with the twisted mass \cite{SVZ06}.}
The term proportional to $a$
 gives a vanishing contribution to the
central charge.
However, the $q$ term (sometimes omitted in the literature) plays an important role.
It combines with the $\xi$ term in the expression for the vortex mass
converting the bare value of $\xi$ into the renormalized one.
In  the problem at hand,
the vortex mass gets renormalized at one loop, and so does the Fayet--Iliopoulos parameter.

\vspace{3mm}

{\em BPS equation for the vortex}

\vspace{2mm}

\begin{figure}[h]
 \centerline{\includegraphics[width=2in]{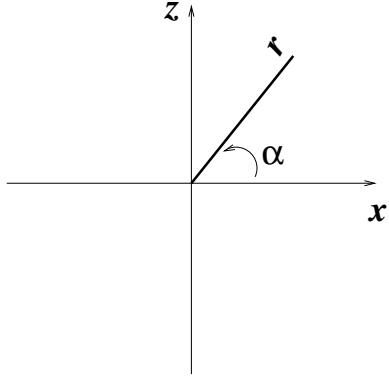}}
 \caption{\small Polar coordinates on the $x,z$ plane. }
 \label{polarc}
 \end{figure}

\noindent
At the classical level the fields $a$ and $\tilde \phi$
play no role. They will be set
\beq
\tilde{q}=0\,, \qquad a= 0\,.
\eeq
The first-order equations describing
 the ANO vortex   in the Bogomol'nyi limit
\cite{B,dvsch,A4}  take the  form
\begin{eqnarray}
&& B - n_e\,e^2 \left(|q|^2- \xi \right)=0\;,
\nonumber\\[2mm]
&&
\left({\mathcal D}_{x}+i{\mathcal D}_{z}\right)\, q=0\,.
\label{ne1}
\end{eqnarray}
with the boundary conditions
\begin{eqnarray}
&&
q\to \sqrt\xi \, e^{ik\alpha}\quad\mbox{at}\quad r\to\infty\,,\nonumber\\[2mm]
&&
q\to 0 \quad\mbox{at}\quad r\to 0\,,
\end{eqnarray}
where $\alpha$ is the polar angle on the $xz$ plane, while $r$ is the distance from the
origin in the same plane (Fig.~\ref{polarc}). Moreover $k$, is an integer,
counting the number of windings.

If Eqs. (\ref{ne1}) are satisfied, the flux of the magnetic field is $2\pi k$
(the winding number $k$ determines the quantized magnetic flux),
and
the vortex mass (string tension) is
\beq
M =\ 2\pi \xi \, k \ ,
\label{vmstension}
\eeq
 The linear
dependence of the
$k$-vortex mass on $k$ implies the absence of their potential  interaction.

For the elementary $k=1$ vortex it is convenient to
introduce two profile functions $\phi (r)$ and  $f(r)$ as follows:
\begin{equation}
q(x) = \phi(r)\,  e^{i\,\alpha}\;,\qquad
A_n(x) =-\frac1{n_e}\varepsilon_{nm}\,\frac{x_m}{r^2}\,[1- f(r)]\ .
\label{profil}
\end{equation}
The {\em ansatz} (\ref{profil})
goes through the set of equations
(\ref{ne1}), and  we get the following two equations
on the profile functions:
\begin{equation}
-\frac1{r}\,\frac{ d f}{ dr} +n_e^2e^2\left(\phi^2-\xi\right) = 0\ ,\qquad
r\, \frac{  d\,\phi}{ dr}- f\,\phi= 0\ .
\label{foe}
\end{equation}

\begin{figure}[h]
\centering
\includegraphics[width=6cm]{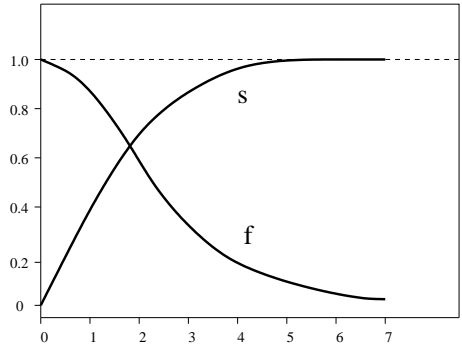}
\caption{\small
Profile functions of the string as functions of the dimensionless
variable $m_{\gamma}r$.
The gauge and scalar profile functions are  given by $f$ and $s\equiv \phi/\sqrt{\xi}$,
respectively.}
\label{figano}
\end{figure}

The boundary conditions for the   profile functions
are rather obvious from the form of the  {\em ansatz} (\ref{profil})
and from our previous discussion. At large distances
\beq
\phi(\infty)=\sqrt{\xi},\qquad f(\infty)= 0\,.
\label{624four}
\eeq
At the same time, at the origin
the smoothness of the field configuration at hand
(the absence of singularities) requires
\begin{equation}
\phi(0)=0,\qquad f(0)= 1\,.
\label{624five}
\end{equation}
These boundary conditions are such that
  the scalar field reaches its vacuum value  at
infinity.
Equations (\ref{foe}) with the above  boundary conditions   lead to a
unique solution for the profile functions, although its analytic form   is
not known. The vortex size is $\sim e^{-1}\,\xi^{-1/2}$.
The solution can be readily obtained numerically.
The profile functions $\phi$ and $f$
which determine the Higgs field and the gauge potential, respectively,
are shown in Fig.~\ref{figano}.

\vspace{3mm}

{\em The fermion zero modes}

\vspace{2mm}

Quantization of vortices  requires the knowledge of the fermion zero modes for the given classical solution. More precisely, since the solution under consideration
is static, we are interested in the zero-eigenvalue solutions of the static fermion
equations which, thus, effectively become two- rather than three-dimensional,
\beq
i\left(\gamma^x\,{\mathcal D}_x + \gamma^z\,{\mathcal D}_z
\right)\psi +n_e\, \sqrt{2}\,\lambda
\,q = 0\,.
\label{n1fzm}
\eeq
This equation is obtained from (\ref{n1bt}) where we dropped the tilted terms
(since $\tilde{q} =0$). The fermion operator
is Hermitean implying that every solution for $\{\psi\,,\,\lambda\}$ is accompanied
by that
for $\{\bar\psi\,,\,\bar\lambda\}$.

Since the  solution to equations (\ref{ne1}) discussed above is 1/2 BPS,
two of the four supercharges annihilate it while the other two generate the fermion zero modes -- superpartners of translational modes.
One can show \cite{RebhanPN} that these are the only normalizable
fermion zero modes in the problem at hand.

\newpage

{\em Short versus long representations}

\vspace{2mm}

The (1+2)-dimensional model under consideration
has four supercharges. The corresponding regular super-representation
is four-dimensional (i.e. contains two bosonic and two fermionic states).

The vortex we discuss has two fermion zero modes.
Hence, viewed as a particle in 1+2 dimensions it forms a super-doublet
(one bosonic state plus one fermionic).
Hence, this is a short multiplet.
This implies, of course, that the BPS bound
must remain saturated when quantum corrections are switched on.
Both, the central charge and the vortex mass get corrections
\cite{vassil,RebhanPN}, but they remain equal
to each other.

\subsubsection{Four-dimensional SQED and the ANO string}
\label{sqed4d}

In this section we will discuss \none SQED.
SQED with extended supersymmetry (i.e. \ntwo)
is also very interesting. This latter model is presented in Appendix B.

The Lagrangian is the same as in Eq.~(\ref{sqed}).
We will consider the simplest case:
one chiral superfield $Q$ with charge $n_e=1/2$, and one
chiral superfield $\tilde Q$ with charge $n_e=-1/2$.
The electric charge of matter is chosen to be half-integer to make contact with
what follows. This normalization is convenient
in the case of non-Abelian models, see Part II.  The Lagrangian in components
can be obtained from Eq.~(\ref{n1bt}) by setting $a=\tilde m =0$.
The scalar potential obviously takes the form
\beq
V=\frac{e^2}{2}\, n_e^2 \big[ \xi -
\left( \bar{q}\,q -  \bar{\tilde q}\,\tilde q\right)
\big]^2\,.
\label{sp4dq}
\eeq
The vacuum manifold is a ``hyperboloid"
\beq
 \bar{q} \,q
- \bar{\tilde{q}}\,\tilde{q}
=\xi \,.
\eeq
Thus, we deal with the Higgs branch of
real dimension two. In fact, the vacuum manifold can be parametrized
by a complex modulus $\tilde{q}q$. On this Higgs branch
the photon field and   superpartners form a massive supermultiplet,
while $\tilde{q}q$ and superpartners form a massless one.

As was shown in \cite{PeninRubak}, no finite-thickness vortices exist
at a generic point on the vacuum manifold, due to the
absence of the mass gap (presence of the massless Higgs excitations).
The moduli fields get involved in the solution at the classical level
generating a logarithmically divergent tail. An infrared regularization can
remove this logarithmic divergence,  and vortices become well-defined,
see \cite{Y99} and Sect.~\ref{sechiggs}.
One of possible infrared regularizations is considering a finite-length
string instead of an infinite string. Then all infrared divergences are cut off at
distances of the
order of the string length. The thickness
of the string is of the order of logarithm of this length.
This is discussed in detail in Sect.~\ref{sechiggs}.
Needless to say, such string is not BPS-saturated.

At the base of the Higgs branch, at $\tilde{q} =0$, the classical solutions
of the BPS equations for $q$ and $A_\mu$ are well-defined.
The form of the solution coincides with that given in Sect.~\ref{sqed3d}.

The fact that there is a flat direction and, hence, massless
particles in the bulk theory does not disappear, of course. Even though at
$\tilde{q} =0$ the classical string solution is well defined, infrared problems arise at
 the
loop level. One can avoid massless particles in the spectrum if one embeds
the theory (\ref{n1bt}) in SQED with eight supercharges, see Sect.~\ref{nonep} and
Appendix B. Then
the Higgs  branch is eliminated, and one is left with isolated vacua. After the embedding
 is done,
one can break ${\mathcal N}=2$ down to ${\mathcal N}=1$, if one so desires.

A simpler framework is provided by the so-called $M$ model.
Its non-Abelian version is considered in Sect.~\ref{secmmodel}.
Here we will outline the construction of this model in the context of \none SQED.

We introduce an extra {\em neutral} chiral superfield $M$, which interacts with $Q$ and $\tilde Q$
through the super-Yukawa coupling,
\beq
{\mathcal L}_M = \int d^2\theta\, d^2\bar\theta \, \, \frac 1h\,\bar M\, M
+\left\{ \int\, d^2\theta\, Q M\tilde Q + \mbox{(H.c)}\right\}\,.
\label{msqed}
\eeq
Here $h$ is a coupling constant. As we will see momentarily the Higgs branch is lifted.
An obvious advantage of this model is that it makes no reference
to \ntwo. This is probably the simplest \none model which supports
BPS-saturated ANO strings without infrared problems.

The scalar potential (\ref{sp4dq}) is now replaced by
\beq
V_M =\frac{e^2}{2}\, n_e^2 \big[ \xi -
\left( \bar{q}\,q -  \bar{\tilde q}\,\tilde q\right)
\big]^2 + h\left| q\,\tilde q
\right|^2 +\left| q\,M
\right|^2 + \left| M \,\tilde q
\right|^2 \,.
\label{sp4dqM}
\eeq
The vacuum is unique modulo gauge transformations,
\beq
q = \bar q = \sqrt\xi\,,\quad \tilde q = 0\,,\quad M =0\,.
\eeq
The classical ANO flux tube solution considered above
remains valid as long as we set, additionally,
$\tilde q = M = 0$. The quantization procedure is straightforward,
since one encounters no infrared problems whatsoever ---
all particles in the bulk are massive. In particular,
there are four {\em normalizable} fermion zero modes (cf. Ref.~\cite{VY}).

For further thorough discussions we refer the reader to Sect.~\ref{aaaa}.

\vspace{2cm}

\centerline{\includegraphics[width=1.3in]{extra3.eps}}

\newpage

\subsection{Monopoles}
\setcounter{equation}{0}

In this section we will discuss  magnetic monopoles --- very interesting objects
which carry magnetic charges. They
emerge as free magnetically charged particles in non-Abelian gauge
theories in which the gauge symmetry is spontaneously broken
down to an Abelian subgroup.\footnote{In the confining regime
monopoles can be obtained in some theories with no adjoint fields,
in which the gauge symmetry is broken completely \cite{us}.
This is a recent development.}
The simplest example was found by 't Hooft
\cite{thooftmon} and
Polyakov \cite{polyakov}. The model they considered had been invented by Georgi and
Glashow \cite{GG} for different purposes. As it often happens,
the Georgi--Glashow model turned out to be more valuable than
the original purpose, which  is long forgotten, while the model itself is alive and well
and is being constantly used by theorists.


\subsubsection{The Georgi--Glashow model:\\[1mm]
vacuum and elementary excitations}

Let us  begin with a brief description of the Georgi--Glashow model.
The gauge group is SU(2) and the matter sector consists of one real scalar field
$\phi^a$ in the adjoint representation (i.e. SU(2) triplet). The Lagrangian of the model is
\begin{equation}
L=-\frac{1}{4g^2}\, F_{\mu\nu}^a\, F^{\mu\nu ,\, a}
+\frac{1}{2} (D_\mu\phi^a )(D^\mu\phi^a )
-\frac{1}{8} \lambda (\phi^a \phi^a - v^2)^2\,,
\label{10.1}
\end{equation}
where the covariant derivative in the adjoint acts as
\begin{equation}
D_\mu\phi^a =\partial_\mu \phi^a +\varepsilon^{abc} A_\mu^b\phi^c\,.
\end{equation}
Below we will focus on the limit of  BPS monopoles.
This limit corresponds to a vanishing scalar coupling, $\lambda\to 0$.
The only role of the last term in Eq. (\ref{10.1}) is to provide a boundary condition for
the scalar field. As is clear from Sect.~\ref{central}
the monopole central charge exists only in ${\mathcal N}=2$ and ${\mathcal N}=4$
superalgebras. Therefore, one should understand the theory (\ref{10.1})
(at $\lambda=0$) as embedded in super-Yang--Mills theories with extended superalgebra. In Part II we will extensively discuss such embeddings in the context
of ${\mathcal N}=2$.

The classical definition  of magnetic charges refers to theories
that support long-range (Coulomb) magnetic field. Therefore,
in consideration of the isolated monopole
 the pattern of the
symmetry breaking should be such that some of the gauge bosons remain  massless.
In the  Georgi--Glashow model (\ref{10.1}) the pattern is as follows:
\begin{equation}
{\rm SU(2)} \to {\rm U(1)}\,.
\label{10.2}
\end{equation}
To see that this is indeed the case let us note the $\phi^a$ self-interaction term
(the last term in Eq. (\ref{10.1})) forces $\phi^a$ to develop a vacuum expectation
value,
\begin{equation}
\langle \phi^a \rangle = v\delta^{3a}\,.
\label{10.3}
\end{equation}
The direction of the vector $\phi^a$ in the SU(2) space (to be referred to as
``color space" or ``isospace") can be chosen arbitrarily. One can always reduce it to the form
(\ref{10.3}) by a global color rotation. Thus, Eq.~(\ref{10.3}) can be viewed as a
(unitary) gauge condition on the field $\phi$.

This gauge is very convenient for discussing the particle content
of the theory, elementary excitations.
Since the color rotation around the third axis does not change the vacuum expectation
value of $\phi^a$,
\begin{equation}
\exp\left\{ i\alpha\frac{\tau_3}{2}\right\}\,\phi_{\rm vac}\exp\left\{
-i\alpha\frac{\tau_3}{2}\right\}= \phi_{\rm vac}\,,\qquad \phi_{\rm
vac}=v\frac{\tau_3}{2}\,,
\label{10wed.1}
\end{equation}
the third component of the gauge field remains massless --- we will call it ``photon,"
\begin{equation}
A_\mu^3\equiv A_\mu\,,\qquad F_{\mu\nu} =\partial_\mu A_\nu -\partial_\nu
A_\mu\,.
\end{equation}
The first and the second components form massive vector bosons,
\begin{equation}
W_\mu^\pm =\frac{1}{\sqrt{2}\,g }\left(A_\mu^1\pm A_\mu^2
\right)\,.
\end{equation}
As usual in the Higgs mechanism, the massive vector bosons eat up
the first and the second components of the scalar field $\phi^a$.
The third component, the physical Higgs field, can be parametrized
as
\begin{equation}
\phi^3 = v +\varphi\,,
\end{equation}
where
$\varphi$ is the physical Higgs field.
In terms of these fields the Lagrangian (\ref{10.1}) can be readily rewritten as
\begin{eqnarray}
L &=&-\frac{1}{4g^2} F_{\mu\nu}\, F_{\mu\nu} +\frac{1}{2}(\partial_\mu\varphi)^2
\nonumber\\[2mm]
&-& \left( D_\alpha W^+_\mu\right) \left(
D_\alpha W^-_\mu\right) + \left( D_\mu W^+_\mu\right)\left(
D_\nu W^-_\nu\right) +
g^2 (v+\phi)^2\, W^+_\mu W^-_\mu\nonumber\\[2mm]
&-& 2\, W_\mu^+ \, F_{\mu\nu}\, W_\nu^- +\frac{g^2}{4}
\left(W_\mu^+\, W_\nu ^- - W_\nu^+\, W_\mu ^-
\right)^2\,,
\end{eqnarray}
where the covariant derivative now includes only the photon field,
\begin{equation}
D_\alpha\,W^\pm = \left(\partial_\alpha \pm i A_\alpha\right)W^\pm \,.
\end{equation}
The last line presents the magnetic moment of the charged (massive)
vector bosons and their self-interaction.
In the limit $\lambda\to 0$ the physical Higgs field is massless. The mass of the
$W^\pm$ bosons is
\begin{equation}
M_W=g\, v\,.
\end{equation}


\subsubsection{Monopoles --- topological argument}
\label{topargu}

Let us explain why this model has a
topologically stable soliton.

Assume that monopole's center is   at the origin and consider a large
sphere ${\mathcal S}_R$ of radius $R$ with the center at the origin.
Since the mass of the monopole is finite, by definition, $\phi^a \phi^a =v^2$ on this sphere. $\phi^a$ is a three-component vector in the isospace subject to the constraint
$\phi^a \phi^a =v^2$ which gives us a two dimensional sphere ${\mathcal S}_G$.
This, we deal here with mappings of ${\mathcal S}_R$ into ${\mathcal S}_G$.
Such mappings split in distinct classes labeled by an integer $n$,
counting how many times the sphere ${\mathcal S}_G$ is swept when we sweep once
the sphere ${\mathcal S}_R$,
since
\begin{equation}
\pi_2 (\mbox{SU(2)/U(1)}) = Z\,.
\label{mtbg}
\end{equation}
${\mathcal S}_G$  =  SU(2)/U(1)  because
for  each given vector $\phi^a$ there is  a U(1) subgroup which does {\em not} rotate it.
The SU(2) group space is a three-dimensional sphere
while that of SU(2)/U(1) is  a two-dimensional sphere.

An isolated monopole field configuration (the 't Hooft--Polyakov monopole) corresponds
to a mapping with $n=1$. Since it is impossible to
continuously deform it to the topologically trivial mapping,
the monopoles are topologically stable.


\subsubsection{Mass and magnetic charge}
\label{maamac}

Classically the monopole mass is given by the energy functional
\begin{equation}
E = \int d^3 x\left\{ \frac{1}{2 \, g^2}\, B_i^a B_i^a +\frac{1}{2}\,
\left( D_i\phi^a\right) \left( D_i\phi^a\right)
\right\}\, ,
\label{fridone}
\end{equation}
where
\begin{equation}
B_i^a = -\frac{1}{2}\, \varepsilon_{ijk} F^a_{jk}\,.
\end{equation}
The fields are assumed to be time-independent, $B_i^a =B_i^a (\vec x )$,
$\phi^a = \phi^a (\vec x )$. For static fields
it is natural to assume that $A_0^a=0$. This assumption will
be verified {\em posteriori}, after we find the field configuration minimizing
the functional (\ref{fridone}). Equation (\ref{fridone}) assumes the limit
$\lambda\to 0$. However, in performing minimization we should keep
in mind the boundary condition $\phi^a (\vec x )\phi^a (\vec x ) \to v^2$
at $|\vec x|\to\infty$.

Equation (\ref{fridone})
can be identically rewritten as follows:
\begin{equation}
E = \int d^3 x  \left\{ \frac{1}{2} \,\left(\frac{1}{g}\, B_i^a-D_i\phi^a\right)\left(\frac{1}{g}\,B_i^a-D_i\phi^a\right)
+\frac{1}{g}\,B_i^a D_i\phi^a
\right\}.
\label{fridthree}
\end{equation}
The last term on the right-hand side is a full derivative.
Indeed, after integrating by parts and using the equation of motion $D_i B_i^a =0$
we get
\begin{eqnarray}
 \int d^3 x\left\{ \frac{1}{g}\,B_i^a D_i\phi^a \right\}&=&\frac{1}{g}\,\int d^3 x\,  \partial_i
\left(B_i^a \phi^a\right)\nonumber\\[2mm]
&=& \frac{1}{g}\,\int_{{\mathcal S}_R} d^2 S_i \,
\left(B_i^a \phi^a\right)\,.
\label{fridtwo}
\end{eqnarray}
In the last line we made use of Gauss' theorem and passed from the volume integration to that over the surface of the large sphere.
Thus, the last term in Eq. (\ref{fridthree}) is topological.

The combination $B_i^a \phi^a$ can be viewed as a gauge invariant definition of the
magnetic field  $\vec {\mathcal B}$. More exactly,
\begin{equation}
{\mathcal B}_i = \frac{1}{v} \, B_i^a \phi^a \,.
\label{fridfour}
\end{equation}
Indeed, far away from the monopole core
one can always assume $\phi^a$ to be aligned in the same way as in the vacuum
(in an appropriate gauge), $\phi^a =v\delta^{3a}$. Then ${\mathcal B}_i=B^3_i$.
The advantage of the definition (\ref{fridfour}) is that
it is gauge independent.

Furthermore,
the  magnetic charge $Q_M$ inside a sphere ${\mathcal S}_R$ can be defined
through the flux of the magnetic field through the surface of the sphere,
\footnote{A remark: Conventions for the charge normalization used in different
books and papers may vary. In his original paper on the magnetic monopole,\cite{Dirac}
Dirac uses the convention $e^2 =\alpha$ and the electromagnetic Hamiltonian
${\mathcal H}= (8\pi)^{-1}\left(\vec E^2 +\vec B^2
\right)$. Then, the electric charge is defined through the flux of the electric field
as $e=(4\pi)^{-1}\int_{{\mathcal S}_R}d^2 S_i {E}_i$, and analogously for the magnetic
charge. We use the convention according to which
$e^2= 4\pi\alpha$,  and the electromagnetic Hamiltonian
${\mathcal H}= (2g^2)^{-1}\left(\vec E^2 +\vec B^2
\right)$. Then $e= g^{-1}\int_{{\mathcal S}_R}d^2 S_i {E}_i$ while
$Q_M=g^{-1}\int_{{\mathcal S}_R}d^2 S_i {B}_i$.
}
\begin{equation}
Q_M= \int_{{\mathcal S}_R} d^2 S_i\, \frac{1}{g} {\mathcal B}_i \,.
\label{fridfive}
\end{equation}
From Eq. (\ref{ten-seven}) (see below) we will
see that
\begin{equation}
{\mathcal B}_i \equiv \frac{1}{v}\, B^a_i\phi^a{\longrightarrow} n^i\,\frac{1}{r^2}\,\,\,\mbox{at $ r\to\infty$} \,,
\end{equation}
and, hence,
\begin{equation}
Q_M= \frac{4\pi}{g}\,.
\end{equation}

Combining Eqs. (\ref{fridfive}), (\ref{fridfour}) and (\ref{fridtwo})
we conclude that
\begin{equation}
E =   v \,  Q_M + \int d^3 x  \left\{ \frac{1}{2} \,\left(\frac{1}{g}\, B_i^a-D_i\phi^a\right)\left(\frac{1}{g}\,B_i^a-D_i\phi^a\right)
\right\}.
\label{fridsix}
\end{equation}
The minimum of the energy functional is attained at
\begin{equation}
\frac{1}{g}\,B_i^a-D_i\phi^a =0\,.
\label{fridseven}
\end{equation}
The mass of the field configuration realizing this minimum
--- the monopole mass --- is obviously equal
\begin{equation}
M_M = \frac{4\pi\, v}{g}\,.
\label{ggmonma}
\end{equation}
Thus, the mass of the critical monopole is in one-to-one relation with its magnetic charge.
Equation (\ref{fridseven}) is nothing but the Bogomol'nyi equation in
the monopole problem.
If it is satisfied, the second order differential equations of motion
are satisfied too.


\subsubsection{Solution of the Bogomol'nyi equation for monopoles}

To solve the Bogomol'nyi equations we need to find an appropriate {\em ansatz} for $\phi^a$.
As one sweeps $S_R$ the vector $\phi^a$ must sweep the group space sphere.
The simplest choice is to identify these two spheres point-by-point,
\begin{equation}
\phi^a = v\,\frac{x^a}{r}= \, v n^a\,, \qquad r\to\infty\,.
\label{ten-one}
\end{equation}
where $n^i\equiv x^i/r$.
This field configuration obviously belongs to the class with $n=1$. The SU(2)
group index $a$ got entangled with the coordinate $\vec x$.
Polyakov proposed to refer to such fields as ``hedgehogs."

Next, observe that finiteness of the monopole energy
requires  the covariant derivative $D_i\phi^a$ to fall off
faster than $r^{-3/2}$  at large $r$, cf. Eq. (\ref{fridone}).
Since
\begin{equation}
\partial_i\phi^a = v\,\frac{1}{r}\left\{\delta^{ai} - n^an^i
\right\}\sim \frac{1}{r}
\label{ten-two}
\end{equation}
one must choose $A^b_i$ in such a way as to cancel (\ref{ten-two}).
It is not difficult to see that
\begin{equation}
A^a_i = \varepsilon^{aij}\, \frac{1}{r}\,  n^j\,, \qquad r\to\infty\,.
\label{ten-three}
\end{equation}
Then the term $1/r$ is canceled in $D_i\phi^a$.

Equations (\ref{ten-one}) and (\ref{ten-three}) determine the index structure
of the field configuration we are going to deal with.
The appropriate {\em ansatz} is perfectly clear now,
\begin{equation}
\phi^a =v\, n^a H (r)\,,\qquad A^a_i = \varepsilon^{aij}\, \frac{1}{r}\,  n^j\, F(r)\,,
\label{ten-six}
\end{equation}
where $H$ and $F$ are functions of $r$ with the boundary conditions
\begin{equation}
H(r) \to 1\,,\qquad F(r) \to 1 \qquad\mbox{at}\,\,\, r\to \infty\,,
\label{ten-four}
\end{equation}
and
\begin{equation}
H(r) \to 0\,,\qquad F(r) \to 0 \qquad\mbox{at}\,\,\, r\to 0\,.
\label{ten-five}
\end{equation}
The boundary condition (\ref{ten-four}) is equivalent to
Eqs. (\ref{ten-one}) and (\ref{ten-three}), while the boundary condition (\ref{ten-five})
guarantees that our solution is nonsingular at $r\to 0$.

After some straightforward algebra we get
\begin{eqnarray}
B^a_i &=&\left(\delta^{ai} -n^an^i
\right)\frac{1}{r}\, F ' + n^an^i\,\frac{1}{r^2}\left(2F-F^2\right)\,,\nonumber\\[3mm]
D_i\phi^a &=&v\left\{ \left(\delta^{ai} -n^an^i
\right)\frac{1}{r}\, H (1-F) +n^an^i H '
\right\}\,,
\label{ten-seven}
\end{eqnarray}
where prime denotes differentiation with respect to $r$.

Let us return now to the Bogomol'nyi equations (\ref{fridseven}).
This is a set of nine first-order differential equations.
Our {\em ansatz} has only two unknown functions. The fact that the {\em ansatz}
goes through and we get two scalar equations on two unknown functions from the
Bogomol'nyi equations is a highly nontrivial check. Comparing Eqs. (\ref{fridseven})
and (\ref{ten-seven}) we get
\begin{eqnarray}
\frac{1}{g} \, F ' &=& v\, H(1-F)\,,\nonumber\\[1mm]
H ' &=&\frac{1}{g\, v} \,\frac{1}{r^2}\left(2F-F^2\right)\,.
\label{ten-eight}
\end{eqnarray}
The functions $H$ and $F$ are dimensionless. It is convenient to make the
radius $r$ dimensionless too. A natural unit of length
in the problem at hand is $(gv)^{-1}$. From now on we will measure $r$ in these units,
\begin{equation}
\rho = r \, (gv)\,.
\end{equation}
The functions $H$ and $F$ are to be considered as functions of $\rho$, while the
prime will denote differentiation over $\rho$. Then the system (\ref{ten-eight})
takes the form
\begin{eqnarray}
  F ' &=&  H(1-F)\,,\nonumber\\[1mm]
H ' &=& \frac{1}{\rho^2}\left(2F-F^2\right)\,.
\label{ten-eighta}
\end{eqnarray}

\begin{figure}[h]
\centering
\includegraphics[width=6cm]{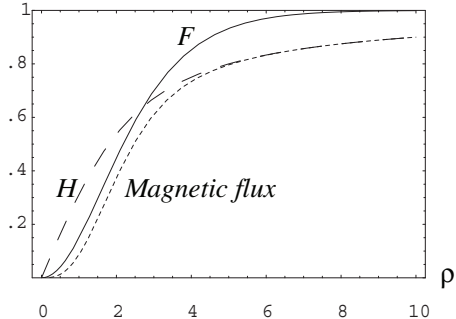}
\caption{\small
The functions $F$ (solid line) and $H$ (long dashes) in the critical monopole solution,
vs. $\rho$. The short-dashed line shows the flux of the magnetic field
${\mathcal B}_i$ (in the units $4\pi$) through the sphere of radius $\rho$.}
\label{10-one}
\end{figure}

These equations have known analytical solutions,
\begin{eqnarray}
  F &=& 1-\frac{\rho}{\mbox{sinh}\rho}\,,\nonumber\\[1mm]
H  &=& \frac{\mbox{cosh}\rho}{\mbox{sinh}\rho}-\frac{1}{\rho}\,.
\label{ten-nine}
\end{eqnarray}
At large $\rho$ the functions  $H$ and $F$ tend to unity
(cf. Eq. (\ref{ten-four}))
while at $\rho\to 0$
$$
F= 0(\rho^2)\,,\qquad H=0(\rho )\,.
$$
They are plotted in Fig. \ref{10-one}. Calculating the flux of the magnetic field
through the large sphere we verify that for the solution at hand
$Q_M=4\pi/g$.

\subsubsection{Collective coordinates (moduli)}
\label{335}

The monopole solution presented in the previous section breaks a number of valid symmetries of the theory, for instance, translational invariance.
As usual, the symmetries are restored after the introduction of the collective coordinates (moduli), which convert a given solution into a family of solutions.

Our first task is to count the number of moduli in the monopole problem.
A straightforward way to count
this number is counting linearly independent zero modes. To this end, one represents
the fields $A_\mu$ and $\phi$ as a sum of the
monopole background plus small deviations,
\begin{equation}
A_\mu ^a = A_\mu ^{a(0)} + a_\mu^a\,,\qquad \phi^a =\phi^{a(0)} +(\delta\phi)^a\,,
\label{10-34}
\end{equation}
where the superscript (0) marks the monopole solution.
At this point it is necessary to impose
a gauge-fixing condition.
A convenient condition is
\begin{equation}
\frac{1}{g}\, D_i a_i^a -\varepsilon^{abc}\phi^b(\delta\phi)^c =0\,,
\label{10-35}
\end{equation}
where the covariant derivative in the first term contains only
the background field.

Substituting the decomposition (\ref{10-34}) in the Lagrangian one finds the
quadratic form for $\{ a, \,\, (\delta\phi)\}$,
 and determines the zero modes of this form (subject to the condition
(\ref{10-35})).

We will not track this procedure in detail, referring the reader to the
original  literature \cite{Mottola}.
Instead,
we suggest a simple heuristic consideration.

Let us ask ourselves what are the valid symmetries of the
model at hand? They are: (i) three translations; (ii)
three spatial rotations; (iii) three rotations in the SU(2) group. Not all these symmetries are independent. It is not difficult to check that the spatial rotations are equivalent to
the SU(2) group rotations for the monopole solution. Thus, we should not count them independently. This leaves us with six symmetry transformations.

One should not forget, however, that two of those six act non-trivially in the
``trivial vacuum."  Indeed, the latter is characterized by the condensate
(\ref{10.3}). While rotations around the third axis in the isospace
leave the condensate intact (see Eq. (\ref{10wed.1})),
the rotations around the first and second axes do not. Thus, the number of moduli
in the monopole problem is $6-2=4$. These four collective coordinates have a very transparent physical interpretation. Three of them correspond to translations.
They are introduced in the solution through the
substitution
\begin{equation}
\vec x \to \vec x - \vec x_0\,.
\end{equation}
The vector $\vec x_0$ now plays the role of the monopole center.
The unit vector $\vec n$ is now defined as
$\vec n = \left(\vec x - \vec x_0\right)/\left| \vec x - \vec x_0\right|$.

The fourth collective coordinate is related to the unbroken U(1) symmetry of the model.
This is the rotation around the direction of alignment of the field $\phi$.
In the ``trivial vacuum" $\phi^a$ is aligned along the third axis.
The monopole generalization of Eq. (\ref{10wed.1}) is
\begin{eqnarray}
A^{(0)} &\to& U^{-1} A^{(0)} U - i U^{-1}\partial U\,,\nonumber\\[2mm]
\phi^{(0)} &\to& U^{-1} \phi^{(0)} U = \phi^{(0)} \,,\nonumber\\[2mm]
U&=& \exp\left\{ i \alpha \phi^{(0)} /v \right\}\,,
\label{10-37}
\end{eqnarray}
where the fields $A^{(0)}$ and $\phi^{(0)}$ are understood here in the matrix
form,
$$
A^{(0)}= A^{a(0)}\,(\tau^a/2)\,,\qquad \phi^{(0)} =\phi^{a(0)}\,(\tau^a/2)\,.
$$
Unlike the vacuum field, which is not changed under
(\ref{10wed.1}), the monopole solution for
the vector field changes its form.  The change looks as a gauge transformation.
Note, however, that the gauge matrix $U$ does not tend to unity at
$r\to\infty$. Thus, this transformation is in fact a global  U(1) rotation.
The physical meaning of the collective coordinate $\alpha$
will become clear shortly. Now let us note that
(i) for small $\alpha$  Eq. (\ref{10-37}) reduces to
\begin{equation}
\delta A^a_i = \alpha\, \frac{1}{v}\,  (D_i\phi^{(0)})^a\,,\qquad\delta\phi = 0\,,
\label{10-38}
\end{equation}
and this is compatible with the gauge condition (\ref{10-35});
(ii) the variable $\alpha$ is compact, since the points $\alpha$ and
$\alpha +2\pi$ can be identified (the transformation of $A^{(0)}$
is identically the same for $\alpha$ and
$\alpha +2\pi$). In other words, $\alpha$ is an angle variable.

Having identified all four moduli relevant to the problem
we can proceed to the quasi-classical quantization.
The task is to obtain quantum mechanics of the moduli.
Let us start from the monopole center coordinate $\vec x_0$.
To this end, as usual, we assume that $\vec x_0$ weakly depends on time $t$,
so that the only time dependence of the solution enters through
$\vec x_0(t)$. The time dependence is important only in time derivatives,
so that the quantum-mechanical Lagrangian of moduli
can be obtained from the following expression:
\begin{eqnarray}
{\mathcal L}_{\rm QM}
&=&  - M_M + \frac{1}{2}\, (\dot x_0)_k (\dot x_0)_j \int\, d^3 x\left\{
\left[ \frac{1}{g}\,F_{ik}^{a(0)} \right]\left[ \frac{1}{g}\,F_{ij}^{a(0)} \right]\right.
\nonumber\\[4mm]
 &+& \left.\left[ D_k \phi^{a(0)} \right]\left[D_j\phi^{a(0)}\right]
\right\}\,,
\label{10-42}
\end{eqnarray}
where $\partial_k A$ and $\partial_k \phi$
where supplemented by appropriate gauge transformations to satisfy the
gauge condition (\ref{10-35}).

Averaging over the angular orientations of $\vec x$ yields
\begin{eqnarray}
{\mathcal L}_{\rm QM}&=& - M_M + \frac{1}{2}\, (\dot{\vec x}_0)^2 \int\, d^3 x
\left\{ \frac{2}{3}\,\frac{1}{g^2}\, B^{a(0)}_iB^{a(0)}_i + \frac{1}{3}\,
D_i\phi^{a(0)} D_i\phi^{a(0)}\right\}\nonumber\\[3mm]
&=& - M_M + \frac{M_M}{2}\, (\dot{\vec x}_0)^2\,.
\label{10-43}
\end{eqnarray}
This last result readily follows if one
combines Eqs. (\ref{fridone}) and (\ref{fridseven}).
Of course, this final answer could have been guessed from the very beginning
since this is nothing but the Lagrangian describing free
non-relativistic  motion of a
particle of mass $M_M$ endowed with the coordinate $\vec x_0$.

Now, having tested the method in the case where the answer was obvious, let us
apply it to the fourth collective coordinate $\alpha$.
Using Eq. (\ref{10-38}) we get
\begin{equation}
{\mathcal L}_{\alpha\rm QM} = \frac{1}{2}\, \frac{M_M}{M_W^2}\,\dot\alpha^2\,,
\label{10-44}
\end{equation}
or, equivalently,
\begin{equation}
{\mathcal H}_\alpha = \frac{1}{2}\, \frac{M_W^2}{M_M}\,p_\alpha^2\,,
\qquad p_\alpha \equiv
-i\, \frac{d}{d\alpha}\,,
\label{10-45}
\end{equation}
where ${\mathcal H}_\alpha$ is the part of the Hamiltonian relevant to $\alpha$.
The full quantum-mechanical Hamiltonian describing the moduli
dynamics is, thus,
\begin{equation}
{\mathcal H} = M_M + \frac{p^2}{2M_M} + \frac{1}{2}\, \frac{M_W^2}{M_M}\,p_\alpha^2\,,\qquad p \equiv -i\, \frac{d}{dx_0}\,.
\label{10-46}
\end{equation}
It describes  free motion of a spinless particle endowed with an internal
(compact) variable $\alpha$. While the spatial part of ${\mathcal H}$ does not raise any questions,
the $\alpha$ dynamics deserves an additional discussion.

The $\alpha$ motion is free, but one should not forget that $\alpha$ is an angle.
Because of the $2\pi$ periodicity,
the corresponding wave functions must  have the form
\begin{equation}
\Psi (\alpha) = e^{ik\alpha}\,,
\label{10-47}
\end{equation}
where $k$ is an integer, $k=0,\pm 1,\pm2,...$.
Strictly speaking, only the ground state, $k=0$, describes the monopole ---
a particle with the magnetic charge $4\pi/g$ and vanishing electric charge.
Excitations with $k\neq 0$ correspond to a particle
with the magnetic charge $4\pi/g$ {\em and} the electric charge $kg$,
the dyon.

To see that this is indeed the case, let us note that for $k\neq 0$
the expectation value of $p_\alpha$ is $k$ and, hence, the expectation value
of $\dot\alpha = (M_W^2/M_M)\, p_\alpha$ is $M_W^2k/M_M$.
Moreover, let us define a gauge-invariant electric field ${\mathcal E}_i$
(analogous to ${\mathcal B}_i$ of Eq. (\ref{fridfour})) as
\begin{equation}
{\mathcal E}_i \equiv \frac{1}{v}\, E_i^a \phi^a = \frac{1}{v}\, \phi^{a(0)} \, {\dot A}^{a(0)}_i
=\frac{1}{v^2}\, \dot\alpha\, \phi^{a(0)} \,  (D_i \phi^{a(0)})\,.
\label{10-48}
\end{equation}
Since for the critical monopole $D_i \phi^{a(0)} =(1/g) B^{a(0)}_i$ we see that
\begin{equation}
{\mathcal E}_i =\dot\alpha\,\frac{1}{M_W}{\mathcal B}_i\,,
\label{10-49}
\end{equation}
and the flux of the  gauge-invariant electric field over the large sphere is
\begin{equation}
\frac{1}{g}\, \int_{{\mathcal S}_R} d^2 S_i \, {\mathcal E}_i
= \frac{M_W^2k}{M_M}\, \frac{1}{M_W}\frac{1}{g}\, \int_{{\mathcal S}_R} d^2 S_i{\mathcal B}_i
\end{equation}
where we replaced $\dot\alpha$ by its expectation value. Thus, the flux
of the electric field reduces to
\begin{equation}
\frac{1}{g}\, \int_{{\mathcal S}_R} d^2 S_i \, {\mathcal E}_i = k g\,,
\label{10-50}
\end{equation}
which proves the above assertion of the electric charge
$kg$.

It is interesting to note that the mass of the  dyon  can be written as
\begin{equation}
M_D = M_M + \frac{1}{2}\, \frac{M_W^2}{M_M}\,k^2\approx \sqrt{M_M^2 + M_W^2\,k^2 }=  v \sqrt{Q_M^2+Q_E^2}\,.
\label{10-51}
\end{equation}
Although from our derivation it might seem that
the square root result is approximate, in fact, the prediction
for the dyon mass
$M_D = v ({Q_M^2+Q_E^2})^{1/2}$
is exact; it follows from the BPS saturation and the central charges
in ${\mathcal N}=2$ model (see Sect.~\ref{central}).

Magnetic monopoles were introduced in theory by Dirac in 1931 \cite{Dirac}.
He considered macroscopic electrodynamics and derived a self-consistency condition
for the product of the magnetic charge of the monopole $Q_M$ and the elementary
electric charge $e$,\footnote{In Dirac's original convention the
charge quantization condition is, in fact, $Q_M  e=(1/2)$.}
\begin{equation}
Q_M \, e = 2\pi\,.
\label{10-54}
\end{equation}
This is known as the Dirac quantization condition.
For the 't Hooft--Polyakov monopole we have just derived that
$Q_M g = 4\pi$, twice larger than in the Dirac quantization condition.
Note, however, that $g$ is the electric charge of the $W$ bosons.
It is {\em not} the minimal possible electric charge
that can be present in the theory at hand. If quarks in the
fundamental (doublet) representation of SU(2) were introduced in the
Georgi--Glashow model, their U(1) charge would be $e=g/2$, and the
Dirac quantization condition would be satisfied for such elementary charges.


\subsubsection{Singular gauge, or how to comb a hedgehog}
\label{singga}

\begin{figure}[h]
\centering
\includegraphics[width=8cm]{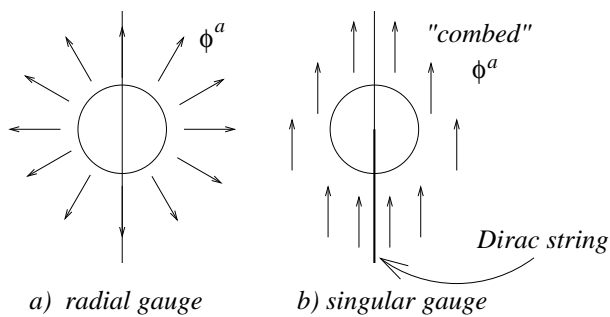}
\caption{\small
Transition from the radial to singular gauge or combing the hedgehog}
\label{comb}
\end{figure}

The {\em ansatz} (\ref{ten-six}) for the monopole solution
we used so far is very convenient for revealing a nontrivial topology
lying behind this solution, i.e. the fact that SU(2)/U(1) $\sim S_2$
in the group space is mapped onto the spatial $S_2$.
However, it is often useful to gauge-transform it in such a way
that the scalar field becomes oriented along the third axis in the color space, $\phi^a\sim \delta^{3a}$, in all space (i.e. at all $x$),
repeating the pattern of the ``plane" vacuum (\ref{10.3}).
Polyakov suggested to refer to this gauge transformation
as to ``combing the hedgehog." Comparison of Figs.~\ref{comb}$a$
and \ref{comb}$b$ shows that this gauge transformation cannot be nonsingular. Indeed, the matrix which combs the
hedgehog,
\begin{equation}
U^\dagger\left( n^a\tau^a\right) U =\tau^3\,,
\end{equation}
has the form
\begin{equation}
U=\frac{1}{\sqrt 2}\left(\sqrt{1+n^3}
+ i\,\frac{\nu^a \tau^a}{\sqrt{1+n^3}}\right)\,,
\end{equation}
where
\begin{equation}
\nu^a = \varepsilon^{3ab}\,n^b\,,\qquad  \nu^a\nu^a
=1-(n^3)^2\,,
\end{equation}
and  $\vec n$ is the unit vector in the direction of
$\vec x$. The matrix $U$ is obviously singular at
$n^3 = -1$ (see Fig.~\ref{comb}). This is a gauge artifact since all physically
measurable quantities
are nonsingular and well-defined. In the ``old" Dirac
description of the monopole \cite{colem}
the singularity of $U$ at $n^3 = -1$ would correspond to the
Dirac string.

In the singular gauge the monopole magnetic field at large
$|\vec x|$ takes the ``color-combed" form
\begin{equation}
B_i \to  \frac{\tau^3}{2}\, \frac{n^i}{r^2}
= 4\pi\,  \frac{\tau^3}{2}\, \frac{n^i}{4\pi\, r^2}\,.
\label{commo}
\end{equation}
The latter equation naturally implies the same magnetic charge
$Q_M= 4\pi/g$, as was derived in Sect.~\ref{topargu}.


\subsubsection{Monopoles in SU($N$)}
\label{moninsun}

Let us return to critical monopoles and extend
the construction presented above from SU(2) to SU($N$)\cite{GodNOlive,We}.
The starting Lagrangian is the same as in Eq.~(\ref{10.1}),
with the replacement of the structure constants $\varepsilon^{abc}$
of SU(2) by the SU($N$) structure constants $f^{abc}$.
The potential of the scalar-field self-interaction can be of a more general
form than in Eq.~(\ref{10.1}).  Details of this potential are unimportant for our purposes since in the critical limit the potential tends to zero;
its only role is to fix the vacuum value of the field $\phi$ at infinity.

Recall that all generators of the Lie algebra can be always divided
into two groups --- the Cartan generators $H_i$, which all commute
with each other, and a set of raising (lowering) operators $E_{\mbox{\boldmath $\alpha$}}$,
\begin{equation}
E_{\mbox{\boldmath $\alpha$}}^\dagger = E_{\mbox{-\boldmath $\alpha$}}\,.
\label{10-100}
\end{equation}
For SU($N$) --- and we will not discuss other groups ---
there are $N-1$ Cartan generators
which can be chosen as
\begin{eqnarray}
H^1 &=& \frac{1}{2}\,{\rm diag} \left\{1, -1, 0,....0  \right\},
\nonumber\\[1mm]
H^2 &=& \frac{1}{2\sqrt 3}\,{\rm diag} \left\{1, 1,-2, 0,....0  \right\},
\nonumber\\[2mm]
&...&..............
\nonumber\\[2mm]
H^m &=& \frac{1}{ \sqrt{2m(m+1)}}\, {\rm diag} \left\{1, 1,1, ..., -m, ..., 0 \right\},
\nonumber\\[2mm]
&...&..............
\\[1mm]
H^{N-1} &=& \frac{1}{ \sqrt{2N(N-1)}}\, {\rm diag} \left\{1, 1,1, ..., 1,  -(N-1)  \right\},
\nonumber
\label{10-103}
\end{eqnarray}
$N(N-1)/2$ raising generators $E_{\mbox{\boldmath $\alpha$}}$, and $N(N-1)/2$ lowering generators
$E_{\mbox{-\boldmath $\alpha$}}$. The Cartan generators are analogs of
$\tau_3/2$ while $E_{\mbox{\boldmath $\pm \alpha$}}$
are analogs of $\tau_\pm/2$.
Moreover, $N(N-1)$ vectors $\mbox{\boldmath $\alpha$},\,\, \mbox{-\boldmath $\alpha$}$ are called   root vectors. They are $(N-1)$-dimensional.

By making an appropriate choice of basis,
any element of SU($N$) algebra can be brought to the Cartan subalgebra.
Correspondingly, the vacuum value of the (matrix) field $\phi \equiv \phi^a T^a$
can always be chosen to be of the form
\begin{equation}
\phi_{\rm vac} = \mbox{\boldmath $h\, H$}\,,
\label{10-101}
\end{equation}
where \boldmath$h$ is an \unboldmath$(N-1)$-component vector,
\begin{equation}
\mbox{\boldmath$ h$} = \{h_1,\,h_2 , ..., \, h_{N-1}\}\,.
\label{10-102}
\end{equation}
For simplicity we will assume that for all simple roots
$ \mbox{\boldmath$ h$}\, \mbox{\boldmath$ \gamma$}>0$
(otherwise, we will just  change the condition defining positive roots to meet this constraint).

Depending on the form of the self-interaction potential
distinct patterns of gauge symmetry breaking can take place.
We will discuss here the case when
the gauge symmetry is maximally broken,
\begin{equation}
{\rm SU}(N) \to {\rm U}(1)^{N-1}\,.
\label{10-103a}
\end{equation}
The unbroken subgroup is Abelian.
This situation is general. In special cases, when $\mbox{\boldmath$ h$}$
is orthogonal to $\mbox{\boldmath$ \alpha$}^m$, for some $m$
(or a set of $m$'s) the unbroken subgroup will contain non-Abelian factors, as
will be explained momentarily.
These cases will not be considered here.

The topological argument proving the existence
of a variety of topologically stable monopoles
in the above set-up parallels that of Sect.~\ref{topargu},
except that Eq.~(\ref{mtbg}) is replaced by
\begin{equation}
\pi_2 \left( {\rm SU}(N)/{\rm U}(1)^{N-1}\right) = \pi_1 \left({\rm U}(1)^{N-1}\right)
= Z^{N-1}\,.
\label{mttbg}
\end{equation}
There are $N-1$ independent windings
in the SU($N$) case.

The gauge field $A_\mu$ (in the matrix form, $A_\mu\equiv A_\mu^a \, T^a$)
can be represented as
\begin{equation}
A_\mu^a \, T^a =\sum_{m=1}^{N-1}\, A_\mu^m\, H^m
+\sum_{\mbox{\boldmath$\alpha$}}\, A_\mu^{\mbox{\boldmath$\alpha$}}\, E_{\mbox{\boldmath$\alpha$}}\,,
\label{10-104}
\end{equation}
where $A_\mu^m$'s ($m=1,..., N-1)$ can be viewed as
``photons," while $A_\mu^{\mbox{\boldmath$\alpha$}}$'s
as ``$W$ bosons."
The mass terms are obtained from the term
$$
{\rm Tr}\, \left( \left[A_\mu, \, \phi\right]
\right)^2
$$
in the Lagrangian.
Substituting here Eqs.~(\ref{10-101}) and (\ref{10-104})
it is easy to see  that the $W$-boson masses are
\begin{equation}
M_{\mbox{\boldmath$\alpha$}}  = g \,
{\mbox{\boldmath$h$}}{\mbox{\boldmath$\alpha$}}\,.
\label{10-105}
\end{equation}
$ N-1$ massive bosons corresponding to simple roots
$\mbox{\boldmath$ \gamma$}$  play a special role:
they can be thought of as {\em fundamental},
in the sense that the quantum numbers and masses of all other
$W$ bosons can be obtained as linear combinations
(with nonnegative integer coefficients)
of those of the fundamental $W$ bosons.
With regards to the masses this is immediately seen from
Eq.~(\ref{10-105}) in conjunction with
\begin{equation}
\mbox{\boldmath $\alpha$} = \sum_{\mbox{\boldmath$ \gamma$}}
k_{\mbox{\boldmath$ \gamma$}} \, \mbox{\boldmath$ \gamma$}\,.
\label{nsrfsr}
\end{equation}

Construction of  SU$(N$) monopoles reduces, in essence,
to that of a SU(2) monopole through various embeddings of
SU(2) in SU$(N$).
Note that each simple root $\mbox{\boldmath$ \gamma$}$
defines an SU(2) subgroup\,\footnote{Generally speaking,
each root $\mbox{\boldmath$ \alpha$}$ defines an SU(2)
subalgebra according to Eq.~(\ref{10-106}), but we will deal only with the simple roots for reasons which will become clear momentarily. }
of SU$(N$)
with the following three generators:
\begin{eqnarray}
t^1 &=& \frac{1}{\sqrt 2}\left(E_{\mbox{\boldmath$\gamma$}}+
E_{-\mbox{\boldmath$\gamma$}}
\right),
\nonumber\\[1mm]
t^2 &=& \frac{1}{\sqrt 2\, i}\left(E_{\mbox{\boldmath$\gamma$}}-
E_{-\mbox{\boldmath$\gamma$}}
\right),
\nonumber\\[2mm]
t^3&=& \mbox{\boldmath$\gamma$}\, \mbox{\boldmath$H$}\,,
\label{10-106}
\end{eqnarray}
with the standard algebra $[t^i,\,t^j] = i\varepsilon^{ijk}\, t^k.$
If the basic SU(2) monopole solution corresponding
to the Higgs vacuum expectation value $v$
is denoted as $\{\phi^a(\mbox{\boldmath$r$};\, v),\,\,
A_i^a  (\mbox{\boldmath$r$};\, v) \}$, see Eq.~(\ref{ten-six}),
the construction of a specific  SU($N$)
monopole
proceeds in three steps: (i) choose a simple root
$\mbox{\boldmath$ \gamma$}$; (ii)
decompose the vector $\mbox{\boldmath$h$}$
in two components, parallel and perpendicular with respect to
$\mbox{\boldmath$ \gamma$}$,
\begin{eqnarray}
\mbox{\boldmath$h$}&=& \mbox{\boldmath$h$}_{\|}+
\mbox{\boldmath$h$}_\perp\,,\nonumber\\[1mm]
\mbox{\boldmath$h$}_{\|} &=&\tilde v  \mbox{\boldmath$\gamma$}\,,
\quad \mbox{\boldmath$h$}_\perp\mbox{\boldmath$\gamma$} =0\,,
\nonumber\\[1mm]
 \tilde v &\equiv&  \mbox{\boldmath$\gamma$}\, \mbox{\boldmath$h$} >0\,;
\label{10-107}
\end{eqnarray}
(iii)
replace $A_i^a  (\mbox{\boldmath$r$};\, v)$
by $A_i^a  (\mbox{\boldmath$r$};\, \tilde v)$ and add a
covariantly constant term to the field $\phi^a(\mbox{\boldmath$r$};\, \tilde v)$ to ensure that at $r\to\infty$
 it  has the correct asymptotic behavior, namely, $2\, {\rm Tr}\, \phi^2
= \mbox{\boldmath$h$}^2$.
Algebraically
the SU($N$) monopole solution takes the form
\begin{equation}
\phi = \phi^a(\mbox{\boldmath$r$};\, \tilde v)\, t^a +
\mbox{\boldmath$h$}_\perp \mbox{\boldmath$H$}\,,
\qquad A_i = A_i^a  (\mbox{\boldmath$r$};\, \tilde v)\,t^a\,.
\label{10-108}
\end{equation}
Note that the mass of the corresponding $W$ boson $M_{\mbox{\boldmath$\gamma$}}= g\tilde v$,
in full parallel with the SU(2) monopole.

It is instructive to  verify that (\ref{10-108})
satisfies the BPS equation (\ref{fridseven}). To this end it is sufficient to
note that $[\mbox{\boldmath$h$}_\perp \mbox{\boldmath$H$}
,\, A_i] =0$, which in turn implies
$$
D_i \left( \mbox{\boldmath$h$}_\perp \mbox{\boldmath$H$}
\right) =0\,.
$$

What remains to be done? We must analyze the magnetic charges of
the SU($N$) monopoles and their masses. In the singular gauge
(Sect.~\ref{singga}) the Higgs field is aligned in  the Cartan subalgebra,
$\phi \sim \mbox{\boldmath $h\, H$}$. The magnetic field at large distances
from the monopole core, being commutative with $\phi$,
also lies in the Cartan subalgebra. In fact, from Eq.~(\ref{10-106})
we infer that combing of the SU($N$) monopole
leads to
\begin{equation}
B_i \to  4\pi
\,\mbox{\boldmath$\gamma$} \mbox{\boldmath$H$}
 \, \frac{n^i}{4\pi\, r^2}\,,
\label{commop}
\end{equation}
which implies, in turn, that the set
of $N-1$ magnetic charges of the
SU($N$) monopole
is given by the components of the $(N-1)$-vector
\begin{equation}
\mbox{\boldmath$Q$}_M =\frac{4\pi}{g}\, \mbox{\boldmath$\gamma$}\,.
\label{su3mc}
\end{equation}
Of course, the very same result is obtained in a
gauge invariant manner from a defining formula
\begin{equation}
 2{\rm Tr}\left( B_i \phi\right) \, \,
 _{\stackrel{\longrightarrow}{\mbox{\tiny{$r\to\infty$}}}\,\,}
 \left(
 \mbox{\boldmath$Q$}_M \mbox{\boldmath$h$}\right)
 \frac{g}{4\pi}\, \frac{n_i}{r^2}\,.
\label{gimc}
\end{equation}
Equation (\ref{fridthree}) implies that
the mass of this monopole is
\begin{equation}
M_{M_{\mbox{\boldmath$\gamma$}}} =  \mbox{\boldmath$Q$}_M \mbox{\boldmath$h$} =\frac{4\pi\,\tilde v}{g}\,,
\end{equation}
to be compared with the mass of the corresponding $W$ bosons,
\begin{equation}
M_{\mbox{\boldmath$\gamma$}}= g \mbox{\boldmath$\gamma$}
\mbox{\boldmath$h$} = g\tilde v\, ,
\end{equation}
in perfect parallel with the SU(2) monopole results of
Sect.~\ref{maamac}. The general
magnetic charge quantization condition
takes the form
\begin{equation}
\exp \left\{i g \, \mbox{\boldmath$Q$}_M \mbox{\boldmath$H$}
\right\}=1\,.
\label{gqmqc}
\end{equation}

Let us ask ourselves what happens if one builds
monopoles on non-simple roots. Such solutions
are in fact composite: they consist of the basic ``simple-root"
monopoles --- the masses and quantum numbers
(magnetic charges) of the composite
monopoles can be obtained by summing up the masses
and quantum numbers of the basic monopoles,
according to Eq.~(\ref{nsrfsr}).


\subsubsection{The $\theta$ term induces a fractional electric charge\\[1mm]
 for the monopole (the Witten effect)}

There is a $P$- and $T$-odd term, the  $\theta$ term, which can be added to the
Lagrangian
for the Yang--Mills theory without spoiling renormalizability. It is given by
\begin{equation}
{\mathcal L}_\theta =  \frac{\theta }{32 \pi^2} F_{\mu \nu}^a
\tilde F^{a \mu \nu } = -  \frac{\theta }{8 \pi^2}\,\, \vec E^a \cdot \vec B^a\,.
\label{threea}
\end{equation}
This interaction violates $P$ and $CP$ but not $C$. As is well known,  this term is a surface term and does not affect the classical equations of motion.  There is however $\theta$ dependence in
instanton effects which involve non-trivial long-range behavior of the gauge fields. As
was realized by Witten \cite{wittheta},  in the presence of magnetic monopoles $\theta$
also has a nontrivial effect, it shifts the allowed values of electric charge in the
monopole sector of the theory.

Since the equations of motions do not change,
the monopole solution obtained above stays intact. What changes is
the effective quantum-mechanical Lagrangian. As usual, we assume an adiabatic
time dependence of moduli.
In the case at hand we must replace the constant phase modulus $\alpha$
by $\alpha(t)$. This generates the electric field
$$E^a_i = \dot\alpha \, (\delta A^a_i/\delta\alpha) = \frac{\dot\alpha}{v}
\left(D_i\phi^{(0)}\right)^a\,,$$
where Eq. (\ref{10-38}) is used. The magnetic field does not change, and can be expressed through $\left(D_i\phi^{(0)}\right)^a$ using Eq. (\ref{fridseven}).
As a result, the quantum-mechanical Lagrangian for $\alpha$
acquires a full derivative term,
\beq
{\mathcal L}_{\alpha\rm QM} = \frac{1}{2\mu }\, \dot\alpha^2
-\frac{\theta}{2\pi}\, \dot \alpha\,,\qquad\quad \mu = \frac{M_W^2}{M_M}\,.
\eeq
This changes the expression for the canonic momentum conjugated to $\alpha$.
If previously $p_\alpha$ was $\dot\alpha/\mu$, now
\beq
p_\alpha =\frac{\dot\alpha}{\mu} -\frac{\theta}{2\pi}\,.
\eeq
Correspondingly,
\beq
\dot\alpha =\mu \left(p_\alpha + \frac{\theta}{2\pi}\right)\,.
\label{33119}
\eeq
From Sect.~\ref{335} we know that the electric charge of the field configuration at hand is (see Eq. (\ref{10-50}))
\begin{equation}
Q_E=\frac{1}{M_W\, g}\,\, \langle\dot\alpha\rangle \,\, \int_{{\mathcal S}_R} d^2 S_i \,
{\mathcal B}_i \,.
\label{10-50a}
\end{equation}
Substituting Eq. (\ref{33119}) and $ \langle p_\alpha\rangle =k $ we arrive at
\beq
Q_E= \left( k +\frac{\theta}{2\pi}\right) g\,.
\eeq
We see that at $\theta\neq 0$ the electric charge of the dyon is non-integer.
As $\theta $ changes from zero to the physically equivalent point
$\theta =2\pi$ the dyon charges shift by one unit.
The dyon spectrum as a whole remains intact.

\vspace{2cm}

\centerline{\includegraphics[width=1.3in]{extra3.eps}}

\newpage

\subsection{Monopoles and fermions}
\label{mafe}
\setcounter{equation}{0}

The critical 't Hooft--Polyakov monopoles
we have just discussed can be embedded
in ${\mathcal N}=2$ super-Yang--Mills.
There are no ${\mathcal N}=1$ models with the
't Hooft--Polyakov monopoles (albeit
${\mathcal N}=1$ theories supporting confined monopoles
are found \cite{us}). The minimal model with the BPS-saturated
't~Hooft--Polyakov monopole
is the ${\mathcal N}=2$ generalization of supersymmetric gluodynamics,
with the gauge group SU(2). In terms of ${\mathcal N}=1$
superfields it contains one vector superfield in the adjoint
describing gluon and gluino, plus one chiral superfield in the adjoint
describing a scalar ${\mathcal N}=2$ superpartner for gluon and
a Weyl spinor, an ${\mathcal N}=2$ superpartner for gluino.

The couplings of the fermion fields to the boson fields are of a special form,
they are fixed by ${\mathcal N}=2$ supersymmetry.
In this section we will first present the Lagrangian of
${\mathcal N}=2$ supersymmetric gluodynamics,
including the part with the adjoint fermions, and then
consider
effects due to the adjoint fermions.
We conclude Sect. \ref{mafe} with a comment
on fermions
in the fundamental representation in the monopole background.

\subsubsection{\boldmath{${\mathcal N}=2$} super-Yang--Mills (without matter)}

Two ${\mathcal N}=1$ superfields are used to build the model,
\beq
W_\alpha = i \left(\lambda_\alpha + i\theta_\alpha \, D -\theta^\beta\, F_{\alpha\beta} - i\theta^2 D_{\alpha\dot \alpha}\bar\lambda^{\dot\alpha}
\right)\,,
\eeq
and
\beq
{\mathcal A} = a + \sqrt{2}\, \psi\,\theta +\theta^2 F\,.
\eeq
Here the notation is spinorial, and all fields are in the adjoint
representation of SU(2). The corresponding generators are
\beq
\big( T^a\big)_{bd} = i \, \varepsilon_{bad}\,.
\eeq
The Lagrangian contains kinetic terms and their supergeneralizations. In components
\begin{eqnarray}
{\mathcal L} &=& \frac{1}{g^2}
\left\{ -\frac{1}{4}F^{a\,\mu\nu}F^a_{\mu\nu} + \lambda^{\alpha,a}\, i\,
{D}_{\alpha\dot\alpha}\, \bar{\lambda}^{\dot\alpha,a}
+ \frac{1}{2}\,  D^a D^a  \right.
\nonumber\\[1mm]
&+& \psi^{\alpha , a}\, i{D}_{\alpha\dot\alpha}\, \bar{\psi}^{\dot\alpha , a} +
({D}^{\mu}\, \bar a ) ({D}_{\mu}\, a )
\nonumber\\[3mm]
 &-& \sqrt{2}\,\varepsilon_{abc}
 \left(\bar{a}^a \, \lambda^{\alpha , b}\, \psi^c_\alpha  + a^a\,
 \bar{\lambda}^b_{\dot\alpha}\,  \bar{\psi}^{\dot\alpha ,c}
\right)
-\left.
i \,\varepsilon_{abc} \,  D^a\; \bar a^b\,   a^c  \right\}\,.
  \label{ldrnm}
\end{eqnarray}
As usual, the $D$ field is auxiliary and can be eliminated
via equation of motion,
\beq
D^a =i\, \varepsilon_{abc}\, \bar a^b a^c\,.
\label{dter}
\eeq
There is a flat direction: if the field $a$ is real
all $D$ terms vanish. If $a$ is chosen to be purely real
or purely imaginary and the
fermion fields ignored we obviously return to the
Georgi--Glashow model discussed above.

Let us perform the Bogomol'nyi completion of the
of the bosonic part of the
Lagrangian (\ref{ldrnm}) for static field configurations.
Neglecting all time derivatives and, as usual,  setting $A_0=0$,
one can write the energy functional as
follows:
\begin{eqnarray}
{\mathcal E}
&=&
\sum_{i=1,2,3;\,a=1,2,3}\,\,
\int{d}^3 x
\left[\frac1{\sqrt{2}g}F^{*a}_{i}
\pm  \frac1{g}D_i a^a\right]^2
\nonumber\\[5mm]
&\mp&
\frac{ \sqrt{2}}{g^2}\, \int d^3 x\, \partial_i \left(F^{*a}_{i}\, a^a
\right) \,,
\label{bogmon}
\end{eqnarray}
where
$$
F_m^*= \frac{1}{2}\, \varepsilon_{mnk}\, F_{nk}\,,
$$
and
the square of the $D$ term (\ref{dter}) is omitted ---
the $D$ term vanishes provided $a$ is real, which we will assume.
This assumption also allows us
to replace the absolute value in the first line by the square brackets.
The term in the second line can be written as an integral over a large sphere,
\beq
\frac{ \sqrt{2}}{g^2}\, \int d^3 x\, \partial_i \left(F^{*a}_{i}\, a^a
\right)=
\frac{ \sqrt{2}}{g^2}\, \int d S_i\left(  a^a \,  F^{*a}_{i}
\right) \,.
\eeq
The Bogomol'nyi equations for the monopole are
\beq
F^{*a}_i \pm \sqrt 2 D_i\,  a^a = 0\,.
\eeq
This coincides with Eq.~(\ref{fridseven}) in the Georgi--Glashow model, up to a normalization.
(The field $a$ is complex, generally speaking, and its kinetic term is normalized
differently.) If the Bogomol'nyi equations are satisfied,
the monopole mass is determined by the surface term (classically).
Assuming that in the ``flat" vacuum $a^a$ is aligned
along the third direction and taking into account
that in our normalization the
magnetic flux is $4\pi$ we get
\beq
M_M =\frac{\sqrt 2\,\,  a^3_{\rm vac}}{g^2}\, \,4\pi\,,
\eeq
where --- we remind ---  $a^3_{\rm vac}$ is assumed to be positive.
This is in full agreement with Eq.~(\ref{ggmonma}).

\subsubsection{Supercurrents and the monopole central charge}
\label{satmcc}

The general classification of central charges in \ntwo theories in four dimensions
is presented in Sect.~\ref{extsc}. Here we will briefly
discuss the Lorentz-scalar central charge $Z$ in the theory (\ref{ldrnm}).
It is this central charge that is saturated by critical monopoles.

The model, being ${\mathcal N}=2$, possesses two conserved supercurrents,
\begin{eqnarray}
J_{\alpha\beta\dot\beta}^I =\frac{2}{g^2}
\left\{ i F_{\beta\alpha}^a\bar\lambda^a_{\dot\beta} + \varepsilon_{\beta\alpha}
D^a \bar\lambda^a_{\dot\beta} + \sqrt{2}\left(D_{\alpha\dot\beta}\bar{a}^a
\right)\psi^a_\beta \right\} +{\rm f.d.}\,,
\nonumber\\[4mm]
J_{\alpha\beta\dot\beta}^{II} =\frac{2}{g^2}
\left\{ i F_{\beta\alpha}^a\bar\psi^a_{\dot\beta} +\varepsilon_{\beta\alpha}
D^a \bar\psi^a_{\dot\beta} - \sqrt{2}\left(D_{\alpha\dot\beta}\bar{a}^a
\right)\lambda^a_\beta \right\}+{\rm f.d.}\,,
\label{scur}
\end{eqnarray}
where f.d. stands for full derivatives. The both expressions can be combined
in one compact formula if we introduce an
SU(2) index $f$ ($f$ =1,2) (to be repeatedly used in  Part II)
in the following way:
\beq
\lambda^f =\left\{
\begin{array}{ll}
\lambda\,,\,\,\, f=1\\
\psi \,,\,\,\, f=2.
\end{array}
\right.
\eeq
Then $\lambda_1 = -\psi$ and $\lambda_2 = \lambda$.
The supercurrent takes the form ($f=1,2$)
\begin{eqnarray}
J_{\alpha\beta\dot\beta ,\,\, f} =\frac{2}{g^2}
\left\{ i F_{\beta\alpha}^a\bar\lambda^a_{\dot\beta ,\,\, f } + \varepsilon_{\beta\alpha}
D^a \bar\lambda^a_{\dot\beta,\,\, f } - \sqrt{2}\left(D_{\alpha\dot\beta}\bar{a}^a
\right)\lambda^a_{\beta ,\,\, f}
\right.
\nonumber\\[3mm]
+\left. \frac{\sqrt 2}{6}
\left[\partial_{\alpha\dot\beta} (\lambda_{\beta ,\, f} \, \bar a)
+
\partial_{\beta\dot\beta} (\lambda_{\alpha,\, f} \, \bar a)
-3 \varepsilon_{\beta\alpha}\partial^\gamma_{\dot\beta} (\lambda_{\gamma ,\, f}
\,\bar a)
\right]
 \right\} .
\label{scurp}
\end{eqnarray}

Classically the commutator of the corresponding supercharges is
\begin{eqnarray}
\{Q^I_\alpha , \, Q^{II}_\beta \}&=&
2\,Z \,\varepsilon_{\alpha\beta}
=
 -\frac{2\sqrt{2} }{g^2}
\,\varepsilon_{\alpha\beta} \int d^3 x\,
{\rm div} \left(
\bar{a}^a \left(\vec E^a - i\vec B^a
\right)
\right)
\nonumber\\[4mm]
&=&
-\frac{2\sqrt{2} }{g^2}\,\varepsilon_{\alpha\beta}\,\int d S_j
 \left(
\bar{a}^a \left(  E^a_j - i\, B^a_j
\right)
\right)\,.
\label{ccc}
\end{eqnarray}
$Z$ in Eq.~(\ref{ccc}) is sometimes referred to as the monopole central charge.
For the BPS-saturated monopoles $M_M=Z$.

Quantum corrections in the monopole central charge and in the mass of the BPS
saturated monopoles were first discussed in
Refs.~\cite{dahodiv,Kaulm,IM} two decades ago.
The monopole central charge is renormalized at one-loop level.
This is obvious due to the fact that the corresponding
quantum correction must convert the bare coupling constant
in  Eq.~(\ref{ccc}) into the renormalized one. The fact
that the logarithmic renormalizations of the monopole mass and the
gauge coupling constant match was established long ago.
However, there is a residual non-logarithmic effect
which cannot be obtained from Eq.~(\ref{ccc}). It was not until
2004 that people realized that the monopole central charge (\ref{ccc})
must be supplemented by an anomalous term \cite{Shifman:2004dr}.

To elucidate the point, let us consider (following \cite{Rebhan:2004vn})
the formula for
the monopole/dyon  mass
obtained in the exact Seiberg--Witten solution \cite{SW1},
\beq
M_{n_e,n_m} =\sqrt{2}  \left|  a  \left( n_e -  \frac{a_D}{a} \, n_m \right)
\right|\,,
\label{cone}
\eeq
where $n_{e,m}$ are integer electric and magnetic numbers (we will consider
here only a particular case when either $n_e =0, 1$ or $n_m=0,1$)
and
\beq
a_D = i\,a\,\left( \frac{4\pi}{g_0^2}- \frac{2}{\pi}\,\ln \frac{M_0}{a}
\right)\,.
\eeq
The subscript 0 is introduced for clarity, it
marks the bare charge.
The renormalized coupling constant is defined
in terms of the ultraviolet parameters as follows:
\beq
\frac{\partial a_D}{\partial a} \equiv   \frac{4\pi i}{g^2}\,.
\eeq
Because of the $a \ln a$ dependence, ${\partial a_D}/{\partial a}$
differs from $a_D/a$ by a constant (non-logarithmic) term, namely,
\beq
\frac{a_D}{a} = i\left( \frac{4\pi }{g^2}-\frac{2}{\pi}
\right).
\label{ctwo}
\eeq
Combining Eq.~(\ref{cone}) and (\ref{ctwo}) we get
\beq
M_{n_e,n_m} =\sqrt{2}  \left|  a  \left( n_e -   i\left( \frac{4\pi }{g^2}-\frac{2}{\pi}
\right) \, n_m \right)
\right|\,,
\label{cthree}
\eeq
This does not match Eq.~(\ref{ccc}) in the non-logarithmic part
(i.e. the term $2\sqrt{2} \, n_m/\pi$).
Since the relative weight of the electric and magnetic parts
in Eq.~(\ref{ccc}) is fixed to be $g^2$,
the presence of the above non-logarithmic term
implies that, in fact, the chiral structure $ E^a_j - i\, B^a_j$
obtained at the canonic commutator level
cannot be maintained once quantum corrections are switched on.
This is a quantum anomaly.

So far no direct calculation of the
anomalous contribution in $\{Q^I_\alpha , \, Q^{II}_\beta \}$
in the operator form has been carried out. However, it is not difficult
to reconstruct it indirectly, using Eq.~(\ref{cthree}),
the fermion term obtained in Ref.~\cite{Shifman:2004dr} and a close parallel
between \ntwo super-Yang--Mills theory and
\ntwo CP$(N-1)$ model with twisted mass
in two dimensions in which a similar problem was solved \cite{SVZ06},
\beq
\left\{Q^I_\alpha , \, Q^{II}_\beta \right\}_{\rm anom}
=
2\, \varepsilon_{\alpha\beta} \,\delta Z_{\rm anom} \,
=
 -
 \left( \varepsilon_{\alpha\beta} \right)
 \,
   {2\sqrt{2} }\,\,  \frac{1}{4\pi^2}\,\,  \int d S_j \, \Sigma^j
 \label{cccprim}
 \eeq
where
\beqn
\Sigma^j &=&
\frac{i}{2}\, \frac{\partial}{\partial\bar\theta^{\dot\beta}}
\left(\bar{\mathcal A}^a \,\bar{W}^a_{\,\dot\alpha}\right)
\left(\sigma^j \right)^{\dot\alpha\dot\beta}\Big|_{\bar\theta=0}
\nonumber\\[5mm]
&=&
\bar{a}^a \left(  \vec E^a + i\, \vec B^a
\right)^j -\frac{\sqrt 2}{2}\, \bar\lambda^a_{\dot\alpha}
\left(\sigma^j \right)^{\dot\alpha\dot\beta}\,\bar\psi^a_{\,\dot\beta}
\,,
\label{cccprimm}
\eeqn
to be added to Eq.~(\ref{ccc}). The (1,0) conversion matrix $\left(\sigma^j \right)^{\dot\alpha\dot\beta}$ is defined in Sect.~\ref{seca3dop}. Equation (\ref{cccprimm})
is to be compared with that obtained at the end of Sect.~\ref{tscl}.

\vspace{2mm}

In the SU($N$) theory
we would have $\frac{N}{8\pi^2}$ instead of $\frac{1}{4\pi^2}$
in Eq.~(\ref{cccprim}).

\vspace{3mm}

Adding the canonic and the anomalous terms
in $\left\{Q^I_\alpha , \, Q^{II}_\beta \right\}$ together
we see that the fluxes generated by color-electric and color-magnetic terms
are now shifted, untied from each other, by a non-logarithmic term
in the magnetic part. Normalizing to the electric term, $M_W =\sqrt{2} a$,
we  get for the magnetic term
\beq
M_M = \sqrt{2}\, a\left(\frac{4\pi}{g^2} - \frac{2}{\pi}
\right)\,,
\label{fnlt}
\eeq
as it is necessary for the consistency with the exact Seiberg--Witten solution.

\subsubsection{Zero modes for adjoint fermions}

Equations for the fermion zero modes can be readily derived from the
Lagrangian (\ref{ldrnm}),
\begin{eqnarray}
&& i D_{\alpha\dot\alpha} \lambda ^{\alpha,\,c} - \sqrt{2}\, \varepsilon_{abc} \, a^a\, \bar \psi^b_{\dot\alpha} =0\nonumber\\[2mm]
&&
i D_{\alpha\dot\alpha} \psi ^{\alpha,\,c} + \sqrt{2}\,  \varepsilon_{abc} \, a^a\, \bar \lambda^b_{\dot\alpha} =0\, ,
\end{eqnarray}
plus Hermitean conjugate. After a brief reflection
we can get two complex (four real) zero modes.\footnote{
This means that the monopole is described by
two complex fermion collective coordinates,
or four real.}
Two of them are obtained if we substitute
\beq
\lambda^\alpha = F^{\alpha\beta}\,,\qquad \bar\psi_{\dot\alpha} = \sqrt{2}\,
 D_{\alpha\dot\alpha}\, \bar a\,.
\eeq
The other two solutions
correspond to the following substitution:
\beq
\psi^\alpha = F^{\alpha\beta}\,,\qquad \bar\lambda_{\dot\alpha} = \sqrt{2}\,
 D_{\alpha\dot\alpha}\, \bar a\,.
\eeq
This result is easy to understand. Our starting theory has eight supercharges.
The classical monopole solution is BPS-saturated, implying that
four of these eight supercharges annihilate the solution
(these are the Bogomol'nyi equations), while the action of the other four
supercharges produces the fermion zero modes.

With four real fermion collective coordinates,
the monopole supermultiplet is four-dimensional:
it includes two bosonic states and two fermionic.
(The above counting refers just to monopole, without its antimonopole
partner. The antimonopole supermultiplet also includes
two bosonic  and two fermionic states.)
From the standpoint of ${\mathcal N}=2$ supersymmetry
in four dimensions this is a short multiplet.
Hence, the monopole states remain BPS saturated
to all orders in perturbation theory (in fact, the criticality of
the monopole supermultiplet is valid
beyond perturbation theory \cite{SW1,SW2}).

\subsubsection{Zero modes for fermions in the fundamental representation}

This topic, being related to an interesting phenomenon of
charge fractionalization,  is marginal for this review.
Therefore, we will limit ourselves to a brief comment.
The interested reader is referred to \cite{harvey,RU,ken}
for further details.
The fermion part of the Lagrangian can be obtained from (\ref{ldrnm})
with the obvious replacement of the adjoint Dirac fermion by the fundamental one,
which we will denote by $\chi$,
\beq
 {\mathcal L}  = \frac{1}{g^2}
\left\{ -\frac{1}{4}F^{a\,\mu\nu}F^a_{\mu\nu} +\frac{1}{2}\, ({D}^{\mu}\, \phi ) ({D}_{\mu}\, \phi )+  \bar\chi \, i\,
\!\!\not{\!\! D}\chi - \bar\chi\,\phi\chi
  \right\} \,.
  \eeq
The Dirac equation then
takes the form
\begin{equation}
( i \gamma^\mu D_\mu - \phi ) \chi = 0\, .
\label{3413}
\end{equation}
Needless to say that the gamma matrices can be now chosen
in any representation we find to be the most convenient.
The one which is indeed convenient here is
\begin{equation}
 \gamma^0 =  \left( \begin{array}{cc}
0 & -i \\
i & 0
\end{array}\right)\,,
\qquad \gamma^i =  \left(\begin{array}{cc}
-i \sigma^i & 0 \\
0 & i \sigma^i  \end{array}\right).
\label{3414}
\end{equation}
For the static 't Hooft--Polyakov monopole configuration (with $A_0=0$)
the zero mode equations reduce to two decoupled equations
\begin{eqnarray}
 \not\!\!{{\mathcal D}}   \chi^-   &\equiv&
        (i \sigma^i D_i +  \phi ) \chi^-   = 0\,.
          \nonumber\\[3mm]
        {\not\!\!{\mathcal D}}^\dagger \chi^+
        &\equiv& ( i \sigma^i D_i -  \phi ) \chi^+ = 0  \,, \quad i=1,2,3\,.
\label{3415}
\end{eqnarray}
provided we parametrize $\chi (\vec x)$  in
terms of the following
two-component spinors:
\begin{equation}
 \chi = \left (\begin{array}{c} \chi^+\\ \chi^- \end{array} \right ) .
\label{3416}
\end{equation}
Now we can use the Callias theorem which says
\beq
 {\rm dim \,\, ker} \not\!\!{\mathcal  D} - {\rm
dim \,\, ker } \,{\not\!\!{\mathcal D}}^\dagger =
n_m \,,
\label{3417}
\eeq
where $n_m$ is the topological number,
$n_m=1$ for the  monopole and $n_m=-1$ for the  antimonopole.
This implies, in turn, that Eq.~(\ref{3415})
has one complex zero mode, i.e. in the case at hand we characterize the
monopole by one complex fermion collective coordinate (and its conjugate, of course).
This fact leads to a drastic consequence:
the monopole acquires a half-integer electric charge.  It becomes a dyon with
charge 1/2 even in the absence of the $\theta$ term.
This phenomenon  -- the charge fractionalization
in the cases with a single  complex fermion collective coordinate --
is well-known in the literature \cite{RU,harvey,SVZ06,ken} and dates
back to Jackiw and Rebbi \cite{jackiw}.

\subsubsection{The monopole supermultiplet: dimension of the BPS representations}

As was first noted by Montonen and Olive \cite{MO}, all states in ${\mathcal N}=2$
model -- $W$ bosons and monopoles alike -- are BPS saturated.
This results in the fact that supermultiplets of this model are short.
Regular (long) supermultiplet would contain
$2^{2{\mathcal N}} = 16$ helicity states, while the short ones
contain $2^{\mathcal N} = 4$ helicity states -- two bosonic and two fermionic. This is in full accord with the fact that
the number of the fermion zero mode on the given monopole solution is four,
resulting in dim-4 representation of the supersymmetry algebra.
If we combine particles and antiparticles together, as is customary in
field theory, we will have one Dirac spinor on the fermion side of the
supermultiplet in  both cases, $W$-bosons/monopoles.
\newpage

\vspace{7mm}

\begin{center}

{\Huge PART II: Long Journey}

\end{center}
\addcontentsline{toc}{section}{\large PART II: Long Journey}

\vspace{6cm}

\centerline{\includegraphics[width=1.8in]{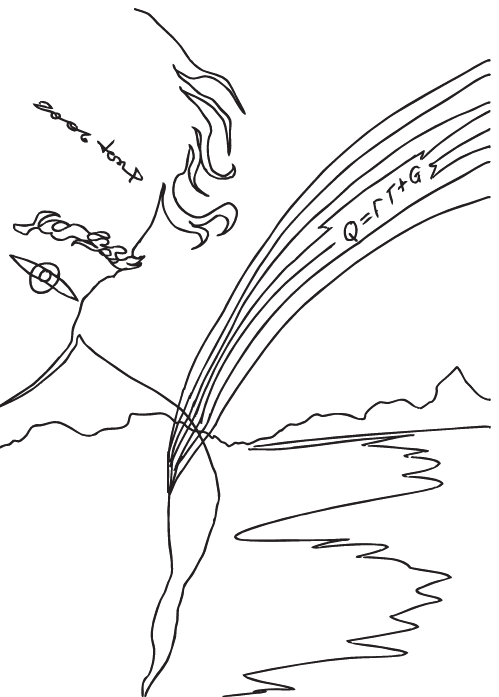}}

\newpage

\mbox{}

\vspace{4cm}

\noindent
 \framebox[1.1in]{
 {\sl Warning:}}

\vspace{4mm}

\noindent
In Part II we switch to Euclidean conventions.
This is appropriate and convenient from the technical point of view,
since we mostly study
static (time-independent) field configurations, and $A_0 =0$.
Then the Euclidean action reduces to the energy functional.
See Appendix A, Sect.~\ref{seca4}.

\newpage

Ever since  't Hooft \cite{thooft}
and Mandelstam \cite{mandelstam}
put forward the hypothesis of the dual Meissner effect
to explain color confinement in non-Abelian gauge
theories  people
were trying to find a controllable
approximation in which one could reliably demonstrate
the occurrence of the dual Meissner effect in these theories.
A breakthrough achievement was the Seiberg--Witten
solution \cite{SW1} of ${\mathcal N}=2$ supersymmetric
Yang--Mills theory. They found massless monopoles and,
adding a small $({\mathcal N}=2)$-breaking deformation,
proved that they condense creating strings carrying
a chromoelectric flux. It was a great success in qualitative
understanding of color confinement.

A more careful examination shows, however, that details of
the Seiberg--Witten confinement are quite different from
those we expect in QCD-like theories. Indeed, a crucial aspect
of Ref.~\cite{SW1} is  that the SU($N$)
gauge symmetry is first broken, at a high scale, down to U(1)$^{N-1}$,
which is then completely broken, at a much lower scale
where condensation of magnetic monopoles occurs. Correspondingly,
the strings in the Seiberg--Witten
solution are, in fact, Abelian strings \cite{ANO}
of the Abrikosov--Nielsen--Olesen
(ANO) type which results, in turn, in confinement
whose structure does not resemble at all that of QCD. In particular, the
``hadronic'' spectrum is much richer than that in QCD
\cite{DS,matt,Strassler,VY,Yrev}. To see this
it is sufficient to observe that, given  the low-energy
gauge group U(1)$^{N-1}$, one has $N-1$ Abelian strings associated with
each of the $N-1$ Abelian factors. Since
\beq
\pi_1 \big(U(1)^{N-1}\big)=Z^{N-1}\,,
\label{towers}
\eeq
the Abelian strings and, therefore, the meson spectrum come in $N-1$ infinite
towers. This feature is not expected in real-world QCD. Moreover,
there is no experimental indications of dynamical Abelization in QCD.

Here in Part II we begin our long journey which covers the advances of the
last decade. Most developments are even fresher, they refer to the last five years or so.
First we dwell on the recent discovery of non-Abelian strings
\cite{HT1,ABEKY,SYmon,HT2}
which appear in certain regimes in \ntwo supersymmetric gauge theories.
Moreover, they were found even in \none theories, the so-called $M$ model,
see Sect.~\ref{secmmodel}.
The most important feature of these strings
is that they acquire orientational zero modes
associated with rotations of their color flux inside a non-Abelian SU$(N)$
subgroup  of the gauge group. The occurrence of these zero modes
makes these strings genuinely non-Abelian.

The flux tubes in non-Abelian theories at weak coupling
were studied in numerous papers previously \cite{VS,HV,Su,SS,KB,KoS,MY}.
These strings are referred to as $Z_N$
strings because they are related to the center of the gauge group SU$(N)$.
Consider, say, the  SU$(N)$ gauge theory with a few (more than one)
scalar fields in the adjoint
representation. Suppose the adjoint scalars condense in such a way that
the SU$(N)$ gauge group is broken
down to its center $Z_N$. Then string solutions are classified according to
\beq
\pi_1 \left(\frac{{\rm SU}(N)}{Z_N}\right)=Z_{N}.
\label{pionezn}
\eeq

In all these previous constructions \cite{VS,HV,Su,SS,KB,KoS,MY} of
the $Z_N$ strings  the flux was always directed in a
fixed group direction (corresponding to a Cartan subalgebra), and no moduli
that would allow to freely rotate its orientation in the group space were ever
obtained. Therefore it is reasonable to call these $Z_N$ strings Abelian, in
contrast the  non-Abelian strings, to be discussed
below,  which have orientational moduli.

Consideration of non-Abelian strings naturally leads us to confined
non-Abelian monopoles. We follow the fate of the classical
't Hooft--Polyakov monopole (classical not in the sense of ``non-quantum"
but rather in the sense of something belonging to textbooks) in the
Higgs ``medium" --- from free monopoles, through a weakly confined
regime, to a highly quantum regime in which confined monopoles manifest themselves
as kinks in the low-energy theory on the string world sheet.
The confined monopoles are sources (sinks) to which the magnetic flux tubes
are attached. We demonstrate that they are dual to quarks
just in the same vein as the magnetic flux tubes are dual to the electric ones.

Our treatise covers the non-Abelian flux tubes both in theories with the minimal
(\none) and extended (\ntwo) supersymmetry.

The third element of the big picture which we explore
is the wall-string junction. From string/D brane theory it is well-known
that the fundamental string can end on the brane. In fact, this is a defining property
of the brane. Since our task is to reveal in gauge
theories all phenomena described by string/D brane theory
we must be able to see the string-wall junctions.
And we do see them! The string-wall junctions which later got the name
{\em boojums} were first observed in an \ntwo gauge theory
in Ref.~\cite{SYwall}. This construction as well as
later advances in the boojum theory are thoroughly discussed in Part II.

The domain walls as brane prototypes must possess another remarkable feature
--- they must localize gauge fields. This localization was first proven to occur
in an \ntwo gauge theory in Ref.~\cite{SYnawall}. The domain-wall
world sheet theory is the theory of three-dimen\-sional
gauge fields after all!

All the above elements combined together lead us
to a thorough understanding of the Meissner effect
in non-Abelian theories. To understand QCD we need
to develop a model of the {\em dual} Meissner effect.
Although this problem is not yet fully solved, we report here
on a significant progress in this direction.

\section{Non-Abelian strings}
\label{strings}
\setcounter{equation}{0}

In this section we discuss a particular class of \ntwo supersym\-metric
gauge theories in which non-Abelian strings were found. One can pose the
question: what is so special about these models that makes an Abelian
$Z_N$ string become non-Abelian? Models we will dwell on below
have
both gauge and flavor symmetries broken by the condensation of scalar
fields. The common feature of these models is that some global diagonal
combination of color and flavor groups survive the breaking. We consider the
case when this diagonal group is SU$(N)_{C+F}$, where the subscript $C+F$
means a combination of global color and flavor groups. The presence of this
unbroken subgroup is responsible for the occurrence of the orientational zero
modes of the string which entail its non-Abelian nature.

Clearly, the presence of supersymmetry is not important for the
construction of non-Abelian strings. In particular, while here we
focus on the  BPS non-Abelian strings in \ntwo  supersymmetric
gauge theories,  in Sect.~\ref{nosusy}  we review non-Abelian strings
in \none supersymmetric theories and in Sect.~\ref{nonBPS} in  non-supersymmetric theories.

\subsection{ Basic model: \ntwo SQCD}
\setcounter{equation}{0}
\label{model}

The model we will deal with derives from
 \ntwo SQCD with the gauge group SU$(N+1)$ and $N_f=N$  flavors of
the fundamental matter  hypermultiplets
which we will call quarks \cite{SW2}. At a generic point on the Coulomb
branch of this theory, the gauge group is broken down to
U(1)$^{N}$.  We will be interested, however,
in a particular subspace of the Coulomb branch, on which
the  gauge group is broken down to SU$(N)\times$U(1). We will
enforce this regime by a special choice of the
quark mass terms.

The breaking SU$(N+1)\to$SU$(N)\times$U(1) occurs at the scale $m$
which is supposed to lie very high, $m\gg\Lambda_{{\rm SU}(N+1)}$,
where $\Lambda_{{\rm SU}(N+1)}$ is the scale of the SU$(N+1)$ theory.
Correspondingly, the masses of the gauge bosons from
SU$(N+1)/$SU$(N)\times$U(1) sector and their superpartners,
are very large --- proportional to $m$ --- and so are the masses
of the $(N+1)$-th color component of the quark fields in the fundamental
representation. We will be interested in the phenomena at
the scales $\ll m$. Therefore, our starting point is in fact
the SU$(N)\times$U(1) model with $N_f=N$ matter fields in
the fundamental representation of SU$(N)$, as it emerges after the
SU$(N+1)\to$SU$(N)\times$U(1) breaking. These matter fields
are also coupled to the U(1) gauge field.

The field content of SU$(N)\times$U(1) \ntwo SQCD with
$N$ flavors is as follows. The \ntwo vector multiplet
consists of the  U(1)
gauge field $A_{\mu}$ and the SU$(N)$  gauge field $A^a_{\mu}$,
(here $a=1,..., N^2-1$), and their Weyl fermion superpartners
($\lambda^{1}_{\alpha}$, $\lambda^{2}_{\alpha}$) and
($\lambda^{1a}_{\alpha}$, $\lambda^{2a}_{\alpha}$), plus
complex scalar fields $a$, and $a^a$. The latter are in the adjoint
representation of SU$(N)$. The spinorial index of $\lambda$'s runs over
$\alpha=1,2$.  In this sector the  global SU(2)$_R$ symmetry inherent to
the \ntwo   model at hand manifests itself through rotations
$\lambda^1 \leftrightarrow \lambda^2$.

The quark multiplets of  the SU$(N)\times$U(1) theory consist
of   the complex scalar fields
$q^{kA}$ and $\tilde{q}_{Ak}$ (squarks) and
the  Weyl fermions $\psi^{kA}$ and
$\tilde{\psi}_{Ak}$, all in the fundamental representation of
the SU$(N)$ gauge group.
Here $k=1,..., N$ is the color index
while $A$ is the flavor index, $A=1,..., N$.
Note that the scalars $q^{kA}$ and ${\bar{\tilde q}}^{\, kA}$
form a doublet under the action of the   global
SU(2)$_R$ group.

The original SU$(N+1)$ theory is perturbed by adding a small
mass term for the adjoint matter,  via the superpotential
${\mathcal W}=\mu\, {\rm Tr}\,\Phi^2$.
Generally speaking, this superpotential breaks
\ntwo down to ${\mathcal N}=1$.
The Coulomb branch shrinks to
a number of  isolated \none vacua \cite{SW1,SW2,DS,APS,CKM}.
In the limit of $\mu\to 0$ these vacua correspond to special
singular points on the Coulomb branch
in which  $N$ monopoles/dyons or
quarks become massless.
The first $(N+1)$ of these points (often referred to as the
Seiberg--Witten vacua) are always at
strong coupling. They correspond to \none vacua of the  pure
SU$(N+1)$ gauge theory.

The massless quark points --- they present vacua of a distinct type,
to be referred to as the quark vacua --- may or may not be at weak
coupling depending on the values of the quark mass parameters  $m_A$.  If
$m_A\gg \Lambda_{{\rm SU}(N+1)}$, the quark vacua do lie at weak coupling.
Below we will be  interested only  in the  quark vacua
assuming that the condition $m_A\gg \Lambda_{{\rm SU}(3)}$ is met.

In the low-energy SU$(N)\times$U(1) theory, which is our starting point,
the perturbation ${\mathcal W}=\mu\, {\rm Tr}\, \Phi^2$ can be truncated, leading
to a crucial simplification. Indeed, since the ${\mathcal A}$
chiral superfield, the ${\mathcal N}=2$ superpartner of the U(1) gauge field,
\footnote{ The superscript 2 in Eq.~(\ref{calasf}) is the global SU(2)$_R$
index of $\lambda$ rather than $\lambda$ squared. }
\beq
{\mathcal A} \equiv  a +\sqrt{2}\lambda^2\theta +F_a\,\theta^2\,,
\label{calasf}
\eeq
it not charged under the gauge group SU$(N)\times$U(1),
one can introduce the superpotential linear in ${\mathcal A}$,
\beq
{\mathcal W}_{{\mathcal A}} =-\frac{N}{2\sqrt{2}}\, \xi\, {\mathcal A}\,.
\label{spla}
\eeq
Here we expand ${\rm Tr}\, \Phi^2$ around its vacuum expectation value (VEV),
and truncate the series keeping only the linear term in ${\mathcal A}$. The truncated
superpotential is a Fayet--Iliopoulos (FI) $F$-term.

Let us explain this in more detail. In  \none
supersymmetric theory with the gauge group SU$(N)\times$U(1)
one can add the following FI term to the action
\cite{FI} (we will call it the FI $D$-term here):
\beq
\xi_3\ D\,
\eeq
where $D$ is the $D$-component of the U(1) gauge field. In \ntwo SUSY
theory the field $D$ belongs to the SU(2)$_R$ triplet, together with the
$F$ components of the chiral field ${\mathcal A}$, $F$ and $\bar F$. Namely,
let us introduce a triplet $F_p$ $(p=1,2,3)$ using
the relations\,\footnote{Attention: The index $p$ is an SU(2)$_R$ index rather than the color index!}
\begin{eqnarray}
D &=& F_3\, , \nonumber\\[1mm]
F &=& \frac1{\sqrt2}\ (F_1+iF_2) \, ,\nonumber\\[1mm]
\bar F &=& \frac1{\sqrt2}\ (F_1-iF_2)\, .
\label{414}
\end{eqnarray}
Now, the generalized FI term can be written as
\beq
S_{\rm FI}\ =- \frac{N}{2}\int d^4x\ \xi_p F_p\, .
\label{fiterm}
\eeq
Comparing this with Eq.~(\ref{spla}) we identify
\begin{eqnarray}
\xi &=& \ (\xi_1-i\xi_2)\, , \nonumber\\[1mm]
\bar\xi &=& \ (\xi_1+i\xi_2)\, .
\label{tripletxi}
\end{eqnarray}
This is the reason why we refer to the  superpotential  (\ref{spla})
as to the FI $F$-term.

A remarkable feature of the FI term  is that it does {\em not}
break \ntwo super\-symmetry
\cite{matt,VY}. Keeping higher order terms of the expansion
of  $\mu\, {\rm Tr}\, \Phi^2$ in powers of ${\mathcal A}$
would inevitably explicitly break \ntwo.
For our purposes it is crucial that the model we will deal with
is {\em exactly} \ntwo super\-symmetric. This ensures that the flux tubes
solutions of the model are BPS-saturated. If higher order terms in
${\mathcal A}$ are taken into account,
\ntwo supersymmetry is broken down to
\none and strings are no longer BPS, generally speaking.
The superconductivity in the model becomes of type I \cite{VY}.

\subsubsection{SU(\boldmath{$N)\times$}U(1) \boldmath{\ntwo} QCD}
\label{action}

The bosonic part of our SU$(N)\times$U(1)
theory has the form \cite{ABEKY}
\beqn
S&=&\int d^4x \left[\frac1{4g^2_2}
\left(F^{a}_{\mu\nu}\right)^2 +
\frac1{4g^2_1}\left(F_{\mu\nu}\right)^2
+
\frac1{g^2_2}\left|D_{\mu}a^a\right|^2 +\frac1{g^2_1}
\left|\partial_{\mu}a\right|^2 \right.
\nonumber\\[4mm]
&+&\left. \left|\nabla_{\mu}
q^{A}\right|^2 + \left|\nabla_{\mu} \bar{\tilde{q}}^{A}\right|^2
+V(q^A,\tilde{q}_A,a^a,a)\right]\,.
\label{qed}
\eeqn
Here $D_{\mu}$ is the covariant derivative in the adjoint representation
of  SU$(N)$, and
\beq
\nabla_\mu=\partial_\mu -\frac{i}{2}\; A_{\mu}
-i A^{a}_{\mu}\, T^a\,.
\label{defnabla}
\eeq
We suppress the color  SU($N$)  indices, and $T^a$ are the
 SU($N$) generators normalized as
$${\rm Tr}\, (T^a T^b)=1/2\;\delta^{ab}\,.$$
The coupling constants $g_1$ and $g_2$
correspond to the U(1)  and  SU$(N)$  sectors, respectively.
With our conventions, the U(1) charges of the fundamental matter fields
are $\pm1/2$.

\vspace{1mm}

The potential $V(q^A,\tilde{q}_A,a^a,a)$ in the action (\ref{qed})
is a sum of  $D$ and  $F$  terms,
\beqn
V(q^A,\tilde{q}_A,a^a,a) &=&
 \frac{g^2_2}{2}
\left( \frac{1}{g^2_2}\,  f^{abc} \bar a^b a^c
 +
 \bar{q}_A\,T^a q^A -
\tilde{q}_A T^a\,\bar{\tilde{q}}^A\right)^2
\nonumber\\[3mm]
&+& \frac{g^2_1}{8}
\left(\bar{q}_A q^A - \tilde{q}_A \bar{\tilde{q}}^A -N \xi_3\right)^2
\nonumber\\[3mm]
&+& 2g^2\left| \tilde{q}_A T^a q^A \right|^2+
\frac{g^2_1}{2}\left| \tilde{q}_A q^A -\frac{N}{2}\,\xi \right|^2
\nonumber\\[3mm]
&+&\frac12\sum_{A=1}^N \left\{ \left|(a+\sqrt{2}m_A +2T^a a^a)q^A
\right|^2\right.
\nonumber\\[3mm]
&+&\left.
\left|(a+\sqrt{2}m_A +2T^a a^a)\bar{\tilde{q}}^A
\right|^2 \right\}\,.
\label{pot}
\eeqn
Here  $f^{abc}$ stand for the structure constants of the SU$(N)$ group, and
the sum over the repeated flavor indices $A$ is implied.

The first and second lines represent   $D$   terms, the third line
the $F_{\mathcal A}$ terms,
while the fourth and the fifth  lines represent the squark $F$ terms.
Using the SU(2)$_R$ rotations we can always direct the FI parameter vector
$\xi_p$ in a given direction. Below in most cases we  will align
the  FI $F$-term to make the parameter $\xi$ real. In other words,
\beq
\xi_3=0, \qquad \xi_2=0, \qquad \xi=\xi_1\,.
\label{fifterm}
\eeq

\subsubsection{The vacuum structure and excitation spectrum}
\label{vacuumstructure}

Now  we briefly review the vacuum structure and
the excitation mass spectrum in  our basic
SU$(N)\times$U(1)  model. The underlying \ntwo SQCD with the gauge group
SU$(N+1)$ has a variety of vacua \cite{APS,CKM,MY}.
In addition to $N$ strong coupling vacua which exist
in pure gauge theory,  there is a number of the so-called $r$ quark vacua, where
$r$ is the number of the quark flavors which develop VEV's in the given vacuum.
Here we will focus\,\footnote{There are singular points on the Coulomb branch
of the underlying SU$(N+1)$ theory where more then $N$ quark flavors become
massless. These singularities are the roots of Higgs branches  \cite{APS,CKM,MY}.}
on a particular isolated vacuum with the maximal possible value
of $r$, $$r=N\,.$$
The  theory (\ref{qed}) is nothing but the low-energy truncation
of the full \mbox{SU$(N+1)$} SQCD which describes physics around this vacuum.

The  vacua of the theory (\ref{qed}) are determined by the zeros of
the potential (\ref{pot}). The adjoint fields develop the following VEV's:
\beq
\langle \Phi\rangle = - \frac1{\sqrt{2}}
 \left(
\begin{array}{ccc}
m_1 & \ldots & 0 \\
\ldots & \ldots & \ldots\\
0 & \ldots & m_N\\
\end{array}
\right),
\label{avev}
\eeq
where we defined the scalar adjoint matrix as
\beq
\Phi = \frac12\, a + T^a\, a^a.
\label{Phidef}
\eeq
For generic values of the quark masses, the  SU$(N)$ subgroup of the gauge group is
broken down to U(1)$^{N-1}$. However, for a special choice
\beq
m_1=m_2=...=m_N,
\label{equalmasses}
\eeq
which we will be mostly interested in in this section, the  SU$(N)\times$U(1)
gauge group remains classically unbroken. In fact, the common value $m$
of the quark masses determines the scale of breaking of the SU$(N+1)$ gauge symmetry of
the underlying theory down to SU$(N)\times$U(1) gauge symmetry
of our benchmark low-energy theory (\ref{qed}).

If the  value of the FI parameter is taken real
we can exploit gauge rotations   to
make the quark VEV's real too. Then
in the case at hand they take the color-flavor locked form
\beqn
\langle q^{kA}\rangle &=&\langle \bar{\tilde{q}}^{kA}\rangle =\sqrt{
\frac{\xi}{2}}\,
\left(
\begin{array}{ccc}
1 & \ldots & 0 \\
\ldots & \ldots & \ldots\\
0 & \ldots & 1\\
\end{array}
\right),
\nonumber\\[4mm]
k&=&1,..., N\qquad A=1,...,N\, ,
\label{qvev}
\eeqn
where we write down the quark fields as an $N\times N$ matrix in
the color and flavor indices.
This particular form of the squark condensates is dictated by the
third line in Eq.~(\ref{pot}). Note that the squark fields stabilize
at non-vanishing values entirely due to the U(1) factor
represented by the second term in the third line.

The  vacuum field (\ref{qvev}) results in  the spontaneous
breaking of both gauge and flavor SU($N$)'s.
A diagonal global SU($N$) survives, however,
\beq
{\rm U}(N)_{\rm gauge}\times {\rm SU}(N)_{\rm flavor}
\to {\rm SU}(N)_{C+F}\,.
\label{c+f}
\eeq
Thus, a color-flavor locking takes place in the vacuum.
A version of this pattern of the symmetry breaking was suggested
long ago \cite{BarH}.

Let us move on to  the issue of the excitation spectrum  in this vacuum
\cite{VY,ABEKY}.
The mass matrix for the gauge fields $(A^{a}_{\mu},
A_{\mu})$ can be read off from the quark kinetic terms  in Eq.~(\ref{qed}).
It shows that all SU$(N)$ gauge bosons become massive, with
one and the same mass
\beq
M_{{\rm SU}(N)} =g_2\,\sqrt\xi\,.
\label{msuN}
\eeq
The equality of the masses is no accident. It is a consequence of the
unbroken SU$(N)_{C+F}$ symmetry (\ref{c+f}).

The mass of the U(1) gauge boson is
\beq
M_{{\rm U}(1)} =g_1\,\sqrt{\frac{N}{2}\,\xi}\,.
\label{mu1}
\eeq
Thus, the theory is fully Higgsed.
The mass spectrum of the adjoint scalar excitations is the same as the one for the
gauge bosons. This is enforced by \ntwo.

What is the mass spectrum of the quark  excitations?
It can be read off from the potential (\ref{pot}).
We have  $4N^2$ real degrees of freedom of quark scalars $q$ and $\tilde q$.
Out of those
$N^2$ are eaten up by the Higgs mechanism.
The remaining $3N^2$ states split in three plus $3(N^2-1)$ states with masses
(\ref{mu1}) and (\ref{msuN}), respectively. Combining these states with
the massive gauge bosons and the adjoint scalar states we get  \cite{VY,ABEKY}
one long \ntwo BPS multiplet (eight real bosonic
plus eight fermionic degrees of freedom) with mass (\ref{mu1})
and $N^2-1$ long \ntwo BPS multiplets with mass (\ref{msuN}).
Note that these supermultiplets come in representations of the unbroken
SU$(N)_{C+F}$ group, namely, the singlet and the adjoint representations.

To conclude this section we want to discuss quantum effects in the theory
(\ref{qed}). At a high scale $m$ the SU$(N+1)$ gauge group is broken down to
SU$(N)\times$U(1) by condensation of the adjoint fields if
the condition (\ref{equalmasses}) is met. The SU$(N)$ sector is
asymptotically free. The running of the corresponding gauge coupling,
if non-interrupted, would drag the theory into the strong coupling regime.
This would invalidate our quasiclassical analysis. Moreover, strong
coupling effects on the Coulomb branch would break SU$(N)$ gauge subgroup
(as well as the  SU$(N)_{C+F}$ group) down to
U(1)$^{N-1}$ by the Seiberg--Witten mechanism \cite{SW1}.
No non-Abelian strings would emerge.

One way out is proposed in \cite{APS,CKM}. One can add more flavors to the
theory making $N_f>2N$. Then the SU$(N)$ sector is not asymptotically free and
does not evolve into the strong coupling
regime. However, the ANO strings in the multiflavor  theory
(on the Higgs branches) become semilocal strings
\cite{AchVas}  and confinement is lost (see Sect.~\ref{semilocal}).
Here we take a different route  assuming the FI parameter $\xi$
to be large \footnote{We discuss this important issue in more details in the
end of Sect.~\ref{confinement}},
\beq
\xi\gg \Lambda_{{\rm SU}(N)}\,.
\label{weakcoupling}
\eeq
This condition ensures weak coupling in the SU$(N)$ sector because
the SU$(N)$ gauge coupling does not run below the scale of the quark VEV's
which is determined by $\xi$. More explicitly,
\beq
\frac{8\pi^2}{g^2_2 (\xi)} =
N\log{\frac{\sqrt{\xi}}{\Lambda_{{\rm SU}(N)}}}\gg 1 \,.
\label{coupling}
\eeq
 Alternatively one can say that
\beq
\Lambda_{{\rm SU}(N)}^{N} = \xi^{N/2}\, \exp\left(-\frac{8\pi^2}{g_2^2 (\xi)}
\right)\ll \xi^{N/2}\,.
\label{4Dlambda}
\eeq

\subsection{ $Z_N$ Abelian strings}
\setcounter{equation}{0}
\label{znstring}

Strictly speaking, \ntwo SQCD with the  gauge group SU($N+1$)  does not have
stable flux tubes. They are unstable due to monopole-antimonopole
pair creation in the SU($N+1$)/SU($N$)$\times$U(1) sector. However, at large $m$
these monopoles become heavy. In fact, there are no such monopoles in
the low-energy theory (\ref{qed}) (where they can be considered as infinitely heavy).
Therefore, the theory (\ref{qed}) has stable string solutions. When
the perturbation $\mu \,{\rm Tr}\, \Phi^2$ is truncated to the FI term (\ref{spla}),
the theory enjoys \ntwo supersymmetry and has BPS string solutions
 \cite{matt,VY,EFMG,MY,ABEKY}. Note that here we  discuss magnetic
flux tubes. They are formed in the Higgs phase of the theory upon condensation
of the squark fields and lead to confinement of monopoles.

Now, let us briefly review the BPS string solutions \cite{MY,HT1,ABEKY} in the model (\ref{qed}). Here we consider the case of equal quark masses
(\ref{equalmasses}) when the global SU($N$)$_{C+F}$ group is unbroken. First we
review the Abelian solutions for $Z_N$ strings and then, in Sect.~\ref{nas} show
that in this limit they acquire orientational moduli.

In fact,
the $Z_N$ Abelian strings considered below are just partial solutions of
the vortex equations (see Eq.~(\ref{gfoes}) below). In the  equal mass limit
(\ref{equalmasses}) the  global SU($N)_{C+F}$ group is restored and the general
solution for the elementary non-Abelian string gets a continuous moduli space
isomorphic to CP$(N-1)$. The $Z_N$ strings are just $N$ discrete points on
this moduli space.

In the generic case of unequal quark masses, the SU($N)_{C+F}$ group is
explicitly broken,
and the continuous moduli space of the string solutions is lifted. Only the $Z_N$
Abelian strings survive this breaking. We will dwell on the case of generic
quark masses in Sect.~\ref{unequalmasses}.

It turns out that the string solutions do not involve the
adjoint fields $a$ and $a^a$. The BPS  strings are ``built'' from gauge and quark
fields only. Therefore, in order to find the classical solution,
in the action (\ref{qed}) we can set the adjoint
fields to their VEV's (\ref{avev}). This is consistent with equations of
motion. Of course, at the quantum level
the adjoint fields start fluctuating, deviating from their VEV's.

We use the {\em ansatz}
\beq
q^{kA}=\bar{\tilde{q}}^{kA}=\frac1{\sqrt{2}}\vp^{kA}
\label{qtilde}
\eeq
reducing the number of the squark degrees of freedom to one complex field
for each color and flavor. With these simplifications the action of the
model (\ref{qed}) becomes
\beqn
S &=& \int {\rm d}^4x\left\{\frac1{4g_2^2}
\left(F^{a}_{\mu\nu}\right)^{2}
+ \frac1{4g_1^2}\left(F_{\mu\nu}\right)^{2}
 \right.
 \nonumber\\[3mm]
&+&
  |\nabla_\mu \vp^A|^2
+\frac{g^2_2}{2}
\left(\bar{\vp}_A T^a \vp^A\right)^2
 +
 \frac{g^2_1}{8}\left(
 | \vp^A|^2 - N\xi \right)^2
,
\label{redmodel}
\eeqn
while the VEV's of the quark fields (\ref{qvev}) are
\beq
\langle\vp \rangle = \sqrt\xi\,{\rm diag}\, \{1,1,...,1\}\,.
\label{diagphi}
\eeq

Since the spontaneously broken gauge U(1) is a part of
the model under consideration, it supports
the conventional ANO strings \cite{ANO},
in which one can discard the SU$(N)_{\rm gauge}$ part
of the action altogether.
The topological stability of the ANO string is due to the fact that
$\pi_1({\rm U(1)}) = Z$.

These are not the strings we are interested in.
At first sight, the triviality of the homotopy group,
$\pi_1 ({\rm SU}(N)) =0$, implies that there are no other topologically
stable strings.
This impression is false. One can
combine the $Z_N$ center of SU($N$) with the elements
$\exp (2\pi i k/N)\in$U(1) to get a topologically stable string solution
possessing both windings, in SU($N$) and U(1). In other words,
\beq
\pi_1 \left({\rm SU}(N)\times {\rm U}(1)/ Z_N
\right)\neq 0\,.
\eeq
It is easy to see that this nontrivial topology amounts to
selecting just one element of $\vp$, say, $\vp^{11}$, or
$\vp^{22}$, etc,
and make it
wind, for instance,\footnote{As explained below,
$\alpha$ is the angle of
the coordinate  $\vec{x}_\perp$ in the perpendicular plane.}
\beq
\vp_{\rm string} = \sqrt{\xi}\,{\rm diag} ( 1,1, ... ,e^{i\alpha  })\,,
\quad x\to\infty \,.
\label{ansa}
\eeq
Such strings can be called elementary;
their tension is $1/N$-th of that of the ANO string.
The ANO string can be viewed as a bound state of
$N$ elementary strings.

More concretely,  the $Z_N$ string solution
(a progenitor of the non-Abelian string) can be written as
follows \cite{ABEKY}:
\beqn
\vp &=&
\left(
\begin{array}{cccc}
\phi_2(r) & 0& ... & 0\\[2mm]
...&...&...&...\\[2mm]
0& ... & \phi_2(r)&  0\\[2mm]
0 & 0& ... & e^{i\alpha}\phi_{1}(r)
\end{array}
\right) ,
\nonumber\\[5mm]
A^{{\rm SU}(N)}_i &=&
\frac1N\left(
\begin{array}{cccc}
1 & ... & 0 & 0\\[2mm]
...&...&...&...\\[2mm]
0&  ... & 1 & 0\\[2mm]
0 & 0& ... & -(N-1)
\end{array}
\right)\, \left( \pt_i \alpha \right) \left[ -1+f_{NA}(r)\right] ,
\nonumber\\[5mm]
A^{{\rm U}(1)}_i &=& \frac{I}{2}\,A_i=\frac{I}{N}\,
\left( \pt_i \alpha \right)\left[1-f(r)\right] ,\;\;\; A^{{\rm U}(1)}_0=
A^{{\rm SU}(N)}_0 =0\,,
\label{znstr}
\eeqn
where $i=1,2$ labels the coordinates in the plane orthogonal to the string
axis, $r$ and $\alpha$ are the polar coordinates in this plane,
and $I$ is the unit $N\times N$ matrix. The profile
functions $\phi_1(r)$ and  $\phi_2(r)$ determine the profiles of the scalar
fields,
while $f_{NA}(r)$ and $f(r)$ determine the SU($N$) and U(1) fields of the
string solutions, respectively. These functions satisfy the following
rather obvious boundary conditions:
\beqn
&& \phi_{1}(0)=0,
\nonumber\\[2mm]
&& f_{NA}(0)=1,\;\;\;f(0)=1\,,
\label{bc0}
\eeqn
at $r=0$, and
\beqn
&& \phi_{1}(\infty)=\sqrt{\xi},\;\;\;\phi_2(\infty)=\sqrt{\xi}\,,
\nonumber\\[2mm]
&& f_{NA}(\infty)=0,\;\;\;\; \; f(\infty) = 0
\label{bcinfty}
\eeqn
at $r=\infty$.

Now, let us derive the first-order equations which determine the profile
functions, making use of the Bogomol'nyi representation \cite{B} of
the model
(\ref{redmodel}). We have
\beqn
T
&=&
\int{d}^2 x   \left\{
\left[\frac1{\sqrt{2}g_2}F^{*a}_{3} +
\frac{g_2}{\sqrt{2}}
\left(\bar{\vp}_A T^a \vp^A\right)\right]^2
\right.
\nonumber\\[3mm]
&+&
\left[\frac1{\sqrt{2}g_1}F^{*}_{3} +
\frac{g_1 }{2\sqrt{2}}
\left(|\vp^A|^2-N\xi \right)\right]^2
\nonumber\\[5mm]
&+&
\left.
 \left|\nabla_1 \,\vp^A +
i\nabla_2\, \vp^A\right|^2
 +\frac{N}{2}\xi\,  F^{*}_3\right\},
\label{bogs}
\eeqn
where $$F_{3}^{*}=\, F_{12}\,\,\,
\mbox{and}\,\,\, F_{3}^{*a}=\, F_{12}^{a}\,,$$
and we assume that the fields in question depend only on
the transverse  coordinates $x_i$, $i=1,2$.

The Bogomol'nyi representation (\ref{bogs})
leads us to the following first-order equations:
\beqn
&& F^{*}_{3}+\frac{g_1^2}{2} \left(\left|
\varphi^{A}\right|^2-N\xi\right)\,=0\, ,
\nonumber\\[3mm]
&& F^{*a}_{3}+ g_2^2 \left(\bar{\vp}_{A}T^a \varphi^{A}\right)
\,=0\, ,
\nonumber\\[3mm]
&& (\nabla_1+i\nabla_2)\varphi^A=0\, .
\label{gfoes}
\eeqn
Once these equations are satisfied, the energy of the BPS object is
given by the last surface term in (\ref{bogs}). Note that the representation
(\ref{bogs}) can be written also with the opposite  sign in front of
the flux terms. Then we would get the Bogomol'nyi equations for the anti-string.

For minimal winding we substitute the {\em ansatz} (\ref{znstr}) in Eqs.
(\ref{gfoes}) to get the first-order equations for the profile functions of the
$Z_N$ string \cite{MY,ABEKY},
\beqn
&&
r\frac{d}{{d}r}\,\phi_1 (r)- \frac1N\left( f(r)
+  (N-1)f_{NA}(r) \right)\phi_1 (r) = 0\, ,
\nonumber\\[4mm]
&&
r\frac{d}{{ d}r}\,\phi_2 (r)- \frac1N\left(f(r)
-  f_{NA}(r)\right)\phi_2 (r) = 0\, ,
\nonumber\\[4mm]
&&
-\frac1r\,\frac{ d}{{ d}r} f(r)+\frac{g^2_1 N}{4}\,
\left[(N-1)\phi_2(r)^2 +\phi_1(r)^2-N\xi\right] = 0\, ,
\nonumber\\[4mm]
&&
-\frac1r\,\frac{d}{{ d}r} f_{NA}(r)+\frac{g^2_2}{2}\,
\left[\phi_1(r)^2 -\phi_2(r)^2\right]  = 0\, .
\label{foes}
\eeqn
These equations present a  $Z_N$-string
generalization of the Bogomol'nyi equations for the ANO string
\cite{B} (see also (\ref{foe}) and (\ref{astringfoe})). They
were solved numerically for the  U(2) case (i.e. $N=2$) in \cite{ABEKY}.
Clearly, the solutions to the first-order equations automatically satisfy the second-order equations of motion.

The tension of this elementary $Z_N$  string is
\beq
T_1=2\pi\,\xi\, .
\label{ten}
\eeq
Since our string is a BPS object, this result is exact and has neither
 perturbative no nonperturbative corrections.
 Note that the tension of
the ANO
string is $N$ times larger; in our normalization
\beq
T_{\rm ANO}=2\pi\,N\,\xi\,.
\label{tenANO}
\eeq

Clearly, the {\em ansatz} (\ref{znstr})
admits permutations,
leading to other $Z_N$ string solutions of type (\ref{znstr}). They can be
obtained by changing the position of the ``winding" field
in Eq.~(\ref{znstr}). Altogether we have $N$ elementary $Z_N$ strings.

Of course, the  first-order equations (\ref{foes}) can be also obtained using
supersymmetry.
We start from  the supersymmetry transformations for
the fermion fields in  the  theory (\ref{qed}),
\begin{eqnarray}
\delta\lambda^{f\alpha}
&=&
\frac12(\sigma_\mu\bar{\sigma}_\nu\epsilon^f)^\alpha
F_{\mu\nu}+\epsilon^{\alpha p}F^m(\tau^m)^f_p\ +\dots,
\nonumber\\[3mm]
\delta\lambda^{af\alpha}
&=&
\frac12(\sigma_\mu\bar{\sigma}_\nu\epsilon^f)^\alpha
F^a_{\mu\nu}+\epsilon^{\alpha p}F^{am}(\tau^m)^f_p\ +\dots,
\nonumber\\[3mm]
\delta\bar{\tilde\psi}_{\dot{\alpha}}^{kA}
&=&
i\sqrt2\
\bar\nabla\hspace{-0.65em}/_{\dot{\alpha}\alpha}q_f^{kA}\epsilon^{\alpha
f}\ +\cdots,
\nonumber\\[4mm]
\delta\bar\psi_{\dot{\alpha}Ak}
&=&
 i\sqrt2\
\bar\nabla\hspace{-0.65em}/_{\dot{\alpha}\alpha}\bar
q_{fAk}\epsilon^{\alpha f}\ +\cdots.
\label{transf}
\end{eqnarray}
Here $f=1,2$ is the SU(2)$_R$ index
and $\lambda^{f\alpha}$  and $\lambda^{af\alpha}$
are the fermions from the ${\mathcal N}=2$ vector supermultiplets of
the U(1) and SU(2) factors,  respectively, while $q^{kAf}$ denotes
the SU(2)$_R$ doublet of the squark fields $q^{kA}$ and $\bar{\tilde{ q}}^{Ak}$ in the quark hypermultiplets.
The parameters of the SUSY transformations
in the microscopic theory are denoted as  $\epsilon^{\alpha f}$.
Furthermore, the $F$ terms in Eq.~(\ref{transf}) are
\beq
F^1+ iF^2 =
i\, \frac{g_1^2}{2}\, \left({\rm Tr} \, |\vp|^2-N\xi\right)\, ,\qquad F^3=0
\label{pdterm}
\eeq
for the U(1) field, and
\beq
F^{a1}+iF^{a2}=
i\, g_2^2\,{\rm Tr}\, \left(\bar{\vp} T^a\vp \right)\,,
\qquad
F^{a3}=0
\label{vdterm}
\eeq
for the SU$(N)$ field. The dots in (\ref{transf})  stand for terms involving
the adjoint scalar fields which vanish on the string solution
(in the equal mass case) because the
adjoint fields are given by their vacuum expectation values (\ref{avev}).

In Ref.~\cite{VY} it was shown that four supercharges selected
by the conditions
\beq
\epsilon^{12}=-\epsilon^{11}\,, \qquad
\epsilon^{21}=\epsilon^{22}
\label{trivQ}
\eeq
act trivially on the BPS string. Imposing the conditions (\ref{trivQ})
and  requiring the left-hand sides of Eqs. (\ref{transf}) to
vanish\,\footnote{If, instead of (\ref{trivQ}), we required other
combinations of the SUSY transformation parameters to vanish
(changing the signs in (\ref{trivQ})) we would get the anti-string equations,
with the opposite direction of the gauge fluxes.} we
get, upon substituting the {\em ansatz}
(\ref{znstr}), the first-order equations (\ref{foes}).

\subsection{Elementary non-Abelian strings}
\setcounter{equation}{0}
\label{nas}

The elementary $Z_N$ strings in the model (\ref{qed})
give rise to {\em bona fide} non-Abelian strings provided the condition
(\ref{equalmasses}) is satisfied \cite{HT1,ABEKY,SYmon,HT2}.
This means that, in addition to trivial translational
moduli, they have extra moduli corresponding to spontaneous
breaking of a non-Abelian symmetry. Indeed, while the ``flat"
vacuum  (\ref{qvev}) is SU($N)_{C+F}$ symmetric,
the solution (\ref{znstr}) breaks this symmetry\,\footnote{At $N=2$ the string solution breaks SU(2) down to U(1).} down to U(1)$\times$SU$(N-1)$ (at $N>2$). This ensures the presence of $2(N-1)$ orientational moduli.

To obtain the non-Abelian string solution from the $Z_N$ string
(\ref{znstr}) we apply the diagonal color-flavor rotation  preserving
the vacuum (\ref{qvev}). To this end
it is convenient to pass to the singular gauge where the scalar fields have
no winding at infinity, while the string flux comes from the vicinity of
the origin. In this gauge we have
\beqn
\vp &=&
U\left(
\begin{array}{cccc}
\phi_2(r) & 0& ... & 0\\[2mm]
...&...&...&...\\[2mm]
0& ... & \phi_2(r)&  0\\[2mm]
0 & 0& ... & \phi_{1}(r)
\end{array}
\right)U^{-1}\, ,
\nonumber\\[5mm]
A^{{\rm SU}(N)}_i &=&
\frac{1}{N} \,U\left(
\begin{array}{cccc}
1 & ... & 0 & 0\\[2mm]
...&...&...&...\\[2mm]
0&  ... & 1 & 0\\[2mm]
0 & 0& ... & -(N-1)
\end{array}
\right)U^{-1}\, \left( \pt_i \alpha\right)  f_{NA}(r)\, ,
\nonumber\\[5mm]
A^{{\rm U}(1)}_i &=& -\frac{1}{N}\,
\left( \pt_i \alpha\right)   f(r)\, , \qquad A^{{\rm U}(1)}_0=
A^{{\rm SU}(N)}_0=0\,,
\label{nastr}
\eeqn
where $U$ is a matrix $\in {\rm SU}(N)_{C+F}$. This matrix parametrizes
orientational zero modes of the string associated with flux rotation
in  SU($N$). The presence of these modes makes the string genuinely
non-Abelian. Since the diagonal color-flavor symmetry is not
broken by the VEV's of the scalar fields
in the bulk (color-flavor locking)
it is physical and has nothing to do
with the gauge rotations eaten by the Higgs mechanism. The orientational moduli
encoded in the matrix $U$ are {\it not} gauge artifacts.

The orientational zero modes of a non-Abelian string were first
observed in \cite{HT1,ABEKY}. In Ref.~\cite{HT1} a general index theorem was
proved which shows that the dimension of elementary string moduli space is
$2N=2(N-1)+2$ where 2 stands for translational moduli while $2(N-1)$ is
the dimension of the internal moduli space.\footnote{The index theorem
in \cite{HT1} deals with more general multiple
strings. It was shown that the dimension of the moduli space of
the $k$-string solution is $2kN$.}
In Ref.~\cite{ABEKY} the explicit
solution for the non-Abelian string which we review here was found and explored.

In fact, non-translational zero modes of strings were discussed earlier
in a U(1)$\times$U(1) model \cite{Wsupercond,Hind}, and
somewhat later, in more contrived models, in Ref.~\cite{Alford}.
(The latter paper is entitled {\em Zero Modes of non-Abelian Vortices}!)
It is worth emphasizing that, along with some apparent similarities, there
are drastic distinctions between the ``non-Abelian strings"
we review here and the strings that were discussed in the 1980's.
In particular, in the example treated  in Ref.~\cite{Alford}
the gauge group is not completely broken in the vacuum, and, therefore, there
are massless gauge fields in the bulk. If the unbroken generator acts
non-trivially on the string flux (which is proportional to a broken generator)
then it can and does create zero modes. Infrared  divergence problems ensue
immediately.

In the case we treat here the gauge group is completely broken
(up to a discrete subgroup $Z_N$). The theory in the bulk is fully Higgsed.
The unbroken group SU$(N)_{C+F}$, a combination of the gauge and flavor
groups, is global. There are no massless fields in the bulk.

It is possible to model the example considered in \cite{Alford}
if we gauge the  unbroken global symmetry
SU$(N)_{C+F}$ of the model (\ref{qed}) with respect to yet another
gauge field  $B_{\mu}$.

Let us also note that a generalization of the non-Abelian
string solutions in six-dimensional gauge theory with eight supercharges
was carried out in \cite{Jstr} while the non-Abelian strings in strongly
coupled vacua were considered in \cite{Bolstr}.

\subsection{The worldsheet effective theory}
\setcounter{equation}{0}
\label{worldsheet}

The non-Abelian string solution (\ref{nastr}) is characterized by two
translational moduli (the position of the string in the (1,2) plane) and
$2(N-1)$ orientational moduli. Below we review the effective
two-dimensional low-energy theory on the string world sheet. As usual,
the translational moduli decouple  and we focus on the internal dynamics
of the orientational moduli. Our string is a 1/2-BPS state in \ntwo
supersymmetric gauge theory with eight supercharges. Thus it has
four supercharges acting in the world sheet theory. This means that we
have extended \ntwo supersymmetric effective theory on the string
world sheet. This theory turns out to be  a two-dimensional
CP$(N-1)$ model \cite{HT1,ABEKY,SYmon,HT2}.
In Sect.~\ref{worldsheet} we will first present a derivation of  this theory
and then discuss underlying physics.

\subsubsection{Derivation of the CP$(N-1)$ model}
\label{kineticterm}

Now, following Refs.~\cite{ABEKY,SYmon,GSY05},  we  will  derive
the effective low-energy
theory for the  moduli  residing in the matrix $U$ in the problem at hand.
As is clear from the string solution (\ref{nastr}),   not each element of
the matrix $U$ will give rise to a modulus. The SU($N-1) \times$U(1)
subgroup remains unbroken
 by the string solution under consideration; therefore
the moduli space is
\beq
\frac{{\rm SU}(N)}{{\rm SU}(N-1)\times {\rm U}(1)}\sim {\rm CP}(N-1)\,.
\label{modulispace}
\eeq
Keeping this in mind we parametrize the matrices entering Eq.~(\ref{nastr})
as follows:
\beq
\frac1N\left\{
U\left(
\begin{array}{cccc}
1 & ... & 0 & 0\\[2mm]
...&...&...&...\\[2mm]
0&  ... & 1 & 0\\[2mm]
0 & 0& ... & -(N-1)
\end{array}
\right)U^{-1}
\right\}^l_p=-n^l n_p^* +\frac1N \delta^l_p\,\, ,
\label{n}
\eeq
where $n^l$ is a complex vector
 in the fundamental representation of SU($N$), and
\beq
 n^*_l n^l =1\,,
\label{unitvec}
\eeq
($l,p=1, ..., N$ are color indices).
As we will show below, one U(1) phase will be gauged away in the effective
sigma model. This gives the correct number of degrees of freedom,
namely, $2(N-1)$.

With this parametrization the string solution (\ref{nastr}) can be
rewritten   as
\beqn
\vp &=& \frac1N[(N-1)\phi_2 +\phi_1] +(\phi_1-\phi_2)\left(
n\,\cdot n^*-\frac1N\right) ,
\nonumber\\[3mm]
A^{{\rm SU}(N)}_i &=& \left( n\,\cdot n^*-\frac{1}{N}\right)
\varepsilon_{ij}\, \frac{x_i}{r^2}
\,
f_{NA}(r) \,,
\nonumber\\[3mm]
A^{{\rm U}(1)}_i &=& \frac1N
\varepsilon_{ij}\, \frac{x_i}{r^2} \, f(r) \, ,
\label{str}
\eeqn
where for brevity we suppress all SU$(N)$  indices. The notation is
self-evident.

Assume  that the orientational moduli
are slowly-varying functions of the string world-sheet coordinates
$x_{k}$, $k=0,3$. Then the moduli $n^l$ become fields of a
(1+1)-dimensional sigma model on the world sheet. Since
$n^l$ parametrize the string zero modes,
there is no potential term in this sigma model.

To obtain the kinetic term  we substitute our solution
(\ref{str}), which depends on the
moduli $ n^l$,
in the action (\ref{redmodel}), assuming  that
the fields acquire a dependence on the coordinates $x_{k}$ via
$n^l(x_{k})$.
In doing so we immediately observe that we have to modify our solution:
we have to include in it the $k=0,3$ components of the gauge potential
which are no more vanishing. In the CP(1) case, as was  shown in
\cite{SYmon},  the potential $A_{k}$ must be orthogonal
(in the SU$(N)$  space)  to the matrix (\ref{n}), as well as to its
derivatives with respect to $x_{k}$. Generalization of these
conditions to the CP$(N-1)$ case leads to the
following {\em ansatz}:
\beq
A^{{\rm SU}(N)}_{k}=-i\,  \big[ \pt_{k} n \,\cdot n^* -n\,\cdot
\pt_{k} n^* -2n\,
\cdot n^*(n^*\pt_{\alpha} n)
\big] \,\rho (r)\, , \quad \alpha=0, 3\,,
\label{An}
\eeq
where we assume the contraction of the color indices inside the parentheses,
$$(n^*\pt_{k} n)\equiv n^*_l\pt_{k} n^l\,, $$
and introduce a new profile function $\rho (r)$.

The function $\rho (r)$ in Eq.~(\ref{An}) is
determined  through a minimization procedure  \cite{ABEKY,SYmon,GSY05}
which generates $\rho$'s own equation of motion. Now we will outline its
derivation. But at first we note that $\rho (r)$ vanishes at infinity,
\beq
\rho (\infty)=0\,.
\label{bcfinfty}
\eeq
The boundary condition at $r=0$ will be determined shortly.

The kinetic term for $n^l$ comes from the gauge and quark kinetic terms
in Eq.~(\ref{redmodel}). Using Eqs.~(\ref{str}) and (\ref{An}) to calculate the
SU($N$)  gauge field strength we find
\beqn
F_{k i}^{{\rm SU}(N)} &=& \left( \pt_{k} n \,\cdot n^*
+n\cdot \,\pt_{k} n^*\right) \, \varepsilon_{ij}\,
\frac{x_j}{r^2}\,
f_{NA}
\left[1-\rho (r)\right]
\nonumber\\[3mm]
&+&
i\left[ \pt_{k} n \,\cdot n^* -n\cdot \,\pt_{k} n^*
 -2n\,\cdot n^*(n^*\pt_{k} n)
\right]
 \, \frac{x_i}{r}\,\, \frac{d\,\rho (r)}{dr}\, .
 \label{Fni}
\eeqn
In order to have a finite contribution  from the term
Tr$\,F_{k i}^2$ in the action we have to impose the constraint
\beq
\rho (0)=1\,.
\label{bcfzero}
\eeq
Substituting the field strength (\ref{Fni}) in the action
(\ref{redmodel}) and including, in addition, the quark kinetic term,
after a rather straightforward but tedious algebra we
arrive at
\beq
S^{(1+1)}= 2 \beta\,   \int d t\, dz \,  \left\{(\pt_{k}\, n^*
\pt_{k}\, n) + (n^*\pt_{k}\, n)^2\right\}\,,
\label{cp}
\eeq
where the coupling constant $\beta$ is given by
\beq
\beta=\frac{2\pi}{g^2_2}\, I \,,
\label{betaI}
\eeq
and $I$ is a basic normalizing integral
\beqn
I & = &
  \int_0^{\infty}
rdr\left\{\left(\frac{d}{dr}\rho (r)\right)^2
+\frac{1}{r^2}\, f_{NA}^2\,\left(1-\rho \right)^2
\right.
\nonumber\\[4mm]
& + &
\left.  g_2^2\left[\frac{\rho^2}{2}\left(\phi_1^2
 +\phi_2^2\right)+
\left(1-\rho \right)\left(\phi_2-\phi_1\right)^2\right]\right\}\, .
\label{I}
\eeqn

The theory in Eq.~(\ref{cp}) is nothing but the two-dimensional CP$(N-1)$ model.
To see that this is indeed the case we can eliminate
the second term in (\ref{cp}) introducing
a non-propagating U(1) gauge field. We review this in Sect.~\ref{Dyn},
and then discuss the underlying physics of the model.

Thus, we obtain the CP$(N-1)$ model as an effective low-energy
theory on the worldsheet of the non-Abelian string. Its coupling constant $\beta$
is related to the four-dimensional coupling $g_2^2$ via the basic normalizing integral
(\ref{I}). This integral must be viewed as an ``action'' for the profile
function $\rho$.

Varying (\ref{I}) with respect to $\rho$
one  obtains the second-order equation which
the function $\rho$ must satisfy, namely,
\beq
-\frac{d^2}{dr^2}\, \rho -\frac1r\, \frac{d}{dr}\, \rho
-\frac{1}{r^2}\, f_{NA}^2 \left(1-\rho\right)
+
\frac{g^2_2}{2}\left(\phi_1^2+\phi_2^2\right)
\rho
-\frac{g_2^2}{2}\left(\phi_1-\phi_2\right)^2=0\, .
\label{rhoeq}
\eeq
After some algebra and extensive use of the first-order equations (\ref{foes})
one can show that the solution of (\ref{rhoeq})  is
\beq
\rho=1-\frac{\phi_1}{\phi_2}\, .
\label{rhosol}
\eeq
This solution  satisfies the boundary conditions (\ref{bcfinfty})
and  (\ref{bcfzero}).
Substituting this solution back in  the expression for the
normalizing integral  (\ref{I}) one can check that
this integral   reduces to a total derivative and is given
by the flux of the string  determined by $f_{NA}(0)=1$.
In this way we arrive at
\beq
I=1\,.
\label{Ieq1}
\eeq

This result can be traced back to the fact that our theory (\ref{redmodel}) is
\mbox{\ntwo} supersymmetric theory, and the string is BPS saturated.
In Sect. \ref{kinkmonopole} we will see
that this fact is crucial for the interpretation of confined monopoles
as sigma-model kinks. Generally speaking, for non-BPS strings, $I$
could  be a certain  function of $N$ (see Ref.~\cite{MMY} for a particular
example).

Equation (\ref{I}) implies
\beq
\beta=\frac{2\pi}{g_2}\, .
\label{beta}
\eeq
The two-dimensional coupling is determined by the four-dimensional non-Abelian
coupling. This relation  is obtained  at the classical level. In quantum theory
both couplings run. Therefore,  we have to specify a scale at which the relation
(\ref{beta}) takes place.
The two-dimensional CP$(N-1)$ model (\ref{cp}) is
an effective low-energy theory appropriate for the description of
internal string dynamics at low energies,  lower than the
inverse thickness of the string which is given by
the masses of the gauge/quark multiplets (\ref{msuN}) and (\ref{mu1})
in the  bulk SU$(N)\times$U(1) theory.
Thus, the parameter $g\sqrt{\xi}$ plays the role of a physical
ultraviolet (UV) cutoff in the action (\ref{cp}).
This is the scale at which Eq.~(\ref{beta}) holds. Below this scale, the
coupling $\beta$ runs according to its two-dimensional renormalization-group
flow, see the  Sect. \ref{Dyn}.

Thus --- we repeat it again --- the model (\ref{cp}) describes the low-energy limit:
all higher order terms in derivatives are neglected. Quartic in
derivatives, sextic, and so on, terms certainly exist. In fact, the
derivative expansion runs in powers of
\beq
\left( g_2\, \sqrt{\xi}\right)^{-1} {\pt_\alpha} \,,
\label{hd}
\eeq
where $g_2\sqrt{\xi}$ gives the order of magnitude of
masses in the bulk theory.
The sigma model (\ref{cp}) is adequate at  scales
below $g_2\sqrt{\xi}$ where the higher-derivative corrections
are negligibly small.

To conclude this subsection let us narrow down the model (\ref{cp}) setting
$N=2$. In this case we deal with the CP(1) model equivalent to the
O(3) sigma model. The action (\ref{cp}) takes the form
\beq
S^{(1+1)}=  \frac{\beta}{2}\,   \int d t\, dz \,  (\pt_{k}\, S^a)^2
\,,
\label{o3}
\eeq
where $S^a$ ($a=1,2,3$) is a real unit vector, $(S^a)^2=1,$
sweeping the two-dimensional sphere $S_2$. It is
defined as
\beq
S^a=-n^*\tau^a n\, .
\label{sn}
\eeq

The model (\ref{o3}), as an effective theory on the worldsheet of
the non-Abelian string in SU(2)$\times$U(1) SQCD
with \ntwo  supersymmetry, was first derived
in \cite{ABEKY} in field-theoretical framework. This derivation was
generalized for arbitrary  $N$ in \cite{GSY05}.
A brane construction of (\ref{cp}) was presented in \cite{HT1}.

\subsubsection{Fermion zero modes}
\label{fermzeromodes}

In Sect.~\ref{kineticterm}  we derived the bosonic part of the effective
\ntwo supersym\-metric CP$(N-1)$ model. Now we
will find fermion zero modes for the
non-Abelian string. Inclusion of these modes into consideration
will demonstrate  that the internal worldsheet dynamics
is given by \ntwo supersymmetric CP$(N-1)$ model.  This program
was carried out in \cite{SYmon} for $N=2$. Here we will dwell on this
construction.

The string solution (\ref{str}) in the  SU(2)$\times$U(1)
theory reduces to
\begin{eqnarray}
\vp
&=&
U \left(
\begin{array}{cc}
\phi_2(r) & 0  \\[2mm]
0 &  \phi_1(r)
\end{array}\right)U^{-1}\, ,
\nonumber \\[4mm]
A^a_{i}(x)
&=&
-S^a \,\varepsilon_{ij}\, \frac{x_j}{r^2}\,
f_{NA}(r)\, ,
\nonumber \\[4mm]
A_{i}(x)
&=&
 \varepsilon_{ij} \, \frac{x_j}{r^2}\,
f(r)\, ,
\label{sna}
\end{eqnarray}
while the parametrization (\ref{n})
reduces to
\beq
S^a\tau^a=U \tau^3 U^{-1},\;\;a=1,2,3.
\label{S}
\eeq
by virtue  of Eq.~(\ref{sn}).

Our string solution is 1/2 BPS-saturated. This means that four
supercharges, out of eight of the four-dimensional theory
(\ref{model}), act trivially
on the string solution
(\ref{sna}). The remaining four supercharges generate four fermion zero
modes which were termed supertranslational modes because they are
superpartners to  two translational zero
modes. The corresponding four fermionic moduli
are superpartners to the coordinates $x_0$ and
$y_0$ of the string center. The  supertranslational
fermion zero modes were found in Ref.~\cite{VY} for the U(1) ANO string in
\ntwo  theory. This is discussed in detail in Appendix B,
see Sect.~\ref{secb3}.
Transition to the non-Abelian model at hand is absolutely
straightforward. We will not dwell on this procedure here.

Instead, we will focus on four {\em additional} fermion zero modes
which arise only for the non-Abelian string, to be referred to
as superorientational. They are superpartners of the bosonic orientational moduli $S^a$.

let us see how one can explicitly construct  these four zero modes (in CP(1))
and study their impact on the string worldsheet.

At $N=2$ the fermionic part of the action  of the model (\ref{qed}) is
\beqn
S_{\rm ferm}
&=&
\int d^4 x\left\{
\frac{i}{g_2^2}\bar{\lambda}_f^a \bar{D}\hspace{-0.65em}/\lambda^{af}+
\frac{i}{g_1^2}\bar{\lambda}_f \bar{\pt}\hspace{-0.65em}/\lambda^{f}
+ {\rm Tr}\left[\bar{\psi} i\bar\nabla\hspace{-0.65em}/ \psi\right]
+ {\rm Tr}\left[\tilde{\psi} i\nabla\hspace{-0.65em}/ \bar{\tilde{\psi}}
\right]\right.
\nonumber\\[3mm]
&+&
\frac{1}{\sqrt{2}}\varepsilon^{abc}\bar{a}^a(\lambda_f^b\lambda^{cf})
+\frac{1}{\sqrt{2}}\varepsilon^{abc}(\bar{\lambda}^{bf}\bar{\lambda}^{c}_f)a^c
\nonumber\\[3mm]
&+&
\frac{i}{\sqrt{2}}\,{\rm Tr}\left[ \bar{q}_f(\lambda^f\psi)+
(\tilde{\psi}\lambda_f)q^f +(\bar{\psi}\bar{\lambda}_f)q^f+
\bar{q}^f(\bar{\lambda}_f\bar{\tilde{\psi}})\right]
\nonumber\\[3mm]
&+&
\frac{i}{\sqrt{2}}\,{\rm Tr}\left[ \bar{q}_f\tau^a(\lambda^{af}\psi)+
(\tilde{\psi}\lambda_f^a)\tau^aq^f +(\bar{\psi}\bar{\lambda}_f^a)\tau^aq^f+
\bar{q}^f\tau^a(\bar{\lambda}^{a}_f\bar{\tilde{\psi}})\right]
\nonumber\\[3mm]
&+&
\left.
\frac{i}{\sqrt{2}}\,{\rm Tr}\left[\tilde{\psi}\left(a+a^a\tau^a\right)\psi
\right]
+\frac{i}{\sqrt{2}}\,{\rm Tr}\left[\bar{\psi}\left(a+a^a\tau^a\right)
\bar{\tilde{\psi}}\right]
\right\}\,,
\label{fermact}
\eeqn
where we use the matrix color-flavor notation for the
matter fermions $(\psi^{\alpha})^{kA}$ and $(\tilde{\psi}^{\alpha})_{Ak}$.
The traces in Eq.~(\ref{fermact}) are performed
over the color-flavor indices. Contraction of spinor indices is assumed
inside the parentheses, say, $(\lambda\psi)\equiv \lambda_{\alpha}\psi^{\alpha}$.

As was mentioned in Sect. \ref{znstring},
the four supercharges selected by the conditions (\ref{trivQ})
act trivially on the BPS string in the theory with the FI term of the $F$ type.
To generate superorientational fermion zero modes the following
method was used in \cite{SYmon}. Assume  the
orientational moduli  $S^a$  in the string solution
(\ref{sna}) to have a slow dependence on the
worldsheet coordinates $x_0$ and $x_3$ (or $t$ and $z$).
Then the four (real) supercharges selected by
the conditions (\ref{trivQ}) no longer  act trivially. Instead,
their action now generates fermion fields proportional to
$x_0$ and $x_3$ derivatives of $S^a$.

This is exactly what one expects from the residual
\ntwo supersymmetry
in the worldsheet  theory. The above four supercharges
generate the worldsheet supersymmetry in the \ntwo
two-dimensional CP(1) model,
\beqn
\delta \chi^a_1
&=&
i\sqrt{2} \left[\left( \pt_0 +i\pt_3\right) n^a \, \varepsilon_2
+\varepsilon^{abc}n^b\left( \pt_0 +i\pt_3\right) n^c\,\eta_2\right]\,,
\nonumber\\[3mm]
\delta \chi^a_2
&=&
i\sqrt{2} \left[\left( \pt_0 -i\pt_3\right)  n^a \, \varepsilon_1
+\varepsilon^{abc}n^b\left( \pt_0 -i\pt_3\right) n^c\,\eta_1\right]\, .
\label{susy2c}
\eeqn
Here $\chi^a_{\alpha}$ ($\alpha=1,2$ is the spinor index) are
real two-dimensional fermions of the  CP(1) model. They are
superpartners of $S^a$ and are subject to the
orthogonality condition
$$S^a\chi^a_{\alpha}=0\,.$$

The real parameters of the \ntwo two-dimensional  SUSY transformations
$\varepsilon_{\alpha}$ and $\eta_{\alpha}$ are identified
with  the parameters of the four-dimensional SUSY
transformations (with the  constraint (\ref{trivQ})) as follows:
\beqn
\varepsilon_1 -i\eta_1
&=&
 \frac1{\sqrt{2}}(\epsilon^{21} +\epsilon^{22})=\sqrt{2}\epsilon^{22}\,,
\nonumber\\[3mm]
\varepsilon_2+i\eta_2
&=&
\frac1{\sqrt{2}}(\epsilon^{11} -\epsilon^{12})=\sqrt{2}\epsilon^{11}\,.
\label{epsilon24}
\eeqn
In this way the worldsheet supersymmetry was used to
re-express the fermion fields obtained upon the action of
these four supercharges
in terms of the (1+1)-dimensional fermions. This procedure  gives
us the superorientational fermion zero modes \cite{SYmon},
\beqn
\bar{\psi}_{Ak\dot{2}}
& = &
\left(\frac{\tau^a}{2}\right)_{Ak}
\frac1{2\phi_2}(\phi_1^2-\phi_2^2)
\left[
 \chi_2^a
+i\varepsilon^{abc}\, n^b\, \chi^c_2\,
\right]\, ,
\nonumber\\[3mm]
\bar{\tilde{\psi}}^{kA}_{\dot{1}}
& = &
\left(\frac{\tau^a}{2}\right)^{kA}
\frac1{2\phi_2}(\phi_1^2-\phi_2^2)
\left[
 \chi_1^a
-i\varepsilon^{abc}\, n^b\, \chi^c_1\,
\right]\, ,
\nonumber\\[5mm]
\bar{\psi}_{Ak\dot{1}}
& = &
0\, , \qquad
\bar{\tilde{\psi}}^{kA}_{\dot{2}}= 0\, ,
\nonumber\\[4mm]
\lambda^{a22}
& = &
\frac{i}{2}\frac{x_1+ix_2}{r^2}
f_{NA}\frac{\phi_1}{\phi_2}
\left[
 \chi^a_1
-i\varepsilon^{abc}\, n^b\, \chi^c_1
 \right]\, ,
\nonumber\\[4mm]
\lambda^{a11}
& = &
\frac{i}{2}\frac{x_1-ix_2}{r^2}
f_{NA}\frac{\phi_1}{\phi_2}
\left[
\chi^a_2
+i\varepsilon^{abc}\, n^b\, \chi^c_2
 \right]\,,
\nonumber\\[4mm]
\lambda^{a12}
& = & \lambda^{a11}
  \, ,\qquad  \lambda^{a21}= \lambda^{a22}\,,
\label{zmodes}
\eeqn
where the dependence on $x_i$ is encoded in the string profile functions, see
Eq.~(\ref{sna}).

Now let us directly check that the zero modes (\ref{zmodes}) satisfy
the Dirac equations of motion. From the fermion action of the model
(\ref{fermact}) we get the Dirac equations for $\lambda^a$,
\beq
\frac{i}{g_2^2} \bar{D}\hspace{-0.65em}/\lambda^{af}
+\frac{i}{\sqrt{2}}\,{\rm Tr}\left(
\bar{\psi}\tau^a q^f+
\bar{q}^f\tau^a\bar{\tilde{\psi}}\right)=0\,.
\label{dirac1}
\eeq
At the same time, for the matter fermions,
\beqn
&&
i\nabla\hspace{-0.65em}/ \bar{\psi}+\frac{i}{\sqrt{2}}\left[\bar{q}_f\lambda^f
-(\tau^a\bar{q}_f)\lambda^{af}+
(a-a^a\tau^a)\tilde{\psi}\right]=0,
\nonumber\\[3mm]
&&
i\nabla\hspace{-0.65em}/ \bar{\tilde{\psi}}+
\frac{i}{\sqrt{2}}\left[\lambda_f q^f
+\lambda^{a}_f(\tau^aq^f)+
(a+a^a\tau^a)\psi\right]=0\, .
\label{dirac2}
\eeqn
Next, we substitute the orientational fermion zero modes (\ref{zmodes})
into these equations. After some algebra one can check
that (\ref{zmodes}) do  satisfy the Dirac equations (\ref{dirac1})
and (\ref{dirac2})
provided the first-order equations for the string profile functions (\ref{foes})
are satisfied.

Furthermore, it is instructive to check that the zero
modes (\ref{zmodes}) do produce
the fermion part of the  \ntwo two-dimensional CP(1) model.
To this end we return to the usual assumption that
the fermion collective coordinates  $\chi^a_{\alpha}$ in
Eq.~(\ref{zmodes}) have an adiabatic dependence on the worldsheet
coordinates $x_k$ ($k=0,3$). This is quite similar to the procedure
of Sect.~\ref{kineticterm} for bosonic moduli.
Substituting Eq.~(\ref{zmodes}) in the fermion kinetic terms
in the bulk  theory (\ref{fermact}),
and taking into account the derivatives of  $\chi^a_{\alpha}$ with
respect to the worldsheet coordinates,  we arrive at
\beq
\beta \int d t d z \left\{\frac12 \, \chi^a_1 \, (\pt_0-i\pt_3)\, \chi^a_1
+\frac12 \, \chi^a_2 \, (\pt_0+i\pt_3)\, \chi^a_2
\right\},
\label{fkin}
\eeq
where $\beta$ is given by the same integral (\ref{beta}) as for
the bosonic kinetic term, see Eq.~(\ref{o3}).

Finally we must discuss the four-fermion interaction term in the CP(1) model.
We can use the worldsheet \ntwo supersymmetry to reconstruct this
term. The SUSY transformations in the CP(1) model
have the form (see e.g. \cite{NSVZsigma} for a review)
\beqn
\delta \chi^a_1
&=&
i\sqrt{2} \left( \pt_1 +i\pt_3\right) n^a \, \varepsilon_2
-\sqrt{2}\varepsilon_1\,n^a(\chi^a_1\chi^a_2)\,,
\nonumber\\[3mm]
\delta \chi^a_2
&=&
i\sqrt{2} \left( \pt_1 -i\pt_3\right)  n^a \, \varepsilon_1
+\sqrt{2}\varepsilon_2\,n^a(\chi^a_1\chi^a_2)\, ,
\nonumber\\[3mm]
\delta n^a
&=&
\sqrt{2}(\varepsilon_1\chi^a_2+\varepsilon_2\chi^a_1)\,,
\label{susy2d}
\eeqn
where we put $\eta_{\alpha}=0$ for simplicity. Imposing this supersymmetry
leads to the following effective theory on the string worldsheet
$$
S_{{\rm CP}(1)}=
\beta \int d t d z \left\{\frac12 (\pt_k S^a)^2+
\frac12 \, \chi^a_1 \, i(\pt_0-i\pt_3)\, \chi^a_1
\right.
$$
\beq
\left.
+\frac12 \, \chi^a_2 \, i(\pt_0+i\pt_3)\, \chi^a_2
-\frac12 (\chi^a_1\chi^a_2)^2
\right\},
\label{ntwocp}
\eeq
This is indeed the action of the \ntwo CP(1) sigma model in its entirety.

\subsubsection{Physics of the CP$(N-1)$ model with \ntwo}
\label{Dyn}

As is quite common in two dimensions, the Lagrangian of our effective theory on the
string worldsheet can be cast in many
different (but equivalent) forms. In particular,
the \ntwo supersymmetric CP$(N-1)$ model  (\ref{cp})  can be understood as
a strong-coupling limit
of a U(1) gauge theory \cite{W93}. Then the  bosonic part
of the  action takes the form
\beqn
S
& =&
\int d^2 x \left\{
 2\beta\,|\nabla_{k} n^{\ell}|^2 +\frac1{4e^2}F^2_{kl} + \frac1{e^2}
|\pt_k\sigma|^2
\right.
\nonumber\\[3mm]
 &+&    4\beta\,|\sigma|^2 |n^{\ell}|^2 + 2e^2 \beta^2(|n^{\ell}|^2 -1)^2
\Big\}\,,
\label{cpg}
\eeqn
where $\nabla_k= \partial_k - i A_k $ while $\sigma$ is a complex scalar
field. The condition (\ref{unitvec}) is
implemented in the limit $e^2\to\infty$. Moreover, in this limit
the gauge field $A_k$  and its \ntwo bosonic superpartner $\sigma$ become
auxiliary and can be eliminated by virtue of the equations of motion,
\beq
A_k =-\frac{i}{2}\, n^*_\ell \stackrel{\leftrightarrow}
{\partial_k} n^\ell \,,\qquad \sigma=0\,.
\label{aandsigma}
\eeq
Substituting Eq.~(\ref{aandsigma}) in the Lagrangian, we can
readily rewrite the action in the form (\ref{cp}).

The coupling constant $\beta$ is asymptotically free \cite{P75}.
The running coupling, as a function of energy $E$,  is given by
the formula
\beq
4\pi\beta=N\,\ln\,{\frac{E}{\Lambda_{\sigma}}}\,,
\label{asyfree}
\eeq
where $\Lambda_{\sigma}$ is a dynamical scale of the sigma model.
The ultraviolet cut-off of the sigma model on the string worldsheet
is determined by  $g_2\sqrt{\xi}$.
Equation~(\ref{beta}) relating the two- and four-dimensional couplings
is valid at this scale.
Hence,
\beq
\Lambda^{N}_{\sigma} = g_2^N\xi^{\frac{N}{2}}
 e^{-\frac{8\pi^2}{g_2^2}} =\Lambda^N_{{\rm SU}(N)}\, .
\label{lambdasig}
\eeq
Here we take into account Eq.~(\ref{4Dlambda}) for the dynamical scale
$\Lambda_{{\rm SU}(N)}$ of the SU$(N)$ factor of the bulk theory.
Note that in the bulk theory {\em per se}, because of the VEV's of
the squark fields, the coupling constant is frozen at
$g_2\sqrt{\xi}$; there are no logarithms below this scale.
The logarithms of the string worldsheet theory take over.
Moreover, the dynamical scales of the bulk and worldsheet
theories turn out to be the same!
We will explain the reason why the dynamical
scale of the (1+1)-dimensional effective theory
on the string worldsheet  is identical to that
of the SU$(N)$ factor of the (3+1)-dimen\-sional gauge theory
later, in Sect.~\ref{dsw}.

The CP$(N-1)$ model was solved by Witten
in the large-$N$ limit \cite{W79}.
We will briefly summarize Witten's results and translate them in terms
of strings in four dimensions \cite{SYmon}.

Classically the  field $n^{\ell}$ can have arbitrary
direction; therefore, one might naively expect
a spontaneous breaking of SU($N$) and
the occurrence of massless Goldstone modes. Well, this cannot happen
in two dimensions. Quantum effects restore the
symmetry. Moreover, the condition
(\ref{unitvec}) gets in effect relaxed. Due to strong coupling
we have more degrees of freedom than in the original Lagrangian,
namely all $N$ fields $n$ become dynamical and acquire
masses $ \Lambda_{\sigma}$.

As was shown by Witten \cite{W79}, the model has $N$ vacua.
These $N$ vacua differ from each other by the expectation value
of the chiral bifermion operator, see e.g. \cite{NSVZsigma}.
At strong coupling  the chiral condensate is the
order parameter. The U(1) chiral symmetry of the CP$(N-1)$ model is
explicitly broken to a discrete $Z_{2N}$ symmetry by the chiral anomaly.
The fermion condensate  breaks $Z_{2N}$ down to $Z_2$. That's the
origin of the $N$-fold degeneracy of the vacuum state.

Physics of the model becomes even more transparent in the mirror
representation which was established  \cite{HoVa} for arbitrary $N$.
In this representation one describes the CP$(N-1)$ model in terms
of the Coulomb gas of instantons (see \cite{FFS} where this was
done for non-supersymmetric CP(1) model) to prove its equivalence to
an affine Toda  theory. The CP$(N-1)$ model (\ref{cpg})
is dual to the following \ntwo affine Toda  model \cite{HoVa,FenInt,cv,EgHoXi},
\begin{eqnarray}
&& S_{\rm mirror}
=
\int d^2 x  d^2 \theta \,
 d^2 \bar{\theta} \,  {\beta}^{-1} \, \sum_{i=1}^{N-1}\bar{Y_i}\, Y_i
\nonumber\\[3mm]
&&
+
 \left\{ \Lambda_{\sigma}\!
\int d^2x   d^2 \theta \, \left(\sum_{i=1}^{N-1} \exp{(Y_i)}
+\prod_{i=1}^{N-1}\exp{(-Y_{i})}\right)\! +\! {\rm H.c.}\right\}\! .
\label{toda}
\end{eqnarray}
Here the last term is a
dual instanton-induced superpotential. In fact the exact form of the
kinetic term in the mirror representation
is not known because it is not protected from quantum correction
in $\beta$. However the superpotential in (\ref{toda})
is {\em exact}. Since the vacuum structure is entirely determined
by the superpotential (\ref{toda}), one immediately confirms
Witten's statement of $N$ vacua.

Indeed, the scalar potential of this affine Toda theory
has $N$ minima. For example, for $N=2$ this theory becomes \ntwo
supersymmetric sine-Gordon theory with scalar potential
\beq
V_{\rm SG}=\frac{\beta }{4\pi^2}\, \Lambda^2_{{\rm CP}(1)}\, \left|
\sinh {y}\right|^2 \, ,
\label{sGpot}
\eeq
which obviously
has two minima, at $y=0$ and $y= \pm i\pi $  (warning: the points $y=   i\pi $ and
$y=   -i\pi $
must be identified; they present one and the same vacuum).

This mirror model explicitly
exhibits  a mass gap of the   order of $\Lambda_{\sigma}$.  It shows
that there are no Goldstone bosons (corresponding to the absence of the
spontaneous breaking of  the  SU$(N)_{C+F}$ symmetry). In terms of
strings in the four-dimensional bulk theory, this means,
in turn, that the string orientation
has no particular direction, it is smeared all over. The strings we
deal with here are genuinely non-Abelian. The
$N$ vacua of the worldsheet theory
(\ref{cpg}) are heirs  of the $N$ ``elementary" non-Abelian
strings of the bulk theory.
Note that these strings are in a highly quantum regime.  They are {\em
not} the $Z_N$ strings of the quasiclassical
U(1)$^{N-1}$ theory since $n^\ell$
is not aligned in the vacuum.

Hori and Vafa originally derived \cite{HoVa} the mirror representation
for the CP$(N-1)$ model in the form of the Toda model.
Since then other useful equivalent representations were obtained, and they were expanded to include the so-called twisted masses of which we will
speak in Sect.~\ref{unequalmasses} and subsequent sections.
A particularly useful mirror representation
of the twisted-mass-deformed CP$(N-1)$ model
was exploited by Dorey \cite{Dorey}.

\subsubsection{Unequal quark masses}
\label{unequalmasses}

The fact that we have $N$ distinct vacua in the worldsheet theory
--- $N$  distinct elementary strings ---
is not quite intuitive in the above consideration.
This is   understandable. At the classical level
the \ntwo two-dimensional CP\mbox{$(N-1)$} sigma model
is characterized by  a continuous vacuum
manifold. This is in one-to-one correspondence with
continuously many strings parametrized by
the moduli $ n^{\ell}$.
The continuous degeneracy is lifted only after quantum effects are taken into
account. These quantum effects become crucial at strong coupling. Gone with this lifting
is the moduli nature of the  fields $n^{\ell}$. They become massive.
This is difficult to grasp.

To make the task easier and to facilitate contact between the bulk  and worldsheet theories,
it is instructive to start from a deformed  bulk theory,
so that the string moduli are lifted already at the classical level.
Then the origin of the $N$-fold degeneracy of the
non-Abelian strings becomes transparent. This will help us
understand, in an intuitive manner, other features
listed above. After this understanding is
achieved, nothing prevents us from returning
to our problem ---  strings with non-Abelian moduli at the classical level ---
by smoothly suppressing the moduli-breaking deformation. The $N$-fold
degeneracy will remain intact as it follows from the Witten index \cite{Windex}.

Thus, let us drop the assumption (\ref{equalmasses}) of equal mass terms
and introduce small mass differences.
With unequal quark masses, the  U$(N)$ gauge group is broken by the condensation
of the adjoint scalars down to U(1)$^{N}$, see (\ref{avev}). Off-diagonal
gauge bosons, as well as the off-diagonal fields of the quark matrix $q^{kA}$,
(together with their fermion superpartners) acquire masses proportional to
various mass differences $(m_A-m_B)$. The effective low-energy theory
contains now only diagonal gauge and quark fields. The reduced action
suitable for the search of the string solution takes the form
\beqn
S &=& \int {\rm d}^4x\left\{\frac1{4g_2^2}
\left(F^{h}_{\mu\nu}\right)^{2}
+ \frac1{4g_1^2}\left(F_{\mu\nu}\right)^{2}
 \right.
 \nonumber\\[3mm]
&+&
  |\nabla_\mu \vp^A|^2
+\frac{g^2_2}{2}
\left(\bar{\vp}_A T^h \vp^A\right)^2
 +
 \frac{g^2_1}{8}\left(
 | \vp^A|^2 - N\xi \right)^2
,
\label{redmodelab}
\eeqn
where the index $h=1,..., (N-1)$ runs over the Cartan generators of the gauge
group SU$(N)$, while the matrix $\vp^{kA}$ is reduced to its diagonal components.

The same steps which previously lead us to Eqs.~(\ref{gfoes}) now give the first-order
string equations in the Abelian model (\ref{redmodelab}),
\beqn
&& F^{*}_{3}+\frac{g_1^2}{2} \left(\left|
\varphi^{A}\right|^2-N\xi\right)\,=0\, ,
\nonumber\\[3mm]
&& F^{*h}_{3}+ g_2^2 \left(\bar{\vp}_{A}T^h \varphi^{A}\right)
\,=0\, ,
\nonumber\\[3mm]
&& (\nabla_1+i\nabla_2)\varphi^A=0\, .
\label{gfoesab}
\eeqn
As soon as the $Z_N$-string solutions (\ref{znstr}) have a diagonal form, they
automatically satisfy the above first-order equations.

However, the Abelian $Z_N$ strings (\ref{znstr}) are  now {\em the only}
solutions to these equations. The family of solutions is discrete.
The global SU$(N)_{C+F}$ group is broken down to U(1)$^{N-1}$,
and the continuous CP$(N-1)$ moduli space of the non-Abelian string
is lifted. In fact, the vector $ n^{\ell}$ gets fixed in $N$ possible positions,
\beq
n^{\ell}=\delta^{\ell \ell_0},\;\;\; \ell_0=1,...,N.
\label{nvac}
\eeq
These $N$ solutions correspond to the
Abelian $Z_N$ strings, see (\ref{znstr}) and (\ref{str}).
If the mass differences are much smaller then
$\sqrt{\xi}$ the set of parameters  $n^{\ell}$ becomes {\em quasi}moduli.

Now, our aim is to derive an effective two-dimensional theory
on the string worldsheet for unequal quark mass terms.
With small mass differences we will still be able introduce orientational
quasimoduli $n^{\ell}$.  In terms of the effective two-dimensional theory on
the string worldsheet, unequal masses lead to a shallow potential
for the quasimoduli  $n^{\ell}$. Let us derive this potential.

Below
we will review the derivation carried out in \cite{SYmon} in the SU(2)$\times$U(1)
model. The case of general $N$ is considered in \cite{HT2}.
In the $N=2$ case two minima of the potential at $S=\{0,0,\pm 1\}$
correspond to two {\em bona fide} $Z_2$ strings.

We start from the
expression for the non-Abelian string
in the singular gauge (\ref{sna}) parametrized by the moduli $S^a$,
and substitute it in the action (\ref{qed}). The only
modification we actually have to make is to supplement our
{\em ansatz} (\ref{sna}) by that for the adjoint scalar field $a^a$;
the neutral scalar field $a$ will stay fixed at its vacuum expectation
value $a=-\sqrt{2}m$.

At large $r$ the field $a^a$ tends to its VEV aligned along
the third axis in the color space,
\beq
\langle a^3 \rangle = -\frac{\Delta m}{\sqrt{2}},\;\;\; \Delta m= m_1-m_2,
\label{N2avev}
\eeq
see Eq.~(\ref{avev}). At the same time, at $r=0$
it must be directed along the vector $S^a$. The reason for this behavior
is easy to understand. The kinetic term for $a^a$  in Eq.~(\ref{qed})
contains the commutator term of the adjoint scalar and the gauge potential.
The gauge potential is singular at the origin, as is seen
from  Eq.~(\ref{sna}). This
implies that $a^a$ must be aligned along $S^a$ at $r=0$. Otherwise,
the string tension would become divergent.
The following {\it ansatz} for $a^a$ ensures this behavior:
\beq
a^a=-\frac{\Delta m}{\sqrt{2}}\, \left[\delta^{a3}\, b +S^a\,  S^3\,
(1-b)\right]\, .
\label{aa}
\eeq
Here we introduced a new profile function $b(r)$ which, as usual, will be
determined from a minimization procedure.
Note that at $S^a=(0,0,\pm 1)$ the field $a^a$ is given by its VEV,
as expected. The boundary conditions for  the function $b(r)$ are
\beq
b(\infty)=1\, ,\qquad  b(0)=0\, .
\label{bbc}
\eeq
Substituting Eq.~(\ref{aa}) in conjunction with (\ref{sna}) in the action
(\ref{qed})
we get the potential
\beq
V_{{\rm CP}(1)}=\gamma\int d^2 x\, \frac{\Delta m^2 }{2} \left(1-S_3^2\right)\, ,
\label{sigpot}
\eeq
where $\gamma$ is given by  the integral
\beqn
\gamma
&=&
 \frac{2\pi}{g_2^2} \, \int_0^{\infty}
r\, dr\, \left\{\left(\frac{d}{dr}b (r)\right)^2
+\frac{1}{r^2}\, f_{NA}^2\,  b^2+\right.
\nonumber\\[4mm]
&+&
\left.
g_2^2\left[\frac{1}{2}\,  (1-b)^2\, \left(\phi_1^2+\phi_2^2\right)
+b\, \left(\phi_1-\phi_2\right)^2\right]\right\}\, .
\label{gamma}
\eeqn
Here two first terms in the integrand come from the kinetic term
of  the adjoint scalar field $a^a$  while the term in the square
brackets comes from the  potential in the action (\ref{qed}).

Minimization with respect to $b(r)$, with the constraint (\ref{bbc}),
yields
\beq
b(r)=1-\rho (r)=\frac{\phi_1}{\phi_2} (r)\,,
\label{bobr}
\eeq
cf. Eqs.~(\ref{I}),  (\ref{rhosol}). Thus,
\beq
\gamma =
I \times \frac{2\pi}{g_2^2}=  \frac{2\pi}{g_2^2}\,.
\eeq
We see that the normalization integrals are the same for both, the kinetic
and the potential terms in the worldsheet sigma model, $\gamma=\beta$.
As a result we arrive at  the following effective theory on the string
worldsheet:
\beq
\label{o3mass}
S_{{\rm CP}(1)}=\beta \int d^2 x \left\{\frac12 \left(\pt_k
S^a\right)^2 +\frac{|\Delta m|^2}{2}\,\left( 1-S_3^2\right)\right\}\, .
\eeq
This is the only functional form that allows \ntwo completion.\footnote{
Note, that although the global SU(2)$_{C+F}$ is broken by $\Delta m$,
the extended ${\mathcal N}=2$  supersymmetry is not. }

The fact that we obtain this form shows that our {\em ansatz} is
fully adequate. The informative aspect of the procedure
is (i) confirmation of the {\em ansatz}  (\ref{aa}) and (ii)
constructive calculation of the constant
in front of $\left(1-S_3^2\right)$ in terms of the bulk parameters.
The mass-splitting parameter $\Delta m$ of the bulk
theory exactly coincides with the twisted mass of the worldsheet  model.

\begin{figure}[h]
\epsfxsize=9.5cm
\centerline{\epsfbox{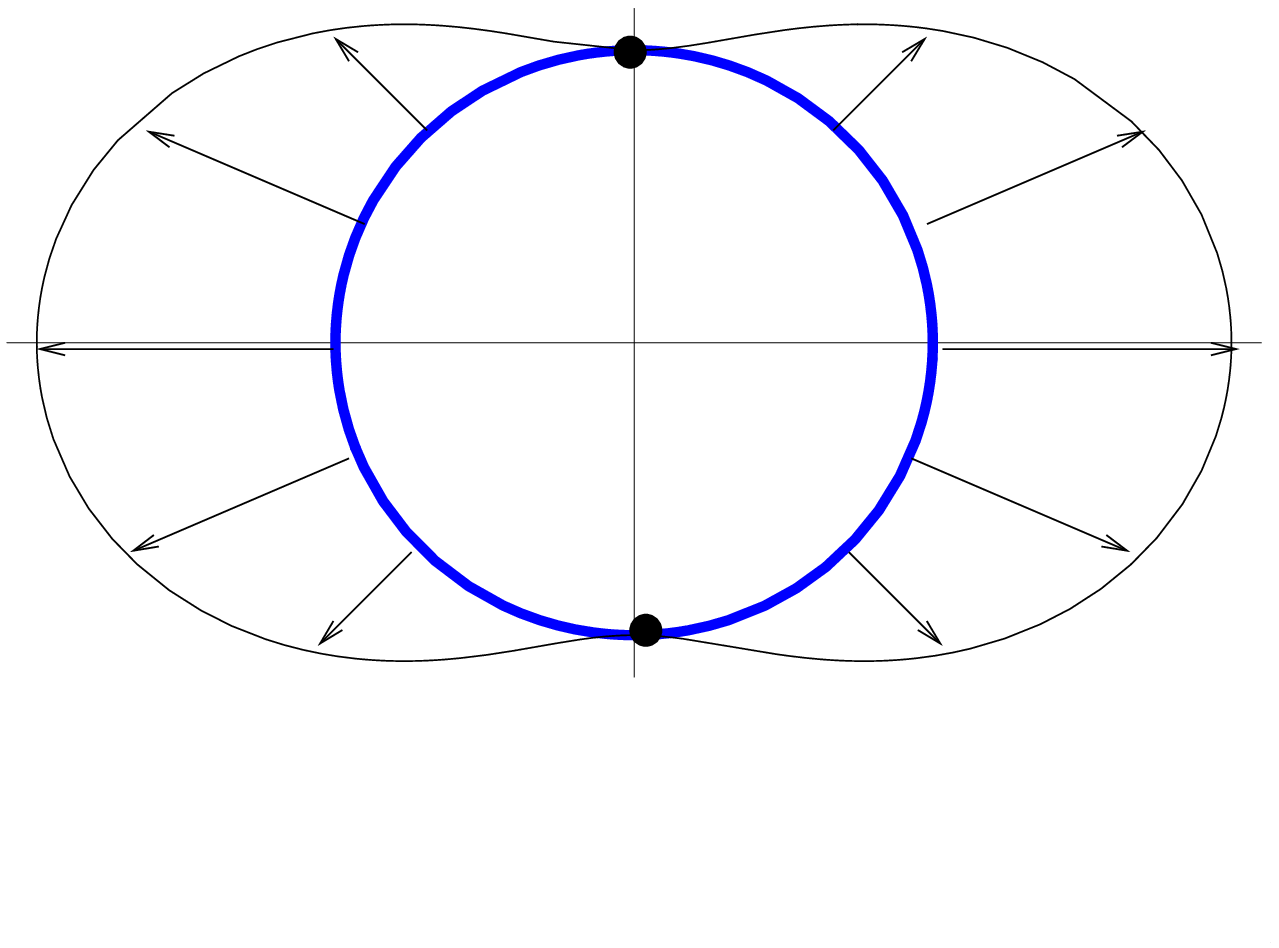}}
 \caption{\small Meridian slice of the target space sphere (thick solid line).
Arrows present the scalar potential in (\ref{o3mass}),
their length being the strength of the potential.
Two vacua of the model are denoted by closed circles.}
\label{f13one}
\end{figure}

The CP(1) model (\ref{o3mass}) has two vacua located at
$S^a=(0,0,\pm 1)$, see Fig.~\ref{f13one}. Clearly these two vacua correspond
to two elementary  $Z_2$ strings.

For generic $N$ the potential in the CP$(N-1)$ model was
obtained in \cite{HT2}. It has the form
\beq
V_{{\rm CP}(N-1)}=2\beta\left\{\sum_{\ell} |\tilde{m}_{\ell}|^2|n^{\ell}|^2-
\left|\sum_{\ell} \tilde{m}_{\ell}|n^{\ell}|^2
\right|^2\right\}\, ,
\label{cppot}
\eeq
where
\beq
\tilde{m}_{\ell}=m_{\ell}-m,\;\;\; m\equiv \frac1N \sum_{\ell}m_{\ell},
\label{tildem}
\eeq
where $\ell =1,...,N$. From the perspective of the bulk theory the index
$\ell$ of the CP$(N-1)$ model coincides with the flavor index, $\ell\equiv A$.
This potential has $N$ vacua (\ref{nvac}) which correspond to $N$ distinct $Z_N$
strings in the bulk theory.

The CP$(N-1)$ model with the potential (\ref{cppot}) is nothing but a bosonic
truncation of the \ntwo two-dimensional sigma model
which was termed the twisted-mass-deformed CP$(N-1)$ model.
This is a generalization of the massless CP$(N-1)$ model which
preserves four supercharges. Twisted chiral superfields in two dimensions
were introduced in \cite{Alvarez} while the twisted mass as an expectation value
of the twisted chiral multiplet was suggested in \cite{HH}. CP$(N-1)$ models
with twisted mass were further studied in \cite{Dorey} and, in particular,
the BPS spectra in these theories were determined exactly.

From the bulk theory standpoint the two-dimensional CP$(N-1)$ model is
an effective worldsheet theory for the non-Abelian string, and the emergence of
\ntwo supersymmetry should be expected.
As we know, the BPS nature of the
strings under consideration does require the worldsheet
theory to have four supercharges.

The twisted-mass-deformed CP$(N-1)$ model can be nicely rewritten as a
strong coupling limit of a U(1) gauge theory \cite{Dorey}. With twisted masses
of the $n^{\ell}$ fields taken into account, the bosonic part of the action
(\ref{cpg}) becomes
\beqn
S
&=&
\int d^2 x \left\{
 2\beta\,|\nabla_{k} n^{\ell}|^2 +\frac1{4e^2}F^2_{kl} + \frac1{e^2}
|\pt_k\sigma|^2
\right.
\nonumber\\[3mm]
 &+& \left. 4\beta\,|\sigma-\frac{\tilde{m}_{\ell}}{\sqrt{2}}|^2 |n^{\ell}|^2 +
 2e^2 \beta^2(|n^{\ell}|^2 -1)^2
\right\}\,.
\label{mcpg}
\eeqn
In the limit $e^2\to \infty$ the $\sigma$ field can be excluded by virtue of an algebraic
equation of motion which leads to the potential (\ref{cppot}).

As was already mentioned, this sigma model gives an effective
description of our non-Abelian string at low energies, i.e. at energies much lower
than the inverse string thickness. Typical momenta in the theory
(\ref{mcpg}) are of the order of $\tilde{m}$. Therefore,
for the action (\ref{mcpg})  to be applicable we must
impose the condition
\beq
\left| \tilde{ m}_{\ell} \right| \ll g_2\sqrt{\xi} \, .
\label{mxi}
\eeq

The description in terms of the twisted-mass-deformed
CP$(N-1)$ model gives us a much
better understanding of dynamics of the non-Abelian strings. If masses
$\tilde{m}_{\ell}$ are much larger than the scale of the CP$(N-1)$ model
$\Lambda_{\sigma}$, the coupling constant $\beta$ is frozen at a large scale
(of the order of masses $\tilde{m}_{\ell}$) and the theory is at weak coupling.
Semiclassical analysis is applicable. The theory (\ref{mcpg}) has
$N$ vacua located at
\beq
n^{\ell}=\delta^{\ell \ell_0},\;\; \sigma=\frac{\tilde{m}_{\ell_0}}{\sqrt{2}},
\;\;\; \ell_0=1,...,N\,.
\label{cpvac}
\eeq
They correspond to the Abelian $Z_N$ strings of the bulk theory, see (\ref{str}).
As we reduce the mass differences $\tilde{m}_{\ell}$ and hit the value
$\Lambda_{\sigma}$, the  CP$(N-1)$ model under consideration
enters the strong coupling regime. At
$\tilde{m}_{\ell}=0$ the global SU$(N)_{C+F}$ symmetry of the bulk theory
is restored. Now $n^{\ell}$  has no particular direction. The condition
(\ref{unitvec}) is relaxed. Still we have $N$ vacua in the worldsheet
theory (Witten's index!). They are seen in the mirror description, see Sect.~\ref{Dyn}.
These vacua  correspond to $N$ elementary non-Abelian
strings in the strong coupling quantum regime.
Thus, we see that for the BPS strings
the transition from the Abelian to non-Abelian regimes is smooth. As we will
discuss in Sect.~\ref{nosusy}, this is not the case for non-BPS strings.
In the latter case the two regimes are separated by a phase transition \cite{GSY05,GSYpt}.

\subsection{Confined monopoles as kinks of the CP$(N-1)$ \\
model}
\setcounter{equation}{0}
\label{kinkmonopole}

Our bulk theory (\ref{qed}) is in the Higgs phase
and therefore the magnetic monopoles of this
theory must be in the confinement phase. If we start from a theory
with the SU$(N+1)$ gauge group  broken to SU$(N)\times$U(1) by condensation
of the adjoint scalar $a$ from which the theory (\ref{qed}) emerges,
the monopoles of the
SU$(N+1)/{\rm SU}(N)\times {\rm U}(1)$ sector can be attached to the endpoints of
the $Z_N$ strings under consideration.
In the bulk theory (\ref{qed}) these monopoles are
infinitely heavy at $m\to\infty$, and hence the $Z_N$ strings are stable.
However, the monopoles residing in the SU$(N)$ gauge group are still present
in the theory (\ref{qed}). As we switch
on the FI parameter $\xi$, the squarks condense triggering confinement of these
monopoles. In this section we will show that these monopoles manifest themselves
as string junctions
of the non-Abelian strings and are seen as kinks in the worldsheet theory
interpolating between distinct vacua of the CP$(N-1)$ model \cite{Tong,SYmon,HT2}.

\begin{figure}
\epsfxsize=9cm
\centerline{\epsfbox{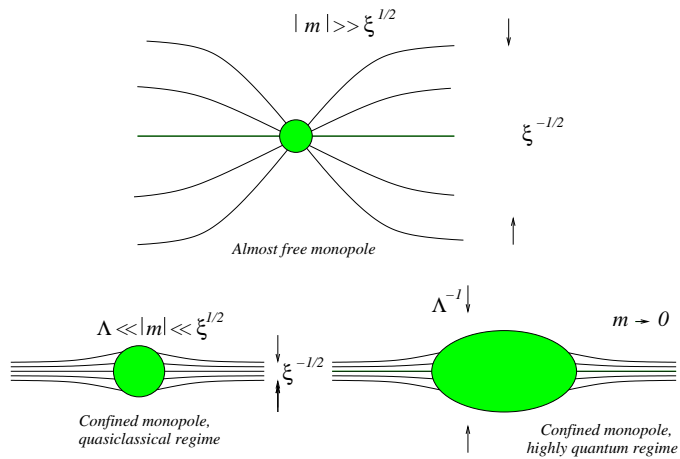}}
\caption{\small
Evolution of the confined monopoles.}
\label{sixf}
\end{figure}

Our task in this section is to trace the evolution of the confined monopoles
starting from the quasiclassical regime, and deep into the quantum regime.
For illustrative purposes it will be even more instructive if we started from the
limit of weakly confined monopoles, when in fact they present
just slightly distorted 't Hooft--Polyakov monopoles (Fig.~\ref{sixf}).

Let us start from the limit $| \Delta{m}_{AB} | \gg \sqrt{\xi }$ and assume all
masses to be of the same order.
In this limit the scalar quark expectation values can be neglected,
and the vacuum structure is determined by VEV's of the adjoint
field $a^a$, see (\ref{avev}). In the non-degenerate case
the gauge symmetry SU($N$) of our bulk model
is broken down to U(1)$^{N-1}$ modulo possible discrete subgroups.
This is the textbook situation for occurrence of the
SU($N$) 't~Hooft--Polyakov monopoles. The monopole core size
is of the order of $| \Delta{m}_{AB}|^{-1}$. The 't Hooft--Polyakov
solution remains valid up to much larger distances, of the order
of $\xi^{-1/2}$. At distances larger than $\sim \xi^{-1/2}$
the quark VEV's become important. As usual, the U(1)
charge condensation leads to the formation of the U(1) magnetic flux tubes,
with the transverse size
of the order of $\xi^{-1/2}$ (see the upper picture in Fig.~\ref{sixf}).
The flux is quantized; the flux tube tension is tiny in
the  scale of the square of the monopole mass. Therefore, what we deal with
in this limit is basically a very weakly confined 't Hooft--Polyakov monopole.

Let us verify that the confined monopole is a junction of two strings.
Consider the junction of two $Z_N$ strings   corresponding to two
``neighboring'' vacua of the CP$(N-1)$ model. For the $\ell_0$-th vacuum $n^{\ell}$
is given by (\ref{cpvac}) while for the $\ell_0 +1$-th vacuum it is given by the
same equations with $\ell_0\to\ell_0 +1$. The flux of this junction is
given by the difference of the fluxes of these two strings. Using (\ref{str})
we get that the flux of the junction is
\beq
  4\pi\,\times \, {\rm diag} \, \frac12\, \left\{  ...\, 0, \,1 ,\,  -1,\, 0
,\, ... \right\} \,,
\label{monflux}
\eeq
with the non-vanishing entries located at
positions $\ell_0$ and $\ell_0 +1$.
These are exactly the  fluxes of $N-1$ distinct 't Hooft--Polyakov
monopoles occurring in the
SU($N$) gauge theory provided that SU($N$)
is spontaneously broken down to U(1)$^{N-1}$. We see that
in the quasiclassical limit of large $| \Delta{m}_{AB} | $ the Abelian
monopoles play the role of junctions
of the Abelian $Z_N$ strings. Note that in various models
the monopole fluxes  and those of
strings were shown  to match each other \cite{Bais,d-one,d-two,PrVi,MY,Kn,Konishi}
so that the monopoles can be confined by strings in the Higgs phase.

Now, let us reduce $\left| \Delta{m}_{AB} \right| $.  If this parameter is
limited inside the interval
\beq
\Lambda \ll \left|  \Delta{m}_{AB} \right|   \ll \sqrt{\xi}\, ,
\label{wcregime}
\eeq
the size of the monopole ($\sim \left| \Delta{m}_{AB} \right|^{-1} $) becomes
larger than the transverse size of the attached strings.
The monopole gets squeezed in earnest by
the strings --- it becomes  a {\em bona fide} confined monopole (the
lower left corner of  Fig.~\ref{sixf}).
A natural question is how this confined monopole is seen in the effective
two-dimensional CP$(N-1)$ model  (\ref{mcpg}) on the string worldsheet.
As soon as the $Z_N$ strings of the bulk theory correspond to $N$ vacua of the
$CP^{N-1}$ model the string junction (confined monopole) is a ``domain wall" ---
kink --- interpolating between these vacua, see Fig.~\ref{f13one}.

Below we will explicitly demonstrate that in the semiclassical
regime (\ref{wcregime}) the solution for the string junction in the bulk
theory is in one-to-one correspondence with
the kink in the worldsheet theory. Then we will show that the
masses of the monopole and kink perfectly match. This was demonstrated
in \cite{SYmon} in the
$N=2$ case.

\subsubsection{The first-order equations for the string junction}
\label{foeqsj}

In this section we derive the first-order equations for the 1/4-BPS
junction of the $Z_N$ strings in the  SU$(N)\times$U(1) theory  in the
quasiclassical limit (\ref{wcregime}).
In this limit $\Delta{m}_{AB} $ is sufficiently small so that we can use
our effective
low-energy description in terms of the twisted-mass-deformed CP$(N-1)$ model (\ref{mcpg}).
On the other hand, $\Delta{m}_{AB} $  is
much larger then the dynamical scale of the CP$(N-1)$ model; hence, the latter is in the weak
coupling regime which  allows one to apply quasiclassical treatment.

The geometry of our junction is shown in the left corner of Fig.~\ref{sixf}.
Both strings are stretched along the $z$ axis. We assume that the monopole
sits near the origin, the $n^{\ell}=\delta^{\ell \ell_0}$-string is at negative
$z$ while the  $n^{\ell}=\delta^{\ell \ell_0+1}$-string is at positive $z$.
The perpendicular plane is parametrized by
$x_1$ and   $x_2$. What is sought for is  a static solution
of the BPS equations, with all relevant fields  depending  only on $x_1$,
$x_2$ and $z$.

Ignoring the time variable
we can represent the energy functional of our theory (\ref{redmodel})
as follows (the Bogomol'nyi representation \cite{B}):
\beqn
E
&=&
\int{d}^3 x   \left\{
\left[\frac1{\sqrt{2}g_2}F^{*a}_{3} +
\frac{g_2}{2\sqrt{2}}
\left(\bar{\vp}_A\tau^a \vp^A\right)
+ \frac1{g_2}D_3 a^a\right]^2
\right.
\nonumber\\[3mm]
&+&
\left[\frac1{\sqrt{2}g_1}F^{*}_{3} +
\frac{g_1 }{2\sqrt{2}}
\left(|\vp^A|^2-2\xi \right)
+ \frac1{g_1}\pt_3 a\right]^2
\nonumber\\[4mm]
&+&
\frac1{g_2^2}\left|\frac1{\sqrt{2}}(F_1^{*a}+iF_2^{*a})+(D_1+iD_2)a^a
\right|^2
\nonumber\\[4mm]
&+&
\frac1{g_1^2}\left|\frac1{\sqrt{2}}(F_1^{*}+iF_2^{*})+(\pt_1+i\pt_2)a
\right|^2
\nonumber\\[5mm]
&+&
 \left|\nabla_1 \,\vp^A +
i\nabla_2\, \vp^A\right|^2
\nonumber\\[4mm]
&+&
\left. \left|\nabla_3 \vp^{A}+\frac1{\sqrt{2}}\left(a^a\tau^a +a
 +\sqrt{2}m_{A}
\right)\vp^A\right|^2\right\}
\label{bogj}
\eeqn
plus surface terms. As compared to the Bogomol'nyi representation (\ref{bogs})
for strings we keep here also terms involving the adjoint fields.
Following our conventions we assume
the quark mass terms to be real implying that  the vacuum expectation values
of  the adjoint scalar fields are real too.
The surface terms mentioned above are
\beq
\left.
E_{\rm surface}=
\xi  \int d^3 x F^{*}_3
+\sqrt{2}\, \xi \int d^2 x \, \langle a\rangle\right|^{z=\infty}_{z=-\infty}
-\sqrt{2}\, \frac{\langle a^a\rangle}{g_2^2}\int d S_n\,  F^{*a}_{n},
\label{surface}
\eeq
where the integral in the last term runs over a large two-dimensional
sphere
at $\vec x^{\, 2} \to \infty$. The first term on the right-hand side
is related to strings, the second to domain walls, while the third to
monopoles (or the string junctions).

The Bogomol'nyi representation (\ref{bogj})
leads us to the following first-order equations:
\beqn
&& F^{*}_1+iF^{*}_2 + \sqrt{2}(\pt_1+i\pt_2)a=0\, ,
\nonumber\\[3mm]
&& F^{*a}_1+iF^{*a}_2 + \sqrt{2}(D_1+iD_2)a^a=0\, ,
\nonumber\\[3mm]
&& F^{*}_{3}+\frac{g_1^2}{2} \left(\left|
\varphi^{A}\right|^2-2\xi\right) +\sqrt{2}\, \pt_3 a =0\, ,
\nonumber\\[3mm]
&& F^{*a}_{3}+\frac{g_2^2}{2} \left(\bar{\vp}_{A}\tau^a \varphi^{A}\right)
+\sqrt{2}\, D_3 a^a =0\, ,
\nonumber\\[3mm]
&& \nabla_3 \vp^A =-\frac1{\sqrt{2}}\left(a^a\tau^a+a+\sqrt{2}m_A\right)
\vp^A\, ,
\nonumber\\[4mm]
&& (\nabla_1+i\nabla_2)\varphi^A=0\, .
\label{foej}
\eeqn
These are our {\it master equations}.
Once these equations are satisfied the energy of the BPS object is
given by Eq.~(\ref{surface}).

Let us discuss the central charges (the surface terms)
of the string, domain wall and
monopole in more detail.  Say, in the
string case, the three-dimensional integral in the first term in
Eq.~(\ref{surface}) gives the length of the string times its flux. In  the
wall case, the two-dimensional integral in the second term in (\ref{surface})
gives the area of the wall times its tension. Finally, in the
monopole case the integral in the last term
in Eq.~(\ref{surface}) gives the magnetic-field flux. This means that
the first-order master equations (\ref{foej}) can be used to study
{\em strings, domain walls,  monopoles and all their possible junctions}.

It is instructive to  check that  the wall, the string and the monopole
solutions, separately, satisfy these equations. For the domain wall  this
check was done in \cite{SYnawall} where we used these equations to
study the string-wall junctions (we review this in Sect. \ref{wallstrj}).
 Let us consider the string solution. Then
the scalar fields $a$ and $a^a$ are  given by their VEV's.
The gauge flux is directed along the $z$ axis, so
that $ F^{*}_1=F^{*}_2= F^{*a}_1=F^{*a}_2=0$.
All fields depend only on the perpendicular
coordinates $x_1$ and $x_2$. As a result,  the first
two equations and the fifth  one  in (\ref{foej}) are trivially
satisfied. The third and the fourth  equations
reduce to the first two equations in Eq.~(\ref{gfoes}). The last equation
in (\ref{foej}) reduces to the last equation in (\ref{gfoes}).

Now, turn to the monopole solution. The 't Hooft--Polyakov monopole
equations \cite{thooftmon,polyakov}  arise from those in Eq.~(\ref{foej}) in
the limit
$\xi=0$. Then all quark fields vanish,  and Eq.~(\ref{foej}) reduces to
the standard first-order equations for the BPS 't~Hooft--Polyakov monopole
( see Sect.~\ref{mafe}),
\beq
F^{*a}_k + \sqrt{2}\, D_k\,  a^a=0 \, .
\label{monopole}
\eeq
The U(1) scalar field $a$ is given by its VEV while the U(1) gauge field
vanishes.

Now, Eq.~ (\ref{surface}) shows that the central charge
of the SU(2) monopole is determined by $\langle a^a\rangle$ which is
proportional to the quark mass difference, see (\ref{avev}).
 Thus, for the monopole
on the Coulomb branch (i.e. at $\xi =0$) Eq.~ (\ref{surface})
yields
\beq
M_M=\frac{4\pi( m_{\ell_0+1}- m_{\ell_0})}{g_2^2}\, .
\label{mmCb}
\eeq
This coincides, of course, with the  Seiberg--Witten result
\cite{SW1} in the weak coupling limit. As we will see shortly,
the same expression continues to hold even if $\Delta{m}_{AB}\ll \sqrt\xi$
(provided that $\Delta{m}_{AB}$ is still much larger than $\Lambda_{{\rm SU}(N)}$).
An explanation will be given in Sect.~\ref{dsw}.

The Abelian version of the first-order equations (\ref{foej})
were   derived in Ref. \cite{SYwall} where they were exploited
to find the 1/4 BPS-saturated solution for the wall-string  junction.
The non-Abelian equations (\ref{foej})
in the  SU(2)$\times$ U(1) theory were derived in \cite{Tong}
where the confined monopoles as a string junctions were considered
at $\Delta m \neq 0$. Then the
non-Abelian equations (\ref{foej}) were  extensively
used  in the  analysis \cite{SYnawall} of the wall-string
junctions in the problem of non-Abelian strings  ending on a stack of domain walls,
see Sect. \ref{wallstrj}. Next, Eqs. (\ref{foej}) for
the confined monopoles as string junctions  were solved in \cite{SYmon} in the
SU(2)$\times$ U(1) theory. Below we will review this solution. Later
all 1/4 BPS solutions for junctions
(in particular, semilocal string junctions) were found in \cite{J14}.

\subsubsection{The string junction solution in the quasiclassical regime}
\label{stringjunctionsolution}

Now we will apply our master equations at $N=2$ in order
to find the junction of the $S^a=(0,0,1)$ and
$S^a=(0,0,-1)$-strings  via an SU(2) monopole
in the quasiclassical limit. We assume that
the $S^a=(0,0,1)$-string is at negative $z$, while the
$S^a=(0,0,-1)$-string is at positive $z$.
We will  show that the solution of
the BPS equations (\ref{foej}) of the four-dimensional  bulk
theory is determined by the kink
solution in the two-dimensional  sigma model (\ref{o3mass}).

To this end   we will look for the solution of equations (\ref{foej}) in the
following {\it ansatz}. Assume that the solution for the string
junction is given, to the leading order in $\Delta m/\sqrt{\xi}$,
by the same string configuration (\ref{sna}), (\ref{An}) and (\ref{aa})
which we dealt with previously
(in the case   $\Delta m \neq 0$)
with $S^a$    slowly-varying
functions of $z$, to be determined below,  replacing the  constant
moduli vector $S^a$.

Now  the functions $S^a(z)$ satisfy the
boundary condition
\beq
S^a(-\infty) = (0,0,1)\,,
\label{-infty}
\eeq
while
\beq
S^a(\infty) = (0,0,-1)\,.
\label{infty}
\eeq
This  {\it ansatz}  corresponds to the non-Abelian string
in which the vector $S^a$ slowly rotates from (\ref{-infty}) at
$z\to-\infty$  to  (\ref{infty}) at  $z\to\infty$. We will show that
the representation (\ref{sna}), (\ref{An}) and (\ref{aa})  solves the master
equations (\ref{foej}) provided the functions $S^a (z)$ are chosen
in a special way.

Note that the first equation in (\ref{foej}) is trivially
satisfied because the field $a$ is  constant and $F^{*}_1=F^{*}_2=0$.
The last equation reduces to the first two  equations in (\ref{foes})
because it does not contain derivatives with respect to $z$
and, therefore, is satisfied for arbitrary functions $S^a(z)$.
The same remark applies also to the third equation in Eq.~(\ref{foej}),
which reduces to  the third equation in  (\ref{foes}).

Let us inspect the fifth equation in  Eq.~(\ref{foej}).
Substituting our {\it ansatz} in this equation and using
the formula (\ref{rhosol}) for $\rho$  we find that this equation
is satisfied provided $S^a (z) $ are chosen to be the  solutions of the
equation
\beq
\pt_3 S^a=\Delta m \left(\delta^{a3}-S^a S^3\right)\, .
\label{kinkeq}
\eeq
Below we will show that these equations are nothing but the
first-order kink equations in the massive CP(1) model.

By the same token, we can consider the second equation in  (\ref{foej}).
Upon substituting there our {\em ansatz},  it reduces to
Eq.~(\ref{kinkeq}) too.
Finally, consider the fourth equation in  (\ref{foej}). One can see that
in fact
it contains an
expansion in the parameter $\Delta m^2/\xi$. This means  that the solution
we have just built  is not exact; it has corrections
of the order of $O(\Delta m^2/\xi)$. To the leading order in this parameter
the fourth equation in (\ref{foej}) reduces to the last equation
in (\ref{foes}). In principle, one could go beyond
the leading order.
Solving the fourth equation in (\ref{foej}) in the next-to-leading
order would allow one  to determine $O(\Delta m^2/\xi)$
corrections to our solution.

Let us dwell on the meaning of Eq. ~(\ref{kinkeq}). This equation
is nothing but
the equation for the kink in the CP(1) model (\ref{o3mass}).
 To see this let us write the Bogomol'nyi representation for kinks in the model
(\ref{o3mass}).  The energy functional can be
rewritten as
\beq
E= \frac{\beta}{2}\int d z \left\{\left| \pt_z S^a-
\Delta m\left (\delta^{a3}-S^a S^3 \right)\right|^2
+2\Delta m\,  \pt_z S^3\right\}\,.
\label{bogcp1}
\eeq
The above  representation implies the first-order equation (\ref{kinkeq})
for the BPS-saturated kink. It also yields $2\beta\Delta m$ for the kink mass.

Thus, we have demonstrated  that the  solution describing the junction
of  the $S^a=(0,0,1)$ and
$S^a=(0,0,-1)$ $Z_2$ strings is given by the non-Abelian string
with a slowly varying orientation vector $S^a$. The variation of $S^a$
is described in terms of the kink solution of the (1+1)-dimensional
CP(1) model with the twisted mass.

In conclusion, we would like to match the masses of the
four-dimensional mono\-pole and two-dimensional
kink. The string mass and that of the string junction is
given by the first and the last
terms in the surface energy (\ref{surface}) (the second term
vanishes). The first term obviously reduces to
\beq
 M_{\rm string}=2\, \pi\, \xi\,\, L,
\label{stringm}
\eeq
i.e.  proportional to the total string length $L$. Note that both the
$S^a=(0,0,1)$ and $S^a=(0,0,-1)$ strings have the same tension (\ref{ten}).
The third term should give the mass of the the monopole.
The surface integral in this term reduces
to the flux of the $S^a=(0,0,-1)$-string  at $z\to\infty$ minus the flux
of the $S^a=(0,0,1)$-string  at $z\to -\infty$. The $F^{*3} $ flux of the
$S^a=(0,0,-1)$-string is $2\pi$ while the $F^{*3} $ flux of the
 $S^a=(0,0,1)$-string is $-2\pi$. Thus, taking into account Eq.~(\ref{avev}),
 we get
\beq
M_M=\frac{4\pi}{g_2^2}\, \Delta m\, .
\label{mm}
\eeq
Note, that although  we discuss the monopole in the confinement
phase  at $\left| \Delta m\right| \ll \sqrt{\xi}$ (in this phase it is
a junction of two strings), nevertheless the $\Delta m$ and $g^2_2$
dependence of its mass coincides with the result (\ref{mmCb})
for the unconfined monopole on the Coulomb branch (i.e. at $\xi=0$).
This is no accident --- there is a deep theoretical reason explaining the
validity of this unified formula. A change occurs only in passing to a
highly quantum regime depicted in the right
lower corner of Fig.~\ref{sixf}.  We will discuss this regime shortly
in Sect.~\ref{tscl}.

It is instructive to compare Eq.~(\ref{mm}) with the kink mass in the effective
CP(1) model on the string worldsheet. As was mentioned,
the surface term in Eq.~(\ref{bogcp1}) gives
\beq
M_{\rm kink}=2\,\beta\, \Delta m  \,.
\label{exhau}
\eeq
Now,
expressing the two-dimensional coupling constant $\beta$ in terms of
the coupling constant of the microscopic theory, see Eq.~(\ref{beta}),
we  obtain
\beq
M_{\rm kink}=\frac{4\pi}{g^2_2}\Delta m\, ,
\label{mk}
\eeq
thus verifying that the four-dimensional
calculation of $M_M$ and the two--dimensional
calculation of $M_{\rm kink}$ yield the same,
\beq
M_M=M_{\rm kink}\, .
\label{monkink}
\eeq
Needless to say, this is in full accordance with the physical picture
that emerged from our analysis, that the two-dimensional CP(1)
model is nothing but the macroscopic description
of the confined monopoles occurring in the four-dimensional
microscopic Yang--Mills theory.
Technically the coincidence of the monopole and kink masses
is based on the fact that the integral in the
definition (\ref{betaI}) of the sigma-model coupling $\beta$ reduces to unity.

\subsubsection{The strong coupling limit}
\label{tscl}

Here we will consider the limit of small $\Delta{m}_{AB}$, when
the effective worldsheet theory develops a strong coupling regime.
For illustrative purposes we will consider the simplest case, $N=2$.
Generalization to generic $N$ is straightforward.

As we further diminish $\left| \Delta m\right|$
approaching $\Lambda_{\sigma}$ and then send $\Delta m$ to zero we
restore the global SU(2)$_{C+F}$ symmetry. In particularly, on the Coulomb
branch, the  SU(2)$\times$U(1) gauge symmetry is restored. The monopole becomes
a truly non-Abelian object. In this limit
the monopole size grows, and, classically, it would explode.
Moreover, the classical formula (\ref{mm}) interpreted
literally  shows that the monopole mass vanishes
(see the discussion of the so-called ``monopole clouds'' in \cite{We} for a
review of the long-standing issue of understanding what becomes of the monopoles
upon restoration of the non-Abelian gauge symmetry).
Thus, classically one would say that the monopoles disappear.

That's where quantum effects on the confining string take over.
As we will explain below,
they materialize the confined non-Abelian monopole as a well defined
stable object \cite{SYmon}.
From the standpoint of the effective worldsheet theory
this $\Delta m$ domain of presents a regime of
highly quantum worldsheet dynamics.

While the string thickness (in the transverse direction) is
$\sim \xi ^{-1/2}$, the
$z$-direction size of the kink representing the confined
monopole in the highly quantum regime is much larger,
$\sim \Lambda^{-1}_{\sigma}$,
see the lower right corner in  Fig.~\ref{sixf}. Still, it
remains finite in the limit $\Delta m\to 0$, stabilized by non-perturbative
effects in the worldsheet CP(1) model. Please, remember that the CP$(N-1)$
models develop a mass gap, and no massless states are present in the spectrum,
see Sect.~\ref{Dyn}. Moreover, the mass of the confined monopole (the CP(1) model kink)
is also determined by the scale $\Lambda_{\sigma}$.
This sets the notion of what the confined  non-Abelian monopole is.
It is a kink in the massless two-dimensional CP$(1)$ model
\cite{SYmon}.

We can get a more quantitative insight in physics of the worldsheet
theory at strong coupling if we invoke the exact BPS spectrum of the
twisted-mass-deformed CP$(N-1)$ model obtained in \cite{Dorey}. It was derived
\cite{Dorey} by  generalizing Witten's analysis \cite{W93} that had been carried out
previously for the massless case. The  BPS states saturate
the central charge $Z$ defined in Eq.~(\ref{compalg}). The
exact formula for this central charge is
\beq
Z_{2d} = i\, \Delta m \, q + m_D\, T\,,
\label{cpcc}
\eeq
where the subscript $2d$ reminds us that the model in question is
two-dimen\-sional. The subscript $D$ in $m_D$ appears for
historical reasons, in parallel with the Seiberg--Witten solution
(it stands for dual). Furthermore,
$T$ is the topological charge of the
kink under consideration, $T=\pm 1$,
while the parameter $q$
\beq
q = 0,\,\,\pm 1,\,\, \pm 2\,, ...
\label{qez}
\eeq
This global U(1) charge of the ``dyonic'' states arises due the presence
of a U(1) group unbroken in (\ref{o3mass}) by the the twisted mass (the
SU(2)$_{C+F}$ symmetry is broken down to U(1) by $\Delta m\neq 0$).

The quantity $m_D$ was introduced \cite{Dorey} in analogy
with $a_D$ of Ref.~\cite{SW1}. In the case $N=2$ it has the form
\beq
m_D =\frac{\Delta m}{\pi}\left[
\frac{1}{2}\,
\ln\frac{\Delta m  + \sqrt{\Delta m^2+4\Lambda_{\sigma}^2}}{\Delta m -
 \sqrt{\Delta m^2+4\Lambda_{\sigma}^2}} -
\sqrt{1+\frac{4\Lambda_{\sigma}^2 }{\Delta m^2}}\,\, \right]\,,
\label{md}
\eeq
where $\Delta m$ is  now assumed to be complex.
The two-dimensional central charge is normalized in such a way that
$M_{\rm kink} = |Z_{2d}|$.

The limit $|\Delta m| /\Lambda_{\sigma} \to \infty$
corresponds to the quasiclassical domain, while corrections
of the type $(\Lambda_{\sigma}/\Delta m)^{2k}$ are induced by instantons.

What happens when one travels from the domain of large $|\Delta m |$ to that
of small $|\Delta m |$?
If $\Delta m =0$ we know (e.g.  from the mirror representation \cite{HoVa})
that there are two degenerate two-dimensional kink supermultiplets,
corresponding to the Cecotti--Fendley--Intriligator--Vafa (CFIV)
index = 2 \cite{CFIV}. They have quantum numbers
$\left\{q,\,\, T\right\} =(0,1)$   and $(1,1)$, respectively.
Away from the point $\Delta m = 0$ the masses of these states are no longer
equal; there is  one singular point with one of
the two states becoming massless \cite{SVZ06}.
The region containing the point $\Delta m = 0$
is separated from the quasiclassical region of large $\Delta m$
by the  curve of marginal stability (CMS)
on which an infinite number of other BPS states, visible
quasiclassically, decay.
Thus, the infinite tower of the $\left\{q,\,\, T\right\} $
BPS states existing in the quasiclassical domain degenerates in
just two stable BPS states in the vicinity of $\Delta m = 0$.

As we outlined above, there are no massless states in the CP(1) model
at $\Delta m =0$.
In particular, the kink (confined monopole) mass   is
 \beq
M_M=\frac{2}{\pi}\Lambda_{\sigma},
\label{strongmm}
\eeq
as it is clear  from  (\ref{md}).
On the other hand, in this limit both the last term in (\ref{surface})
and the surface term in (\ref{bogcp1}) vanish for
the monopole and kink masses, respectively. What's wrong?

This puzzle was solved by the following observation:
anomalous terms in the central charges of both four-dimensional
and two-dimensional SUSY algebras are present in these theories.
In two dimensions the anomalous terms were obtained in \cite{LosSh,SVZ06}.
In four dimensions the bifermion anomalous term was discovered
in \cite{SYmon}. We refer the reader to Sect.~\ref{satmcc}
for a more detailed discussion.

In the bulk
theory the central charge associated with the monopole
is defined through the anticommutator
\beq
\{\bar{Q}_{\dot{\alpha}}^f\,  \bar{Q}_{\dot{\beta}}^g\} =2\,
\varepsilon_{\dot{\alpha}\dot{\beta}}\,
\varepsilon^{fg}\,  \bar{Z}_{4d}\,,
\eeq
where $\bar{Z}_{4d}$ is an SU(2)$_R$ singlet; the subscript $4d$
will remind us of four dimensions. It is most convenient to write
$\bar{Z}_{4d}$ as a topological charge (i.e. the integral over a topological
density),
\beq
\bar{Z}_{4d} = \int \, d^3 x \, \bar{\zeta}^0 (x)\,.
\label{4dcc}
\eeq

In the model at hand
\beqn
&&\bar{\zeta}^\mu
=
\frac{1}{\sqrt{2}}\varepsilon^{\mu\nu\rho\sigma}
\,\partial_\nu\left(
\frac{i}{g_2^2}\, a^a F^a_{\rho\sigma} + \frac{i}{g_1^2}\, a F_{\rho\sigma}
-\frac{i}{2\pi^2}\, a^a F^a_{\rho\sigma}
\right.
\nonumber\\[4mm]
&&+
\left.
\frac{i}{8\sqrt{2}\pi^2}\left[
\lambda^a_{f\alpha}(\sigma_{\rho})^{\alpha\dot\alpha}
(\bar{\sigma}_{\sigma})_{\dot\alpha\beta} \lambda^{af\beta}+
2g_2^2\tilde\psi_{A\alpha}(\sigma_{\rho})^{\alpha\dot\alpha}
(\bar{\sigma}_{\sigma})_{\dot\alpha\beta}\psi^{A\,\beta}
\right]
\right)\,.
\nonumber\\
\label{4anom}
\eeqn
Note that the  general structure of the operator in the
square brackets is unambiguously fixed by
dimensional arguments, the Lorentz symmetry and other symmetries
of the bulk theory. The numerical coefficient was first found in \cite{SYmon}
by matching the monopole and kink masses at $\Delta m=0$.
We also include the bosonic part of the anomaly term associated with
magnetic field here (last term in the first line), see  Sect.~\ref{satmcc}.

The above expression is an operator equality.
In the low-energy limit, the Seiberg--Witten exact solution
allows one to obtain the full matrix element of the operator on the right-hand side
(which includes
all perturbative and non-perturbative corrections)
by replacing $a$ by $a_D$.

The fermion part of the anomalous term plays a crucial role in the Higgs phase
for the confined monopole.
On the Coulomb branch it does not contribute to the
mass of the monopole due to a fast fall off of
the fermion fields at infinity.
On the Coulomb branch the  bosonic   anomalous terms become important.
The relationship between the 't~Hooft--Polyakov monopole mass
and the ${\mathcal N}=2$ central charge is analyzed
in  \cite{Rebhan:2004vn}, which identifies
an anomaly in the central charge explaining
a constant (i.e. non-logarithmic) term in the monopole mass
on the Coulomb branch. The result of Ref.~\cite{Rebhan:2004vn} is in
agreement with the  Seiberg--Witten formula for the monopole mass.
In Sect.~\ref{satmcc} we presented  the operator form of the
central charge anomaly. Note, that the coefficient in front of fermionic
term involving $\lambda$-fermions in (\ref{4anom})coincides with the
one in (\ref{cccprimm}) obtained
by supersymmetrization of the bosonic anomalous term.

\subsection{Two-dimensional kink and
four-dimensional Seiberg--Witten solution}
\setcounter{equation}{0}
\label{dsw}

Why  the 't Hooft--Polyakov  monopole mass (i.e. that on the Coulomb branch
at $\xi=0$) is given by the same formula (\ref{mmCb}) as  the mass
(\ref{mm})  of
the strongly confined  large-$\xi$ monopole
(subject to condition $\sqrt{\xi}\gg\Delta m$)?

This fact was noted  in
Sect.~\ref{stringjunctionsolution}. Now we will explain the reason lying
behind this observation \cite{SYmon,HT2}.
 {\em En route}, we will explain another striking
observation made in Ref.~\cite{Dorey}. A remarkably close parallel between
the
four-dimensional Yang--Mills theory with $N_f=2$ and the two-dimensional
CP(1) model was noted, at an observational level,
by virtue of comparison of  the corresponding central charges.
The observation was made on the Coulomb branch
of the Seiberg--Witten theory, with unconfined monopoles/dyons
of the 't Hooft--Polyakov type. Valuable as it is,
the parallel was quite puzzling since the solution of the  CP(1) model
seemed to have no physics connection to the Seiberg--Witten solution.
The latter gives the mass of the unconfined
monopole in the Coulomb regime  at $\xi=0$
while the CP(1) model emerges only in the  Higgs regime of the bulk theory.

We want to show, that the reason for the correspondence
mentioned above is that in the BPS sector
(and {\em only} in this sector) the parameter $\xi$, in fact, does not
appear in relevant formulae. Therefore, one can vary $\xi$ at will, in particular,
making it less than  $| \Delta m | $ or even tending to zero,
where CP(1) is no more the string worldsheet theory for our bulk model.
Nevertheless, the parallel expressions for the central charges
and other BPS data in four dimensions and two dimensions, trivially
established at $| \Delta m | \ll \xi$, will continue to hold
even on the Coulomb branch. The  ``strange coincidence" we observed in
Sect.~\ref{stringjunctionsolution} is no accident. We deal with
an exact relation which stays valid including both
perturbative and non-perturbative corrections.

Physically the monopole in the Coulomb phase is very different from the one
in the confinement phase, see Fig.~\ref{sixf}. In the Coulomb phase it is
a 't~Hooft--Polyakov monopole, while in the confinement phase it becomes
related to a junction of two non-Abelian strings. Still let us show
that the masses of these two objects are given by the same expression,
\beq
M_M^{\rm Coulomb }= M_M^{\rm confinement }
\label{mmCc}
\eeq
provided that $\Delta m$ and the gauge couplings are kept fixed.
The superscripts refer to the Coulomb and monopole-confining phases,
respectively.

The crucial observation here is that the mass of the monopole cannot depend
on the FI parameter $\xi$. Start from the monopole in the Coulomb phase
at $\xi=0$. Its mass is given by  the exact Seiberg--Witten formula \cite{SW2}
\beqn
M_M^{\rm Coulomb}
&=&
\sqrt{2}\left| a_D^3\, \left(a^3=-\frac{\Delta m}{\sqrt{2}} \right) \right|
\nonumber\\[4mm]
&=&
\left|\frac{\Delta m}{\pi}\ln{\frac{\Delta m}{\Lambda_{{\rm SU}(2)}}}
+\Delta m\sum_{k=1}^{\infty}
c_k\left(\frac{\Lambda}{\Delta m}\right)^{2k}\right|\, ,
\label{mmSW}
\eeqn
where $a_D^3$ is the dual Seiberg--Witten  potential for the SU(2)
gauge group. We take into account the fact that for $N_f=2$ the first
coefficient of the $\beta$ function is  2.

In Eq.~(\ref{mmSW})
$a^3=- {\Delta m}/\sqrt{2}$ is the
argument of $a_D^3$, the logarithmic term takes into account the
one-loop result  (\ref{coupling}) for the
SU(2) gauge coupling at the scale $\Delta m$, while the power
series represents instanton-induced terms ---  small corrections at large $a$.

Now, if we switch on a small FI parameter $\xi\neq 0$ in the theory,
on dimensional grounds we could expect  corrections to
the monopole mass  in powers of $\sqrt{\xi}/\Lambda_{{\rm SU}(2)}$
and/or $\sqrt{\xi}/\Delta m$ in Eq.~(\ref{mmSW}).

But ... these corrections are {\em forbidden}
by the U(1)$_R$ charges. Namely, the U(1)$_R$ charges of
$\Lambda_{{\rm SU}(2)}$ and
$\Delta m$ are equal to 2 (and so is the U(1)$_R$ charge of the
central charge under consideration)
while $\xi$ has a vanishing U(1)$_R$ charge.
For convenience, the U(1)$_R$ charges of different fields and parameters
of the microscopic theory are collected in Table~\ref{table2}. Thus, neither
$(\sqrt{\xi}/\Lambda_{{\rm SU}(2)} )^k $ nor $(\sqrt{\xi}/\Delta m)^k$ can appear.

\begin{table}
\begin{center}
\begin{tabular}{|c|c | c| c|c| c | c | c| c|}
\hline
Field/parameter  & $a$ & $a^a$ & $\lambda^{\alpha}$ & $q$ &
$\psi^{\alpha}$ & $m_{A}$ & $\Lambda_{SU_{N}})$ & $\xi$
\\[3mm]
\hline
U(1)$_R$ charge & 2 & 2 & 1 & 0 & $-1$
& 2 & 2 & 0
\\[2mm]
\hline
\end{tabular}
\end{center}
\caption{\small The U(1)$_R$
charges of fields and parameters of the bulk theory.}
\label{table2}
\end{table}

By the same token, we could start from the confined monopole at large $\xi$,
and study the dependence of the monopole (string junction) mass
as a function of $\xi$ as we reduce $\xi$. Again, the above  arguments
based on the U(1)$_R$  charges tell us  that no corrections in powers
of  $\Lambda_{{\rm SU}(2)}/\sqrt{\xi}$ and $\Delta m/\sqrt{\xi}$  can appear.
This leads us to  Eq. (\ref{mmCc}).

Another way to arrive at the same conclusion is to observe that
the monopole mass depends on $a$ (anti)holomorphically,
cf. Seiberg--Witten's formula
(\ref{mmSW}). Thus, it
cannot depend on the FI parameter $\xi$ which is not holomorphic
(it is a component of the SU(2)$_R$ triplet \cite{matt,VY}).

Now let us turn to the fact that the mass of the monopole in the confinement phase
is given by the  kink mass   in the CP(1) model, see (\ref{monkink}). In this way we obtain
\beq
M_M^{\rm Coulomb }\leftrightarrow M_M^{\rm confinement }
\leftrightarrow M_{\rm kink}\,.
\label{mmk}
\eeq
In particular, at one loop, the kink
mass is determined by renormalization of the CP(1)-model coupling
constant $\beta$, while the monopole mass on the Coulomb branch is determined
by the renormalization of $g^2$.
This leads to the relation $$\Lambda_{\sigma}=\Lambda_{{\rm SU}(2)}$$
between the two- and four-dimensional dynamical scales. It was
noted earlier as a ``strange coincidence," see Eq. (\ref{lambdasig}).
The first coefficient of the
$\beta$ functions is two ($N$ for generic $N$) for both theories.
Now we know the physical reason behind this coincidence.

Clearly, the above relation can be generalized (cf.~\cite{Dorey,HT2}) to
cover the SU($N$)$\times$U(1) case
with $N_f=N$  flavors on the four-dimensional
side, and CP$(N-1)$ sigma models on the two-dimensional side.

This correspondence can be seen in more quantitative terms \cite{Dorey,HT2}.
Four-dimensional  U$(N)$ SQCD with \ntwo and $N_f=N$
flavors is described by the degenerate Seiberg--Witten curve
\beq
y^2=\frac14\left[\prod_{i=1}^{N}(x+\tilde{m}_i)-\Lambda_{{\rm SU}(N)}^N\right]^2
\label{SWcurve}
\eeq
in the special point (\ref{avev}) on the Coulomb branch which becomes
a quark vacuum upon the $\xi$ deformation. The periods of this curve give
the BPS spectrum of the two-dimensional CP$(N-1)$ model \cite{Dorey}.
We quoted
this spectrum for CP(1) in Eqs.~(\ref{cpcc}) and (\ref{md}).

In fact,  Dorey demonstrated
\cite{Dorey} that the BPS spectra of the two-dimensio\-nal
CP$(N-1)$ model and four-dimensional
SU$(N)$ SQCD coincide with each other if one chooses a point on the Coulomb branch
corresponding to the baryonic Higgs branch defined by the condition
$\sum m_A=0$ (in the SU(2) case the gauge equivalent choice is to   set $m_1=m_2$).

At the same time, we observe that the BPS
spectra of the {\em massive} states in the SU(2)
and U(2) theories, respectively,
coincide in the corresponding quark vacua upon identification of $m_A$ of
the SU$(N)$ theory with $\tilde{m}_{A}$ of
the U$(N)$ theory.
In particular, in the $N=2$ case one must
identify $m_1=m_2$ of the SU(2) theory with $\Delta m/2$ of the U(2) theory.
Note that that the
vacuum (\ref{avev}) and  (\ref{qvev}) of the U$(N)$ theory is an isolated vacuum rather
than a root of a Higgs branch. There are no massless states in the U$(N)$
bulk theory in this vacuum, see Sect.\ref{vacuumstructure} for more detail.

Note also that the BPS spectra of both theories include not only
the monopole/kink
and ``dyonic'' states but elementary excitations with $T=0$ as well.
On the two-dimensional side they correspond to elementary fields $n^{\ell}$
in the large $\Delta{m}_{AB}$ limit.
On the four-dimensional side they correspond to non-topological
(i.e. $T=0$ and $q=\pm 1$) BPS excitations
of the string  with masses proportional to  $\Delta{m}_{AB}$ confined to the string.

The latter can be interpreted
as follows.  Inside the string the squark profiles
vanish, effectively bringing us into the Coulomb branch ($\xi =0$)
where the $W$ bosons and
quarks would become BPS-saturated
states in the bulk. Say, for $N=2$
on the Coulomb branch, the $W$ boson
and off-diagonal quark mass would reduce to $\Delta m$.
Hence, the  $T=0$ BPS excitation
of the string is a wave of such $W$ bosons/quarks propagating along the
string.  One could term  it a ``confined $W$ boson/quark." It is
localized in the perpendicular but not in the transverse direction.
What is important, it has no connection with the bulk Higgs phase $W$
bosons, which are non-BPS and are much heavier than $\Delta m$. Neither these
non-topological excitations have connection to the bulk quarks of our
bulk model, which are not BPS-saturated too.

To conclude, let us mention that recently Tong
compared \cite{Tad} a conformal theory
with massless quarks and monopoles arising
on the Coulomb branch of the  four-dimensional \ntwo SQCD (upon a special
choice of the mass parameters $\Delta{m}_{AB}$), at the so-called
Argyres--Douglas point \cite{AD}, with the twisted-mass-deformed
two-dimensional CP$(N-1)$
model. The coincidence of the monopole and kink masses explained
above
ensures that the CP$(N-1)$ model flows to a non-trivial conformal point at these values
of $\Delta{m}_{AB}$.  The scaling dimensions of
the chiral primary operators in four- and two-dimensional conformal theories
were shown to agree \cite{Tad}, a very nice result, indeed.

\subsection{ More quark flavors}
\setcounter{equation}{0}
\label{semilocal}

In this section we will abandon the assumption $N_F=N$ and consider
the theory (\ref{qed}) with more fundamental
flavors, $N_F>N$. In this case we have a number of  isolated vacua such as
(\ref{avev}), (\ref{qvev}), in which $N$ squarks out of $N_f$ develop VEV's,
while the adjoint VEV's are determined by the masses of these quarks, as in
Eq.~(\ref{avev}).

Now, let us focus on the equal mass case. Then the isolated
vacua coalesce, and a Higgs branch develops from the common root whose location
on the Coulomb branch is given by Eq.~(\ref{avev}) (with all masses set equal). The
dimension of this branch is $4N(N_f-N)$, see \cite{APS,MY}. The Higgs branch is
non-compact and has a hyper-K\"ahler geometry \cite{SW2,APS}. It
has a compact base manifold defined by the condition
\beq
\bar{\tilde{q}}^{kA}=q^{kA}\, .
\label{basehiggs}
\eeq
The dimension of this manifold
is twice less than the total dimension of the Higgs branch,
$2N(N_f-N)$, which implies 4 for $N_f=3$ and 8 for $N_f=4$
in the simplest $N=2$ case.
The BPS string solutions exist only on the base manifold of the Higgs
branch. The flux tubes become non-BPS-saturated
if we move in non-compact directions \cite{EY}.
Therefore, we will limit ourselves to the vacua which belong
to the base manifold.

Those strings that emerge in multiflavor
theories, i.e. $N_f>N$,  (typically on the
Higgs branches) as a rule are are not conventional ANO strings.
Rather, they become the so-called semilocal strings (for a comprehensive
review see \cite{AchVas}). The simplest model where the
semilocal strings appear is the Abelian Higgs
model with two complex flavors
\begin{equation}
\label{ah2fl}
S_{AH}=\int d^4x\left\{\frac1{4g^2}\,F^2_{\mu\nu}+|\nabla_\mu
q^A|^2+\frac{g^2}{8}(|q^A|^2-\xi)^2\right\},
\end{equation}
where $A=1,2$ is the flavor index.

If  $\xi \neq 0$ the scalar fields develop VEV's breaking the U(1) gauge group.
The photon field becomes massive, together with one real scalar field.

In fact,  for the particular  choice of the quartic coupling made
in Eq.~(\ref{ah2fl}) this scalar field has the same mass as the photon. Then
the model (\ref{ah2fl}) is the bosonic part of a supersymmetric theory;
the flux tubes are BPS-saturated.
The topological reason for the existence of the ANO flux tubes is that
$$\pi_1[{\rm U}(1)]=Z$$
for the U(1) gauge group. On the other hand,  in Eq.~(\ref{ah2fl}) we can go to
the low-energy limit
integrating out the massive photon and its scalar counterpart.
This will lead us to a four-dimensional sigma model on the manifold
\beq
|q^A|^2=\xi\,.
\label{vmsamol}
\eeq
The vacuum manifold (\ref{vmsamol}) has
dimension $4-1-1=2$, where we subtract
one real condition mentioned above,  as well as one phase that is gauged away.
Thus, the manifold (\ref{vmsamol})
represents a two-dimensional sphere $S_2$.  The low-energy limit
of the theory  (\ref{ah2fl}) is the $O(3)$ sigma model.

We remember that
$$
\pi_2[S_2]=\pi_1[{\rm U}(1)]=Z\,,$$
and this is the topological reason for
the existence of instantons in the two-dimensional
$O(3)$ sigma model \cite{PolBel}.  Uplifted in four dimensions the instantons
become string-like objects (lumps).

Just as the O(3) sigma-model instantons, the semilocal strings
possess two additional zero modes associated with its complexified
size modulus $\rho$ in the model (\ref{ah2fl}). Hence, the
semilocal strings
interpolate between the ANO strings and two-dimensional  sigma-model instantons
uplifted in four dimensions. At $\rho =0$ we have the ANO string while
at $\rho \to\infty$ the string becomes nothing but the
two-dimensional instanton elevated in four dimensions.
At generic $\rho \neq 0$ the semilocal
string is characterized by a power fall-off of the profile functions at infinity,
instead of the exponential fall-off characteristic of the  ANO string.

Now, if we turn to our {\em non}-Abelian theory (\ref{qed}),  we will see that
the semilocal strings in this theory have ``size'' moduli
in addition to the $2(N-1)$ orientational
moduli $n^{\ell}$. The total dimension of the
moduli space of the semilocal string was shown \cite{HT1} to be
\beq
2N_f=2+2(N-1)+2(N_f-N)\,,
\label{dimsemiloc}
\eeq
where the first, the second and the third term above correspond to the translational,
orientational and the size moduli.

No study of geometry of the moduli space of the semilocal
strings were carried out for quite some time due to infrared problems.
It was  known \cite{Ward,LeSa} that the size zero modes are logarithmically
non-normalizable in the infrared, as is the case for the sigma-model instantons
in  two dimensions.  This problem was addressed in \cite{SYseml} where
non-Abelian strings in the  U(2) gauge theory were treated.
The effective theory on the
string worldsheet was shown  to have the form
\beq
S^{(1+1)}=  \beta \, M_W\,   \int d t\, dz \left\{
\frac{\rho^2}{4}\, (\pt_{k}\, S^a)^2 +|\pt_{k}\, \rho_i|^2\, \right\}\,
\ln\, {\frac{1}{|\rho|\,\, \delta\, m}}
\,,
\label{toric}
\eeq
where $M_W$ is the $W$-boson mass, see Eq.~(\ref{msuN}). The subscript
$i=3,...N_f$, while $\rho_i$ stand for $(N_f-2)$ complex
fields associated with the size moduli. The parameter $\delta m$ here
measures small quark mass differences. It is necessary to introduce
this infrared parameter, slightly lifting the size moduli $\rho_i$, in order to
regularize the infrared logarithmic divergence.

The metric (\ref{toric}) is derived in \cite{SYseml} for
large --- but not too large --- values of $|\rho|^2\equiv |\rho_i^2|$ lying
inside the window
\beq
\frac{1}{M_W}\ll |\rho|\ll\frac{1}{\delta m}\,.
\label{window}
\eeq
The inequality on the left-hand side refers to the limit in which the semilocal
string becomes an O(3) sigma-model lump.
The inequality on the right-hand side ensures the validity of the logarithmic approximation. The action (\ref{toric})
was obtained in the logarithmic approximation.

 For $\rho_i$'s lying inside the window (\ref{window}),
with a logarithmic accuracy, one can introduce new variables
\beq
z_i=\rho_i\left[M_W^2\, \ln\, {\frac{1}{|\rho|\,\delta m}}\, \right]^{1/2}\, .
\label{zrho}
\eeq
In terms of these new variables the metric
of the worldsheet theory (\ref{toric}) was shown to become flat \cite{SYseml}.
Corrections
to this flat metric run in powers of
$$
\frac{1}{M_W\, |\rho|}\,\,\,\mbox{and} \,\,\,
\left( \ln \,{\frac{1}{|\rho|\, \delta m}}
\right)^{-1}\,.
$$
These corrections have not yet been calculated within the field-theory approach.

On the other hand, the very same problem
was analyzed from the D-brane theory side.
Using brane-based arguments Hanany and Tong conjectured \cite{HT1,HT2}
(see also Ref.~\cite{Jstr})
that the effective theory on the worldsheet of the non-Abelian
semilocal string is given by the strong-coupling limit ($e^2\to\infty$)
of the following two-dimensional gauge theory:
\beqn
S
& =&
 \int d^2 x \left\{
 2\beta\,|\nabla_{k} n^{\ell}|^2 +2\beta\,|\nabla_{k} z_i|^2
 +\frac1{4e^2}F^2_{kl} + \frac1{e^2}
|\pt_k\sigma|^2
\right.
\nonumber\\[3mm]
 &+&  4\beta\,
 \left|
 \sigma-\frac{\tilde{m}_{\ell}}{\sqrt{2}}\right|^2
 \, |n^{\ell}|^2 +
 4\beta\,
 \left|
 \sigma-\frac{\tilde{m}_{i}}{\sqrt{2}}\right|^2
 |z_i|^2
\nonumber\\[4mm]
&+&
2e^2\,  \beta^2\left[\,
|n^{\ell}|^2-|z_i|^2 -1\right]^2
\Big \}\,,
\label{sl2d}
\eeqn
where $\ell=1,...,N$ and $i=N+1,...N_f$. Furthermore, $z_i$ denote $(N_f-N)$
complex fields associated with the size moduli.  The
fields $n^{\ell}$ and $z_i$ have the charges  +1 and $-1$
with respect to the U(1) gauge field in Eq.~(\ref{sl2d}).
This theory is similar to the model (\ref{mcpg}) for the
$N_f=N$ non-Abelian string.

The Hanany--Tong
conjecture is supported by yet another argument. As was discussed
Sect.~\ref{dsw}, the BPS spectrum of dyons on the Coulomb branch of
the four-dimensional
theory must coincide with the BPS spectrum in
the two-dimensional theory on the
string worldsheet. We expect that this correspondence extend  to
theories with $N_f>N$. The two-dimensional theory (\ref{sl2d})
was studied in \cite{DorHolT} where
it was shown that its BPS spectrum agrees with the spectrum of
four-dimensional U$(N)$ SQCD with $N_f$ flavors.
In particular, the one-loop coefficient of the $\beta$ function is
$2N-N_f$ in both theories. This leads to the identification of their
scales, see Eq.~(\ref{lambdasig}).
As a matter of fact,
Ref.~\cite{DorHolT} deals with the SU$(N)$
theory at the root of the baryon Higgs branch, much in the
same vein as \cite{Dorey}. However, as was
explained in Sect.~\ref{dsw},
the  BPS spectra of the massive states in  these four-dimensional
theories are the same.

The above argument shows that the two-dimensional theory (\ref{sl2d})
is a promising candidate for an effective theory on the
semilocal string worldsheet. In particular, the metric in (\ref{sl2d})
is asymptotically flat. The variables $z_i$ in (\ref{sl2d}) should be
identified with the ones in Eq.~(\ref{zrho}) introduced within the field-theory
framework in Ref.~\cite{SYseml}.
It is quite plausible that corrections
to the flat metric in powers $1/( M_W\,  |\rho|)$ are properly reproduced by the
worldsheet theory (\ref{sl2d}). Nevertheless, the results of \cite{SYseml}
clearly demonstrate the approximate nature of the
worldsheet theory (\ref{sl2d}).
Namely,
corrections at large $\rho_i$ suppressed by large infrared logarithms
$\left( \ln\, (1/ |\rho|\, \delta m) \right)^{-1} $ are
certainly not captured in Eq.~(\ref{sl2d}).

The
implication of the ``semilocal nature"
of the semilocal strings which is
most important
from the physical standpoint is the  loss
of the monopole confinement \cite{EY,SYseml} i.e. the loss
of the Meissner effect. To study
the monopole confinement as a result of the
squark condensation we must consider
a string of a finite length $L$ stretched between a
heavy probe monopole and antimonopole from the
${\rm SU}(N+1)/{\rm SU}(N)\times{\rm  U}(1)$ sector.
The ANO string has a typical
transverse size $( g\sqrt{\xi})^{-1}$. If $L$ is much larger than this
size the energy of this probe configuration is
\beq
V(L)=TL\, ,
\label{conf}
\eeq
where $T$ is the string tension. The
linear potential in Eq.~(\ref{conf}) ensures confinement of monopoles.

For semilocal strings this conventional picture drastically changes.
Now the transverse string size can be arbitrarily large. Imagine a configuration
in which the string size becomes much larger than $L$. Then we will
clearly deal with the  three-dimensional rather than
two-dimensional problem.  The
monopole flux is no longer trapped in a narrow flux tube. Instead, it
freely spreads over a large three-dimensional volume,  of the size of order of
$L$ in all directions. Obviously,
this will give rise to a Coulomb-type potential between the probe monopoles,
\beq
V(L)\sim 1/L\, ,
\label{coulomb}
\eeq
possibly augmented by logarithms of $L$.
At large $L$ the energy of this configuration is lower than the one of the flux-tube
configuration (\ref{conf});  therefore, it is energetically favorable.

To summarize,
semilocal strings can indefinitely  increase their transverse
size and effectively disintegrate, so that the linear potential (\ref{conf})
gives place to the Coulomb potential (\ref{coulomb}). In fact, lattice studies
unambiguously
show that the semilocal string thickness always increases upon small
perturbations \cite{Leese}.
Formation of semilocal strings  on the Higgs
branches leads to a dramatic physical effect --- deconfinement.

\subsection{Non-Abelian $k$-strings}
\setcounter{equation}{0}

In this section we will briefly review how multi-strings,  with the
winding number $k>1$, can be constructed.
One can consider them as bound states of $k$ BPS
elementary strings. The Bogomol'nyi representation (\ref{bogs})
implies that the tension of the BPS-saturated $k$-string is
determined by its total U(1) flux, $2\pi \, k$. This entails, in turn, that
in \ntwo SQCD, see Eq.~(\ref{qed}),  the $k$-string tension
has the form
(\ref{qed})
\beq
T_k=2\pi\,k\,\xi\,.
\label{tenkstr}
\eeq
Equation (\ref{tenkstr})
implies that the elementary strings
that form composite $k$-strings do not
interact.

If one considers $k$ elementary strings, forming the given $k$-string, at large separations the corresponding moduli space
obviously factorizes into $k$ copies of the moduli spaces of the
elementary strings. This suggests that the dimension of the total moduli space is
\beq
2kN_f= 2k +2k(N-2)+ 2k(N_f-N)\,,
\label{dimmodkstr}
\eeq
see (\ref{dimsemiloc}). The total dimension is written as a sum of
dimensions of the translational, orientational and size moduli spaces.
This result was
confirmed by the Hanany--Tong
index theorem \cite{HT1} which implies (\ref{dimmodkstr})
at any separations. The moduli space
of well separated elementary strings forming the given $k$-string, say,
at $N_f=N$ is
\beq
\frac{\big[ C\times {\rm CP}(N-1)\big]^k}{S_k}\,,
\eeq
where $S_k$ stands for permutations of the
elementary string positions.

An explicit solution for a non-Abelian 2-string at zero separation
in the simplest bulk theory with $N=N_f=2$ was constructed in \cite{ASY2str}.
It has a peculiar feature. If the orientation vectors of the two strings $S^a_1$
and $S^a_2$ are opposite,  the composite 2-string becomes an Abelian ANO string.
It carries no non-Abelian flux. Therefore,  SU(2)$_{C+F}$ rotations act trivially on this particular string. This means that the internal moduli space
of this string is singular \cite{HashT,ASY2str}.
The section of the orientational moduli
space corresponding to $S^a_1=-S^a_2$ degenerates into a point. In \cite{ASY2str}
it was argued that the internal moduli of the 2-string at zero separation
is equivalent to ${\rm CP}(2)/Z_2$. This differs by a discrete quotient from the
result CP(2) obtained in \cite{HashT}. Later results obtained in \cite{Jrecon,Jreconsemi}
confirm the ${\rm CP}(2)/Z_2$ metric.

The metric on the $k$-string moduli space for generic $k$ is not known. For Abelian
$k$-strings exponential corrections to the flat metric were calculated
in \cite{ManSpe}. Exponentially small corrections are natural since
in this case the vortices are characterized by an exponential fall off of their profile functions at large distances.

Hanany and Tong exploited \cite{HT1,HT2} a D-brane construction to
obtain the $k$-string metric in terms of the Higgs branch of a
two-dimensional gauge theory, see (\ref{mcpg}) and (\ref{sl2d}).
What they came up with
is an \mbox{\ntwo} supersymmetric U$(k)$ gauge theory with $N$ fundamental
and $(N_f-N)$ anti-fundamental flavors $n^{\ell}$ and $\rho_i$,
respectively,  ($\ell=1,...,N$ and $i=N,...,N_f$),  plus an
adjoint chiral multiplet $Z$. The $D$-term condition for this theory
is
\beq
\frac1{2\beta}[\bar{Z},Z] +n^{\ell} \bar{n}_{\ell}-\rho_i\bar{\rho}^i =1\, .
\label{Dtermkstrings}
\eeq
The metric defined by this Higgs branch has corrections
to the factorized metric which run in (inverse) powers of separations
between the elementary strings. Thus, it exhibits a
dramatic disagreement with the field-theory expectations.  Still the
metric  is believed
\cite{HT1,HT2,ob1,ob2} to correctly
reproduce some data protected by supersymmetry, such as the BPS spectrum.

To derive all moduli of the general $k$-string solution
the so-called {\em moduli matrix method} was
developed in  \cite{Jk}. It was observed
that the
substitution
\beq
\vp=S(z,\bar{z})\, H_0(z),\qquad A_1+iA_2= S^{-1}\bar{\pt}_z S\, ,
\label{modmat}
\eeq
solves the
 last of the first-order equations (\ref{gfoes}).
Here $z=x_1+ix_2$ and $H_0$ is an $N\times N_f$ matrix with a holomorphic
dependence on $z$.

Then the equations for the gauge field strength in (\ref{gfoes})
yield an equation on $S(z,\bar{z})$ which is rather hard to solve in the general
case.  It was argued, however,  that the factor $S$ involves no new
moduli parameters \cite{Jk}.
Therefore,  all moduli parameters reside in the moduli matrix $H_0(z)$.
Determining $H_0(z)$ gives one a moduli space which agrees with the moduli space
corresponding to the Higgs branch (\ref{Dtermkstrings}).

\subsection{A physical picture of the monopole confinement}
\setcounter{equation}{0}
\label{confinement}

\begin{figure}[h]
\epsfxsize=8cm
\centerline{\epsfbox{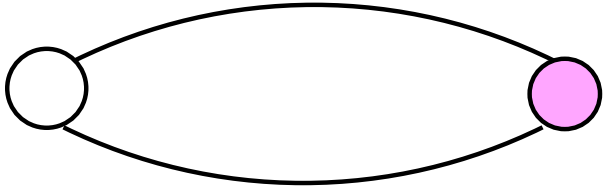}}
\caption{\small
Monopole and antimonopole bound  into a ``meson.''
The binding is due to strings.  Open and closed
circles denote the monopole and antimonopole, respectively. }
\label{figmmeson}
\end{figure}

In this section we will return to our basic \ntwo SQCD with the U$(N)$ gauge group
and $N_f=N$ flavors (\ref{qed}) and discuss an emerging physical picture of
the monopole confinement. As was reviewed in detail in Sect.~\ref{kinkmonopole}, elementary confined monopoles can be viewed as
junctions of two elementary strings. Therefore,  the physical spectrum of the
theory includes monopole-antimonopole ``mesons''
formed by two elementary  strings in a loop configuration shown in
Fig.~\ref{figmmeson}.

If  spins of such ``mesons'' of order one, their mass is of the order of
the square root of the  string tension $\sqrt{\xi}$.
Deep in the quantum non-Abelian regime ($\tilde{m}^l=0$), the
CP$(N-1)$-model strings
carry no average SU($N$) magnetic flux \cite{W79},
\beq
\langle n^l \rangle =0\,,
\label{zeronl}
\eeq
see Eq.~(\ref{str}). What they do carry
is the  U(1) magnetic flux which determines their
tension.

Monopoles are seen in the worldsheet theory as CP$(N-1)$ kinks. At $\tilde{m}^l=0$
they become  non-Abelian too,
much in the same way as strings.  They  carry no  average   SU($N$) magnetic flux.
(Unlike strings, even in the classical regime they do not carry the U(1) magnetic flux, see
(\ref{monflux}).)

Moreover, the monopoles acquire global flavor quantum numbers.
We know that the CP$(N-1)$ model kinks
at strong coupling are described by the $n^l$ fields  \cite{W79,HoVa} and, therefore,
in fact, they belong to the fundamental representation
of the global SU($N$)$_{C+F}$ group.
This means that the
monopole-antimonopole ``mesons'' formed by the string configuration shown in
Fig.~\ref{figmmeson} can belong either to singlet or to adjoint representations
of the global ``flavor'' group SU($N$)$_{C+F}$, in
full  accordance with our expectations.

Singlets resemble glueballs. In weakly coupled bulk theory ($g_1^2\ll 1$)
the singlet mesons can decay into massive vector multiplets
formed by gauge and quark fields, with mass (\ref{mu1}), see
Sect.~\ref{vacuumstructure}.
The monopole-antimonopole mesons with the adjoint flavor quantum numbers are also
metastable in weakly coupled bulk theory ($g_2^2\ll 1$),
they decay into massive gauge/quark multiplets which carry the adjoint quantum
numbers with respect to the global unbroken SU($N$)$_{C+F}$ group and have masses determined by Eq.~(\ref{msuN}).

Two elementary strings of the monopole-antimonopole meson shown in
Fig.~\ref{figmmeson} can form a non-BPS bound state. Hence, in practice the
composite meson looks as if the monopole
was connected to the antimonopole by a single string. In fact, there are
indications that
this is what happens in the theory at hand.  Interactions of
elementary $Z_2$ strings were studied in \cite{MY} in the
simplest case $N=2$. An interaction
potential for the elementary $Z_2$ strings with $S^a=(0,0,+1)$ and $S^a=(0,0,-1)$
was found to be attractive at large distances,
\beq
U \sim\, -\,
\Big\{  M_{{\rm SU}(2)}\,R \Big\} ^{-1/2}\,\, e^{
-M_{{\rm SU}(2)}R
}
\, ,
\label{intpot}
\eeq
where $R$ stands for the
distance between two parallel strings. The gauge boson mass is
given in Eq.~(\ref{msuN}). This attractive potential leads to formation of a bound state
a composite string.

Note that we have $N$ distinct elementary strings.
As was discussed in Sect.~\ref{Dyn}, $N$ elementary
strings differ from each other  in quantum regime by the value of
the bifermion condensate
of the CP$(N-1)$ model fermions \cite{NSVZsigma}.
Therefore, the physical picture of the monopole confinement
is not absolutely similar to what we expect in QCD, see
the discussion in the beginning of this
section. Namely, we have $N$   different degenerate ``mesons''  ( at $N>2$) of
the type discussed above, associated with $N$ different elementary strings.

In QCD (and in nature)
we have instead a single meson with the
given quantum numbers, plus its radial excitations
which have higher masses.
This is typical for BPS strings in supersymmetric gauge theories. We will
see in Sect.~\ref{nonBPS} that in non-supersymmetric theories the situation is different:
elementary strings are split and, therefore, different ``mesons'' become split too.

\begin{figure}[h]
\epsfxsize=8cm
\centerline{\epsfbox{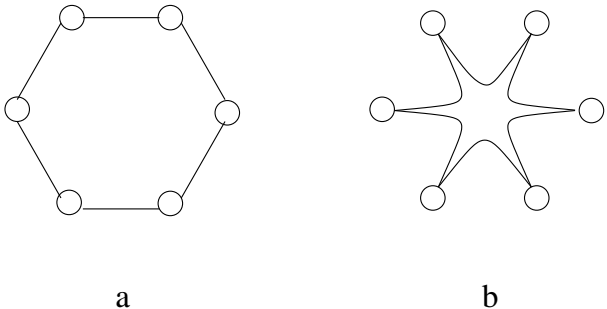}}
\caption{\small
a). A schematic picture of  the ``baryon'' formed by monopoles and strings for $N=6$; b).
The ``baryon'' acquires the shape of a star once the neighboring strings form non-BPS bound states.
 }
\label{figbaryon}
\end{figure}

In addition to the
``mesons'' and gauge/quark multiplets, the physical spectrum contains also
``baryons'' built of  $N$ elementary monopoles connected to each other  by
elementary strings forming a closed ``necklace configuration," see
Fig.~\ref{figbaryon}a. In the classical limit $\tilde{m}^l\gg \Lambda_{\sigma}$
all strings carry the SU($N$) magnetic fluxes given by
\beq
\int d^2 x \, F^{*}_{\,\,\,{\rm SU}(N)}=2\pi \left( n\,\cdot n^*-\frac{1}{N}\right)\,,
\label{strflux}
\eeq
with $n^l=\delta^{ll_0}$, $l_0=1,...,N$ for $N$ elementary strings forming the
``baryon.'' The monopoles carry the SU($N$) magnetic fluxes
given in Eq.~ (\ref{monflux}) and,
therefore, can be located at the corners of the polygon in Fig.~\ref{figbaryon}a.

In highly quantum regime, at $\tilde{m}^l =0$, both strings and monopoles carry
no average SU($N$) magnetic flux, see (\ref{zeronl}). The confined monopoles
are seen as kinks interpolating between the ``neighboring'' quantum vacua
of the CP$(N-1)$ model (a.k.a. strings) in the closed necklace configuration in
Fig.~\ref{figbaryon}a.

As was mentioned,
the monopoles/kinks acquire flavor global quantum numbers.
They become fundamentals in  SU($N)_{C+F}$. Thus, the ``baryon'' is in the
$$\prod_1^N (N)$$ representation of SU($N)_{C+F}$. Note that both quarks and
monopoles do not carry
baryon numbers. Therefore,  our ``baryon''  has no baryon number too.
The reason for this is that the U(1) baryon current is coupled to a
 gauge boson in the U($N$) gauge
theory that we consider here. This means,
in particular, that the ``baryons''  can decay into the  monopole ``mesons''
or gauge/quark multiplets.

We mentioned that  the ``neighboring'' elementary strings can form a
non-BPS bound state, a composite string.
It is plausible then that in practice the monopole ``baryon'' actually resembles a
configuration shown in Fig.~\ref{figbaryon}b.

Let us emphasize that all states seen in the physical
spectrum of the theory are gauge singlets. This goes without saying.
While color charges of the
gauge/quark multiplets are screened by the Higgs mechanism,
the monopoles are confined by non-Abelian strings.

Let us also stress in conclusion that in the limit $\tilde{m}^l=0$ the global
group SU($N$)$_{C+F}$ is restored
in the bulk and both strings and confined monopoles become non-Abelian. One might argue
 that this restoration could happen  only at the classical level.
 One could suspect that in quantum theory a
``dynamical Abelization'' ( i.e. a cascade
breaking of the gauge symmetry U($N$)$\to$U(1)$^{N} \to {\rm discrete\; subgroup}$ )
might occur. This could have happened if the  adjoint VEV's  that are classically
vanish at $\tilde{m}^l=0$
(see (\ref{avev}))  could have developed  dynamically in quantum theory.

 At $\tilde{m}^l\neq 0$ the global SU($N$)$_{C+F}$
group is explicitly broken down  to U(1)$^{N-1}$ by quark masses. At $\tilde{m}^l=0$ this group is classically restored. If it could be proven to be dynamically broken
at $\tilde{m}^l=0$,
this would mean a spontaneous symmetry breaking, with obvious consequences,
such as the corresponding Goldstone modes.

We want to explain why  this cannot and does not happen in the
theory at hand. First of all,
 if a global symmetry is not spontaneously broken at the tree level then
it cannot be broken
by quantum effects at  week coupling in ``isolated'' vacua. Second, if the
global group SU($N$)$_{C+F}$
were
broken spontaneously at $\tilde{m}^l=0$ this would ensure the presence of massless Goldstone
bosons. However, we know that there are no massless states in the spectrum of the bulk
theory, see Sect.~\ref{model}.

Finally, the breaking of SU($N$)$_{C+F}$
in the $\tilde{m}^l=0$ limit would mean that the twisted masses of the worldsheet
CP$(N-1)$ model
would  not be given by $\tilde{m}^l$; instead they would be shifted,
$$\tilde{m}_{(tw)}^l=\tilde{m}^l +c^l \Lambda_{CP(N-1)}\,,$$
where $c^l$ are  some coefficients.
In Sect.~\ref{dsw}  it was
shown \cite{SYmon,HT2} that the BPS spectrum of the CP$(N-1)$ model on
the string should coincide with
the BPS spectrum of the four-dimensional bulk theory on the Coulomb branch.
The BPS spectrum of the CP$(N-1)$ model is determined by $\tilde{m}_{(tw)}^l$
while the BPS spectrum of the bulk theory on the Coulomb branch is determined by
$\tilde{m}^l$. In \cite{Dorey} it was shown that the BPS spectra of both theories coincide
at $\tilde{m}_{(tw)}^l=\tilde{m}^l$. Thus, we conclude that $c^l=0$,
and the twisted
masses vanishes in
the $\tilde{m}^l=0$ limit.

Hence, the global SU($N$)$_{C+F}$ group is not
broken in the bulk and both
strings and confined monopoles become non-Abelian at $\tilde{m}^l=0$.

\vspace{20mm}

\centerline{\includegraphics[width=1.3in]{extra3.eps}}

\newpage

\section{Less supersymmetry}
\label{nosusy}

Let us move towards less supersymmetric theories.
In this section we will review non-Abelian strings in four-dimensional
gauge theories with  ${\mathcal N}=1$. In Sect.~\ref{nonBPS} we will deal
with  ${\mathcal N}=0$.

As was discussed in the beginning of Sect.~\ref{strings},
the Seiberg--Witten mechanism of confinement \cite{SW1,SW2} relies
on a cascade gauge symmetry breaking: the
non-Abelian gauge group breaks down to an Abelian subgroup at a higher scale
by condensation
of the adjoint scalars, and  at a lower scale the Abelian subgroup
 breaks down to a discrete
subgroup by condensation of quarks ( or monopoles depending on the type
of vacuum considered). This leads to formation
of the ANO flux tubes and ensures an Abelian nature of confinement
of the monopoles (  or quarks respectively).

On the other hand, non-supersymmetric QCD-like theories as well as \mbox{\none}
SQCD have no adjoint scalars and, as a result, no cascade
gauge symmetry breaking occurs. Confinement in these theories is believed to
be essentially non-Abelian. This poses a problem of understanding
confinement in theories of this type. Apparently, a straightforward extrapolation
of the Seiberg--Witten  confinement scenario to these theories does not work.

The discovery of the non-Abelian strings \cite{HT1,ABEKY,SYmon,HT2}
suggests a novel
possibility of solving this problem. In the \ntwo gauge theory (\ref{qed})
the SU$(N)$ subgroup of the U$(N)$ gauge group remains unbroken
after condensation
the vacuum expectation value  $\langle a^a \rangle =0$, see (\ref{avev}).
This circumstance demonstrates
that the formation of the non-Abelian strings does not rely on the presence
of adjoint VEV's. This suggests, in turn, that  we can give  masses to
the adjoint fields, make them heavy,
and eventually decoupling the adjoint fields  altogether, without loosing qualitative
features of the non-Abelian confinement mechanism
reviewed previously.

This program --- moving towards less
supersymmetry --- was initiated in Ref.~\cite{SYnone} which we will discuss
in this section. In \cite{SYnone} we considered
\ntwo gauge theory (\ref{qed}),  with the  gauge group U$(N)$.
This theory was deformed by a
mass term $\mu$ for the adjoint matter  fields  breaking \mbox{\ntwo} supersymmetry
down to ${\mathcal N}=1$. The breaking terms do not affect classical solutions
for the non-Abelian strings. The latter are still 1/2 BPS-saturated.
However, at the quantum level the
strings ``feel'' the presence of \ntwo supersymmetry breaking terms.
Effects generated by  these terms first show up in  the sector of
the fermion zero modes.

If the adjoint mass parameter
$\mu$ is kept finite, the non-Abelian string in the \none model at hand
is well defined and supports confined
monopoles. However, at $\mu\to \infty$,
as the adjoint superfield becomes very heavy --- we approach the limit of \none SQCD
---
an infrared problem develops. This is due to the fact that in \none SQCD defined
in a standard way the vacuum manifold is no longer an isolated point.
Rather,  a flat direction develops (a Higgs branch). The presence of
the massless states obscure physics of the non-Abelian strings.
In particular, the strings become infinitely thick \cite{SYnone}.
Thus, one arrives at a dilemma: either one must abandon the attempt
to decouple  the adjoint superfield,
or, if this decoupling is performed, confining non-Abelian strings cease to exist
\cite{SYnone}.

A way out was suggested recently in \cite{us}.
A relatively insignificant modification
of the benchmark \ntwo model takes care of the infrared problem. All we have to do is
to add a neutral meson superfield $M$ coupled to the quark superfields
through a superpotential term.  Acting together with the mass term of the adjoint
superfield, $M$ breaks \ntwo down to \none. The limit $\mu\to\infty$
in which the adjoint superfield completely decouples, becomes well defined.
No flat directions emerge. The limiting theory is \none SQCD
supplemented by the meson superfield. It supports
non-Abelian strings. The junctions of these strings present confined monopoles,
or, better to say, what becomes of the monopoles in the theory
where there are no adjoint scalar fields. There is a continuous path following which one
can trace the monopole
evolution in its entirety: from the 't Hooft--Polyakov monopoles which
do not exist without the adjoint scalars to the confined monopoles
in the adjoint-free environment.

\subsection{Breaking \ntwo supersymmetry down to \none}
\setcounter{equation}{0}
\label{nonep}

In Sect.~\ref{nonep} we will outline main
results of Ref.~\cite{SYnone} where non-Abelian strings
were considered in an \none  gauge theory obtained as a deformation of
the \ntwo theory (\ref{qed}) by mass terms of the adjoint matter.

\subsubsection{Deformed theory and string solutions}
\label{dtss}

Let us  add a superpotential mass term to our \ntwo SQCD,
 \beq
{\mathcal W}_{{\mathcal N}=1}=\sqrt{\frac{N}{2}}\,
\frac{\mu_1}{2} \, {\mathcal A}^2
+ \frac{\mu_2}{2}\, \left({\mathcal A}^a\right)^2\,,
\label{superpotbr}
\eeq
where $\mu_1$ and $\mu_2$ are mass parameters for the chiral
superfields belonging to  \ntwo U(1) and SU$(N)$ gauge supermultiplets,
respectively,
while the factor $\sqrt{\frac{N}{2}}$ is included for convenience. Clearly,
the mass term  (\ref{superpotbr}) splits these supermultiplets,
breaking \ntwo supersymmetry down to \none.

The bosonic part of the  SU$(N)\times$U(1) theory has  the form
(\ref{qed}) with the potential
\beqn
V(q^A,\tilde{q}_A,a^a,a)_{{\mathcal N}=1} &=&
 \frac{g^2_2}{2}
\left( \frac{1}{g^2_2}\,  f^{abc} \bar a^b a^c
 +
 \bar{q}_A\,T^a \,q^A -
\tilde{q}_A T^a\,\bar{\tilde{q}}^A\right)^2
\nonumber\\[3mm]
&+& \frac{g^2_1}{8}
\left(\bar{q}_A q^A - \tilde{q}_A \bar{\tilde{q}}^A-N\xi\right)^2
\nonumber\\[3mm]
&+& \frac{g^2_2}{2}\left| 2\tilde{q}_A T^a q^A +\sqrt{2}\mu_2a^a\right|^2+
\frac{g^2_1}{2}\left| \tilde{q}_A q^A +\sqrt{N}\mu_1 a \right|^2
\nonumber\\[3mm]
&+&
\frac12\sum_{A=1}^N \left\{ \left|(a +2T^a a^a)q^A\right|^2
\right.
\nonumber\\[3mm]
&+&
\left.
\left|(a +2T^a a^a)\bar{\tilde{q}}_A
\right|^2 \right\}\,,
\label{N1pot}
\eeqn
where the sum over repeated flavor indices $A$ is implied.
The potential (\ref{N1pot}) differs from the one in (\ref{pot})
in two ways. First, we use SU(2)$_R$ invariance of the original \ntwo
theory with the potential (\ref{pot}) to rotate the FI term. In
Eq.~(\ref{N1pot}) it is the FI $D$ term,
while in Sect. \ref{strings} we considered the FI $F$ term.

Second, there
are \ntwo supersymmetry breaking contributions from  $F$ terms in
Eq.~(\ref{N1pot}) proportional to the mass parameters $\mu_1$ and $\mu_2$.
Note that we set the quark mass differences  at zero and redefine $a$
to absorb the average value of quark masses.

As in Eq.~(\ref{pot}), the FI term triggers the spontaneous breaking
of the gauge symmetry. The vacuum expectation values
of the squark fields can be chosen as
\beqn
\langle q^{kA}\rangle &=&\sqrt{
\xi}\, \left(
\begin{array}{ccc}
1 & 0 & ...\\
... & ... & ... \\
... & 0 & 1  \\
\end{array}
\right),\,\,\,\,\,\,\langle \bar{\tilde{q}}^{kA}\rangle =0,
\nonumber\\[3mm]
k&=&1,...N,\qquad A=1,... N\,,
\label{N1qvev}
\eeqn
while the adjoint field VEV's are
\beq
\langle a^a\rangle =0,\,\,\,\,\langle a\rangle =0,
\label{N1avev}
\eeq
see (\ref{avev}).

We see that the quark VEV's has the color-flavor locked form (see (\ref{qvev}))
implying that the SU$(N)_{C+F}$ global symmetry is unbroken in the vacuum.
Much in the same way as in \ntwo SQCD,
this symmetry leads to the emergence of the
orientational zero modes of the $Z_N$ strings.

Note that VEV's (\ref{N1qvev}) and  (\ref{N1avev}) do not depend on
supersymmetry breaking parameters $\mu_1$ and $\mu_2$. This
is due to the fact that our choice of parameters in (\ref{N1pot}) ensures
vanishing of the adjoint VEV's, see (\ref{N1avev}). In particular, we have
the same pattern of symmetry breaking all the way down to very large
$\mu_1$ and $\mu_2$, where the adjoint fields decouple. As in \ntwo SQCD
we assume $\sqrt{\xi}\gg \Lambda_{{\rm SU}(2)}$ to ensure weak coupling.

Now, let us discuss the mass spectrum in the \none  theory at hand.
Since both U(1) and SU$(N)$ gauge groups are broken by squark condensation, all
gauge bosons become massive. Their masses are given in Eqs. (\ref{msuN}) and
(\ref{mu1}).

To obtain the scalar boson masses we expand the potential (\ref{N1pot})
near the vacuum (\ref{N1qvev}), (\ref{N1avev}) and diagonalize the
corresponding mass matrix. Then, $N^2$ components of $2N^2$  (real)
component scalar field $q^{kA}$ are eaten by the Higgs mechanism
for the U(1) and
SU$(N)$
gauge groups, respectively. Other $N^2$  components are split as follows:
one component acquires mass (\ref{mu1}). It becomes
the scalar component of  a massive \none vector U(1) gauge multiplet.
Moreover, $N^2-1$  components acquire masses (\ref{msuN}) and become
scalar superpartners of the SU$(N)$ gauge bosons in \none massive gauge
supermultiplets.

Other $4N^2$ real scalar components of fields $\tilde{q}_{Ak}$, $a^a$ and $a$
produce the following states: two states acquire mass
\beq
m_{{\rm U}(1)}^{+}=g_1\sqrt{\frac{N}{2}\xi\,\lambda_1^{+}}\, ,
\label{u1m1}
\eeq
while the mass of other two states is given by
\beq
m_{{\rm U}(1)}^{-}=g_1\sqrt{\frac{N}{2}\xi\,\lambda_1^{-}}\, ,
\label{u1m2}
\eeq
where $\lambda_1^{\pm}$ are two roots of the quadratic equation
\beq
\lambda_i^2-\lambda_i(2+\omega^2_i) +1=0\,,
\label{queq}
\eeq
for $i=1$, where we introduced two \ntwo supersymmetry breaking
parameters associated with the U(1) and SU$(N)$ gauge groups, respectively,
\beq
\omega_1=\frac{g_1\mu_1}{\sqrt{\xi}}\,,\qquad
\omega_2=\frac{g_2\mu_2}{\sqrt{\xi}}\,.
\label{omega}
\eeq
Other $2(N^2-1)$ states acquire mass
\beq
m_{{\rm SU}(N)}^{+}=g_2\sqrt{\xi\lambda_2^{+}}\,,
\label{su2m1}
\eeq
while the remaining  $2(N^2-1)$ states become massive, with mass
\beq
m_{{\rm SU}(N)}^{-}=g_2\sqrt{\xi\lambda_2^{-}}\, ,
\label{su2m2}
\eeq
where $\lambda_2^{\pm}$ are two roots of the quadratic equation
(\ref{queq}) for $i=2$. Note that all states come either as singlets
or adjoints with respect to the unbroken SU$(N)_{C+F}$.

When the SUSY breaking parameters $\omega_{i}$ vanish,
the masses (\ref{u1m1}) and (\ref{u1m2}) coincide with the U(1) gauge
boson mass (\ref{mu1}). The corresponding states form a bosonic part of a long
\ntwo
massive U(1) vector supermultiplet \cite{VY}, see also
Sect.~\ref{vacuumstructure}.

If
$\omega_1\neq 0$ this supermultiplet splits into a  \none vector multiplet,
with mass (\ref{mu1}),  and two chiral multiplets, with masses
(\ref{u1m1}) and (\ref{u1m2}). The same happens with the states with masses
(\ref{su2m1}) and (\ref{su2m2}). With vanishing $\omega$'s they combine
into  bosonic parts of $(N^2-1)$  \ntwo vector supermultiplets with mass
(\ref{msuN}). If $\omega_i \neq 0$ these multiplets split into $(N^2-1)$
\none vector multiplets (for the SU$(N)$ group) with mass (\ref{msuN})
and $2(N^2-1)$ chiral multiplets with masses (\ref{su2m1}) and (\ref{su2m2}).
Note that the same splitting pattern was found in \cite{VY} in the
Abelian case.

Now let us take a closer look at the spectrum obtained above in the limit
of large \ntwo supersymmetry breaking parameters $\omega_i$, $\omega_i\gg 1$.
In this limit
the larger masses $m_{{\rm U}(1)}^{+}$ and $m_{{\rm SU}(N)}^{+}$ become
\beq
m_{{\rm U}(1)}^{+}= m_{{\rm U}(1)}\omega_1=g_1^2\sqrt{\frac{N}{2}}\mu_1
\,,\quad
m_{{\rm SU}(N)}^{+}= m_{{\rm SU}(N)}\omega_2=g_2^2\mu_2\, .
\label{amass}
\eeq
In the limit $\mu_i\to \infty$ these are the masses of the heavy adjoint
scalars $a$ and $a^a$. At $\omega_i\gg 1$ these fields decouple and
can be integrated out.

The low-energy theory in this limit
contains massive gauge \none multiplets and chiral multiplets with
the lower masses $m^{-}$. Equation (\ref{queq}) gives for these masses
\beq
m_{{\rm U}(1)}^{-}= \frac{m_{{\rm U}(1)}}{\omega_1}=\sqrt{\frac{N}{2}}\,\frac{\xi}{\mu_1}
\,,\quad
m_{{\rm SU}(2)}^{-}= \frac{m_{{\rm SU}(2)}}{\omega_2}=\frac{\xi}{\mu_2}\,.
\label{light}
\eeq
In particular, in the limit of infinite $\mu_i$ these masses tend to zero.
This reflects the presence of a Higgs branch in \none SQCD.
To see the Higgs branch and calculate its
dimension, please,  observe that our theory (\ref{qed}) with
the potential (\ref{N1pot})  in the limit
$\mu_i\to \infty$ flows to \none
SQCD with the gauge group SU$(N)\times$U(1) and the FI $D$ term.
The bosonic part of the action of the latter theory is
\beqn
S&=&\int d^4x \left\{\frac1{4g^2_2}
\left(F^{a}_{\mu\nu}\right)^2 +
\frac1{4g^2_1}\left(F_{\mu\nu}\right)^2
+ \left|\nabla_{\mu}
q^{A}\right|^2 + \left|\nabla_{\mu} \bar{\tilde{q}}^{A}\right|^2
\right.
\nonumber\\[4mm]
& + &
\left.
\frac{g^2_2}{2}\left(
 \bar{q}_A\,T^a q^A -
\tilde{q}_A T^a\,\bar{\tilde{q}}^A\right)^2
+\!\! \frac{g^2_1}{8}
\left(\bar{q}_A q^A - \tilde{q}_A \bar{\tilde{q}}^A-N\xi\right)^2
\right\}.\nonumber\\
\label{noneqcd}
\eeqn
All $F$ terms disappear in this limit, and we are left only with the $D$ terms.
We have $4N^2$ real components of $q$ and $\tilde{q}$ fields while the
number of the $D$ term constraints in (\ref{noneqcd})
is $N^2$. Moreover, $N^2$ phases are
eaten by the Higgs mechanism. Thus, dimension of the Higgs branch in
Eq.~(\ref{noneqcd}) is $$4N^2-N^2-N^2=2N^2\,.$$

The vacuum (\ref{N1qvev}) corresponds to the base point of the
Higgs branch with $\tilde{q}=0$. In other words, flowing from \ntwo
theory (\ref{qed}), we do not recover the entire Higgs branch
of \none SQCD. Instead,  we arrive at a single
vacuum ---  a base point of a Higgs branch.

The scale of \none SQCD $$\Lambda^{{\mathcal N}=1}_{{\,\rm SU}(N)}$$
is expressed in terms of the scale $\Lambda_{{\rm SU}(N)}$
of the deformed \ntwo theory as follows:
\beq
\big(\Lambda^{{\mathcal N}=1}_{{\,\rm SU}(N)}\big)^{2N}=\mu_2^{N}\,
\Lambda^{N}_{{\rm SU}(N)}\,.
\label{lambdanone}
\eeq
To keep the bulk theory at weak coupling in the limit
of large $\mu_i$ we assume that
\beq
\sqrt{\xi}\gg \Lambda^{{\mathcal N}=1}_{{\,\rm SU}(N)}\,.
\label{N1xilam}
\eeq

Now,
considering the theory (\ref{qed})
with the potential (\ref{N1pot}),
let us return to the case of arbitrary $\mu_i$ and discuss non-Abelian
string solutions. BPS-saturation is maintained.
By the same token as for the BPS strings in \ntwo
we use the {\em ansatz}
\beq
q^{kA}\equiv \vp^{kA}\,,\qquad \tilde{q}_{Ak}=0\,.
\label{tildeq0}
\eeq
The adjoint fields are set to zero. Note that Eq.~(\ref{tildeq0})
is an SU(2)$_{R}$-rotated version of (\ref{qtilde}).
The  FI $F$ term considered in Sect.~\ref{strings} is
rotated into the FI $D$ term
in (\ref{N1pot}).

With these simplifications the \none model (\ref{qed}) with the
potential (\ref{N1pot}) reduces to the model (\ref{redmodel}) which was
exploited in Sect.~\ref{strings}
to obtain non-Abelian string solutions.
The reason for this is that the adjoint fields play no role in the
string solutions, and we let them vanish, see Eq.~(\ref{N1avev}). Then
\ntwo  breaking terms vanish, and the
potential (\ref{N1pot})
reduces to the one in Eq.~(\ref{pot}) (up to an SU(2)$_{R}$ rotation).

This allows us to parallel the construction of the non-Abelian strings carried out
in Sect. \ref{nas}. In particular, the elementary
string solution is given by Eq.~(\ref{str}). Moreover, the {\em bosonic} part
of the worldsheet theory is nothing but the CP$(N-1)$ sigma model (\ref{cp}),
with the coupling constant $\beta$ determined by the  coupling $g_2$ of the
bulk theory via Eq.~(\ref{beta}) at the scale $\sqrt{\xi}$.
The latter scale plays
the role of the UV  cutoff in the worldsheet theory.

At small values of the deformation parameter,
$$\mu_2\ll\sqrt{\xi}\,,$$ the coupling constant $g_2$ of
the four-dimensional bulk theory is determined by the scale
$\Lambda_{{\rm SU}(N)}$ of
the \ntwo theory. Then Eq.~(\ref{beta}) implies (see (\ref{lambdasig}))
\beq
\Lambda_{\sigma}=\Lambda_{{\rm SU}(N)}\, ,
\label{cpscale1}
\eeq
where we take into account  that the first coefficient of the $\beta$ function
is $N$ both in the \ntwo limit of the four-dimensional bulk theory and
in the two-dimensional CP$(N-1)$ model, see (\ref{asyfree}).

Instead, in the limit of large $\mu_2$, $$\mu_2\gg\sqrt{\xi}\,,$$ the coupling
constant $g_2$ of
the  bulk theory is determined by the scale
$\Lambda^{{\mathcal N}=1}_{{\,\rm SU}(N)}$ of
\none SQCD (\ref{noneqcd}), see (\ref{lambdanone}). In this limit
Eq.~(\ref{beta}) gives
\beq
\Lambda_{\sigma}=\frac{\big(\Lambda^{{\mathcal N}=1}_{{\,\rm SU}(N)}\big)^2}{\sqrt{\xi}}\,,
\label{cpscale2}
\eeq
where we take into account the fact that the first coefficient of the $\beta$ function
in \none  SQCD is $2N$.

\subsubsection{Limits of applicability}
\label{limits}

As was discussed above,  both the string solution and the bosonic part of the
worldsheet  theory for the non-Abelian strings in
\none   with the potential (\ref{N1pot}) are identical to those in
\ntwo. However, the occurrence of the Higgs branch in the limit
$\mu\to\infty$ manifests itself at
the quantum level \cite{SYnone}. At the classical level
light fields appearing in the bulk theory in the large-$\mu$ limit do not enter
the string solution. The string is ``built'' of heavy fields. However, at the quantum
level couplings to the light fields lead to a dramatic effect: an effective string
thickness becomes large due to  long range tails
of the string profile functions associated with the light fields. As a matter of fact,
we demonstrated \cite{SYnone} that  in
the fermion sector
this effect is seen already at the classical level. Some of the fermion zero modes on the string solution
acquire long-range tails and become non-normalizable in the limit $\mu\to\infty$.

\begin{figure}
\epsfxsize=9cm
\centerline{\epsfbox{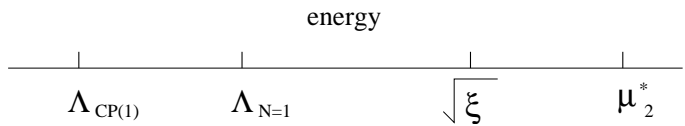}}
\caption{\small
Relevant scale hierarchy in the limit  $\mu_2\gg \sqrt{\xi}$.}
\label{scales}
\end{figure}

Below we will estimate the range of validity of the description of non-Abelian string
dynamics  by the
CP$(N-1)$ model (\ref{cp}). To this end let us note that higher derivative
corrections to (\ref{cp}) run in powers of
\beq
\Delta\, \pt_k\,,
\label{hder}
\eeq
where $\Delta$ is the string transverse size (thickness). At small $\mu_2$
it is quite clear that
$\Delta\sim1/\sqrt{\xi}\,.$
A typical energy scale on the string
worldsheet is given by the scale $\Lambda_{\sigma}$ of the CP$(N-1)$ model which,
in turn,
is given by (\ref{cpscale1}) at small $\mu_2$. Thus, $\pt \to\Lambda_{\sigma}$
and
higher-derivative corrections run in powers of $\Lambda_{\sigma}/\sqrt{\xi}$.
At small $\mu_2$ the higher-derivative corrections are suppressed by powers
of
$ \Lambda_{\sigma}/\sqrt{\xi}\ll 1$ and can be ignored.
However, with $\mu_2$ increasing, the fermion zero modes
acquire long-range tails \cite{SYnone}. This means that an
effective thickness of the string grows. The thickness is determined by masses of lightest
states (\ref{light}) of the bulk theory,
\beq
\Delta\sim \frac{1}{m^{-}}=\frac{\mu_2}{\xi}\, .
\label{delta}
\eeq
The higher-derivative terms are small if $\left( \Delta\,\Lambda_{\sigma}\right) \ll 1$.
Substituting
here the scale of the CP$(N-1)$ model given at large $\mu_2$ by (\ref{cpscale2})
and the scale of \none SQCD (\ref{lambdanone}) we arrive at
\beq
\mu_2\ll\mu_2^{*}\,,
\label{cond}
\eeq
where the critical value of   $\mu_2$ is
\beq
\mu_2^{*} = \frac{\xi}{\Lambda_{\sigma}}=\frac{\xi^{3/2}}
{\big(\Lambda_{{\,\rm SU}(N)}^{{\mathcal N}=1}\big)^2}\,.
\label{mucrit}
\eeq
If the  condition (\ref{cond}) is met,
the \ntwo CP$(N-1)$ model gives  a good description of worldsheet physics.
A hierarchy of relevant scales in our theory is displayed in Fig.~\ref{scales}.

If we increase $\mu_2$ above the critical value (\ref{mucrit})
the non-Abelian strings
become thick and their  worldsheet dynamics is
no  longer described   by \ntwo CP$(N-1)$ sigma model. The
higher-derivative corrections on the worldsheet explode.
Note that the physical reason for the growth of the string thickness $\Delta$
is the presence of the Higgs branch in \none SQCD. Although
the classical string solution (\ref{sna}) retains a finite transverse size,
the Higgs branch manifests itself at the quantum level. In particular,
the fermion zero modes feel  the Higgs branch and acquire long-range
logarithmic tails.

\subsection{The $M$ model}
\setcounter{equation}{0}
\label{secmmodel}

In Sect.~\ref{nonep} we learned that the occurrence of the Higgs branch in
\none SQCD obscures physics of the  non-Abelian strings.
Thus, it is highly desirable to get rid of the Higgs branch, keeping \none.
This was done in \cite{us}. Below we will review some results pertinent
to the issue.

To eliminate light states we will introduce a particular \ntwo breaking deformation
in the U$(N)$ theory with the potential (\ref{N1pot}). Namely,
we uplift the quark mass matrix $m_A^B$ (see Eq.~(\ref{pot}) where this
matrix is assumed to be diagonal) to the superfield status,
$$
m_A^B \to M_A^B\,,
$$
and introduce the superpotential
\beq
{\mathcal W}_M = Q M \tilde Q\,.
\eeq
The matrix $M$ represents $N^2$ chiral superfields of the mesonic type
(they are color singlets). Needless to say,  we have to
add, in addition,  a kinetic term for $ M_A^B$,
\beq
 S_{M\rm kin} = \int d^4x \, d^2\theta \, d^2\bar{\theta}\; \;\frac{2}{h}\;
{\rm Tr}\,\bar{M}M
\,,
\label{mkin}
\eeq
where $h$ is a new coupling constant
(it supplements the set of the gauge couplings).
In particular, the kinetic term for the scalar components of $M$ takes the form
\beq
\int d^4x \left\{\frac1h \left|\pt_{\mu} M^0\right|^2
+\frac1h \left|\pt_{\mu} M^a\right|^2\right\},
\label{kinM}
\eeq
where we use the decomposition
\beq
M^A_B=\frac12\, \delta_B^A\;M^0 +(T^a)^A_B\;M^a\,.
\label{adjointM}
\eeq

At $h=0$ the matrix field $M$ becomes sterile, it is frozen and in essence
returns to the status
of a constant numerical matrix. The theory
acquires flat directions (a moduli space).
With non-vanishing $h$ these flat directions are lifted, and $M$ is
determined by the minimum of the scalar potential, see below.

The uplift of the quark mass matrix to superfield is a crucial step
which allows us to lift the Higgs branch which would develop in this
theory in the large $\mu$ limit if $M$ were a constant matrix.
We will refer to this theory as to the $M$ model.

The potential $V(q^A,\tilde{q}_A,a^a,a,M^0,M^a)$ of the $M$ model
is
\beqn
& & V(q^A,\tilde{q}_A,a^a,a,M^0,M^a) =
 \frac{g^2_2}{2}
\left( \frac{1}{g^2_2}\,  f^{abc} \,\bar a^b a^c
 +
 {\rm Tr}\,\bar{q}\,T^a q -
{\rm Tr}\,\tilde{q}\, T^a\,\bar{\tilde{q}}\right)^2
\nonumber\\[3mm]
&+& \frac{g^2_1}{8}
\left({\rm Tr}\,\bar{q} q - {\rm Tr}\,\tilde{q} \bar{\tilde{q}}-
N\xi\right)^2+
\frac{g^2_2}{2}\left| 2{\rm Tr}\,\tilde{q}T^a q +\sqrt{2}\mu_2a^a\right|^2
\nonumber\\[3mm]
&+&
\frac{g^2_1}{2}\left| {\rm Tr}\,\tilde{q} q +\sqrt{N}\mu_1 a \right|^2
+\frac12 \,{\rm Tr}\, \left\{ \left|(a +\,2\,T^a\, a^a)q +
\frac1{\sqrt{2}}q(M^0 +2T^a M^a)
\right.
\right|^2
\nonumber\\[3mm]
&+&
\left.
\left|(a +\,2\,T^a\, a^a)\bar{\tilde{q}}
+\frac1{\sqrt{2}}\bar{\tilde{q}}(M^0 +2T^a M^a)
\right|^2 \right\}
+\frac{h}{4}\left|{\rm Tr}\,\tilde{q}q\right|^2
+h\left|{\rm Tr}\,qT^a\tilde{q}\right|^2
\,.
\nonumber\\[3mm]
&&\mbox{}
\label{Mapot}
\eeqn
The two last terms here are $F$ terms of the $M$ field.
In Eq.~(\ref{Mapot}) we also introduced the FI $D$-term for the U(1) field,
with the FI parameter $\xi$.

The FI term triggers the spontaneous breaking
of the gauge symmetry. The VEV's
of the squark fields and adjoint fields are given by (\ref{N1qvev}) and
(\ref{N1avev}), respectively, while the VEV's of $M$ field vanish,
\beq
\langle M^a\rangle =0,\,\,\,\,\langle M^0\rangle =0\,.
\label{Mvev}
\eeq

\vspace{2mm}

The color-flavor locked form of the quark VEV's in
Eq.~(\ref{N1qvev}) and the absence of VEV's of the adjoint field $a^a$
and the meson fields $M^a$ in
Eqs.~(\ref{N1avev}) and (\ref{Mvev})
result in the fact that, while the theory is fully Higgsed, a diagonal
SU($N$)$_{C+F}$ symmetry survives as a global symmetry,
much in the same way as in $\mu$-deformations of \ntwo QCD. Namely, the global rotation
\beq
q\to UqU^{-1},\qquad a^aT^a\to Ua^aT^aU^{-1},\qquad M\to U^{-1}MU,
\label{c+fmm}
\eeq
is not broken by the VEV's (\ref{N1qvev}), (\ref{N1avev}) and (\ref{Mvev}).
Here $U$ is a matrix from SU($N$).
As usual, this symmetry  leads to the emergence of
orientational zero modes  of the $Z_N$ strings in the theory with the potential
(\ref{Mapot}).

At large $\mu$ one can readily integrate out the adjoint fields ${\mathcal A}^a$ and
${\mathcal A}$.  The bosonic part of the action of the $M$ model
takes the form
\beqn
S&=&\int d^4x \left[\frac1{4g^2_2}
\left(F^{a}_{\mu\nu}\right)^2 +
\frac1{4g^2_1}\left(F_{\mu\nu}\right)^2+
{\rm Tr}\,\left|\nabla_{\mu}
q\right|^2 + {\rm Tr}\,\left|\nabla_{\mu} \bar{\tilde{q}}\right|^2
\right.
\nonumber\\[4mm]
&+& \frac1h \left|\pt_{\mu} M^0\right|^2+
\frac1h \left|\pt_{\mu} M^a\right|^2
+\frac{g^2_2}{2}
\left(
 {\rm Tr}\,\bar{q}\,T^a q -
{\rm Tr}\,\tilde{q} T^a\,\bar{\tilde{q}}\right)^2
\nonumber\\[3mm]
&+& \frac{g^2_1}{8}
\left({\rm Tr}\,\bar{q} q - {\rm Tr}\,\tilde{q} \bar{\tilde{q}}-
N\xi\right)^2+
 {\rm Tr}|qM|^2 +{\rm Tr}|\bar{\tilde{q}}M|^2
\nonumber\\[3mm]
&+&
\left.
\frac{h}{4}\left|{\rm Tr}\,\tilde{q}q\right|^2
+h\left|{\rm Tr}\,qT^a\tilde{q}\right|^2
\right\}
\,.
\label{mmodel}
\eeqn

The vacuum of this theory is given by Eqs. (\ref{N1qvev}) and (\ref{Mvev}).
The mass spectrum of elementary excitations over
this vacuum consists of the \none gauge multiplets for
the U(1) and SU($N$) sectors, with masses given in Eqs.~(\ref{msuN}) and
(\ref{mu1}). In addition, we have chiral multiplets $\tilde{q}$ and $M$, with masses
\beq
m_{{\rm U}(1)} =\sqrt{\frac{hN\xi}{4}}
\label{U1mass}
\eeq
for the U(1) sector, and
\beqn
m_{{\rm SU}(N)} = \sqrt{\frac{h\xi}{2}}
\label{SUNmass}
\eeqn
for the SU($N$) sector.

\vspace{2mm}

It is worth emphasizing  that there are no massless states in the bulk theory.
At $h=0$ the theory with the potential (\ref{Mapot}) develops a Higgs branch
in the large-$\mu$ limit (see Sect.~\ref{nonep}). If $h\ne 0$, $M$ becomes a
fully dynamical field.  The Higgs branch is lifted, as follows from Eqs.~(\ref{U1mass})
and (\ref{SUNmass}).

The \none SQCD with the $M$ field, the $M$ model, belongs to the class of theories
introduced by Seiberg \cite{Sdual} to provide a dual description of
conventional \none SQCD
with the SU($N_c$) gauge group and $N_f$ flavors of fundamental matter, where
$$N_c=N_f - N$$  (for reviews see Refs.~\cite{IS,MS}).
There are significant distinctions, however.

Let us point out the main
differences of the $M$ model (\ref{mmodel}) from those introduced \cite{Sdual}
by Seiberg:

(i) The  theory (\ref{mmodel}) has the U($N$) gauge group rather than SU($N$);

(ii) It  has the FI $D$ term instead of a linear in $M$ superpotential
in Seiberg's models;

(iii) Following \cite{us} we consider the case $N_f=N$ which would correspond to
Seiberg's $N_c=0$ in the original SQCD.
The theory (\ref{mmodel}) is asymptotically free,
while Seiberg's dual theories are most meaningful (i.e. have
the maximal predictive power with regards to the original strongly coupled \none SQCD)
below the left edge of the conformal
window, in the range
$N_f<(3/2)\,N_c$, which would correspond to $N_f >3N$ rather
than $N_f = N$. Note that at $N_f >3N$ the theory
(\ref{mmodel}) is not asymptotically free and is thus uninteresting from our standpoint.

In addition, it is worth noting that at $N_f>N$ the vacuum (\ref{N1qvev}), (\ref{Mvev})
becomes metastable: supersymmetry is broken \cite{ISS}. The $N_c=N_f - N$
supersymmetry-preserving vacua have vanishing VEV's of the quark fields and
a non-vanishing VEV of the $M$ field
\footnote{This is correct for the version of the theory with
$\xi$-parameter introduced via superpotential.}. The latter vacua are
associated with the gluino
condensation in pure SU($N$) theory,
$\langle\lambda\lambda\rangle \neq 0$, arising  upon decoupling $N_f$ flavors
\cite{IS}. In the case $N_f=N$ to which we limit ourselves
the vacuum (\ref{N1qvev}), (\ref{Mvev}) preserves supersymmetry. Thus, despite
a conceptual similarity between Seiberg's models and ours,
dynamical details are radically different.

Now, it is time pass to solutions for non-Abelian
BPS strings in the $M$ model \cite{us}.  Much in the same way as in Sect.~\ref{nonep}
we use the {\em ansatz} (\ref{tildeq0}). Moreover,
we set the adjoint fields and the $M$ fields at zero. With these simplifications
the \none model  with the
potential (\ref{Mapot}) reduces to the model (\ref{redmodel}) which we used
previously in the original construction of the non-Abelian strings.

In particular, the solution for the elementary
string is given by (\ref{str}). Moreover, the bosonic part
of the effective worldsheet theory is again described by
the CP$(N-1)$ sigma model (\ref{cp})
with the coupling constant $\beta$ determined by (\ref{beta}).
The scale of this CP$(N-1)$ model is given by Eq.~(\ref{cpscale2})
in the limit of large $\mu$.

To conclude this section let us note a somewhat related development:
{\em non}-BPS non-Abelian strings were
considered in metastable vacua of a dual description of \none SQCD at $N_f>N$ in
Ref.~\cite{Jmeta}.

\subsection{Confined non-Abelian monopoles}
\label{evol}
\setcounter{equation}{0}

Since supersymmetric CP$(N-1)$ model is an  effective
low-energy theory describing worldsheet physics  of the non-Abelian string
in the $M$ model \footnote{This statement
is not quite correct. We will discuss
supersymmetry of the effective worldsheet theory in Sect.~\ref{susyenh}.
Fine details of worldsheet physics may slightly deviate from those
one obtains in  supersymmetric CP$(N-1)$.},
consequences of this model ensue, in particular, $N$ degenerate vacua and kinks
that interpolate between them, similar to the kinks
that emerge in \ntwo SQCD. These kinks are interpreted as
(confined) non-Abelian monopoles \cite{Tong,SYmon,HT2}, the descendants
of the 't Hooft--Polyakov monopole (see Sect.~\ref{kinkmonopole}).

Let us  discuss what happens with these monopoles
as we deform our theory and eventually end up with the $M$ model.
It is convenient to split this deformation into several distinct stages.
We will describe what happens with the monopoles as one passes
from one stage to another. Some of these steps involving deformations
of \ntwo QCD were already discussed in Sect.~\ref{kinkmonopole}. Here
we focus on deformations of \ntwo QCD leading to the $M$ model.

A qualitative evolution of the monopoles under consideration
as a function of the  relevant parameters is presented in
Fig.~\ref{sixf}.

\vspace{1mm}

(i) We start from \ntwo SQCD turning off the \ntwo
breaking parameters $h$ and $\mu$'s
as well as the FI parameter in the potential (\ref{Mapot}), i.e. we start from the Coulomb
branch of the theory,
\beq
\mu_1=\mu_2=0,\qquad h=0, \qquad \xi=0, \qquad M\neq 0\, .
\label{stage1}
\eeq
As was explained in Sect.~\ref{secmmodel}, the field $M$
is frozen in this limit and can take arbitrary values (the notorious flat direction).
The matrix  $M^A_B$ plays the role of a fixed mass parameter matrix for
the quark fields.
As a first step let us consider the diagonal matrix $M$, with distinct diagonal entries,
\beq
M^A_B ={\rm diag}\,\{M_1,...,M_N\}\,.
\label{diffmasses}
\eeq
Shifting the field $a$ one can always make $\sum_{A}M_A=0$ in the limit $\mu_1=0$.
Therefore $M^0=0$. If all $M_A$'s are different the gauge group SU($N$) is broken down
to U(1)$^{(N-1)}$ by a VEV of the SU($N$) adjoint scalar (see (\ref{avev})),
\beq
\langle a^k_l\rangle = -\frac{1}{\sqrt{2}} \,\delta^k_l  M_l  \,.
\label{adjvev}
\eeq
Thus, as was already discussed in Sect.~\ref{kinkmonopole},
there are 't Hooft--Polyakov monopoles embedded in the broken gauge SU($N$).
Classically, on the Coulomb branch
the masses of $(N-1)$ elementary monopoles are proportional to
$$|(M_A-M_{A+1}) \,| /g_2^2\,. $$

In  the limit $(M_A-M_{A+1})\to 0$ the monopoles tend to
become massless, formally, in the classical
approximation. Simultaneously their size becomes infinite \cite{We}.
The mass and size are stabilized by  highly quantum confinement effects. The monopole confinement occurs in the Higgs phase, at $\xi\neq 0$.

\vspace{1mm}

(ii) Now let us make the FI parameter $\xi$ nonvanishing.
This triggers the squark condensation. The
theory is in the Higgs phase. We still keep \ntwo breaking parameters $h$ and $\mu$'s
vanishing,
\beq
\mu_1=\mu_2=0,\qquad h=0, \qquad \xi\neq 0, \qquad M\neq 0.
\label{stage2}
\eeq
The squark condensation leads to formation of the $Z_N$ strings.
Monopoles become confined by these strings. As we discussed in
Sect.~\ref{kinkmonopole}
$(N-1)$ elementary monopoles become junctions of pairs of elementary strings.

Now, if we  reduce $|\Delta M_A| $,
\beq
\Lambda_{{\rm CP}(N-1)} \ll \left|\Delta M_A\right|   \ll \sqrt{\xi}\, ,
\label{ququr}
\eeq
the size of the monopole along the string $\sim \left|(M_A-M_{A+1}) \,\right| ^{-1}\,$
becomes larger than the transverse size of the attached strings.
The monopole becomes  a {\em bona fide} confined
monopole (the  lower left corner of  Fig.~\ref{sixf}).  At nonvanishing $\Delta M_A$
the effective theory on the string worldsheet is CP$(N-1)$ model
with twisted mass terms \cite{Tong,SYmon,HT2}, see Sect.~\ref{unequalmasses}.
Two $Z_N$ strings attached to a elementary monopole correspond to two
``neighboring'' vacua of the CP$(N-1)$ model. The monopole
(aka the string junction of two $Z_N$ strings) manifests itself
as a kink interpolating between these two vacua.

\vspace{1mm}

(iii)  Next, we switch off the mass differences
$\Delta M_A$ still keeping the \ntwo breaking
parameters vanishing,
\beq
\mu_1=\mu_2=0,\qquad h=0, \qquad \xi\neq 0, \qquad M = 0 \,.
\label{stage3}
\eeq

The values of the twisted masses in CP$(N-1)$ model coincide
with $\Delta M_A$ while the
size of the twisted-mass sigma-model kink/confined monopole
is of the order of  $\sim \left|(M_A-M_{A+1}) \,\right| ^{-1}\,$.

As we diminish $\Delta M_A$
approaching $\Lambda_{{\rm CP}(N-1)}$ and then getting  below
$\Lambda_{{\rm CP}(N-1)}$,
the monopole size grows, and, classically, it would explode.
This is where quantum effects in the worldsheet theory take over.
It is natural to refer to this domain of parameters as the ``regime of
highly quantum dynamics."
While the thickness of the string (in the transverse direction) is
$\sim \xi ^{-1/2}$, the
$z$-direction size of the kink  representing the confined
monopole in the highly quantum regime is much larger, $\sim
\Lambda_{{\rm CP}(N-1)}^{-1}$, see the  lower right corner in  Fig.~\ref{sixf}.

In this regime the confined monopoles become truly
non-Abelian. They  no longer carry average magnetic flux since
\beq
\langle n^l\rangle =0\, ,
\label{nvev}
\eeq
in the strong coupling limit of the CP$(N-1)$ model \cite{W79}. The kink/monopole
belongs to the fundamental representation of the SU($N$)$_{C+F}$
group \cite{W79,HoVa}.

\vspace{1mm}

(iv) Thus, with vanishing $\Delta M_A$ we still have confined ``monopoles"
(interpreted as  kinks) stabilized
by quantum effects in the worldsheet CP$(N-1)$ model. Now we can
finally switch on the \ntwo breaking parameters $\mu_i$
and $h$,
\beq
\mu_i\neq 0,\qquad h\neq 0, \qquad \xi\neq 0, \qquad M = 0\, .
\label{stage4}
\eeq
Note that the last equality here is automatically satisfied in the vacuum, see
Eq.~(\ref{Mvev}).

As was discussed in Sect.~\ref{secmmodel}
the effective worldsheet description of the non-Abelian string is
still given by a deformation of the supersymmetric  CP$(N-1)$ model
(for a more careful discussion of the worldsheet
supersymmetry see Sect.~\ref{susyenh}). This
model still has $N$ vacua (Witten's index!)
which should be interpreted as $N$ elementary non-Abelian
strings in the quantum regime, with kinks
interpolating between these vacua. These kinks
should still be interpreted as non-Abelian
confined monopoles/string junctions.

Note that although the adjoint fields are still
present in the theory (\ref{Mapot}) their VEV's vanish (see (\ref{N1avev}))
and the monopoles cannot be seen in the semiclassical approximation.
They are seen solely as worldsheet kinks.
Their mass and inverse size are determined by $\Lambda_{\sigma}$ which
in the limit of large $\mu_i$ is given by Eq.~(\ref{cpscale2}).

\vspace{1mm}

(v) At the last stage,  we take the limit of large masses of the adjoint fields
in order to eliminate them from the physical spectrum altogether,
\beq
\mu_i\to \infty,\qquad h\neq 0, \qquad \xi\neq 0, \qquad M = 0\, .
\label{stage5}
\eeq
The theory flows to \none SQCD extended by the  $M$ field.

In this limit we get a remarkable result: although the adjoint fields
are eliminated from our theory
and the monopoles cannot be seen in any semiclassical description,
our analysis shows
that confined non-Abelian monopoles still exist in
the theory (\ref{mmodel}). They are seen
as  kinks in the effective worldsheet theory on the non-Abelian string.

\vspace{1mm}

(vi) The  confined monopoles are in the highly quantum regime, so they
carry no average magnetic flux (see Eq.~(\ref{nvev})). They are genuinely non-Abelian.
Moreover, they acquire global flavor quantum numbers. In fact, they
belong to the fundamental representation of the global SU($N$)$_{C+F}$ group
(see Refs.~\cite{W79,HoVa} where this phenomenon
is discussed in the context of the CP$(N-1)$-model kinks).

It is quite plausible that
the emergence of these non-Abelian monopoles can
shed light on mysterious objects introduced by Seiberg: ``dual magnetic''
quarks which play an important role in the description of \none SQCD at strong
coupling \cite{Sdual,IS}.

\subsection{Fermionic sector of the worldsheet theory}
\label{susyenh}
\setcounter{equation}{0}

In this section we will discuss the fermionic sector of the low-energy effective
theory on the worldsheet of the non-Abelian string in  the $M$ model,
as well as supersymmetry of the worldsheet theory.
First, we note that our string is 1/2 BPS-saturated.
Therefore, in the \ntwo limit (with \ntwo  breaking
parameters $\mu_i$ and $h$ vanishing)
four supercharges out of eight present in the bulk
theory are automatically
preserved on the string worldsheet. They become supercharges
in the  CP$(N-1)$ model.

For simplicity let us discuss the simplest case of $N=2$ thus limiting ourselves
to the CP(1) model. Generalization to arbitrary $N$ is rather straightforward.
The action of this model is given by (\ref{ntwocp})
where we used the fact that CP(1) is equivalent to the $O(3)$ sigma model
defined in terms of a unit real vector $S^a$, see (\ref{sn}).

This model has two real bosonic degrees of freedom. Two real fermion fields
$\chi_1^a$ and $\chi_2^a$ are subject to constrains
\beq
\chi_1^aS^a=0, \qquad \chi_2^aS^a=0\,.
\label{fermconstr}
\eeq
Altogether we have four real fermion fields in the model (\ref{ntwocp}).

What happens when we break \ntwo supersymmetry of the bulk model
by switching on parameters
 $\mu_i$ and $h$? The 1/2 ``BPS-ness" of the string solution
requires only two supercharges. However, as we will show below, the number of the
fermion zero modes in the string
background does not change. This number is fixed by the index theorem.
Thus, the number of (classically) massless fermion fields in the worldsheet
model does not change.

It was shown in \cite{SYnone} that the (2,2)
supersymmetric
sigma model with the CP$(N-1)$ target space does not admit (0,2) supersymmetric
deformations. Therefore, it was concluded in \cite{SYnone} that the world sheet theory has ``accidental'' supersymmetry enhancement. A similar phenomenon
of supersymmetry enhancement was found earlier
in \cite{svritz2} in the worldsheet theory on domain walls.

While supersymmetry enhancement certainly takes place
for domain walls, Edalati and Tong suggested \cite{EdT}
that on the string worldsheet (2,2) SUSY is ruined through
a mixing of superorientational and supertranslational zero modes
resulting in  a ``natural" un-enhanced (0,2) supersymmetry
on the string worldsheet.
It was shown \cite{EdT} that the sigma model with the $C\times {\rm CP}(1)$ target space does admit (0,2)
supersymmetric deformations. It is not clear at the moment whether or not
this mixing actually occurs in the $M$ model. There are no general reasons
that would forbid it.

If it actually occurs then the worldsheet
(0,2) supersymmetric $C\times {\rm CP}(1)$ model
must have a $\mu$-deformed four-fermion interaction of the form
\beqn
S^{(0,2)}_{{\rm CP}(1)}
&=&
\beta \int d t d z \left\{\frac12 (\pt_k S^a)^2+
\frac12 \, \chi^a_1 \, i(\pt_0-i\pt_3)\, \chi^a_1
\right.
\nonumber\\[3mm]
&+& \left.
\frac12 \, \chi^a_2 \, i(\pt_0+i\pt_3)\, \chi^a_2
-\frac12\,\frac1{1+c|\mu_2|^2/(g^2_2\xi)} \,(\chi^a_1\chi^a_2)^2
\right\},
\label{02o3}
\eeqn
where $c$ is an unknown coefficient. Also the first constraint in
Eq.~(\ref{fermconstr}) must be
replaced by
\beq
\chi_1^a S^a=c/2\,(\mu_2\zeta_1 + \bar{\mu}_2\bar{\zeta}_1)
\,,
\eeq
where $\zeta_1$ is the right-moving two-dimensional fermion field associated with
the supertranslational zero modes.
If the Edalati--Tong  conjecture  \cite{EdT} is correct
the four-fermion term disappears in the large-$\mu$ limit. To confirm or rule
out this scenario one has to calculate the coefficient in front of the four-fermion term in
(\ref{02o3}). This is not a simple calculation, and
the issue calls for further studies.

In any case, the worldsheet supersymmetric model must have $N$
degenerate vacua to be interpreted as
$N$ elementary strings of the bulk theory. This number is protected by Witten's index and survives possible \ntwo breaking deformations.
We used this fact in Sect.~\ref{evol}
to demonstrate the presence of kinks in the string worldsheet theory.
The kinks which interpolate between these vacua are confined monopoles.
Below we  will show that the occurrence  of four ($4(N-1)$ in the general case)
superorientational fermion zero modes on the non-Abelian strings follows from an
index theorem.

\subsubsection{Index theorem}
\label{indext}

In this section we will
discuss an index theorem establishing the number of
the fermion zero modes on the string. For definiteness we will consider the $M$ model
\cite{us}.
Similar theorems can be easily proved for ordinary \none QCD (\ref{noneqcd}) as well
as for theories at intermediate $\mu$'s with potentials (\ref{N1pot}) or (\ref{Mapot}).
They generalize index theorems obtained long ago for simple non-supersymmetric
models \cite{Indexth}.

The  fermionic part of the action  of the model (\ref{mmodel}) is
\beqn
S_{\rm ferm}
&=&
\int d^4 x\left\{
\frac{i}{g_2^2}\bar{\lambda}^a \bar{D}\hspace{-0.65em}/\lambda^{a}+
\frac{i}{g_1^2}\bar{\lambda} \bar{\pt}\hspace{-0.55em}/\lambda
+ {\rm Tr}\left[\bar{\psi} i\bar\nabla\hspace{-0.75em}/ \psi\right]
+ {\rm Tr}\left[\tilde{\psi} i\nabla\hspace{-0.75em}/ \bar{\tilde{\psi}}
\right]
\right.
\nonumber\\[3mm]
&+&
\frac{2i}{h}{\rm Tr}\left[\bar{\zeta} \bar{\pt}\hspace{-0.55em}/\zeta\right]
+\frac{i}{\sqrt{2}}\,{\rm Tr}\left[ \bar{q}(\lambda\psi)-
(\tilde{\psi}\lambda)\bar{\tilde{q}} +(\bar{\psi}\bar{\lambda})q-
\tilde{q}(\bar{\lambda}\bar{\tilde{\psi}})\right]
\nonumber\\[3mm]
&+&
\frac{i}{\sqrt{2}}\,{\rm Tr}\left[ \bar{q}\, 2T^a\, (\lambda^{a}\psi)-
(\tilde{\psi}\lambda^a)\, 2T^a\, \bar{\tilde{q}}
 +(\bar{\psi}\bar{\lambda}^a)\, 2T^a\, q -
\tilde{q}\,2T^a\, (\bar{\lambda}^{a}\bar{\tilde{\psi}})\right]
\nonumber\\[3mm]
&+&
i\,{\rm Tr}\left[ \tilde{q}(\psi\zeta)+
(\tilde{\psi}q\zeta) +(\bar{\psi}\bar{\tilde{q}}\bar{\zeta})+
\bar{q}(\bar{\tilde{\psi}}\bar{\zeta})\right]
\nonumber\\[3mm]
&+&
\left.
i\,{\rm Tr}\left(\tilde{\psi}\psi M+
\bar{\psi}
\bar{\tilde{\psi}}\bar{M}\right)
\right\}\,,
\label{fermactmm}
\eeqn
where the matrix color-flavor notation is used for the
matter fermions $(\psi^{\alpha})^{kA}$ and $(\tilde{\psi}^{\alpha})_{Ak}$,
and the traces are performed
over the color-flavor indices. Moreover, $\zeta$ denotes
the fermion component of the matrix $M$ superfield,
\beq
\zeta^A_B=\frac12 \delta^A_B\, \zeta^0 + (T^a)^A_B\,\zeta^a\,.
\rule{0mm}{7mm}
\label{matrixzeta}
\eeq

In order to find the number of the fermion zero modes in the background
of the non-Abelian string solution (\ref{str}) we have to carry out the following
program. Since our string solution depends only on two coordinates
$x_i$ ($i=1,2$), we can reduce our theory to two dimensions. Given the theory
defined on the $(x_1,x_2)$ plane we have to identify an axial current and derive
the anomalous divergence for this current. In two dimensions the axial current anomaly
takes the form
\beq
\pt_ij_{i5}\sim F^{*}\,,
\label{appranom}
\eeq
where $F^{*}=(1/2)\varepsilon_{ij}F_{ij}$ is the dual U(1) field strength in
two dimensions.

Then, the integral over the left-hand side over the $(x_1,x_2)$ plane gives us the index
of the 2D Dirac operator $\nu$ coinciding with the number of the 2D left-handed minus
2D right-handed zero modes of this
operator in the given background field. The integral over the right-hand side is
proportional to the string flux. This will fix the number of the chiral fermion zero
modes\,\footnote{Chirality is understood here as the two-dimensional chirality.} on
the string with the given flux. Note that the reduction of the theory to two dimensions
is an important step in this program. The anomaly relation in four dimensions
involves the instanton charge $F^{*}F$ rather than the string flux and is therefore
useless for our purposes.

The reduction of \none gauge theories to two dimensions is discussed in
detail in \cite{W93} and here we will be brief. Following \cite{W93} we
use the rules
\beqn
&& \psi^{\alpha}\to(\psi^{-},\psi^{+}), \qquad
\tilde{\psi}^{\alpha}\to(\tilde{\psi}^{-},\tilde{\psi}^{+}),
\nonumber\\[3mm]
&& \lambda^{\alpha}\to(\lambda^{-},\lambda^{+}),\,\qquad
\zeta^{\alpha}\to(\zeta^{-},\zeta^{+}).
\label{2dreduc}
\eeqn
With these rules the Yukawa interactions in (\ref{fermact}) take the form
\beqn
{\mathcal L}_{\rm Yukawa} &=&
i\sqrt{2}\,{\rm Tr}\left[ -\bar{q}(\hat{\lambda}_{-}\psi_{+}
-\hat{\lambda}_{+}\psi_{-})+
(\tilde{\psi}_{-}\hat{\lambda}_{+}
-\tilde{\psi}_{+}\hat{\lambda}_{-})\bar{\tilde{q}}
+ {\rm c.c.}\right]
\nonumber\\[3mm]
&- &  i\,{\rm Tr}\left[ \tilde{q}(\psi_{-}\zeta_{+}-\psi_{+}\zeta_{-})+
(\tilde{\psi}_{-}q \zeta_{+}-
\tilde{\psi}_{+}q \zeta_{-})
+ {\rm c.c.}\right],
\label{yukawa}
\eeqn
where the color matrix $\hat{\lambda} = (1/2)\,\lambda +T^a\lambda^a$.

\begin{table}
\begin{center}
\begin{tabular}{|c|c | c| c|c| c | c | c|c | c | c |}
\hline
$\rule{0mm}{6mm}$ Field  & $\psi_{+}$ & $\psi_{-}$ & $\tilde{\psi}_{+}$ & $\tilde{\psi}_{-}$
& $\lambda_{+}$ & $\lambda_{-}$ & $\zeta_{+}$ & $\zeta_{-}$ & $q$ & $\tilde{q}$
\\[3mm]
\hline
$\rule{0mm}{5mm}$ U(1)$_R$ charge & $-1$ & 1 & $-1$ & 1 & $-1$
& 1 & $-1$ & 1 & 0 & 0
\\[2mm]
\hline
$\rule{0mm}{5mm}$ U(1)$_{\tilde{R}}$ charge & $-1$ & 1 & 1 & $-1$ & $-1$
& 1 &  1 & $-1$ & 0 & 0
\\[2mm]
\hline
\end{tabular}
\end{center}
\caption{{\small  The U(1)$_R$ and U(1)$_{\tilde{R}}$
charges of fields  of the two-dimensional reduction of the  theory.}}
\label{tableu1ch}
\end{table}

\begin{figure}[h]
\epsfxsize=8cm
\centerline{\epsfbox{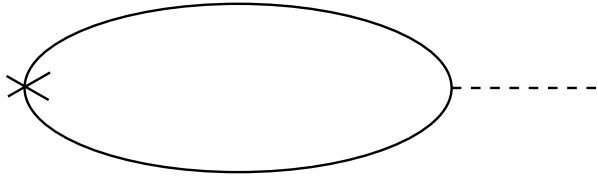}}
\caption{\small
Diagram for the chiral anomaly in two dimensions. The solid lines denote fermions
$\psi$, $\tilde{\psi}$, the dashed line denotes photon, while the cross denotes
insertion of the axial current.}
\label{figanom}
\end{figure}

It is easy to see that ${\mathcal L}_{\rm Yukawa}$ is classically invariant under the
chiral  U(1)$_{R}$ transformations with the U(1)$_{R}$ charges presented in
Table~\ref{tableu1ch}.
The axial current associated with this  U(1)$_{R}$ is not anomalous \cite{W93}.
This is easy to understand. In two dimensions the chiral anomaly comes from
the diagram shown in Fig.~\ref{figanom}. The U(1)$_{R}$ chiral
charges of the fields $\psi$ and $\tilde{\psi}$ are the same while their electric
charges
are opposite. This leads to cancellation of their contributions to this
diagram.

It turns out that for the   particular string solution we are interested in
the classical two-dimensional action has more symmetries than generically, for
a general background. To see this, please,  note that the field $\tilde{q}$ vanishes
on the string solution (\ref{str}), see (\ref{tildeq0}). Then the Yukawa
interactions (\ref{yukawa}) reduce to
\beq
i\sqrt{2}\,{\rm Tr}\left[ -\bar{q}(\hat{\lambda}_{-}\psi_{+}
-\hat{\lambda}_{+}\psi_{-})
\right]
-i\,{\rm Tr}\left[
\tilde{\psi}_{-}q \zeta_{+}-
\tilde{\psi}_{+}q \zeta_{-}
\right]+ {\rm c.c.}
\label{redyukawa}
\eeq
The fermion $\psi$ interacts only with $\lambda$'s while
the fermion $\tilde{\psi}$ interacts only with $\zeta$. Note also that
the interaction in the last line in (\ref{fermactmm}) is absent because
$M=0$ on the string solution.
This property allows us to introduce another chiral symmetry in the theory,
the one which is relevant for the string solution. We will refer to this
extra chiral symmetry as  U(1)$_{\tilde{R}}$.

The U(1)$_{\tilde{R}}$ charges of our set of fields  are also shown in Table~\ref{tableu1ch}.
Note that  $\psi$
and $\tilde{\psi}$ have the opposite charges under this symmetry. The corresponding
current then has the form
\beq
\tilde{j}_{i5}=
\left(
\begin{array}{c}
\rule{0mm}{6mm} \bar{\psi_{-}}\psi_{-}-\bar{\psi_{+}}\psi_{+}
-\bar{\tilde{\psi}}_{-}\tilde{\psi}_{-} +\bar{\tilde{\psi}}_{+}\tilde{\psi}_{+}
+\cdots\\[3mm]
-i\bar{\psi_{-}}\psi_{-}-i\bar{\psi}_{+}\psi_{+}
+i\bar{\tilde{\psi}}_{-}\tilde{\psi}_{-} +i\bar{\tilde{\psi}}_{+}\tilde{\psi}_{+}
+\cdots
\rule{0mm}{6mm}\\
\end{array}
\right),
\label{current}
\eeq
where the ellipses stand for terms associated with  the $\lambda$ and $\zeta$
fields which do not contribute to the anomaly relation.

It is clear that the U(1)$_{\tilde{R}}$ symmetry  is anomalous
in  quantum theory. The contributions
of the fermions $\psi$  and  $\tilde{\psi}$ double in the diagram in
Fig.~\ref{figanom} rather than cancel. It is not difficult
to find the coefficient in the anomaly formula
\beq
\pt_i\tilde{j}_{i5} = \frac{N^2}{\pi} F^{*} \,,
\label{anom}
\eeq
which can be normalized e.g. from \cite{ShVa}.  The factor $N^2$ appears
due to the presence of $2N^2$ fermions $\psi^{kA}$ and $\tilde{\psi}_{Ak}$.

Now, taking into account the fact that the flux of the $Z_N$ string under consideration is
\beq
\int d^2 x \,F^{*}=\frac{4\pi}{N}\, ,
\label{flux}
\eeq
(see the  expression for the U(1) gauge field for  the solution (\ref{znstr})
or (\ref{str})) we conclude that the total number of the fermion zero modes
in the string background
\beq
\nu\,= \,4N \,.
\label{number}
\eeq
This number can be decomposed as
\beq
\nu\,= \,4N= \, 4(N-1)+4\,,
\label{splitnumber}
\eeq
where 4 is the number of the supertranslational modes while
$4(N-1)$ is the number of the superorientational modes. Four supertranslational modes
are associated with four fermion fields in the two-dimensional effective
theory on the string worldsheet, which are superpartners of the bosonic translational
moduli $x_0$ and $y_0$. Furthermore, $4(N-1)$ corresponds to $4(N-1)$ fermion fields in
the  \ntwo CP$(N-1)$ model on the string worldsheet.  CP$(N-1)$ describes
dynamics of the orientational moduli of the string.
For $N=2$ the latter number ($4(N-1)=4$) counts four fermion fields $\chi_1^a$,
$\chi_2^a$ in the model (\ref{ntwocp}).

A similar theorem can be formulated for \none theory with
the potential  (\ref{N1pot}) as well; it implies
$4(N-1)$ zero modes in this case too, i.e.
the doubling of the number of the fermion zero modes on the string as
compared with the one which follows from ``BPS-ness.''

In \cite{SYnone} and \cite{us} four orientational fermion zero modes
were found  explicitly in \none QCD and the $M$ model, by
solving the Dirac equations in the string background.
Note that these fermion zero modes in the $M$ model
are perfectly normalizable provided we keep the coupling constant
$h$ non-vanishing. Instead,
in conventional
\none SQCD without the $M$ field the second pair of the fermion zero modes
(proportional to $\chi_1^a$) become non-normalizable in the
large-$\mu$ limit \cite{SYnone}.
This is related to the presence of the Higgs branch and
massless bulk states in conventional
\none SQCD. As was already mentioned more than once,
in the $M$ model, Eq.~ (\ref{mmodel}), we have no massless states in the bulk.

Note that in both translational and orientational sectors the number of the fermion
zero modes is twice larger than the one dictated by 1/2 ``BPS-ness."

\vspace{2cm}

\centerline{\includegraphics[width=1.3in]{extra3.eps}}

\newpage

\section{ Non-BPS Non-Abelian strings}
\setcounter{equation}{0}
\label{nonBPS}

In this section we will review non-BPS non-Abelian strings.
In particular, they appear in non-supersymmetric
theories. We will see that, although for BPS strings in supersymmetric
theories the transition from quasiclassical to quantum regimes in the
worldsheet theory on the string goes smoothly (see Sect. \ref{unequalmasses}),
for the non-Abelian strings in non-supersymmetric theories
these two regimes are separated by a phase transition.

Next, we will show that the same behavior is typical for non-BPS strings
in supersymmetric gauge theories. As an example we consider non-Abelian strings
in the so-called \none$^{*}$ theory which is a deformed \nfour supersymmetric theory
with supersymmetry broken down to \none in a special way.

\subsection{ Non-Abelian strings in non-supersymmetric theories}
\setcounter{equation}{0}
\label{nonsusy}

In this section we will review some results reported in
\cite{GSY05,GSYpt} treating
non-Abelian strings in non-supersymmetric gauge theories. The theory
studied in \cite{GSY05} is just a bosonic part of  \ntwo
supersymmetric QCD with the gauge group SU$(N)\times$U(1) described
in Sect.~\ref{strings} in the supersymmetric setting. The action of
this model is
\beqn
S &=& \int {\rm d}^4x\left\{\frac1{4g_2^2}
\left(F^{a}_{\mu\nu}\right)^{2}
+ \frac1{4g_1^2}\left(F_{\mu\nu}\right)^{2}
+\frac1{g^2_2}|D_{\mu}a^a|^2
 \right.
 \nonumber\\[3mm]
&+&
  |\nabla^\mu \vp^A |^2
+\frac{g^2_2}{2}
\left(\bar{\vp}_A T^a \vp^A \right)^2
 +
 \frac{g^2_1}{8}\left(|\vp^A|^2- N\xi \right)^2
 \nonumber\\[3mm]
 &+&\left.
\frac12 \left|\left(a^a T^a +\sqrt{2}m_A\right)\vp^A\right|^2
+\frac{i\,\theta}{32\,\pi^2} \, F_{\mu\nu}^a F^{*a}_{\mu\nu}
 \right\}\,,
\label{nsusymodel}
\eeqn
where  $ F^{*a}_{\mu\nu}=(1/2)\,\ve_{\mu\nu\alpha\beta}F_{\alpha\beta}$.
This model is a bosonic part of \ntwo supersym\-metric theory
(\ref{qed}) where, instead of two squark fields $q^k$ and $\tilde{q}_k$,
only one fundamental scalar $\vp^k$ is introduced for each flavor
$A=1,...,N_f$, see the reduced model (\ref{redmodel}) in
Sect.~\ref{znstring}. We also limit ourselves to the case $N_f=N$ and drop
the neutral scalar field $a$ present in (\ref{qed}) as it plays no role in the string
solutions.  To keep the theory at weak coupling we consider large
values of the  parameter $\xi$ in (\ref{nsusymodel}),
$\xi\gg\Lambda_{{\rm SU}(N)}$.

We  assume here that
\beq
\sum_{A=1}^N m_A =0\,.
\label{vsde}
\eeq
Later on it will be convenient to make a specific choice of the
parameters $m_A$, namely,
\beq
m^A= m \times {\rm diag} \left\{  e^{2\pi i/N} , \,e^{4\pi i/N},\, ... ,
 \,e^{2(N-1)\pi i/N} , \, 1\right\} \,,
\label{spch}
\eeq
where $m$ is a single common parameter. Then the constraint (\ref{vsde})
is automatically satisfied. We can (and will) assume $m$ to be real and
positive.   We also introduce a $\theta$ term in the model
(\ref{nsusymodel}).

Clearly the vacuum structure of the model (\ref{nsusymodel}) is
the same as of the theory (\ref{qed}), see Sect.~\ref{model}.
Moreover, the $Z_N$ string solutions are the same; they are given in
Eq.~(\ref{znstr}). The adjoint field plays no role in this solution and
is given by its VEV (\ref{avev}). The tensions of these strings
are given classically by Eq.~(\ref{ten}). However, in contrast with
supersymmetric theory, now the tensions of $Z_N$ strings
acquire quantum corrections in loops.

If masses of the fundamental matter vanish in (\ref{nsusymodel})
this theory has unbroken SU$(N)_{C+F}$ much in the same  way as
the theory (\ref{qed}). In this limit the $Z_N$ strings acquire
orientational zero modes and become non-Abelian. The corresponding solution
for the elementary non-Abelian string is  given by Eq. (\ref{nastr}).
Below we will consider two-dimensional effective low-energy theory
on the worldsheet of such non-Abelian string. Its physics appears
to be quite different as compared with the one in the supersymmetric
case.

\subsubsection{Worldsheet theory}

Derivation of the  effective worldsheet theory  for the
non-Abelian string in the model (\ref{nsusymodel}) can be
carried out much in
the same way as in the supersymmetric case \cite{GSY05}, see
Sect.~\ref{worldsheet}. The worldsheet theory now is two-dimensional
non-supersymmetric CP$(N-1)$ model (\ref{cp}). Its coupling constant
$\beta$ is given by the coupling constant $g^2_2$ of the bulk theory
via the relation (\ref{betaI}). Classically the normalization integral
$I$ is given by (\ref{I}). Then it follows that
$I=1$ as in supersymmetric case. However, now we expect quantum
corrections to modify this result. In particular, $I$ can become a function
of $N$ in quantum theory.

Now, let us discuss the impact of  the $\theta$ term which we introduced
in our bulk theory (\ref{nsusymodel}).
At first sight, seemingly  it cannot produce any effect because our
string is magnetic.
However, if one  allows for slow variations of $n^l$ in $z$
and $t$, one immediately observes that
the electric field is generated via $A_{0,3}$ in Eq.~(\ref{An}). Substituting
$F_{k i}$ from (\ref{Fni}) into the $\theta$ term in the
action (\ref{nsusymodel})
and taking into account the contribution from $F_{kn}$
times $F_{ij}$
($k,n=0,3$ and $i,j=1,2$) we get  the topological term
in the effective CP$(N-1)$ model (\ref{cp}) in the form
\beq
S^{(1+1)}=   \int d t\, dz \,\left\{2 \beta\,
\left[(\pt_{\alpha}\, n^*\pt_{\alpha}\, n) + (n^*\pt_{\alpha}\, n)^2\right]-
\frac{\theta}{2\pi}\, I_{\theta}\,\, \varepsilon_{\alpha\gamma}\,
(\pt_{\alpha}\, n^*\pt_{\gamma}\, n )
\right\}\,,
\label{cpN}
\eeq
where $I_{\theta}$ is another normalizing integral given by
the formula
\beqn
I_{\theta}
&=&
- \int dr \left\{2f_{NA}(1-\rho) \, \frac{d\rho}{dr}
 +(2\rho-\rho^2) \, \frac{df}{dr}\right\}
\nonumber\\[3mm]
&=&
\int dr \frac{d}{dr}\left\{2f_{NA}\, \rho -\rho^2 \, f_{NA}\right\}.
\label{Itheta}
\eeqn
As is clearly  seen,  the integrand here reduces to a total
derivative,  and is determined by the boundary conditions for
the profile functions $\rho$ and $f_{NA}$. Substituting (\ref{bcfinfty}),
(\ref{bcfzero}), and (\ref{bcinfty}), (\ref{bc0}) we get
\beq
I_{\theta}=1\,,
\label{Ithetaeq1}
\eeq
independently of the form of the  profile functions. This latter circumstance
is perfectly natural for the topological term.

The additional term  in the CP$(N-1)$ model (\ref{cpN})
we have just derived is the $\theta$ term in the
standard normalization. The result (\ref{Ithetaeq1}) could have been
expected since physics is $2\pi$-periodic with respect to
$\theta$ both in the
four-dimensional bulk gauge theory and in the
two-dimensional worldsheet CP$(N-1)$ model. The result (\ref{Ithetaeq1})
is not  sensitive to the presence of supersymmetry. It will hold in
supersymmetric models as well. Note that the complexified bulk
coupling constant converts into the complexified
worldsheet coupling constant,
$$
\tau = \frac{4\pi}{g^2_2} + i\frac{\theta}{2\pi}
\to 2\beta+ i\frac{\theta}{2\pi}\,.
$$

Now let us introduce small masses for the fundamental matter in
(\ref{nsusymodel}).
Clearly the diagonal
color-flavor group SU$(N)_{C+F}$ is now broken by adjoint VEV's
down to U(1)$^{N-1}\times Z_N$.
Still, the solutions for the Abelian (or $Z_N$) strings are the same
as was discussed in Sect.~\ref{unequalmasses} since the adjoint
field does not enter these solutions. In particular,  we  have $N$
distinct $Z_N$   string solutions depending on what particular squark winds
at infinity, see Sect.~\ref{unequalmasses}. Say, the string solution
with the winding last flavor  is still given by Eq.~(\ref{znstr}).

What is changed with the color-flavor   SU$(N)_{C+F}$
explicitly broken by $m_A\neq 0$,
the rotations (\ref{nastr}) no more generate zero modes.
In other words, the fields
$n^{\ell}$ become quasi-moduli:
a shallow potential  (\ref{cppot})
for the quasi-moduli $n^l$ on the string worldsheet
is generated \cite{SYmon,HT2,GSY05}. Note that we can replace
$\tilde{m}_A$ by $m_A$ due to the condition (\ref{vsde}).
This potential is shallow as long as $m_A\ll \sqrt\xi$.

The potential simplifies if the mass terms are chosen according to
(\ref{spch}),
\beq
V_{{\rm CP}(N-1)}=2\beta\, m^2 \left\{
1
-\left|
\sum_{\ell=1}^N \, e^{2\pi i\, \ell/N }\,  | n^\ell |^2
\right|^2\right\}\,.
\label{pots}
\eeq
This potential is obviously invariant under the cyclic
$Z_N$ substitution
\beq
\ell\to\ell +k\,,\qquad    n^\ell \to n^{\ell + k}\,,
\qquad \forall\,\,\ell\,,
\label{cycle}
\eeq
with $k$ fixed. This property will be exploited below.

Now our effective two-dimensional theory on the string worldsheet becomes
a massive  CP$(N-1)$ model. As in the supersymmetric case
the potential  (\ref{pots})
has $N$ vacua
at
\beq
n^\ell=\delta^{\ell\ell_0}\,, \qquad \ell_0 =1, 2, ... , N\,.
\label{nell}
\eeq
These vacua correspond
to $N$ distinct Abelian $Z_N$ strings with $\vp^{\ell_0 \ell_0}$
winding at infinity, see Eq.~(\ref{str}).

\subsubsection{Physics in the large-$N$ limit}
\label{largeN}

The massless non-supersymmetric CP$(N-1)$  model (\ref{cpN})
was solved a long time ago by Witten in the large-$N$ limit
\cite{W79}. The massive case with the potential (\ref{pots}) was
considered at large $N$ in \cite{GSY05,GSYpt} in relation with the
non-Abelian strings.  Here we will briefly review this analysis.

As was discussed in Sect. \ref{unequalmasses}, the CP$(N-1)$ model can be
understood as a
strong coupling limit of the U(1) gauge theory. The action has the form
\beqn
S
&=&
\int d^2 x \left\{
 2\beta\,|\nabla_{k} n^{\ell}|^2 +\frac1{4e^2}F^2_{kp} + \frac1{e^2}
|\pt_k\sigma|^2 -\frac{\theta}{2\pi}\ve_{kp}\pt_k A_p
\right.
\nonumber\\[3mm]
&+&
\left.
 4\beta\,|\sigma-\frac{\tilde{m}_{\ell}}{\sqrt{2}}|^2 |n^{\ell}|^2 +
 2e^2 \beta^2(|n^{\ell}|^2 -1)^2
\right\}\,,
\label{nscpg}
\eeqn
where we also included the $\theta$ term. As in the supersymmetric case,
in the limit $e^2\to \infty$ the $\sigma$ field can be eliminated via
the algebraic equation of motion which leads to the
theory (\ref{cpN}) with the potential  (\ref{cppot}).

The $Z_N$-cyclic symmetry (\ref{cycle}) now takes the form
\beq
\sigma\to e^{i\frac{2\pi k}{N}}\sigma\,,\qquad    n^\ell \to
n^{\ell +k}\,, \qquad \forall\,\,\ell\,,
\label{cycleg}
\eeq
where $k$ is fixed.

It turns out  that the non-supersymmetric version
of the massive  CP$(N-1)$  model (\ref{nscpg}) has two phases separated by
a phase transition \cite{GSY05,GSYpt}. At large values of the mass
parameter $m$ we have the Higgs phase while at small $m$ the theory is in
the Coulomb/confining phase.

\begin{center}

{\bf The Higgs phase}

\end{center}

At large $m$, $m\gg\Lambda_{\sigma}$, the renormalization group flow of
the coupling constant $\beta$ in (\ref{nscpg}) is frozen at the scale
$m$.  Thus, the model at hand is at weak coupling and  the
quasiclassical analysis is applicable.
The potential  (\ref{pots}) have $N$ degenerate vacua which are
labeled by the order parameter $\langle\sigma\rangle$,
the vacuum configuration being
\beq
n^{\ell}=\delta^{\ell \ell_0},\;\;
\sigma=\frac{\tilde{m}_{\ell_0}}{\sqrt{2}},
\;\;\; \ell_0=1,...,N\,,
\label{nscpvac}
\eeq
as in the supersymmetric case, see (\ref{cpvac}).
In each given vacuum the $Z_N$ symmetry (\ref{cycleg}) is spontaneously
broken.

These vacua  correspond to Abelian $Z_N$ strings of the bulk theory.
$N$ vacua of the worldsheet theory have strictly degenerate vacuum
energies.  From the four-dimensional point of view this means that we
have $N$ strictly degenerate  $Z_N$ strings.

There are $2(N-1)$ elementary excitations. Here we count
real degrees of
freedom. The action (\ref{nscpg}) contains $N$ complex fields
$n^\ell$.
The common phase of $n^{\ell_0}$ is gauged away.
  The condition $|n^\ell|^2 = 1$ eliminates one more
field.   These elementary excitations have  physical masses
\beq
M_\ell =
|m_\ell-m_{\ell_0}|\,,\qquad \ell\neq\ell_0\,.
\label{elmass}
\eeq

Besides, there are kinks (domain ``walls" which are particles in two
dimensions) interpolating between these vacua. Their masses scale as
\beq
M^{\rm kink}_{\ell} \sim \beta\,M_\ell \,.
\label{kinkmass}
\eeq
The kinks  are much  heavier than elementary
excitations at weak coupling. Note that they have nothing to do
with Witten's $n$ solitons \cite{W79} identified as solitons at
strong coupling. The point of phase transition separates these
two classes of solitons.

As was already discussed in the supersymmetric case
(see Sect.~\ref{kinkmonopole})
the flux of the Abelian 't Hooft--Polyakov monopole
is the difference of the fluxes of two ``neighboring'' strings,
see (\ref{monflux}).  Therefore, the confined monopole
in this regime is obviously  a junction of two distinct
$Z_N$ strings. It is seen as a quasiclassical kink
interpolating between the ``neighboring'' $\ell_0$-th and $(\ell_0 +1)$-th
vacua  of the
effective massive CP$(N-1)$ model on the string worldsheet.
A  monopole can move freely along the string as both
attached strings are tension-degenerate.

\begin{center}

{\bf The Coulomb/confining phase}

\end{center}

Now let us discuss the Coulomb/confining phase of the theory
occurring at small $m$.
As was mentioned, at $m=0$ the CP$(N-1)$  model was solved by Witten
in the
large-$N$ limit \cite{W79}. The model at small $m$ is very similar to
Witten's solution. (In fact, in the large-$N$ limit it is just the same.)
The paper \cite{GSYpt} presents a generalization of
Witten's analysis to the massive case which is then used to study
the phase transition between the $Z_N$ asymmetric and symmetric phases.
Here we will briefly summarize Witten's results for the massless model.

The non-supersymmetric CP$(N-1)$  model is asymptotically free
(as its supersymmetric version) and develops its own scale
$\Lambda_{\sigma}$.
If $m=0$, classically the  field $n^{\ell}$ can have arbitrary
direction; therefore, one might naively expect
spontaneous breaking of SU$(N)$ and
the occurrence of massless Goldstone modes.
This cannot happen
in two dimensions. Quantum effects restore the
full symmetry making the vacuum unique. Moreover, the condition
$|n^{\ell}|^2=1$ gets in effect relaxed. Due to strong coupling
we have more degrees of freedom than in the original Lagrangian,
namely all $N$ fields $n$ become dynamical and acquire
masses $ \Lambda_{\sigma}$.

\begin{figure}
\epsfxsize=8cm
\centerline{\epsfbox{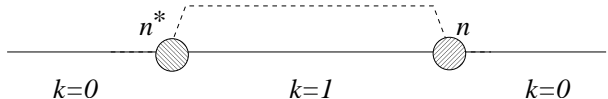}}
\caption{\small
Linear confinement of the $n$-$n^*$ pair.
The solid straight line represents the string.
The dashed line shows
the vacuum energy density (normalizing ${\mathcal E}_0$ to zero).}
\label{fig:conf}
\end{figure}

This is not the end of the story, however. In addition, one gets
another
composite degree of freedom.
The  U(1) gauge field $A_{k}$ acquires a standard
kinetic term at  one-loop level,\footnote{By loops here we mean
perturbative expansion in $1/N$ perturbation theory.
} of the form
\beq
{N}\,\Lambda^{-2}\,\,  F_{kp}  F_{kp}\, .
\label{gkinterm}
\eeq
Comparing Eq.~(\ref{gkinterm}) with (\ref{nscpg})
we see that the charge of the $n$ fields with respect to this photon
is $1/\sqrt{ N}$.
The Coulomb potential between two charges in
two dimensions is linear in separation  between these charges.
The linear potential scales as
\beq
V(R) \sim \frac{\Lambda_{\sigma}^2}{N}\, R\, ,
\label{nine}
\eeq
where $R$ is  separation. The force is attractive for pairs
$\bar n$ and $n$, leading to the formation
of weakly coupled bound states (weak coupling is the
manifestation of the $1/N$ suppression of the confining potential).
Charged states are eliminated from the spectrum. This is the reason
why the
$n$ fields were called ``quarks" by Witten. The spectrum
of the theory consists of $\bar{n} n$-``mesons.'' The picture of
confinement of $n$'s is shown in Fig.~\ref{fig:conf}.

The validity of the above consideration rests on large $N$.
If $N$ is not large Witten's solution \cite{W79}
ceases to be applicable. It remains  valid in the qualitative sense,
however. Indeed, at $N=2$ the model was solved exactly \cite{ZZ,Zam}
(see also
\cite{Coleman}).
Zamolodchikovs found that the spectrum of the
O(3) model consists of a triplet of degenerate states
(with mass $\sim \Lambda_{\sigma}$).
At $N=2$ the action (\ref{nscpg}) is built of doublets.
In this sense one can
say that Zamolodchikov's solution exhibits confinement of
doublets. This is in qualitative accord with the large-$N$ solution
\cite{W79}.

Inside the $\bar n\,n$ mesons, we have  a constant electric field,
see Fig.~\ref{fig:conf}. Therefore the spatial interval between $\bar n$
and $n$ has a higher energy density than the domains outside the meson.

Modern understanding of the vacuum structure of the massless
CP$(N-1)$ model
\cite{Wtheta} (see also \cite{Stheta})
allows one to reinterpret  confining dynamics of the
$n$ fields in different terms \cite{MMY,GSY05}. Indeed, at large $N$,
along with the unique ground state,
the model has $\sim N$ quasi-stable local minima, quasi-vacua,
which become absolutely stable at $N=\infty$. The relative
splittings between the values of the energy density in the adjacent
minima
is of the order of $1/N$, while
the probability of the false vacuum decay is proportional to
$N^{-1}\exp (-N)$ \cite{Wtheta,Stheta}. The $n$
quanta ($n$ quarks-solitons) interpolate between
the adjacent minima.

The existence of a large family of quasi-vacua
can be inferred from the study of the $\theta$ evolution of the
theory.
 Consider the topological susceptibility, i.e. the correlation
function of two topological densities
\beq
\int d^2 x \, \langle Q(x),\,\, Q(0)\rangle\,,
\label{correlator}
\eeq
where
\beq
Q=\frac{i}{2\pi }\, \varepsilon_{kp}
\partial_k A_p
=\, \frac{1}{2\pi }\, \varepsilon_{kp}
\left(\partial_k
n_{\ell}^*\,\, \partial_p n^{\ell}
\right).
\label{topdens}
\eeq
The correlation function (\ref{correlator}) is proportional to the
second derivative
of the vacuum energy with respect to the $\theta$ angle. From
(\ref{topdens})
it is not difficult to deduce that this correlation function  scales
as $1/N$ in the large $N$
limit. The vacuum energy by itself scales as $N$. Thus, we conclude
that, in fact, the
vacuum energy should be a function of $\theta/N$.

\begin{figure}
\epsfxsize=6cm
\centerline{\epsfbox{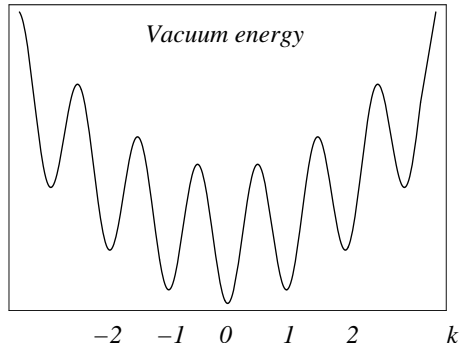}}
\caption{\small
The vacuum structure of
CP$(N-1)$ model at $\theta =0$.}
\label{odin}
\end{figure}

On the other hand, on general grounds,  the vacuum energy must be
a $2\pi$-periodic function of
$\theta$. These two requirements are seemingly self-contradictory.
A way out reconciling the above facts is as follows. Assume that we
have a family of quasi-vacua
 with energies
\beq
 E_k (\theta) \sim \, N\, \Lambda_{\sigma}^2
\left\{1 + {\rm const}\left(\frac{2\pi k +\theta}{N}
\right)^2
\right\}
\,,\,\,\,\, k=0\ldots, N-1\,.
\label{split}
\eeq
A schematic picture of these vacua
is given in Fig. ~\ref{odin}. All  these minima are entangled in the
$\theta$ evolution.
If we vary $\theta$ continuously from $0$ to $2\pi$ the
depths of the minima ``breathe." At $\theta =\pi$ two vacua become
degenerate, while for larger values of $\theta$ the former
global minimum
becomes local while the adjacent local minimum becomes global.
It is obvious that for the neighboring vacua which are not too far
from the global minimum
\beq
E_{k+1}-E_k \sim \frac{\Lambda_{\sigma}^2}{N}\,.
\label{distance}
\eeq
This is also the confining force acting between $n$ and $\bar n$.

One could introduce  order parameters that would
distinguish between distinct vacua from the vacuum family.
An obvious choice is the expectation value of the topological charge.
The kinks $n^{\ell}$ interpolate, say, between the global minimum
and the first local one on the right-hand side. Then $\bar n$'s
interpolate between the
first local minimum and the global one. Note that the vacuum energy
splitting
is an effect suppressed by $1/N$.
At the same time, kinks  have masses which scale
as $N^0$,
\beq
M^{\rm kink}_{\ell}\sim \Lambda_{\sigma}\,.
\label{kinkmasscoulomb}
\eeq
The multiplicity of
such kinks is $N$ \cite{Achar}, they form an $N$-plet of SU$(N)$.
This is in full accord with the fact that the large-$N$
solution of (\ref{nscpg}) exhibits $N$ quanta of the complex field
$n^{\ell}$.

Thus we see that the  CP$(N-1)$
model has a fine structure of ``vacua'' which are split,
with the splitting of the order of $\Lambda_{\sigma}^2/N$. In
four-dimensional bulk
theory these ``vacua'' correspond to elementary non-Abelian strings.
Classically all these strings have the same tension (\ref{ten}). Due to
quantum effects in the worldsheet theory
the degeneracy is lifted:
the elementary strings become split, with the tensions
\beq
T  = 2\pi\xi +c_1\, N\, \Lambda_{\sigma}^2
\left\{1 + c_2\left(\frac{2\pi k +\theta}{N}
\right)^2
\right\}
\,,
\label{splitten}
\eeq
where $c_1$ and $c_2$ are numerical coefficients.
Note that (i) the splitting does not appear to any finite order
in the   coupling constants; (ii) since
$\xi\gg \Lambda_{\sigma}$,  the splitting is suppressed in both
parameters, $\Lambda_{\sigma}/\sqrt{\xi}$ and $1/N$.

Kinks of the worldsheet theory represent confined monopoles
(string junctions) in the four-dimensional bulk theory. Therefore
kink confinement in CP$(N-1)$ model can be interpreted as follows
\cite{MMY,GSY05}.
The non-Abelian monopoles,
in addition to  the four-dimensional confinement
(which ensures that the monopoles
are attached to the strings) acquire a two-dimensional confinement along
the string:
a monopole--antimonopole   forms a meson-like configuration,
with necessity, see Fig.~\ref{fig:conf}.

In  summary, the  CP$(N-1)$
model in the Coulomb/confining phase,
at small $m$, has a
vacuum family with a fine structure.
For each given $\theta$ (except $\theta =\pi, \,\, 3\pi$, etc.)
the true ground state is unique, but there is a large number
of ``almost" degenerate ground states.
 The $Z_N$
symmetry is unbroken. The classical condition (\ref{unitvec})
is replaced by $\langle n^{\ell}\rangle =0$.
The spectrum of physically observable states
consists of kink-anti-kink  mesons which form  the   adjoint
representation of SU$(N)$.

Instead, at large
$m$ the theory is in the Higgs phase; it has $N$ strictly degenerate
vacua (\ref{nscpvac}); the $Z_N$ symmetry is broken. We have
$N-1$ elementary excitations $n^{\ell}$  with  masses
given by Eq.~(\ref{elmass}).
Thus we conclude that these two regimes
should be separated by a phase transition  at some critical value $m_*$
\cite{GSY05,GSYpt}. This phase
transition is associated with the $Z_N$ symmetry breaking: in the
Higgs phase the
$Z_N$ symmetry is spontaneously broken, while in the Coulomb phase it
is restored. For $N=2$ we deal with $Z_2$ which makes the situation
akin to the Ising model.

\begin{figure}
\epsfxsize=11cm
\centerline{\epsfbox{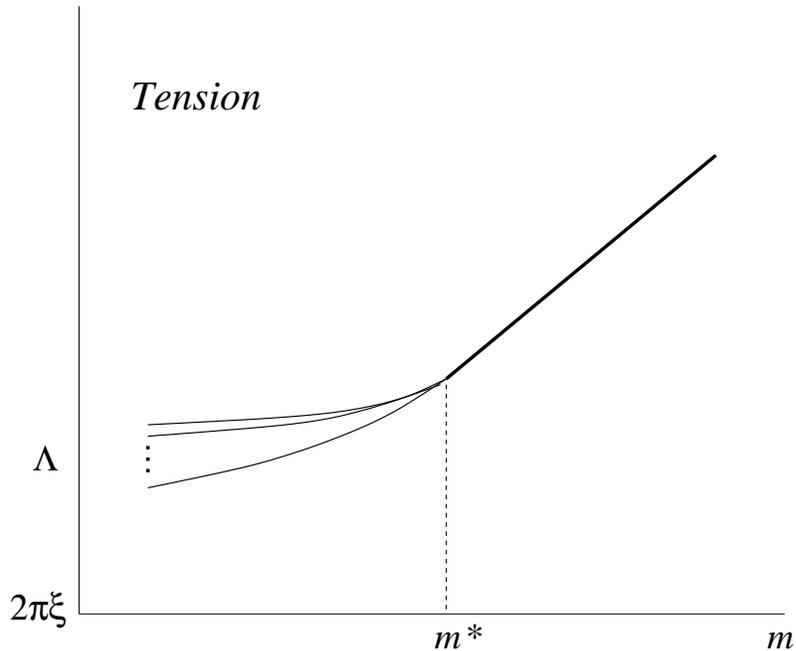}}
\caption{\small Schematic dependence of string tensions on the mass parameter $m$.
At small $m$ in the non-Abelian confinement phase the tensions are split
while in the Abelian confinement phase at large $m$ they are degenerative.}
\label{phtrans}
\end{figure}

In the worldsheet theory this is a phase transition between the
Higgs and Coulomb/confining phase. In the bulk theory it can be
 interpreted  as a phase transition
between the Abelian and non-Abelian confinement. In the Abelian confinement
phase
at large $m$, the $Z_N$ symmetry is spontaneously broken, all $N$ strings
are strictly degenerate, and there is no two-dimensional confinement of
the 4D-confined monopoles. In contrast,
in the non-Abelian confinement phase occurring at small $m$, the
$Z_N$ symmetry is fully restored, all $N$ elementary strings are split, and
the 4D-confined monopoles combine with antimonopoles to form a
meson-like configuration
on the string, see Fig.~\ref{fig:conf}. We show schematically
the dependence of the string tensions on $m$ in these two phases in
Fig.~\ref{phtrans}.

In \cite{GSYpt} the  phase transition point is found using large-$N$
methods developed by Witten in \cite{W79}. It turns out that the
critical point
\beq
m_{*}=\Lambda_{\sigma}.
\label{critmass}
\eeq
The vacuum energy is calculated in both phases and is shown to be
continuous at the critical point. If one approaches the critical point,
say, from the Higgs phase some composite states of the worldsheet
theory (\ref{nscpg}), such as the photon and the kinks, become light.
One is tempted to believe that these states become massless at the
critical point (\ref{critmass}).
However, this happen only in the very narrow vicinity of the phase
transition point where $1/N$ expansion fails. Thus, the large-$N$
approximation is not powerful enough to determine
the critical behavior.

To conclude this section we would like to stress that
we encounter a crucial difference between the non-Abelian confinement
in supersymmetric and non-supersymmetric gauge theories.
For BPS strings in
supersymmetric theories we have no phase transition separating
the phase of the non-Abelian strings from that of the Abelian strings
\cite{SYmon,HT2}. Even for small values of
the mass parameters supersymmetric theory
strings are strictly degenerate, and the $Z_N$ symmetry is
spontaneously broken. In particular, at $\Delta m_A=0$ the order parameter
for the broken $Z_N$, which
differentiates the $N$ degenerate vacua of the supersymmetric CP$(N-1)$
model, is the bifermion condensate of two-dimensional fermions living
on the string worldsheet of the non-Abelian BPS string,
see Sect.~\ref{Dyn}.

Moreover, the presence of the phase transition between Abelian and
non-Abelian confinement in non-supersymmetric theories suggests a
solution for the problem of enrichment of the hadronic spectrum
mentioned in the beginning of Sect.~\ref{strings}, see
also a more detailed discussion in Sect.~\ref{confinement}. In the phase
of Abelian confinement we have $N$ strictly degenerative
Abelian $Z_N$ strings which give rise to too many ``hadron" states,
not present in actual QCD. Therefore Abelian $Z_N$ strings
can hardly play a role of prototypes for QCD confining strings.
Although the BPS strings in supersymmetric theories become
non-Abelian as we tune the mass parameters $m_A$ to a
common value, still there are $N$ strictly degenerative  non-Abelian
strings and, therefore, still too many ``hadron" states in the spectrum.

As was explained in this section, the situation in non-supersymmetric
theories is quite different. As we make mass parameters $m_A$ equal
we enter the non-Abelian confinement phase. In this phase $N$
elementary non-Abelian strings are split. Say, at $\theta=0$ we have
only one lightest elementary string producing a single two-particle
meson with the given flavor quantum numbers and spin, exactly as observed
in nature. If $N$ is large, the splitting is small, however. If $N$ is
not-so-large the splitting is of the order of $\Lambda_{\sigma}^2$.
Therefore, the mesons produced by  excited strings are unstable and
may appear invisible experimentally.

\subsection{Non-Abelian strings in \none$^{*}$ theory}
\setcounter{equation}{0}
\label{nfour}

So far
in our quest
for the non-Abelian strings we focused on a particular model, with the
SU$(N)\times$U(1) gauge group and fundamental matter. However, it is
known that solutions for $Z_N$ strings were first found in
simpler models,  with the SU$(N)$ gauge group and adjoint matter
\cite{VS,HV,Su,SS} (in fact, the gauge group becomes
SU$(N)/Z_N$ if only adjoint matter is present in the theory).
A natural question which immediately comes to one's mind is:
can these $Z_N$ strings under some special conditions develop
orientational zero modes and become non-Abelian? The answer
to this question is yes. Solutions for non-Abelian strings
in the simplest theory with the SU(2) gauge group and adjoint
matter were found in \cite{MMY} (actually, the gauge group of this theory is SO(3)).
Here we will briefly review the
results of this paper.

Although the model considered in \cite{MMY} is  supersymmetric
the price one has to pay for its simplicity
is that the strings which appear in this model are not BPS. The reason
is easy to understand. One cannot introduce the FI term
in the theory with the gauge group SU$(N)$ and, therefore, one cannot
construct the string central charge \cite{gorskys}.

The model considered in \cite{MMY} is the
so-called  \none$^*$ supersymmetric theory with the gauge
group SU(2). It is a deformed \nfour theory with the mass
terms for three \none chiral superfields.
Let us take two equal masses, say $m_1=m_2= m$, while the third mass
$m_3$ is assumed to be distinct. Generally speaking,
\nfour supersymmetry is broken
down to \none, unless $m_3=0$.  If $m_3=0$ the theory has
\ntwo supersymmetry. It exemplifies \ntwo gauge theory with the
adjoint matter (two \none
flavors of adjoint matter with equal masses).

Classically the vacuum structure of this theory was studied in
\cite{VafaW}, while quantum effects were taken into account in
\cite{DonagiW}.
The \nfour theory with the SU(2) gauge
group has three vacua, and if the coupling constant of the \nfour
theory is small\,\footnote{Note that the coupling of the
unbroken \nfour theory $g^2$ does not run since the \nfour theory
is conformal.}, $g^2\ll 1$, one of these vacua is at weak coupling.
All three adjoint scalars condense in this vacuum. Therefore, it
is called the Higgs vacuum  \cite{VafaW,DonagiW}. Two other vacua of the
theory are always at strong coupling. For small
$m_3$ they corresponds to the monopole and dyon vacua of the
perturbed \ntwo theory \cite{SW1}. Here we will concentrate
on the Higgs vacuum in the weak coupling regime.

In this vacuum the gauge group SU(2) is broken down
to $Z_2$ by the adjoint scalar
VEV's. Therefore there are stable $Z_2$
non-BPS strings associated with
\beq
\pi_1({\rm SU}(2)/Z_2)=Z_2\,.
\eeq

If we choose a special value of $m_3$,
$$m_3=m\,,$$
there is a
diagonal $O(3)_{C+F}$ subgroup of the global gauge group SU(2),
and the flavor $O(3)$ group, unbroken by vacuum condensates.
In parallel with Refs.~\cite{ABEKY,HT1,SYmon,HT2} (see also
Sect.~\ref{strings}), the presence
of this group leads to emergence of orientational zero modes
of the $Z_2$-strings associated with rotation of the color magnetic
flux of the string inside the SU(2) gauge group, which converts the $Z_2$
string into non-Abelian.

Let us discuss this model in more detail.
In terms of \none supermulti\-plets, the \nfour supersymmetric
gauge theory with the SU(2) gauge group contains a vector
multiplet, consisting of the gauge field $A_{\mu}^a$ and gaugino
$\lambda^{\alpha a}$, and three chiral multiplets $\Phi^a_A$, $A=1,2,3$,
all in the adjoint representation of the gauge group, with $a=1,2,3$ being
the SU(2) color index. The superpotential of the \nfour gauge theory
is
\beq
W_{{\mathcal N}=4}= -\frac{\sqrt{2}}{g^2}
\varepsilon_{abc}\Phi_1^a\Phi_2^b\Phi_3^c\,.
\label{n4sup}
\eeq
One can deform this theory, breaking \nfour supersymmetry down to
\ntwo,
by adding two equal mass terms
$m$, say, for the first two flavors of the adjoint matter,
\beq
\label{N2mass}
W_{{\mathcal N}=2} = {m\over 2g^2}\sum_{A=1,2} \left(\Phi_A^a\right)^2\,.
\eeq
Then, the third flavor combines with the vector
multiplet to form an $\mbox{\ntwo}$ vector supermultiplet, while the first two
flavors (\ref{N2mass}) can be treated as \ntwo massive adjoint matter.
If one wishes, one can further break supersymmetry down to \none, by adding a mass
term  to the $\Phi_3$ multiplet,
\beq
\label{N1mass}
W_{{\mathcal N}=1^*} = {m_3\over 2g^2} \left(\Phi_3^a\right)^2\,.
\eeq

The bosonic part of the action
is
\beqn
S_{{\mathcal N}=1^*}
&=&
 {1\over g^2}\int d^4 x\left(
\frac14\left(F_{\mu\nu}^a\right)^2 +
\sum_A \left| D_\mu\ \Phi_A^a\right|^2+
\right.
\nonumber\\[3mm]
&+&
\frac12\sum_{A,B}\left[(\bar{\Phi}_A\bar{\Phi}_B)(\Phi_A\Phi_B)-
(\bar{\Phi}_A\Phi_B)(\bar{\Phi}_B\Phi_A)\right]+
\nonumber\\[3mm]
&+&
\left.
\sum_{A}\left|\frac1{\sqrt{2}}
\varepsilon_{abc}\varepsilon^{ABC}\Phi^b_B\Phi^c_C
-m_A\Phi^a_A\right|^2\right) \, ,
\label{su2}
\eeqn
where
$D_\mu\ \Phi_A^a =
\pt_\mu \Phi^a_A + \varepsilon^{abc}A_\mu^b\Phi^c_A$,
and we use the same notation $\Phi^a_A$ for the scalar components
of the corresponding chiral superfields.

As was mentioned above, we are going to study the so-called Higgs
vacuum of the
theory (\ref{su2}), when all three adjoint scalars develop the
VEV's of the order of $m,\, \sqrt{mm_3}$.
 The scalar condensates $\Phi^a_A$ can be
written in the form of the following 3$\times$3
color-flavor matrix (convenient for the SU(2) gauge group and three
and three chiral flavor superfields)
\beq
\label{gvac}
\langle\Phi_A^a\rangle =\frac{1}{\sqrt{2}}\left(
\begin{array}{ccc}
  \sqrt{mm_3} &
 0& 0 \\
 0 &
\sqrt{mm_3} & 0 \\
  0 & 0 & m
\end{array}
\right)\,.
\eeq
These VEV's break the SU(2) gauge group completely. The $W$-bosons masses
are
\beq
m^2_{1,2}=m^2 +mm_3
\label{Wmass}
\eeq
for $A^{1,2}_\mu$,
while the mass of the photon fields $A_{\mu}^3$ is
\beq
m^2_{\gamma}=2mm_3\,.
\label{phmass}
\eeq

In what follows,
we will be especially interested in a particular point
in the parameter space: $m_3=m$.
For this value of $m_3$, (\ref{gvac}) presents a
symmetric color-flavor locked  vacuum
\beq
\label{svac}
\langle\Phi_A^a\rangle =\frac{m}{\sqrt{2}}\left(
\begin{array}{ccc}
  1 &
 0& 0 \\
 0 &
1 & 0 \\
  0 & 0 & 1
\end{array}
\right)\,.
\eeq
This symmetric vacuum respects the global $O(3)_{C+F}$ symmetry,
\beq
\Phi\rightarrow O\Phi O^{-1},\;\;\; A_{\mu}^a\rightarrow
O^{ab}A_{\mu}^b\,,
 \label{c+fmmy}
\eeq
which combines transformations from the
global color and flavor groups, similarly to the SU$(N)_{C+F}$
group of the U(N) theories, see Sect.~\ref{strings}.
It is this symmetry that is responsible for the presence of the non-Abelian strings in
the vacuum (\ref{svac}).

Note that at $m_3=m$
masses of all gauge bosons are equal,
\beq
m^2_{g}=2m^2\,,
\label{gmassmmy}
\eeq
as is clearly seen
from (\ref{Wmass}) and (\ref{phmass}).
This means, in particular, that in the point $m_3=m$ we loose all  traces of
the ``Abelization" in our theory, which are otherwise present at
generic values of $m_3$.

Let us also emphasize that the coupling
$g^2$ in Eq.~(\ref{su2}) is the \nfour coupling constant.
It does not run in the \nfour theory at scales above $m$,
and we assume it to be small,
\beq
g^2\ll 1\,.
\label{n4coupling}
\eeq
At the scale $m$ the gauge group SU(2) is broken
in the vacuum (\ref{svac}) by the scalar VEV's.
Much in the same was as in the U$(N)$
theory (see Sect.~\ref{strings}) the running of the coupling constant
below the scale $m$
is determined by the $\beta$ function of the effective
two-dimensional sigma model on the worldsheet of the
non-Abelian string.

Skipping details we present here the solution
for the non-Abelian string in the model (\ref{su2}) found in
\cite{MMY}.
When $m_3$ approaches $m$, the theory acquires additional symmetry.
In this case the scalar VEV's take the form (\ref{svac}), preserving
the global combined color-flavor symmetry (\ref{c+f}).
On the other hand, the $Z_2$ string solution itself is not invariant
under this symmetry.
The symmetry  (\ref{c+f})  generates orientational
zero modes of the string.
The string solution in the singular gauge is
\beqn
\Phi_A^a &=&O\left(
\begin{array}{ccc}
  \frac{g}{\sqrt{2}}\phi & 0 & 0 \\
 0 & \frac{g}{\sqrt{2}}\phi & 0 \\
  0 & 0 & a_0
\end{array}\right)O^{-1}
\nonumber\\[3mm]
&=&
 \frac{g}{\sqrt{2}}\phi  \delta^a_A
+S^a S^A \left(a_0-\frac{g}{\sqrt{2}}\phi\right)\,,
\nonumber\\[5mm]
A^a_i
&=&
 S^a\;\frac{\varepsilon_{ij}x_j}{r^2}\, f(r),\qquad i,j=1,2\,,
\label{n4str}
\eeqn
where we introduced  the unit orientational vector $S^a$,
\beq
S^a=O^a_b\delta^{b3}=O^a_3\, .
\label{Smmy}
\eeq
It is easy to see that the orientational vector $S^a$ defined above
coincides with the one we introduced in Sect.~\ref{strings},
see Eq.~(\ref{S}).
The solution (\ref{n4str}) interpolates between the Abelian
$Z_2$ strings for which $\vec S=\{ 0,0,\pm 1\}$.
We see that the string flux is determined now by an arbitrary vector
$S^a$ in the color space,  much in the same way
as for the non-Abelian strings in the U$(N)$ theories.

Since this string is not BPS-saturated,
the profile functions in (\ref{n4str}) satisfy now the {\em
second}-order differential equations,
\beqn
&&
\phi''+{1\over
r}\phi'-{1\over r^2}f^2\phi =
\phi\left(g^2\phi^2-\sqrt{2}m_3a_0\right)+2\phi\left(a_0-
\frac{m}{\sqrt{2}}\right)^2 ,
\nonumber\\[4mm]
&&
a_0''+ {1\over r}a_0' = -\frac{m_3}{\sqrt{2}}\left(
g^2\phi^2-\sqrt{2}m_3 a_0\right)
+2g^2\phi^2\left(a_0-\frac{m}{\sqrt{2}}\right),
\nonumber\\[4mm]
&&
f''- {1\over r}f' = 2g^2f\phi^2 \,,
\label{n4streq}
\eeqn
where the primes stand for derivative with respect to $r$, and
the boundary conditions are
\beqn
&&
\phi(0) = 0, \ \ \ \ \phi(\infty) =\frac{\sqrt{mm_3}}{g},
\nonumber\\[4mm]
&&
a_0'(0)=0, \ \ \ \ a(\infty) = \frac{m}{\sqrt{2}},
\nonumber\\[4mm]
&&
f(0) = 1, \ \ \ \ f(\infty) = 0\,.
\label{n4bc}
\eeqn
The string tension is
\beqn
&&
T=2\pi\int_0^{\infty} r dr \left[
\frac{f'^2}{2g^2r^2} +\phi'^2+\frac{a_0'^2}{g^2}
+\frac{f^2\phi^2}{r^2}
\right.
\nonumber\\[4mm]
&&
+
\left.
\frac{g^2}{2}\left(\phi^2-\frac{\sqrt{2}m_3}{g^2}a_0\right)^2
+2\phi^2\left(a_0-\frac{m}{\sqrt{2}}\right)^2\right].
\label{n4ten}
\eeqn

The second-order equations for the string profile
functions were solved in \cite{MMY}  numerically and the string tension was found
as a function of the mass ratio $m_3/m$. Note that for the BPS string
(which appears in the limit $m_3\to 0$) the tension is
$$T_{\rm BPS}=2\pi\, mm_3/g^2\,.$$

The effective worldsheet theory for the non-Abelian string (\ref{n4str}) was shown
to be the non-supersymmetric CP(1) model \cite{MMY}.
Its coupling constant $\beta$ is related to the coupling constant
$g^2$ of the bulk theory via (\ref{betaI}), where now
 the normalization integral $$I\sim 0.78\,.$$
In this theory  there is a 't Hooft--Polyakov
monopole  with the unit magnetic charge.
Since the $Z_2$-string charge is $1/2$, it cannot end on the
monopole, much in the same way as for the monopoles in the U$(N)$ theories,
see Sect.~\ref{kinkmonopole}.
Instead, the confined monopole appears to be a
junction of the $Z_2$ string and anti-string. In the worldsheet
CP(1) model it is seen as a kink interpolating between the
two vacua.

At small values of the mass difference $m_3-m$ the worldsheet theory is
in the  Coulomb/confining phase, see Sect.~\ref{largeN},
although, strictly speaking, the large-$N$ analysis  is
not applicable in this case. Still, the
monopoles, in addition to four-dimensional confinement  ensuring that
they are attached to a string,  also experience confinement in two dimensions,
along the string \cite{MMY}. This means that each
monopole on the string
must be accompanied by an antimonopole,
with a linear potential between them along the string.
As a result, they form a
meson-like configuration, see Fig.~\ref{fig:conf}.
As was mentioned in Sect.~\ref{largeN},  this follows from the
exact solution of the CP(1) model \cite{ZZ,Zam}:
only the triplets of SU(2)$_{C+F}$ are seen in the spectrum.

\vspace{2cm}

\centerline{\includegraphics[width=1.3in]{extra3.eps}}

\newpage

\section{Strings on the Higgs branches}
\label{sechiggs}

One common feature of supersymmetric gauge theories is the presence of moduli
spaces ---  manifolds on which scalar fields can develop arbitrary VEV's
without violating the zero energy condition. If on these vacuum manifolds
the gauge group is broken, either completely or partially,
down to a discrete subgroup, these
manifolds are referred to as the Higgs branches.

One may pose a question: what happens with the flux tubes and confinement
in theories with the Higgs branches? The Higgs branch represents an extreme case
of type-I superconductivity, with vanishing Higgs mass. One may ask oneself
whether or not the ANO strings still exist in
this case, and if yes, whether they provide confinement for external
heavy sources.

This question was posed
and studied first in \cite{PeninRubak} where the authors concluded that
the vortices do not exist on the
Higgs branches due to infrared problems.
In Ref.~\cite{AchDPUrr,PUrr} the \none SQED  vortices were further
analyzed. It was found that at a generic point on the Higgs branch
strings are unstable.  The only vacuum which supports string
solutions is the base point of the Higgs branch where
the strings become BPS-saturated. The so-called
``vacuum selection rule" was put forward in \cite{AchDPUrr,PUrr}
to ensure this property.

On the other hand, in \cite{Y99,EY} it was shown that infrared problems can be
avoided provided certain infrared regularizations are applied. Say, in \cite{Y99,EY}
the infrared divergences were regularized through embedding
of  \none SQED in
softly broken \ntwo SQED. Another alternative is to consider
a finite length-$L$ string instead of an infinitely long string. In this case
the impact of the Higgs branch was shown to ``roughen" the string,
making it
logarithmically ``thick.'' Still, the string solutions do exist and produce
confinement for heavy trial sources. However, now the confining potential
is not linear in separation; rather it
behaves as
\beq
V(L) \sim \frac{L}{\ln{L}}
\label{confpot}
\eeq
at large $L$. Below we will briefly review the string solutions
on the Higgs branches, starting from the simplest case of the flat Higgs branch
and then considering a more common scenario, when the Higgs branch is curved by
the FI term.

\subsection{Extreme type-I strings}
\setcounter{equation}{0}

In this section we will review the classical solutions for
the ANO vortices (flux tubes) in the theories with the flat Higgs potential which
arises in supersymmetric settings \cite{Y99}.
Let us start from the Abelian Higgs model,
\begin{equation}
\label{ah}
S_{\rm AH}=\int d^4x\left\{\frac1{4g^2}\,F^2_{\mu\nu}+|\nabla_\mu
q|^2+\lambda\left(|q|^2-v^2\right)^2\right\},
\end{equation}
for a single complex field $q$  with the quartic coupling $\lambda\to 0$.
Here $$\nabla_{\mu}=\partial_{\mu}-in_eA_{\mu}\,,$$ where $n_e$ is
the electric charge of the field $q$.
Following \cite{Y99}, we will first consider this model  with a small
but nonvanishing $\lambda$ and
then take the limit $\lambda = 0$.

Obviously, the field $q$ develops a VEV, $q=v$,
spontaneously breaking the U(1) gauge group.
The photon acquires the mass
\beq
\label{mgamma}
m^2_{\gamma}=2n_e^2g^2v^2,
\eeq
while the Higgs particle mass is
\beq
\label{Hmass}
m_q^2=4\lambda v^2\,.
\eeq

The model (\ref{ah}) is the standard Abelian Higgs model which supports
the ANO strings \cite{ANO}.
For generic values of $\lambda$ the Higgs mass
differs from that of the photon. The ratio of the photon mass to the
Higgs mass is an
important parameter --- in the theory of superconductivity it characterizes
the type of the
superconductor in question.
Namely, for $m_q<m_\gamma$ we have the type-I
superconductor in which two well-separated ANO strings attract each
other.  On the other hand, for $m_q > m_\gamma$ we have the type-II
superconductor in which two
well-separated strings repel each other. This is due to the fact
that the scalar field gives rise to attraction between two vortices,
while the electromagnetic field gives rise to repulsion.

Now, let us  consider the extreme type-I limit in which
\beq
\label{gammahiggs}
 m_q\ll m_\gamma \,.
\eeq
We will assume the week coupling regime in the model  (\ref{ah}),
$\lambda\ll g^2 \ll 1$.

The general guiding idea which will
lead us in the search for the string solution in the extreme type-I limit
is a separation of
different fields at distinct scales which are
obviously present in the problem at hand due
to the ``extremality" condition (\ref{gammahiggs}).
This method goes back to the original paper
by Abrikosov \cite{ANO} in which the tension of the type-II string
 had been calculated under the condition $m_{q}\gg m_{\gamma}$.
A similar idea was used in \cite{Y99} to calculate the tension
of the type-I string under the condition $m_{q}\ll m_{\gamma}$.

To the leading order in $\ln m_\gamma/m_q$ the vortex
solution has the following structure in
the plane orthogonal to the string axis: The
electromagnetic field is confined in a core with the radius
\begin{equation}
\label{rad}
R_g\ \sim\ \frac1{m_\gamma}\, \ln\,{\frac{m_\gamma}{m_q}}\ .
\end{equation}
At the same time, the scalar field is close to zero inside the core.
Outside the core the electromagnetic field is vanishingly
small, while the scalar field behaves as
\beq
\label{scsol}
q=v\left\{ 1-\frac{K_0(m_q r)}{\ln (1/m_q R_g)}\right\}\,
e^{i\alpha}\, ,
\eeq
where $r$ and $\alpha$ are  polar coordinates
in the orthogonal plane (Fig.~\ref{polarc}).
Here $K_0$ is the (imaginary argument) Bessel function
with the exponential fall-off at infinity and logarithmic
behavior at small arguments, $$K_0(x)\sim \ln (1/x )\,\,\,\mbox{at}\,\,\, x\to 0\,.$$
The reason for this behavior is that in the absence of the
electromagnetic field outside the core the scalar field satisfies the free
equation  of motion, and (\ref{scsol}) presents the
appropriate solution to this equation.
From (\ref{scsol}) we see that the scalar field slowly (logarithmically)
approaches its boundary value $v$.

The  tension of this string  is \cite{Y99}
\begin{equation}
\label{rten}
T \ =\ \frac{2\pi v^2}{\ln \left( m_\gamma/m_q\right) }\, .
\end{equation}
The main contribution to the tension in (\ref{rten})
 comes from the logarithmic ``tail'' of the scalar field $q$.
It is given by the kinetic term for the scalar field in (\ref{ah}).
This term contains a logarithmic integral over $r$. Other terms
in the action are suppressed by
inverse powers of $\ln\left( m_\gamma/m_q\right)$
as compared with the contribution quoted in (\ref{rten}).

The results in Eqs. (\ref{rad}) and  (\ref{rten}) imply that if we naively take the
limit $m_q\to 0$ the string becomes infinitely thick and its
tension tends to zero \cite{Y99}. This apparently means that there are no
strings in the limit $m_q =0$.  As was mentioned above, the absence of the ANO
strings in the theories with the flat Higgs potential was first noticed in
\cite{PeninRubak}.

One might think that the
absence of ANO strings means that there is no
confinement of monopoles in the theories
with the Higgs branches.

So, do the ANO strings exist, or don't they?

As we will see shortly confinement does not disappear \cite{Y99}.
It is the formulation of the problem that has to be changed a little bit
in the case at hand.

So far we considered infinitely long ANO strings. However,
an appropriate  setup in the confinement problem is in fact slightly different \cite{Y99}.
We have to consider a monopole--antimonopole pair at
a large but finite separation $L$. Our aim is to take the limit
$m_q\to 0$. This limit will be perfectly smooth provided
we  consider the ANO string of a finite
length $L$, such that
\begin{equation}
\label{L}
\frac1{m_\gamma}\ \ll\ L\ \ll\ \frac1{m_q}\, .
\end{equation}
Then it turns out  \cite{Y99} that $1/L$ plays the role of the $IR$ cutoff
in Eqs. (\ref{rad}) and (\ref{rten}), rather than $m_q$.
The reason for this is that for $r\ll L$ the problem is two-dimensional
and the solution of the  two-dimensional free equation of motion for the scalar field
given by (\ref{scsol}) is logarithmic. If we naively put $m_q=0$ in this solution
the Bessel function reduces to the logarithmic function which cannot
reach a finite boundary value at infinity. Thus, as we mentioned
above,  infinitely long flux tubes do not exist.

However, for $r\gg L$, the problem becomes three-dimensional. The solution to the three-dimensional free scalar
equation of motion behaves as $$(q-v) \sim 1/|\vec x| \,$$
where $x_n$  ($n=1,2,3$) are the spatial coordinates in the three-dimensional
space.

With this behavior the scalar field reaches its
boundary value at infinity. Clearly $1/L$ plays a role of the IR
cutoff for the logarithmic behavior of the scalar field.

Now we can safely put $m_q=0$.
The formula for the radius of the electromagnetic core of the vortex
takes the form
\begin{equation}
\label{ts}
R_g\ \sim\ \frac1{m_\gamma}\ln \left(m_\gamma L\right)\, ,
\end{equation}
while the string tension now becomes \cite{Y99}
\begin{equation}
\label{ct}
T\ =\ \frac{2\pi v^2}{\ln \left( m_\gamma L\right) }\, .
\end{equation}
The ANO string becomes ``thick." Nevertheless, its
transverse size $R_g$ is much smaller than its length $L$, $$R_g\ll
L\,,$$
so that the string-like structure is clearly identifiable.
As a result, the potential acting between the probe well-separated
monopole and antimonopole  confines but is
no longer linear in $L$. At large $L$ \cite{Y99}
\begin{equation}
V(L)\ =\ 2\pi v^2\, \frac L{\ln\left( m_\gamma L\right) }\, .
\end{equation}

The potential $V(L)$ is an order parameter which distinguishes
different phases of a given gauge theory
(see, for example, \cite{seiobz}). We conclude that
on the Higgs branches one deals
with a
new confining phase, which had never been observed previously. It is clear that
this phase can arise only in supersymmetric theories
because we have no Higgs branches without supersymmetry.

\subsection{Example: \none SQED with the FI term}
\label{aaaa}

Initial comments regarding this model are presented in Part I,
see Sect.~\ref{sqed4d}.
The SQED Lagrangian in terms of superfields
is presented in Eq.~(\ref{sqed}), while the component expression can be found
in (\ref{n1bt}). For convenience we reiterate here
crucial features of \none SQED, to be exploited below.

The field content of \none SQED is as follows. The vector multiplet
contains the U(1) gauge field $A_{\mu}$ and the Weyl fermion
$\lambda^{\alpha}$, $\alpha=1,2$. The chiral matter multiplet contains
two complex scalar fields $q$ and $\qt$ as well as two complex Weyl
fermions $\psi^{\alpha}$ and $\tilde{\psi}_{\alpha}$. The bosonic part
of the action is
\beq
\label{qedN1}
S_{QED} =\int d^4 x \left\{ \frac1{4 g^2}F_{\mu\nu}^2 +
\bar{\nabla}_{\mu}\bar{q}\nabla_{\mu}q+
\bar{\nabla}_{\mu}\tilde{q}\nabla_{\mu}\bar{\tilde{q}}
+V(q,\qt)\right\},
\eeq
where
$$
\nabla_{\mu}=\partial_{\mu} -\frac{i}{2}A_{\mu}\,,\qquad
\bar{\nabla}_{\mu}=\partial_{\mu} +\frac{i}{2}A_{\mu}\,.
$$
Thus, we assume the matter fields to have electric charges $n_e=\pm 1/2$.
The potential of this theory comes from the $D$ term and
reduces to
\beq
\label{pot1}
V(q,\qt)=\frac{g^2}{8}\left( |q|^2 -|\tilde{q}|^2
-\xi\right)^2\,.
\eeq
The  parameter $\xi$ is the Fayet--Iliopoulos parameter
introduced through $\xi_3$.

The vacuum manifold of the
theory (\ref{qedN1}) is the Higgs branch determined
by the condition
\beq
\label{1hb}
|q|^2 - |\qt |^2
          = \xi.
\eeq
The dimension of this Higgs branch is two. To see this please observe that
in the problem at hand we
have two complex scalars (four real variables) subject to
one constraint (\ref{1hb}). In addition, we have to subtract one gauge
phase; thus, we have $4-1-1=2$.

The mass spectrum of this theory consists of one massive vector
$\mbox{\none}$ multiplet, with mass
\beq
\label{mgamma1}
m^2_{\gamma}=\frac12g^2v^2\, ,
\eeq
(four bosonic + four fermionic states) and one chiral massless field associated with
fluctuations along the Higgs branch.
The VEV of the scalar field above
is given by
\beq
\label{vevhb}
v^2= |\langle q \rangle |^2 + |\langle \qt \rangle |^2\,.
\eeq

Next, following \cite{EY}, let us consider strings
supported by this theory. First we will choose
the VEV of the scalar field to lie on the base point of the Higgs branch,
\beq
q=\sqrt{\xi},\qquad \qt=0\,.
\label{basepoint}
\eeq
Then the massless field $\qt$ plays no role in the string solution and can be
set to zero. This case is similar to the case of non-Abelian strings in
\none SQCD described in detail in Sect.~\ref{nonep}. On the base of the Higgs
branch we do have (classically) BPS ANO strings with the tension given by (\ref{ten}).
In particular, their profile functions are determined by (\ref{profil}) and
satisfy the first-order equations (\ref{foe}).

Now consider a generic vacuum on the Higgs branch. The string solution has the
following structure \cite{EY}. The electromagnetic field, together with the
massive scalar, form a string core of  size $\sim
1/( g\sqrt{\xi})$. The solution for this core is essentially given by the BPS
profile functions for the gauge field and massive scalar $q$. Outside the core
the massive fields almost vanish, while the light (massless) fields living on
the Higgs branch produce a logarithmic ``tail.'' Inside this ``tail''
the light scalar fields interpolate between the base point (\ref{basepoint})
and the VEV's of scalars $q$ and $\qt$  on the Higgs branch
(\ref{1hb}). The tension of the string is given by the sum of tensions
coming from the core and ``tail'' regions,
\beq
\label{tenL}
T= 2\pi\xi + \frac{2\pi\xi}{\ln\, {(g\sqrt{\xi}\,L)}}\,\, l^2,
\eeq
where $l$  the length of the geodesic line on the Higgs branch between
the base point and VEV,
\beq
\label{length}
l=
\int_0^1 d t \, \sqrt{g_{MN}\, \left(
\partial_{t}\vp^N \right)\left(  \partial_{t}\vp^N\right)}\,,
\eeq
where $g_{MN}$ is the metric on the Higgs branch, while $\vp^N$ stand
for massless scalars living on the Higgs branch. For example,
for $v^2\gg\xi$ $$l^2=v^2/\xi \,,$$ and
the ``tail'' contribution in (\ref{tenL}) matches
the result (\ref{ct}) for the string tension on the flat Higgs branch.

In (\ref{tenL}) we consider the string of a finite length $L$
to ensure an infrared regularization. It is also possible \cite{EY} to embed
\none SQED (\ref{qedN1}) in softly broken \ntwo SQED much in the same
way as it was done in Sect.~\ref{nonep} for non-Abelian strings. This
procedure slightly lifts the Higgs branch making even infinitely long strings
well defined. Note, however, that within this procedure the string
is {\em not} BPS-saturated at a generic point on the Higgs branch.

\vspace{2cm}

\centerline{\includegraphics[width=1.3in]{extra3.eps}}

\newpage

\section{Domain walls as $D$-brane prototypes}
\setcounter{equation}{0}
\label{walls}

D branes are extended objects in string theory on which strings can
end \cite{P}.
Moreover, the gauge fields are the lowest excitations
of open superstrings, with the endpoints attached to  D branes.
 SU$(N)$ gauge theories are obtained as a field-theoretic reduction
of a string theory  on the  worldvolume of a stack of $N$ D branes.

Our task is to see how the above assertions are implemented in
field theory. We have already thoroughly discussed
field-theoretic strings.
Solitonic objects of the domain wall type
were also extensively studied in supersymmetric gauge theories in 1+3 dimensions
in the last decade. The original impetus was provided by the Dvali--Shifman
observation \cite{DvSh} of the
critical (BPS-saturated) domain walls
in \none  gluodynamics, with the tension scaling as
$N\Lambda^3$.
The peculiar $N$ dependence of the tension prompted \cite{Witten:1997ep}
a D brane interpretation of such walls. Ideas as to how
flux tubes can end on the BPS walls were analyzed \cite{KoganKS}
at the qualitative level shortly thereafter.
Later on, BPS saturated domain-walls and their junctions with strings
were discussed \cite{AbrTow,GPTT} in a more quantitative aspect
in \ntwo sigma models.
Some remarkable parallels between field-theoretical critical solitons
and the D-brane string theory construction were discovered.

In this and subsequent sections we will review the parallel  found
between the field-theoretical  BPS domain walls in gauge theories
and D branes/strings. In other words, we will discuss BPS domain walls
with the emphasis on localization of the  gauge fields on their world volume.
In this sense the BPS domain walls
become $D$ brane prototypes in field theory.

As was mentioned,
research on field-theoretic mechanisms of gauge field localization
on the domain walls attracted much attention in recent years.
The only viable mechanism of gauge field localization
was put forward in Ref. \cite{DvSh}
where it was noted that if a gauge field is confined in the bulk and
is unconfined (or less confined) on the brane, this naturally gives rise
to a gauge field on the wall (for further developments see Refs.
\cite{DRubak,DVil}).
Although this idea seems easy to implement, in fact
it requires a careful consideration of quantum effects
(confinement is certainly such an effect)
which is hard to do at strong coupling.

Building on these initial proposals models with localization of gauge
fields on the worldvolume of domain walls at weak coupling
in \ntwo supersym\-metric gauge theories were
suggested in \cite{SYwall,SYnawall,SakaiT}.
Using a dual language,
the basic idea can be expressed as follows:
the gauge group is completely Higgsed in the bulk while inside the wall
the charged scalar fields almost vanish.
In the bulk magnetic flux tubes are formed while inside the wall
the magnetic fields can propagate
freely. In Ref.~\cite{SYwall} domain walls in the simplest \ntwo QED
theory were considered while Refs.~\cite{SakaiT,SYnawall} deal with the
domain walls in non-Abelian \ntwo gauge theories (\ref{qed}), with the
gauge group U$(N)$. Below we will review some results obtained in these papers.

The moduli space of the multiple domain walls in \ntwo supersymmetric
gauge theories  and sigma models
were studied in \cite{GTT,Tw,Jwms,Jw,Jwms2}.
Note that the domain walls can intersect \cite{GTTi,KakimotoSak,Jtrian}.
In particular, in \cite{Jweb,Jnaweb} a honeycomb webs of walls
were obtained in Abelian and non-Abelian gauge theories, respectively.

We start our discussion of the BPS domain walls as $D$ brane prototypes in
the simplest Abelian theory
-- \ntwo SQED with 2 flavors \cite{SYwall}. It supports  both,
the BPS-saturated domain walls and the
BPS-saturated ANO strings if the Fayet--Iliopoulos term is added
to the theory.

\subsection{ \ntwo supersymmetric QED}
\setcounter{equation}{0}
\label{n2qed}

\none SQED (four supercharges) was discussed in Sect.~\ref{vortandflt}.
Now we will extend supersymmetry to \ntwo (eight supercharges).
Some relevant features of this model are summarized in Appendix B.

The field content of \ntwo SQED is as follows.
In the gauge sector we have the U(1) vector
\ntwo multiplet. In the matter sector we have $N_f$  matter hypermultiplets.
In this section we will limit ourselves to $N_f=2$.  This is the simplest case
which admits domain wall interpolating between quark vacua.
The bosonic part of the action of this
theory is
\beqn
&&
S=\int d^4 x \left\{ \frac{1}{4 g^2} F_{\mu \nu}^2 + \frac{1}{g^2}
|\partial_\mu a|^2 +\bar{\nabla}_\mu \bar{q}_A \nabla_\mu q^A +
\bar{\nabla}_\mu \tilde{q}_A \nabla_\mu \bar{\tilde{q}}^A
\right.\nonumber\\[3mm]
 &&
 +
 \left.\frac{g^2}{8}\left(|q^A|^2-|\tilde{q}_A|^2-\xi\right)^2+\frac{g^2}{2}
\left|\tilde{q}_A q^A\right|^2
+\frac{1}{2}(|q^A|^2+|\tilde{q}^A|^2)\left| a+\sqrt{2}m_A\right|^2 \right\}, \nonumber\\
\label{n2sqed}
\eeqn
where
\beq
\nabla_\mu=\partial_\mu-\frac{i}{2}A_\mu\,,\qquad
\bar{\nabla}_\mu=\partial_\mu+\frac{i}{2}A_\mu\,.
\eeq
With this convention the electric charges of the matter fields
are $\pm1/2$ (in the units of $g$).
Parameter $\xi$  in Eq.~(\ref{n2sqed}) is the coefficient in front of the Fayet--Iliopoulos term.
It is introduced as in Eq.~(\ref {fiterm}) with
$F_3 = D$ and $F_{1,2}=0$. In other words,
here we introduce the Fayet--Iliopoulos term as the $D$ term.
Furthermore, $g$ is the U(1)
gauge coupling. The
index $A=1,\,2$ is the flavor index.

The mass parameters $m_1,m_2$ are
assumed to be  real. In addition we will assume
\beq
\Delta m \equiv
m_1-m_2\gg g\sqrt{\xi} \,.
\label{mggxi}
\eeq
Simultaneously, $\Delta m \ll  (m_1+m_2)/2$.
There are two vacua in this theory: in the first vacuum
\beq
a=-\sqrt{2} m_1,\qquad q_1=\sqrt{\xi}, \qquad  q_2=0\,,
\label{fv}
\eeq
and in the second one
\beq
a=-\sqrt{2} m_2, \qquad q_1=0,  \qquad q_2=\sqrt{\xi}\,.
\label{sv}
\eeq
The vacuum expectation value (VEV) of the
field $\tilde{q}$ vanishes in both  vacua.
Hereafter in search for domain wall solutions
we will stick to the {\em ansatz} $\tilde{q}=0$.

Now let us discuss the mass spectrum in both quark vacua.
Consider for definiteness  the first vacuum, Eq.~(\ref{fv}).
 The spectrum  can be obtained by diagonalizing  the quadratic form
in  (\ref{n2sqed}). This is done in Ref.~\cite{VY};  the result is as
follows:  one real component of the field $q^1$ is eaten up by the Higgs
 mechanism to become the third components of the massive photon. Three
components of the massive photon, one remaining component of $q^1$ and
four real components of  the fields
$\tilde{q}_1$ and $a$ form one long \ntwo
multiplet (8 boson states + 8 fermion states), with mass
\beq
\label{mgammaw}
m_{\gamma}^2=\frac12\, g^2\,\xi.
\eeq

The second flavor $q^2$,  $\tilde{q}_2$
 (which does not condense in this vacuum)
forms one short \ntwo multiplet (4 boson states + 4 fermion states), with
mass $\Delta m$ which is heavier than
the mass of the vector supermultiplet.
The latter assertion
applies to the regime (\ref{mggxi}).
In the second vacuum the mass spectrum is   similar --- the
roles of the first and the second flavors are interchanged.

If we consider the limit opposite to that in Eq.~(\ref{mggxi}) and tend
$\Delta m\to 0$,
the ``photonic" supermultiplet becomes heavier than that of $q^2$,
the second flavor field.
Therefore, it can be integrated out, leaving us with
the theory of massless moduli from  $q^2,\,\tilde{q}_2$,
which interact through a nonlinear sigma model with the K\"ahler term
corresponding to the Eguchi--Hanson metric. The manifold parametrized by these (nearly) massless fields
is obviously four-dimensional. Both vacua discussed above lie at the base of this
manifold. Therefore, in considering the domain wall solutions
in the sigma model limit $\Delta m\to 0$ \cite{GTT,Tw,GPTT}
one can limit oneself to the base manifold, which is, in fact, a two-dimensional sphere.
In other words, classically,  it is sufficient to
consider the domain wall in the CP(1) model
deformed by a twisted mass term (related to a nonvanishing $\Delta m$),
see Fig.~\ref{f13one}.
This was first done in \cite{Tw}. A more general analysis of the
domain walls on the Eguchi--Hanson manifold can be found in \cite{arai}.
An interesting \none deformation of the model (\ref{n2sqed})
which was treated in the literature \cite{rauzzi}
in the quest for ``confinement on the wall"
automatically requires construction of the wall
on the Eguchi--Hanson manifold, rather than the CP(1) wall,
since in this case the two vacua of the model
between which the wall interpolates do not lie
on the base.

\subsection{Domain walls in \ntwo SQED}
\setcounter{equation}{0}
\label{wallsinqed}

A BPS domain wall
interpolating between the two vacua of the  bulk
theory (\ref{n2sqed}) was explicitly constructed in Ref. \cite{SYwall}.
Assuming that all fields depend only on the coordinate
$z=x_3$, it is possible to write the energy in the Bogomol'nyi form
\cite{B},
\beqn
  E &=&
 \int dx_3 \left\{ \left|\nabla_3 q^A \pm \frac{1}{\sqrt{2}}q^A
 (a+\sqrt{2}m_A)\right|^2\right.
 \nonumber\\[3mm]
&+&\left. \left|\frac{1}{g}\partial_3 a \pm \frac{g}{2 \sqrt{2}}
\left(|q^A|^2-\xi\right)\right|^2
\pm \frac{1}{\sqrt{2}} \xi \partial_3 a\right\}.
\label{bogfw}
\eeqn
Requiring the first two terms above to vanish gives us the BPS equations
for the wall. Assuming that $\Delta m >0$ we choose the upper sign in
(\ref{bogfw}) to get
\begin{eqnarray}
\nabla_z q^A &=& -\frac1{\sqrt{2}}q^A\left(a+\sqrt{2}m_A\right),
\nonumber\\[3mm]
\label{wfoe}
\pt_z a &=&- \frac{g^2}{2\sqrt{2}}\left(|q^A|^2-\xi\right).
\end{eqnarray}
These first-order equations should be supplemented by the following boundary
conditions:
\begin{eqnarray}
q^1(-\infty) &=&\sqrt{\xi},\quad q^2(-\infty)=0,
\quad
a(-\infty)=-\sqrt{2}m_1\,;
\nonumber\\[3mm]
\label{wbc}
q^1(\infty)
&=& 0,\quad
|q^2(\infty)|=\sqrt{\xi},\quad a(\infty)=-\sqrt{2}m_2,
\end{eqnarray}
which show  that our wall interpolates between the two quark vacua.
Here we use a U(1) gauge rotation  to make $q^1$ in the left vacuum
real.

 The tension is given by the total derivative term
(the last one in Eq.~(\ref{bogfw}))
which can be identified as the $(1,0)$ central charge
of the supersymmetry algebra,
\beq
T_{\rm w}=\xi \, \Delta m\,.
\label{wten}
\eeq

WE can find the solution to the first-order equations (\ref{wfoe})
compatible with the boundary
conditions (\ref{mggxi}). The range of variation of the field
$a$ inside the wall is of the order of $\Delta m$ (see Eq.~(\ref{wbc})).
Minimization of  its kinetic energy
implies this field is slowly varying.
Therefore, we may safely assume that the wall is thick; its
size $R\gg 1/g\sqrt{\xi}$. This fact will be confirmed shortly.

\begin{figure}[h]
\epsfxsize=9cm
\centerline{\epsfbox{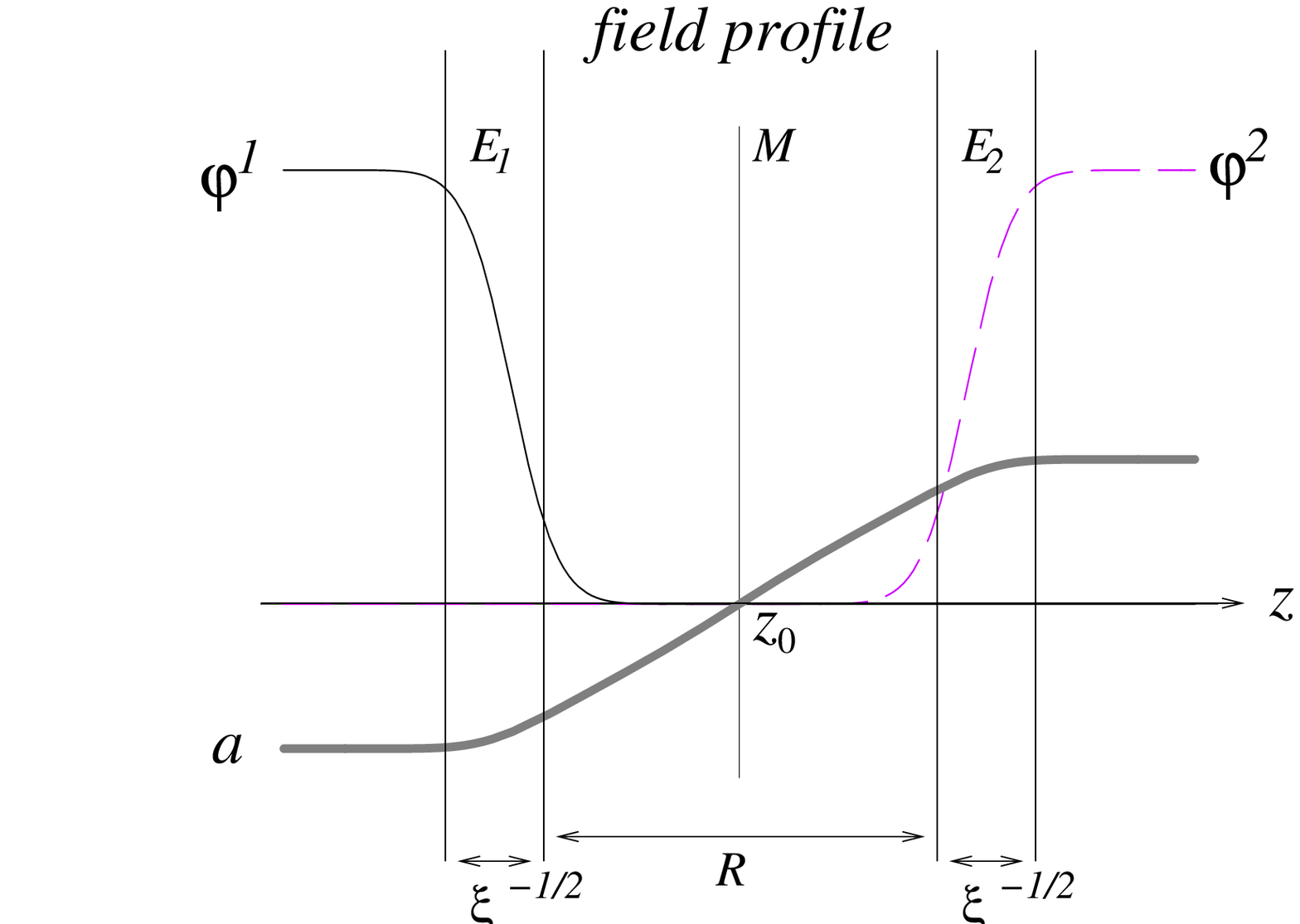}}
 \caption{\small Internal structure of the domain wall:
two edges (domains $E_{1,2}$) of the width $\sim (g\sqrt{\xi})^{-1}$
are separated by a broad middle band (domain $M$) of the width $R$,
see Eq. (\ref{R}).}
\label{syfigthree}
\end{figure}

We arrive at the following picture of the domain wall at hand.
The wall solution has a three-layer structure \cite{SYwall},
see Fig.~\ref{syfigthree}.
In the two outer layers --- let us call them edges, they  have thickness
${O}(({g \sqrt{\xi}})^{-1}$) which means that they are thin  ---  the squark fields drop
to zero exponentially;  in the inner layer
the field $a$ interpolates between its two vacuum values.

Then to the leading order we can put the quark fields to zero
in (\ref{wfoe}) inside the inner layer.
The second equation in (\ref{wfoe}) tells us that $a$ is a linear
function of $z$. The solution for $a$ takes the form
\beq
\label{a}
a=-\sqrt{2}\left( m-\Delta m \frac{z-z_0}{R}\right),
\eeq
where the collective coordinate $z_0$ is the position  of the wall center
(and $\Delta m$ is assumed positive).
The solution  is valid in a
wide domain of $z$
\beq
\label{inside}
\left| z-z_0\right| < \frac{R}{2}\,,
\eeq
except narrow areas of size $\sim 1/g\sqrt{\xi}$ near the edges of the
wall at $z-z_0=\pm R/2$.

Substituting the solution (\ref{a}) in the second equation in (\ref{wfoe})
we get
\beq
\label{R}
R=\frac{4\Delta m}{g^2 \xi}= \frac{2\Delta m}{m_\gamma^2}\,.
\eeq
Since $\Delta m /g\sqrt\xi \gg 1$, see Eq. (\ref{mggxi}), this result
shows that $R\gg1/g\sqrt{\xi}$, which justifies our approximation.
This approximation will be referred to as the thin-edge approximation.

Furthermore, we can now use the first relation in Eq.~(\ref{wfoe})
to determine   tails
of the quark fields inside the wall. As was
mentioned above, we  fix the gauge imposing
the condition that $q^1$ is real at $z\to -\infty$, see more detail
discussion in \cite{SYwall}.

Consider first the left edge (domain $E_1$ in Fig. \ref{syfigthree})
at
$z-z_0=-R/2$. Substituting the above solution for $a$ in the equation for
$q^1$ we get
\beq
\label{wq1}
q^1=\sqrt{\xi}\,
e^{-\frac{m_{\gamma}^2}{4}\left(z-z_0+\frac{R}{2}\right)^2
}\,\, ,
\eeq
where $m_{\gamma}$ is given by  (\ref{mgammaw}).
This behavior is valid in the domain $M$, at
$(z-z_0+R/2)\gg1/g\sqrt{\xi}$, and shows
that the field of the first quark flavor tends to zero  exponentially
inside the wall, as was expected.

By the same token,  we can consider the behavior of the second quark flavor
near the right edge of the wall at  $z-z_0=R/2$. The
first equation  in (\ref{wfoe})   for $A=2$
implies
\beq
\label{wq2}
q^2=\sqrt{\xi}\,
e^{-\frac{m_{\gamma}^2}{4}\left(z-z_0-\frac{R}{2}\right)^2 -i\sigma
}\,\,,
\eeq
which is valid in the domain $M$ provided that
$(R/2-z+z_0)\gg1/g\sqrt{\xi}$.
Here $\sigma$ is an arbitrary phase which cannot be gauged away.
Inside the wall the second
quark flavor tends to zero  exponentially too.

It is not difficult to check that the main contribution to the wall
tension comes from the middle layer while the edge domains produce contributions
of the order of $\xi^{3/2}$ which makes them
negligibly small.

Now let us comment on the   phase factor in  (\ref{wq2}).
Its origin
is as follows \cite{SYwall}.
The bulk theory at $\Delta m\neq 0$ has the U(1)$\times$U(1) flavor symmetry
corresponding to two independent rotations of two quark flavors. In both
vacua only one quark develops a  VEV. Therefore, in both vacua only one
of these
two U(1)'s is broken. The corresponding phase is eaten by the Higgs mechanism.
However, on the wall both quarks have non-vanishing values, breaking both U(1)
groups. Only one of corresponding  two phases is eaten by the Higgs mechanism.
The other one becomes a Goldstone mode living on the wall.

Thus, we have two collective coordinates characterizing our wall solution,
the position of the center $z_0$ and the phase $\sigma$. In the effective
low-energy   theory on the wall
they become scalar fields of the worldvolume (2+1)-dimensional theory,
$z_0 (t,x,y)$ and $\sigma (t,x,y)$, respectively.
The target space of the second field is $S_1$.

This wall is an $1/2$ BPS solution of the Bogomol'nyi equations.
In other words, four out of eight supersymmetry
generators of the \ntwo bulk theory are broken.
As was shown in \cite{SYwall}, the four supercharges selected
by the conditions
\beqn
\bar{\ve}^2_{\dot{2}}&=& -i\ve^{21}\,,\qquad
\bar{\ve}^1_{\dot{2}}=-i\ve^{22}\,,
\nonumber\\[3mm]
\bar{\ve}^1_{\dot{1}}&=& i\ve^{12}\, ,\qquad
\bar{\ve}^2_{\dot{1}}=i\ve^{11}\,,
\label{wsusy}
\eeqn
act trivially on the wall solution. They become the four supersymmetries
acting in the (2+1)-dimensional effective  worldvolume theory on the
wall.  Here $\ve^{\alpha f}$ and $\bar{\ve}^f_{\dot{\alpha}}$ are eight
supertransformation parameters.

\subsection{Effective field theory on the wall}
\setcounter{equation}{0}
\label{wkinterms}

In this section we will review the  (2+1)-dimensional  effective
low-energy theory of the moduli on the wall \cite{SYwall}.
To this end we will make the wall
collective coordinates $z_0$ and  $\sigma$ (together with their
fermionic superpartners) slowly varying fields
depending on $x_n$  ($n=0,1,2$),
 For simplicity let us consider the
bosonic fields $z_0 (x_n)$ and  $\sigma  (x_n)$;
the  residual supersymmetry will allow us to
readily reconstruct the fermion part of the effective action.

Because $z_0 (x_n)$ and  $\sigma  (x_n)$ correspond to zero modes of
the wall, they have no potential terms in the worldsheet theory.
Therefore, in fact our task is to derive kinetic terms,
much in the same way as it was done for strings,
see Sect.~\ref{worldsheet}.  For $z_0 (x_n)$ this procedure is very simple.
Substituting the wall solution
(\ref{a}), (\ref{wq1}), and (\ref{wq2}) in  the action
(\ref{n2sqed}) and taking into account the $x_n$ dependence of this
modulus we immediately get
\beq
\label{kinz0}
\frac{T_{\rm
w}}{2} \, \int d^3 x \; (\pt_n z_0   )^2\, .
\eeq
As far as the
kinetic term for $\sigma  (x_n)$ is concerned more effort is needed. We
start from Eqs. (\ref{wq1}) and  (\ref{wq2}) for the quark
fields.
Then we will have to modify our {\em ansatz} introducing
nonvanishing components
of the gauge field,
\beq
\label{kingpot}
A_n=\chi(z)\, \pt_{n}\sigma(x_n)\,.
\eeq
These components of the gauge field are needed
to make worldvolume action well-defined.
They are introduced
in order to cancel  the $x$ dependence of the quark fields
far away from the wall (in the
quark vacua at $z\to \infty$) emerging through the  $x$ dependence
of $\sigma (x_n)$, see  Eq.~(\ref{wq2}).

Thus, we introduce a new profile function  $\chi(z)$.
It has no role in the construction of the static wall solution
{\em per se}. It is unavoidable, however,
in constructing the kinetic part of the worldsheet theory
of the moduli. This new profile function
is described by its own action, which will be subject to
minimization. This is quite similar to derivation
of the worldsheet effective theory for non-Abelian strings,
see Sect.~\ref{worldsheet}.

The gauge potential in  Eq.~(\ref{kingpot}) is  pure gauge
far away from the wall and is not pure gauge inside the wall. It does lead to
a non-vanishing field strength.

To ensure proper vacua at $z\to\pm \infty$ we
impose the following boundary conditions on the function $\chi (z)$
\beqn
&&
\chi(z)\to 0,\;\,\,  z\to - \infty \,,
\nonumber\\[2mm]
&&
\chi(z)\to -2,\;\,\, z\to + \infty\, .
\label{bchi}
\eeqn
Remember the
electric charge of the quark fields is $\pm 1/2$.

Next, substituting Eqs.~(\ref{wq1}), (\ref{wq2}) and (\ref{kingpot})
in the  action (\ref{n2sqed}) we arrive at
\begin{eqnarray}
S_{2+1}^{\sigma} &=&\left[\int d^3 x \;\frac12(\pt_n \sigma )^2\right]
\nonumber\\[4mm]
&\times& \int dz\,
\left\{\frac1{g^2}(\pt_z\chi)^2 +
\chi^2|q^1|^2 +(2+\chi)^2|q^2|^2\right\}.
\label{kinsigint}
\end{eqnarray}
The expression in the second line must be considered as an ``action" for the $\chi$ profile function.

Our next task is to explicitly find the function $\chi$. To
this end we have to minimize (\ref{kinsigint}) with
respect to $\chi$. This gives the following equation:
\beq
-\pt_z^2\chi+g^2\chi |q^1|^2+g^2(2+\chi)|q^2|^2=0\, .
\label{eqonchi}
\eeq
The equation for $\chi$ is of the second order.
This is because the domain wall is no longer BPS state once we
switch on the dependence of the moduli  on the ``longitudinal"
variables $x_n$.

To the leading order in $g\sqrt{\xi}/{\Delta m}$ the solution of
Eq.~(\ref{eqonchi}) can be obtained in the same manner as
it was done  previously for other profile functions.
Let us first discuss what happens outside the inner part
of the wall. Say, at $z-z_0 \gg R/2$ the profile $|q^1|$ vanishes
while $|q^2|$ is exponentially close to
$\sqrt{\xi}$ and, hence,
\beq
\chi \to - 2 +{\rm const}\, e^{-m_\gamma (z-z_0)}\,.
\label{bvfchi}
\eeq
At $z_0 -z \gg R/2$
the profile function $\chi$ falls off exponentially to zero.
Thus,
outside the inner part of the  wall, at $|z-z_0|\gg R/2$, the function
$\chi$ approaches its
boundary values with the exponential rate of approach.

Of most interest, however, is the inside part,
the middle domain $M$ (see Fig.~\ref{syfigthree}).
Here both quark profile functions vanish, and Eq.~(\ref{eqonchi})
degenerates into $\pt_z^2 \chi =0$. As a result, the solution takes
the form
\beq
\chi=-1-2 \frac{z-z_0}{R}\, .
\label{sfchiinm}
\eeq
In the narrow edge domains $E_{1,2}$ the exact $\chi$
profile smoothly interpolates between the boundary
values, see Eq. (\ref{bvfchi}),  and the linear  behavior (\ref{sfchiinm})
inside the wall. These
edge domains give small corrections to the leading term   in the
action.

Substituting the solution (\ref{sfchiinm})  in  the $\chi$
action, the second line in Eq. (\ref{kinsigint}),  we finally arrive at
\beq
\label{kinsig}
S_{2+1}^{\sigma}=\frac{\xi}{\Delta m}\, \int d^3 x \;\frac12 \,
(\pt_n \sigma )^2\, .
\eeq

\vspace{2mm}

As well-known \cite{P77}, the compact scalar field $\sigma (t,x,y)$
can be reinterpreted to be
dual to the (2+1)-dimensional Abelian gauge field living on the wall.
The emergence of the gauge field on the wall is easy to
understand. The quark fields   almost vanish inside the wall.
Therefore the  U(1)  gauge group is restored inside the wall while it
is Higgsed
in the bulk. The dual U(1) is in the confinement regime in the bulk.
Hence, the dual U(1) gauge field is localized on the wall, in full accordance with the
general argument of Ref.~\cite{DvSh}.
The compact scalar field $\sigma (x_n)$ living on the wall is a manifestation of this
magnetic localization.

The action in Eq.~(\ref{kinsig}) implies that the coupling constant of our
effective  U(1)  theory on the wall is given by
\beq
\label{21coupling}
e^2=4\pi^2\,  \frac{\xi}{\Delta m}\, .
\eeq
In particular, the definition of the (2+1)-dimensional gauge field
 takes the form
\beq
\label{21gaugenorm}
F^{(2+1)}_{nm}=\frac{e^2}{2\pi} \,
\varepsilon_{nmk}\, \partial^k \sigma\, .
\eeq
This finally
 leads us   to the following
effective low-energy theory of the moduli fields on the wall:
\beq
\label{21theory}
S_{2+1}=\int d^3 x \, \left\{\frac{T_w}{2}\,\,  (\pt_n z_0 )^2+
\frac{1}{4\, e^2}\, (F_{nm}^{(2+1)})^2 +\mbox{fermion terms} \right\}.
\eeq
The fermion content of
the worldvolume theory
is given by two three-dimen\-sional Majorana spinors, as
is required by ${\mathcal N}=2$ in three dimensions
(four supercharges, see (\ref{wsusy})).
The full worldvolume theory is a
 U(1) gauge theory in $(2+1)$ dimensions, with
four supercharges. The Lagrangian and the
corresponding superalgebra
can be obtained by reducing four-dimensional
\none SQED (with no matter)
to three dimensions.

The field $z_0$ in (\ref{21theory}) is the \ntwo superpartner of the
gauge field $A_n$. To make it more transparent we make a rescaling,
introducing a new field
\beq
a_{2+1}=2\pi\xi\,z_0\,.
\label{az}
\eeq
In terms of $a_{2+1}$ the action
(\ref{21theory}) takes the form
\beq
S_{2+1}=\int d^3 x \left\{ \frac{1}{2e^2}
\left(\partial_n a_{2+1}\right)^2
+\frac{1}{4 e^2} \left( F_{mn}^{(2+1)} \right)^2
+\mbox{fermions}\right\}.
\label{pure}
\eeq
The gauge coupling constant $e^2$ has dimension of mass in three
dimensions. A characteristic scale of massive excitations on the
worldvolume theory is of the order of the inverse thickness of the wall $1/R$,
see (\ref{R}). Thus, the dimensionless parameter that characterizes the
coupling strength in the worldvolume theory is $e^2 R$,
\beq
e^2 R=\frac{16\pi^2}{g^2}.
\label{duality}
\eeq
This can be interpreted
as a feature of the bulk--wall duality:
the weak coupling regime in the bulk theory
corresponds to strong coupling on the wall and {\em vice versa}
\cite{SYwall,SYdual}.
Of course, finding explicit domain wall solutions and deriving the effective
theory on the wall assumes weak coupling in the bulk, $g^2\ll 1$.
In this limit the worldvolume theory is in the strong coupling regime
and is not very useful.

The fact that each domain wall has two bosonic collective coordinates
--- its center and the phase --- in the sigma model limit was noted
in \cite{AbrTow,Tw}.

To  summarize, we showed that the  worldvolume theory on the domain
wall is the U(1) gauge theory (\ref{pure}) with extended supersymmetry,
\ntwo. Thus, the domain wall
in the theory (\ref{n2sqed})
presents an example of a field-theoretic $D$ brane: it localizes
a gauge field on its worldvolume. In string theory gauge fields
are localized on $D$ branes because fundamental open strings can
end on $D$ branes. It turns out that this is also true for
field-theoretic ``$D$ branes."
In fact, various junctions of
field-theoretic strings (flux tubes)  with domain walls
were found explicitly \cite{GPTT,SYwall,SYnawall}.
We will review 1/4-BPS  junctions in Sect.~\ref{wallstrj}.
Meanwhile, in the remainder of this section we will consider  non-Abelian generalizations
of the localization effect for the gauge fields.

\subsection{Domain walls in the U$(N)$ gauge theories}
\setcounter{equation}{0}
\label{uNwalls}

In this section we will review the domain walls in \ntwo QCD
(see Eq.~(\ref{qed}))
with the  U$(N)$ gauge group. We assume that the number of the quark flavors
in this theory $N_f>N$, so the theory has many vacua of the type
(\ref{avev}), (\ref{qvev}) depending on which $N$ quarks out of
$N_f$ develop VEV's. We can denote different vacua as
$(A_1,A_2,...,A_N)$ specifying which quark flavors develop VEV's.
First, we will consider a general case assuming all quark masses
to be different.

\subsubsection{Non-degenerate masses}

Let us arrange the quark masses as follows:
\beq
m_1>m_2>...>m_{N_f}.
\label{morder}
\eeq
In this case the theory (\ref{qed})  has
\beq
\frac{N_f!}{N!(N_f-N)!}
\label{numvac}
\eeq
isolated vacua.

Domain walls interpolating between these vacua were classified
in \cite{SakaiT}. Below we will review this classification.

The
Bogomol'nyi representation of the action (\ref{qed}) leads to the
first-order equations for the wall configurations \cite{LambertTong},
see also \cite{SYnawall},
\begin{eqnarray}
\pt_z
\vp^A &=&
-\frac1{\sqrt{2}}\left(a_a\tau^a+a+
\sqrt{2}m_A\right)\vp^A,
\nonumber\\[3mm]
\label{nawfoe}
\pt_z a^a &=&-
\frac{g_2^2}{2\sqrt{2}}\left(\bar{\varphi}_A\tau^a \vp^A
\right),
\nonumber\\[3mm]
\pt_z a &=&-
\frac{g_1^2}{2\sqrt{2}}\left(|\vp^A|^2
-2\xi
\right),
\end{eqnarray}
where we used the {\em ansatz} (\ref{qtilde}) and introduced a single quark
field $\vp^{kA}$ instead of two fields $q^{kA}$ and $\tilde{q}_{Ak}$.
These walls are 1/2 BPS.
The wall tensions are
given by the surface term
\beq
T_{\rm w}=\sqrt{2}\xi\int d z\,\pt_z a \,.
\label{wtension}
\eeq
They can be written as \cite{SakaiT}
\beq
T_{\rm w}=\xi\,\vec{g}\,\vec{m}\, ,
\label{wtenst}
\eeq
where we use (\ref{avev}) and $\vec{m}=(m_1,...,m_{N_f})$ while
\beq
\vec{g}=\sum_{i=1}^{N_f-1} k_i\alpha_i\, .
\eeq
Here $k_i$ are integers while $\alpha_i$ are simple roots of the
SU($_{N_f}$) algebra,
\beqn
\alpha_1
&=&
(1,-1,0,...,0)\, ,
\nonumber\\[1mm]
\alpha_2
&=&
(0,1,-1,...,0)\, ,... \,,
\nonumber\\[1mm]
\alpha_{N_f-1}
&=&
(0,...,0,1,-1)\, .
\label{roots}
\eeqn
Elementary walls arise if one of $k_i$'s reduces to unity while  all
other integers in the set vanish. The tensions of the
elementary walls are
\beq
T_{\rm w}^{i}=\xi\,(m_i-m_{i+1})\,.
\label{wtenstel}
\eeq
The $i$-th elementary wall interpolates between the vacua $(...,i,...)$
and $(...,i+1,...)$. All other walls can be considered as a composite
states of elementary walls.

As an example  let us consider the theory (\ref{qed}) with the U(2) gauge group
and $N_f=4$.
Explicit solutions for the elementary  walls in the limit
\beq
(m_i-m_{i+1}) \gg g\sqrt{\xi}
\label{largemassdiff}
\eeq
where obtained in \cite{SYnawall}. They have the same three-layer
structure as in the Abelian case, see Sect.~\ref{wallsinqed}. Say, the
elementary wall interpolating between the vacua $(1,2)$ and $(1,3)$
has the following structure. In the left edge domain quark $\vp^2$
varies from its VEV $\sqrt{\xi}$  to zero exponentially, while
in the right edge domain the quark $\vp^3$ evolves from zero to its  VEV
$\sqrt{\xi}$. In the broad middle domain the fields $a$ and $a^3$
linearly interpolate between their VEV's in two vacua. A novel
feature of the domain wall solution as compared to the Abelian case
(see Sect.~\ref{wallsinqed}) is that the quark field $q^1$ does not vanish
both outside and inside the wall.

The solution for the elementary wall has two real moduli
much in the same way as in the Abelian
case: the wall center $z_0$ and a compact phase. The phase can be
rewritten as  a U(1) gauge field. Therefore, the effective theory
on the elementary wall is of the type (\ref{pure}), as in
the Abelian case. The physical reason behind the localization of the
U(1) gauge field on the wall worldvolume is easy to understand.
Since the quark $q^1$ does not vanish  inside the wall only an appropriately chosen
U(1) field, namely $(A_{\mu}-A^3_{\mu})$,
which does not interact with this quark field can propagate freely inside the
wall.

In the case of generic quark masses for composite domain walls
the effective  worldvolume theory contains U(1) gauge fields
associated with each elementary wall. However the metric on the moduli
space can be more complicated. For example the metric for the
$\alpha_1+\alpha_2$ composite wall was shown  \cite{Tw,Jcigar} to have a cigar-like
geometry.

\subsubsection{Degenerate masses}

Now let us  consider the case of degenerate quark masses. As we know
from Sect.~\ref{model} the non-Abelian gauge group is not broken in
this case by adjoint fields. It turns out that in this case certain
composite domain walls can localize non-Abelian gauge fields
\cite{SYnawall}.

Let us consider  the following choice of quark mass parameters:
\beqn
&&
m_1=m_2\, ,
\nonumber\\[2mm]
&&
m_3=m_4\,,
\nonumber\\[2mm]
&&
\Delta m \equiv m_1-m_3>0\,,
\label{mchoice}
\eeqn
assuming the condition $\Delta m\gg g\sqrt{\xi}$
to be satisfied for both the Abelian and non-Abelian coupling constants
$g_1$ and $g_2$. For this degenerate choice of masses four out of six
isolated vacua of the theory with non-degenerate masses coalesce.

This theory
has two isolated vacua $(1,2)$ and $(3,4)$ with unbroken SU(2)$_{C+F}$
symmetry while other four vacua coalesce
and a Higgs branch is developed from the common root. We will denote
this Higgs branch as $(A,B)$, where $A=1,2$ and $B=3,4$.

The elementary domain walls interpolating between the $(1,2)$ and $(A,B)$
vacua (or between the $(A,B)$ and $(3,4)$) have the same structure
as elementary walls described above for the theory with non-degenerate
masses. Following \cite{SYnawall}  we will be mostly interested
in the composite  $(1,2)\to (3,4)$ wall which is a bound
state of two elementary walls mentioned above.

The solution of the first-order equations (\ref{nawfoe}) for this wall
has a familiar three-ayer structure similar to the solutions for the
elementary walls.
Now all quark fields vanish inside the wall. The solution
for the $a$ fields in the middle domain $M$
is given by
\beqn
a
&=&
-\sqrt{2}\left( m_1-\Delta m\,
\frac{z-z_0+\tilde{R}/2}{\tilde{R}}\right),
\nonumber\\[3mm]
a^3 &=& 0\,,
\label{ac}
\eeqn
where we introduce the thickness $\tilde{R}$ of the composite wall,  to be
considered large, $\tilde{R}\gg 1/g\sqrt{\xi}$, see below.
The equation for $a^3$ in (\ref{nawfoe}) is trivially satisfied,
while the equation for $a$ yields
\beq
\tilde{R}=\frac{2\Delta m}{g_1^2\,  \xi}\,,
\label{tR}
\eeq
demonstrating that indeed   $\tilde{R}\gg 1/g\sqrt{\xi}$.

Substituting the above solutions in the first  two equations
in (\ref{nawfoe}) we determine the fall-off of the quark
fields inside the wall. Namely, near the left edge at
$(z-z_0 + R/2)\gg 1/g\sqrt{\xi}$
\beq
\vp^{kA}=
\sqrt{\xi}\,
\left(
\begin{array}{cc}
1 & 0\\[2mm]
0 & 1
\end{array}
\right)
e^{-\frac{\Delta m}{2\tilde{R}}\left(z-z_0+\frac{\tilde{R}}
{2}\right)^2} \,,\qquad A=1,2\, ,
\label{wq12}
\eeq
while near the right edge at $(R/2-z+z_0)\gg 1/g\sqrt{\xi}$
\beq
\vp^{kB}= \sqrt{\xi}\,(\tilde{U})^{kB}
e^{-\frac{\Delta m}{2\tilde{R}}\left(z-z_0-\frac{\tilde{R}}
{2}\right)^2}\,,\qquad B=3,4\, ,
\label{wq34}
\eeq
where  $\tilde{U}$ is a matrix from the U(2) global flavor group,
which takes into account possible flavor rotations  inside the flavor pair
$B=3,4$. It can be represented as product of a U(1) phase factor and
a matrix $ U$  from  SU(2)
\beq
\label{flmat}
\tilde{U}=e^{i\sigma_0}\;U\,.
\eeq
This matrix is parametrized by four phases, $\sigma_0$ plus
three phases residing in the matrix $U$.

The occurrence of these four wall moduli --- one related to U(1)
and three to SU(2) --- is quite similar to the occurrence of one U(1)
phase $\sigma$ for the domain wall in the Abelian theory, see
Sect.~\ref{wallsinqed}.
In \cite{SYnawall}   these four moduli
were identified with (2+1)-dimensional
gauge fields living on the wall worldvolume. Namely, the
phase $\sigma_0$ is identified with the U(1) gauge field while the
SU(2) matrix $U$ gives rise to a non-Abelian SU(2) gauge field.

Thus, we get four gauge fields localized on the wall.
The physical interpretation of this result is as follows. The quark fields
are condensed outside the $12  \to  34$ wall while inside they
(almost) vanish. Therefore both the U(1) and SU(2) gauge fields of
the bulk theory
are Higgsed in the bulk while they can freely propagate inside the wall.

The worldvolume theory derived in \cite{SYnawall} for the
composite wall has the form
\beqn
S_{2+1}
&=&
\int d^3 x \, \left\{\frac1{2e^2_{2+1}} (\pt_n a_{2+1} )^2+
\frac1{2g^2_{2+1}} (D_n a^a_{2+1} )^2 \right.
\nonumber\\[3mm]
&+&
\left. \frac{1}{4\, e^2_{2+1}}\, \left[ F_{nm}^{(2+1)}\right]^2
+\frac{1}{4\, g^2_{2+1}}\, \left[ F_{nm}^{(2+1)a}\right]^2
+\mbox{fermions} \right\},
\label{21theoryna}
\eeqn
where (2+1)-dimensional couplings  are given by
 \beqn
e^2_{2+1}
&=& 2\pi^2\,  \frac{\xi}{\Delta m}\, ,
\nonumber\\[3mm]
g^2_{2+1}
&=&
2\pi^2\,\,  \frac{g_1^2}{g_2^2}\,\frac{\xi}{\Delta m}\, ,
\label{21couplings}
\eeqn
in terms of the parameters of the
bulk theory.
The domain wall is a 1/2-BPS object so it preserves four supercharges
on its worldvolume. Thus, we must have the extended \ntwo supersymmetry
with four supercharges in the (2+1)-dimensional worldvolume theory.
This is in accord with Eq.~(\ref{21theoryna}) in which
the U(1)  and  SU(2)  gauge fields are combined with the scalar fields
$a_{2+1}$ and $a^a_{2+1}$ to form the bosonic parts of \ntwo vector
multiplets.

A few comments are in order here.
The first comment refers to   four non-compact   moduli
$a,\, a^a$ which emerged in Eq.~(\ref{21theoryna}).
We can exploit gauge transformations in the  worldvolume theory
to put two of them to zero, say $a^{1,2}_{2+1}=0$. The other two
$a^3_{2+1}$ and $a_{2+1}$ should  be identified
with (linear combinations of) two centers of the elementary walls comprising
our composite wall.
More exactly, as $a_{2+1}$ has no
interactions whatsoever it is to be  identified with the center of mass
of the composite wall,
\beq
\label{centerid}
a_{2+1}=
\pi\xi\;(z_1+z_2),
\eeq
where $z_1$ and $z_2$ are the positions of the elementary walls forming the
composite wall,
while $a_{2+1}^3$ can be identified with the relative separation between
the elementary walls,
\beq
\label{separationid}
a_{2+1}^3=\pi\xi\;\frac{g_1}{g_2}\;(z_1-z_2).
\eeq

The second comment is devoted to a technical
element of the derivation of Eq.~(\ref{21theoryna})
in \cite{SYnawall}. In fact, this
worldvolume action was obtained by
a procedure
similar to the one described in Sect~\ref{wkinterms}
for the Abelian case, only
at the quadratic level (i.e. omitting non-Abelian nonlinearities).
To recover cubic and quartic (truly non-Abelian) terms in Eq.~(\ref{21theoryna})
and thus rigorously prove the non-Abelian nature of the worldvolume
theory one needs to go beyond the quadratic approximation.

Nevertheless, there are rather convincing general arguments showing that the
suggestion made in \cite{SYnawall} is correct. First, the number of fields
matches. We have four compact phases and two non-compact centers. Upon
dualization, they fit into a vector multiplet of 3D \ntwo
theory with the SU(2)$\times$U(1) gauge group. Say, if the gauge group was
U(1)$^4$ (as in the case of non-degenerate masses, see \cite{SakaiT})
we would have four phases and four non-compact coordinates.\,\footnote{
In the case of non-degenerate masses the composite $(1,2)\to
(3,4)$ wall is a bound state of four elementary walls, see
\cite{SakaiT}. However, two of them disappear in the limit
(\ref{mchoice}).} Thus, the  non-Abelian gauge symmetry of the
 worldvolume theory, in effect,  is supported by supersymmetry.
Second, there are only two distinct coupling constants in
(\ref{21theoryna}) rather then four. This also indicates that three
phases, upon dualization, should be united in the SU(2) gauge theory.

A 1/4-BPS solution
for a junction of the non-Abelian string (discussed in
Sect.~\ref{strings}) with the composite wall described
above is found in \cite{SYnawall}.
Thus, the non-Abelian string can end on the
composite $(1,2)\to (3,4)$ wall in the theory with degenerate masses,
see (\ref{mchoice}).
Since the non-Abelian string carries a non-Abelian flux with
arbitrary direction of the magnetic flux inside the SU(2)
gauge group this becomes still
another argument in favor of localization of the non-Abelian
gauge fields on the  worldvolume of the composite wall.

\vspace{2cm}

\centerline{\includegraphics[width=1.3in]{extra3.eps}}

\newpage

\section{Wall-string junctions}
\label{wallstrj}

In Sect.~\ref{walls} we reviewed the construction of $D$-brane prototypes
in field theory. In string theory $D$ branes are extended objects on which
fundamental strings can end. To make contact with this string/brane picture
one may address a question whether or not
solitonic strings can end on a domain wall
which localizes gauge fields. The answer to this question is yes. Moreover,
the string endpoint plays a role of a charge with respect to the gauge field
localized on the wall surface. This issue was studied in \cite{GPTT}
in the sigma-model setup and in \cite{DVil} for gauge theories at strong
coupling. A solution for a 1/4-BPS wall-string junction in the
\ntwo supersymmetric U(1) gauge theory at week coupling was found in
\cite{SYwall}, while \cite{SYnawall} deals with its non-Abelian generalization.
Further studies of the wall-string junctions was carried out in
\cite{J14} where all 1/4-BPS solution to Eqs. (\ref{foej}) were obtained,
and in
\cite{SakaiT,ASYboo} where the energy associated with the wall-string
junction (boojum) was calculated, and in \cite{Tbrane,SYdual} where
a quantum version of the effective theory on the domain wall worldvolume
which takes into account charged matter (strings of the bulk theory) was derived.
Below we will review the wall-string junction solutions and then briefly discuss
how the presence of strings in the bulk modifies the effective theory on the wall.

\subsection{Strings ending on the wall}
\setcounter{equation}{0}
\label{junctions}

To begin with, let us review the solution for the simplest   1/4-BPS  wall-string
junction in
\ntwo SQED  obtained in \cite{SYwall}. As was discussed in
Sect.~\ref{walls}
in both vacua of the theory the gauge field is Higgsed
while it can spread freely inside the wall. This is the physical reason why
the ANO string  carrying a magnetic flux can end on the wall.

Assume that at large distances from the string endpoint which lies at $r=0$, $z=0$
the wall is almost parallel to the $(x_1,x_2)$ plane while the string
is stretched along the $z$ axis. As usual, we look for a static solution
assuming that all relevant fields can depend only on $x_n$, ($n=1,2,3$).
The Abelian version of the first-order equations (\ref{foej}) in
various 1/4-BPS junctions
for the theory (\ref{n2sqed}) is \cite{SYwall}
\beqn
&& F^{*}_1-iF^{*}_2 - \sqrt{2}(\pt_1-i\pt_2)a=0\, ,
\nonumber\\[3mm]
&& F^{*}_{3}-\frac{g^2}{2} \left(\left| q^{A}\right|^2-\xi\right)
-\sqrt{2}\, \pt_3 a =0\, ,
\nonumber\\[3mm]
&& \nabla_3 q^A =-\frac1{\sqrt{2}}q^A(a+\sqrt{2}m_A)\, ,
\nonumber\\[4mm]
&& (\nabla_1-i\nabla_2)q^A=0\, .
\label{foeaj}
\eeqn
These equations generalize the first-order equations for the wall
(\ref{wfoe})
and the Abelian ANO  string.

Needless to say, the solution of the first-order equations (\ref{foeaj}) for a
string
ending on the wall can be found only numerically especially near the
end point of
the string  where both the string and the wall profiles are heavily deformed.
However, far away from the string endpoint, deformations are
weak and we can find the asymptotic behavior analytically.

\begin{figure}[h]
\epsfxsize=7cm
\centerline{\epsfbox{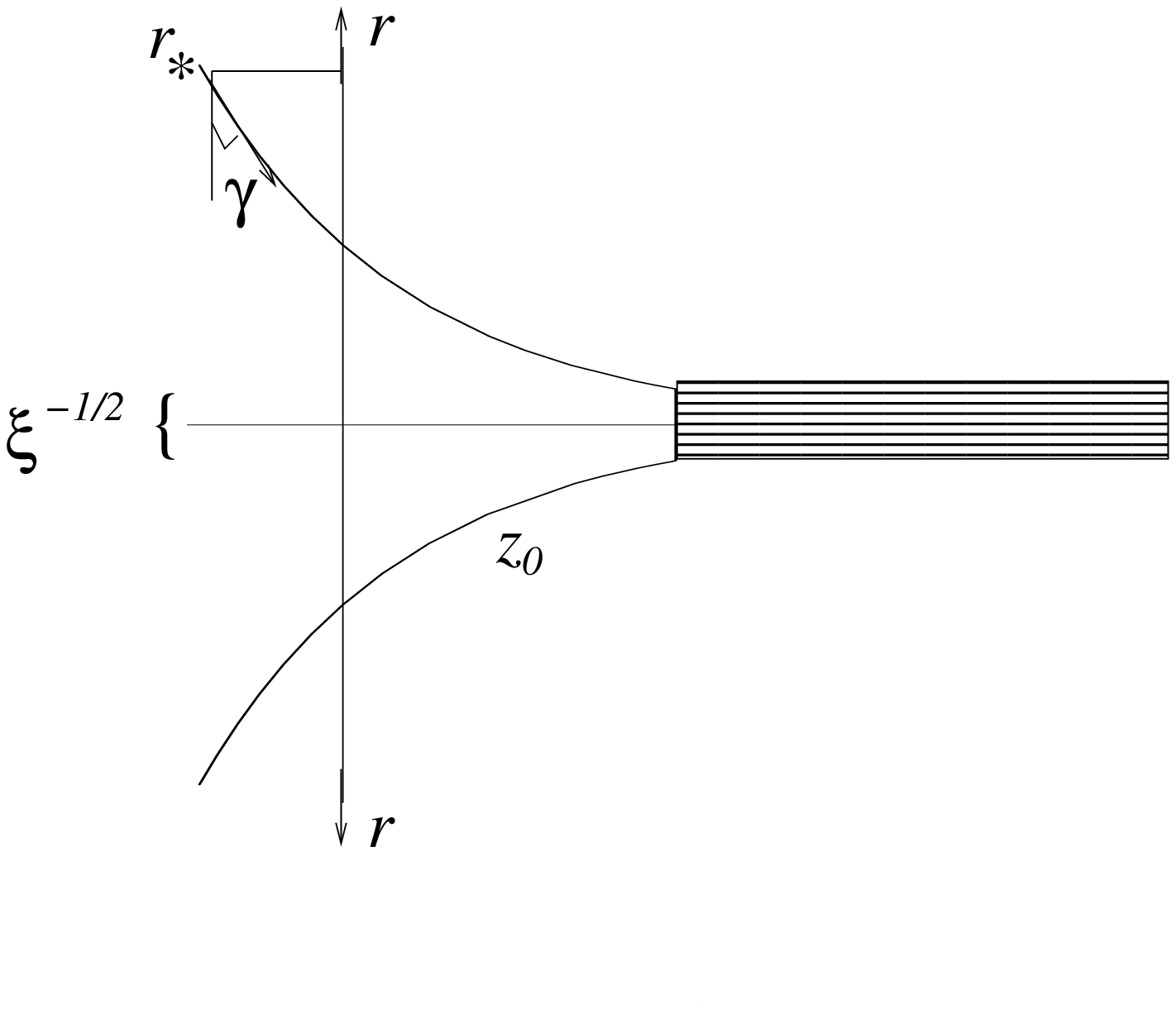}}
 \caption{\small Bending of the wall due to the string-wall junction.
The flux tube extends to the right infinity. The wall
profile is logarithmic at transverse distances larger than
$\xi^{-1/2}$ from the string axis. At smaller distances
the adiabatic approximation fails.}
\label{syfigfive}
\end{figure}

Let the string be on the $z>0$ side of the wall, where the vacuum
is given by Eq.~(\ref{sv}).
 First note that in the region $z\to \infty$ far away
from the string endpoint  at  $z\sim 0$ the solution to
(\ref{foeaj}) is given by an almost unperturbed ANO string.
Now consider the domain $r\to\infty$ at small $z$. In this domain the solution
to (\ref{foeaj}) is given by a  perturbation of the wall solution. Let us
use the {\em ansatz} in which the solutions for the fields $a$ and $q^{A}$ are
given by
the same equations (\ref{a}), (\ref{wq1}) and (\ref{wq2})  in which the
size of the wall is still given by  (\ref{R}), and
{\em the only modification} is
that the position of the wall $z_0$ and the phase $\sigma$  now become
slowly-varying functions of $r$ and $\alpha$, the polar coordinates on
the $(x_1,x_2)$ plane. It is quite obvious that $z_0$ will depend only on $r$,
as schematically depicted in Fig. \ref{syfigfive}.

Substituting this {\em ansatz} into the first-order equations (\ref{foeaj})
one arrives at the equations which determine the adiabatic dependence
of the moduli $z_0$ and $\sigma$ on $r$ and $\alpha$
\cite{SYwall},
\beqn
\label{derz}
&&\pt_{r}z_0=-\frac1{\Delta m r}\,,
\\[3mm]
\label{dersig}
&&\frac{\pt \sigma}{\pt \alpha}=1,\qquad \frac{\pt \sigma}{\pt r}=0\, .
\eeqn

Needless to say  our adiabatic approximation holds
only provided the $r$ derivative is small, i.e.
sufficiently far from the string,
 $\sqrt{\xi} r\gg1$.
The solution to Eq.~(\ref{derz}) is straightforward,
\beq
\label{zbend}
z_0=-\frac1{\Delta m }\ln {r} + {\rm const}.
\eeq
We see that the wall is logarithmically bent according to the Coulomb law
in 2+1 dimensions
(see Fig. \ref{syfigfive}).
 This bending produces a balance of forces
between the string and the wall in the $z$ direction so that the whole
configuration is
static.
The solution to Eq.~(\ref{dersig}) is
\beq
\label{sigmaalpha}
\sigma =\alpha\, .
\eeq
This vortex solution is certainly expected and welcome.
One can identify the
compact scalar field $\sigma$ with the electric  field living
on the domain wall  worldvolume via (\ref{21gaugenorm}). Equation
(\ref{sigmaalpha}) implies
  \beq
\label{elf}
F_{0i}^{(2+1)}=\frac{e^2}{2\pi}
 \,\,\frac{x_i}{r^2}
\eeq
for this electric field,
where the (2+1)-dimensional coupling is given by (\ref{21coupling}).

This is the field of a point-like electric charge
in 2+1 dimensions placed at $x_i=0$.
The interpretation of this result is that the string endpoint
 on the wall plays a role of the electric charge in
the  dual U(1) theory on the wall.
From the standpoint of the bulk theory, when the string ends on the wall,
the magnetic flux it brings with it spreads out inside the wall
via the Coulomb law in (2+1) dimensions.

From the above discussion it is clear that
in the worldvolume theory (\ref{pure}), the fields (\ref{zbend})
and (\ref{elf}) can be
considered as produced by classical point-like charges which interact in a
standard way with the electromagnetic field $A_n$ and the scalar field
$a_{2+1}$,
\beqn
S_{2+1}&=&\int d^3 x \left\{ \frac{1}{2e^2}
(\partial_n a_{2+1})^2+\frac{1}{4 e^2} ( F_{mn}^{(2+1)} )^2
\right.
\nonumber\\[4mm]
&+& A_n\,j_n - a_{2+1}\,\rho\Big\},
\label{cl}
\eeqn
where the classical electromagnetic current and the charge density of
static  charges are
\beq
j_n (x)=n_e\{ \delta^3 (x),0,0\}\,,\qquad\rho(x)=n_s\delta^3 (x)\,.
\label{clcur}
\eeq
Here $n_e$ and $n_s$ are electric and scalar charges
 associated with the string endpoint
with respect to the electromagnetic field $A_n$ and the scalar field $a$,
respectively \cite{SYdual},
\beqn
&&
n_e=+1,\;\;\; \mbox{incoming flux},
\nonumber\\[2mm]
&&
n_e=-1,\;\;\; \mbox{outgoing flux},
\label{elch}
\eeqn
while their   scalar charges are
\beqn
&&
n_s=+1,\;\;\; \mbox{string from the right},
\nonumber\\[2mm]
&&
n_s=-1,\;\;\; \mbox{string from the  left}.
\label{scch}
\eeqn
These rules are quite obvious from the perspective of the bulk
theory. The anti-string carries the opposite flux to that
of a string in (\ref{elf}) and  the bending of the wall
produced by the string coming from the left is opposite
to the one in (\ref{zbend}) associated with the string coming from the right.

\subsection{Boojum energy}
\setcounter{equation}{0}
\label{secboojum}

Let us now  calculate the energy of the wall-string junction,
the boojum.
There are two distinct contributions to this energy \cite{ASYboo}.
The first contribution is due to the gauge field (\ref{elf}),
\beqn
E^G_{(2+1)} &=& \int
\frac{1}{2 e^2_{2+1}} (F_{0i})^2\,  2 \pi r\, dr
\nonumber\\[4mm]
&=&\frac{\pi \xi}{\Delta m
}\int  \frac{ dr}{r}=\frac{\pi \xi}{\Delta m
} \ln {\left(g\sqrt{\xi}L\right)}.
\label{gfcc}
\eeqn
The integral $\int dr/r$ is logarithmically divergent both in the
ultraviolet and
infrared. It is clear that the UV divergence is cut off at the
transverse size of the string $\sim 1/g\sqrt{\xi}$
 and presents no problem. However, the infrared divergence is
much more serious. We introduced a large size $L$ to regularize it in
(\ref{gfcc}).

The second contribution, due to the $z_0$  field (\ref{zbend}), is
proportional to $\int dr/r$ too,
\beqn
E^H_{(2+1)} &=& \int \, \frac{T_{\rm w}}{2} \left(\partial_r
\, z_0 \right)^2
\, 2 \pi r\,  dr
\nonumber\\[4mm]
&=&
\frac{\pi \xi}{\Delta m
}\ln{\left(g\sqrt{\xi}L\right)}\,.
\label{zfcc}
\eeqn
Both contributions are logarithmically divergent in the infrared.
Their  occurrence is an obvious
feature of charged objects coupled to massless fields
in $(2+1)$ dimensions due to the
fact that the fields $A_n$ and $a_{2+1}$
do not die off at infinity, which means infinite energy.

The above two contributions are equal (with the logarithmic
accuracy), even though their physical
interpretation is different. The total  energy of the string junction is
\beq
E^{G+H}=\frac{2\pi \xi}{\Delta m
}\ln{\left(g\sqrt{\xi}L\right)}\,.
\label{eto}
\eeq

We see that in our attempt to include  strings  as point-like
charges in the worldvolume theory (\ref{cl}) we encounter problems
 already at the classical level. The energy of a single charge is
IR divergent.
It is clear that the infrared problems will become even more severe in
quantum theory.

A way out was suggested in \cite{ASYboo,SYdual}. For the infrared
divergences to cancel
we should consider strings and anti-strings with incoming and outgoing
fluxes as well as strings
coming from the right and from the left.
Clearly, only configurations with total both electric and scalar
charges zero have finite energy (see (\ref{elch}) and (\ref{scch})).

\begin{figure}[h]
 \centerline{\includegraphics[width=2in]{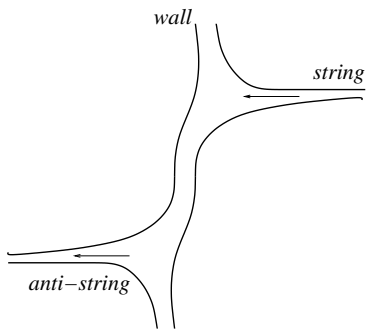}}
 \caption{\small  String and anti-string ending on the wall
from different sides. Arrows denote
the direction of the magnetic flux.
}
\label{fig:strtstr}
\end{figure}

In fact, it was shown in \cite{ASYboo}
that the configuration depicted in Fig.~\ref{fig:strtstr} is a
non-interacting 1/4-BPS configuration. All logarithmic contributions are
canceled; the
junction energy in this geometry is given by a finite negative contribution
\beq
E=-\frac{8\pi}{g^2}\,\Delta m\,,
\label{boojum}
\eeq
which is called the {\sl boojum energy}. In fact this energy
was first calculated in  \cite{SakaiT}. A procedure allowing one to
separate this finite energy from logarithmic contributions described above (and make
it well-defined) was discussed in \cite{ASYboo}.

\subsection{Finite-size rigid strings stretched between the walls.
Quantizing string endpoints}
\setcounter{equation}{0}
\label{quantum}

Now, after familiarizing ourselves
with the junctions of the BPS walls with the semi-infinite strings,
the boojums, we can ask whether or not the junction can acquire
a dynamical role. Is there a formulation of the
 problem in which one can speak of a junction
as of a particle sliding on the wall?

The  string energy is  its tension (\ref{ten}) times its
length. If we have a single wall, all strings attached to it have half-infinite
length; therefore, they are infinitely heavy.  In the wall worldvolume
theory  (\ref{cl}) they may be seen as classical infinitely heavy
point-like charges. The junctions are certainly non-dynamical objects in  this case.

In order to be able to treat junctions as ``particles"
we need to make strings ``light" and deprive them of their
internal dynamics, i.e. switch off all string excitations.
It turns out a domain in the parameter space is likely to exist
where these goals can be
achieved.

In this section we will review a quantum version of the worldvolume theory
(\ref{cl}) with additional charged matter fields. The latter will
represent the junctions on the wall worldvolume \cite{Tbrane,SYdual} (of course,
the junctions have  strings of the bulk theory attached to them; these strings will be rigid).

Making string masses finite is prerequisite. To this end
one  needs at least two domain walls at a finite distance from each other
with strings stretched
between them. This scenario was suggested in \cite{Tbrane}. A quantum
version of the wall worldvolume theory in which the strings were
represented by a charged chiral matter superfield in 1+2 dimensions
was worked out. However,  in the above scenario the strings
were attached to each wall from one side.
From the discussion in Sect.~\ref{secboojum} it must be clear  that this theory
is not free from infrared problems.
The masses of (1+2) dimensional charged fields are infinite.

\begin{figure}[h]
\centerline{\includegraphics[width=2in]{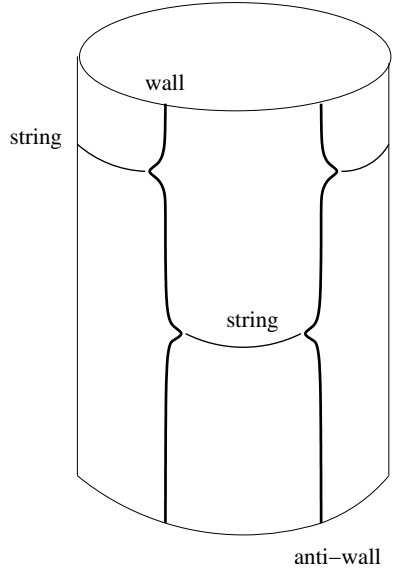}}
 \caption{\small A wall and antiwall connected by strings on
the cylinder. The circumference of the circle (the transverse slice of the
cylinder)
is $L$.
}
\label{fig:cylinder}
\end{figure}

To avoid these infinities we need a configuration with
strings coming both from the right and from the left sides of each wall.
This configuration was suggested in \cite{SYdual}, see Fig.~\ref{fig:cylinder}.

Let us describe this set-up in more detail.
First, we compactify the $x_3=z$ direction in our bulk
theory (\ref{n2sqed}) on a circle of length  $L$. Then we
consider a pair ``wall plus antiwall"  oriented
in the $\{x_1,\,x_2\}$ plane,  separated by a distance $l$ in the
perpendicular direction, as shown in
see Fig.~\ref{fig:cylinder}. The wall and antiwall
experience attractive forces. Strictly speaking,
this is not a BPS configuration ---
supersymmetry in the worldvolume theory is broken. However,  the wall-antiwall
interaction due to overlap of their profile functions is
exponentially suppressed at large separations,
\beq
L\sim l\gg R\,,
\label{largesep}
\eeq
where $R$ is the wall size (see Eq.~(\ref{R})). In what follows we will neglect
exponentially suppressed effects. If so, we neglect the effects which
break supersymmetry in our (2+1)-dimensional worldvolume
theory. Thus, it continues to have four conserved supercharges
(\ntwo supersymmetry
in (2+1) dimensions) as was the case for the isolated single wall.

Let us denote the wall position as  $z_1$
while that of the  antiwall as $z_2$. Then
$$l=z_2-z_1\,.$$
On the wall worldvolume $z_{1,2}$ become scalar fields.
The kinetic terms for these fields in the worldvolume theory are obvious
(see (\ref{21theory})),
\beq
 \frac{T_{\rm w}}{2}
\left[
(\partial_n z_1)^2+ (\partial_n z_1)^2\right]
=\frac{1}{2e^2}
\left[(\partial_n a^{(1)}_{2+1})^2 +(\partial_n a^{(2)}_{2+1})^2\right]
,
\label{kinetic}
\eeq
where we use (\ref{az}) to define the fields $a_{2+1}^{(1,2)}$. The sum
of these
fields,
$$
a_{+}\equiv \frac{1}{\sqrt{2}}\left(a^{(2)}_{2+1} + a^{(1)}_{2+1}\right),
$$
with the corresponding superpartners, decouples from other fields
forming a free field theory describing dynamics of the center-of-mass of
our construction. This is a trivial part which will not concern us here.

An interesting part is associated with the field
\beq
a_{-}\equiv \frac{1}{\sqrt{2}}\left(a^{(2)}_{2+1} - a^{(1)}_{2+1}\right)\,.
\label{aminus}
\eeq
The factor $1/\sqrt 2$ ensures that $a_{-}$
has a canonically normalized kinetic term. By definition, it is related to the
relative wall-antiwall
separation, namely,
\beq
a_{-}=\frac{2\pi\xi}{\sqrt{2}}\,l  \, .
\label{al}
\eeq
Needless to say, $a_{-}$ has all necessary ${\mathcal N}=2$ superpartners.
In the bosonic sector we introduce the gauge field
\beq
A_n^{-}\equiv \frac{1}{\sqrt{2}}\left( A_n^{(1)}-A_n^{(2)}\right),
\label{Aminus}
\eeq
with the canonically normalized kinetic term. The strings stretched between
the wall and antiwall, on both sides, will be represented by two chiral
superfields,
$S$ and $\tilde S$, respectively. We will denote the corresponding
bosonic components by $s$ and $\tilde s$.

In terms of these fields the quantum version of the theory (\ref{cl})
is completely determined by the charge assignments (\ref{elch}) and (\ref{scch})
and $\mbox{\ntwo}$ supersymmetry. The charged matter fields
have the opposite electric charges and distinct mass terms, see below.
A mass term for one of them is introduced by virtue of a ``real mass,"
a three-dimensional generalization of the twisted mass in two dimensions
\cite{twisted}.
It is necessary due to the fact that there are
two inter-wall distances, $l$ and $L-l$.
The real mass breaks parity.
The bosonic part of the action has the form
\beqn
S_{\rm bos} &=&\int d^3 x\,
\left\{\frac{1}{4e^2}\, F_{mn}^-\,F^{-}_{mn} +\frac{1}{2e^2}\,
\left(\partial_n\,a_-\right)^2+ \left|  {\mathcal D}_n s\right|^2 +\left|
\tilde{\mathcal D}_n \tilde s\right|^2\right.
\nonumber\\[3mm]
&+&
2  a^2_- \, \bar s \,s +2(m-a_-)^2\, \bar{\tilde s} \,\tilde s
+e^2\left(|s|^2-|\tilde{s}|^2\right)^2\Big\}\,.
\label{qu}
\eeqn
According to our discussion in Sect.~\ref{junctions},
the fields $s$ and $\tilde s$  have charges +1 and $-1$ with respect to the
gauge fields $A_n^{(1)}$
and $A_n^{(2)}$, respectively. Hence,
\beqn
{\mathcal D}_n &=& \pt_n -i\left( A_n ^{(1)}-A_n^{(2)}\right)=
\pt_n-i\sqrt{2}A_n^{-}\,,
\nonumber\\[3mm]
\tilde{\mathcal D}_n
&=&
\pt_n +i\left( A_n^{(1)}-A_n^{(2)}\right)=
\pt_n+i\sqrt{2}A_n^{-}\,.
\label{nabla}
\eeqn
The electric charges of boojums with respect to the field $A_n^{-}$ are
$\pm \sqrt{2}$. The last term in (\ref{qu}) is the $D$ term dictated by
supersymmetry.

So far,
$m$ is a free parameter whose relation to $L$ will be determined shortly.
Moreover, $F_{mn}^-= \pt_m \, A_n^ - - \pt_n \, A_m^ -$.
The theory (\ref{qu}) with the pair of chiral multiplets
$S$ and $\tilde{S}$ is free of IR divergences and global $Z_2$ anomalies
\cite{AHISS,BHO}. At the classical level it is clear from our discussion
in Sect.~\ref{secboojum}. A version of the
worldvolume theory (\ref{qu}) with a single supermultiplet
$S$ was considered in Ref. \cite{Tbrane} but, as was mentioned,
this version is not free of IR divergences.

It is in order to perform a crucial test of our theory (\ref{qu}) by
calculating the masses
of the charged matter multiplets $S$ and $\tilde{S}$. From (\ref{qu})
we see that the mass of $S$ is
\beq
m_s=\sqrt{2}\,\,\bra a_{-}\ket\,.
\label{Ms}
\eeq
Substituting here the relation (\ref{al}) we get
\beq
m_s= 2\pi\xi\,l \,.
\label{Mstr}
\eeq
The mass of the charged matter field $S$ reduces to the mass of the string
of the bulk theory
stretched between the wall and antiwall at separation $l$, see (\ref{ten}).
Great success! Of course, this was expected. Note that this is a nontrivial check
of consistency between the worldvolume theory and the bulk theory. Indeed,
the charges of the strings endpoints (\ref{elch}) and (\ref{scch}) are unambiguously
fixed by the classical solution for the wall-string junction.

Now, imposing the relation between the free mass parameter $m$ in
(\ref{qu}) and
the length of the compactified $z$-direction $L$ in the form
\beq
m=\frac{2\pi\xi}{\sqrt{2}}\,L
\label{mL}
\eeq
we get the mass of the chiral field $\tilde{S}$ to be
\beq
m_{\tilde{s}}= 2\pi\xi\,(L-l)\, .
\label{Mtstr}
\eeq
The mass of the string $\tilde{S}$ connecting the wall with the  antiwall from
the other side of the cylinder is the string tension times $(L-l)$, in full
accordance with our expectations, see Fig.~\ref{fig:cylinder}.

\begin{figure}[h]
 \centerline{\includegraphics[width=4in]{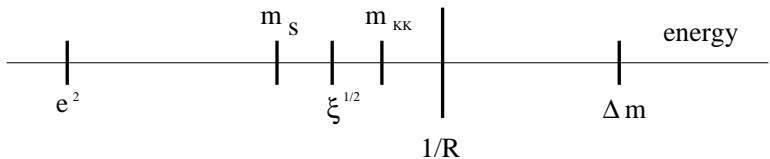}}
 \caption{\small  Mass scales of the bulk and worldvolume theories.
}
\label{fig:masses}
\end{figure}

The theory (\ref{qu}) can be considered as an effective low-energy
(2+1)-dimensional description of the wall-antiwall system dual to
the (3+1)-dimen\-sional bulk theory (\ref{n2sqed}) under the choice of
parameters specified below (Fig.~\ref{fig:masses}).
Most importantly, we use the quasiclassical
approximation in our bulk theory (\ref{n2sqed}) to find the solution
for the string-wall junction \cite{SYwall} and derive the wall-antiwall
worldvolume effective theory
(\ref{qu}). This assumes weak coupling in the bulk, $g^2\ll 1$.
According to the
duality relation (\ref{duality}) this implies strong coupling in the
worldvolume theory.

In order to be able to work with the worldvolume theory
we want to continue  the theory (\ref{qu}) to the
weak coupling regime,
\beq
e^2\ll \frac1{R}\, ,
\label{bwc}
\eeq
which means strong coupling in the bulk theory, $g^2\gg 1$. The  general
idea is that at $g^2\ll 1$ we can use the bulk theory (\ref{n2sqed}) to
describe our wall-antiwall system while at $g^2\gg 1$ we better
use the worldvolume theory (\ref{qu}). In \cite{SYdual}
this set-up was termed
{\sl  bulk--brane duality}.
In spirit --- albeit not in detail --- it is similar to the AdS/CFT correspondence.

In order for the theory (\ref{qu}) to give a correct low-energy description of
the wall-antiwall system the masses of strings (including boojums)
in this theory should be
much less than the masses of both the wall and string excitations.
These masses are of
order of $ 1/R$ and $m_{KK}=k/l\sim k/(L-l)$, respectively,
where $k$ is an integer. The high mass gap for the string excitations make
strings rigid.

These constraints were studied in \cite{SYdual}. It was found that
for the constraints to be satisfied
different scales of the theory must have a hierarchy
shown in Fig.~\ref{fig:masses}.

The scales $\Delta m$, $\sqrt{\xi}$ and
$e^2_{2+1}\sim \xi/\Delta m$
are determined by the string and wall tensions in our bulk theory, see
(\ref{ten}) and (\ref{wten}).  In particular, the (2+1)-dimensional
coupling $e^2$
is determined by the ratio of the wall tension to the square of the string
tension, as follows from Eqs.~(\ref{21theory}) and (\ref{az}).
Since the strings and walls in the bulk theory
are BPS-saturated, they receive
no quantum corrections. Equations (\ref{ten}) and  (\ref{wten}) can be
continued to
the strong coupling regime in the bulk theory. Therefore, we always can
take such values
of
parameters
$\Delta m$ and  $\sqrt{\xi}$  that the conditions
\beq
e^2 \ll \sqrt{\xi}\ll \Delta m
\label{conditions}
\eeq
are satisfied.

To actually prove duality
between the bulk theory (\ref{n2sqed})and the worldvolume theory (\ref{qu})
we only need to prove the condition
\beq
\sqrt{\xi}\ll \frac1R \,,
\label{xiR}
\eeq
which ensures that strings are lighter than the wall excitations.
This will give us the hierarchy of the
mass scales shown in Fig.~\ref{fig:masses}. With the given values of the
parameters
$\Delta m$ and  $\sqrt{\xi}$ we have another free parameter of the bulk theory
to ensure (\ref{xiR}), namely, the coupling constant $g^2$.
However,  the
scale $1/R$ (the mass scale of various massive excitations living on the wall)
is not protected by supersymmetry and we cannot prove
that the regime (\ref{xiR}) can be reached
at strong coupling in the bulk theory. Thus, the above   bulk--brane duality
is in fact a
conjecture  essentially equivalent to the statement that the regime
 (\ref{xiR}) is  attainable under a certain choice of parameters.
Note, that if the condition (\ref{xiR}) is not met, the wall excitations
become lighter than the strings under consideration, and the theory (\ref{qu})
does not correctly
describe low-energy physics of the theory on the walls.

\subsection{Quantum boojums. Physics of the  worldvolume theory}
\setcounter{equation}{0}
\label{wvphys}

 What is a boojum loop?

Let us integrate out the string multiplets $S$ and $\tilde{S}$ and study
the effective theory for the U(1) gauge supermultiplet at scales below
 $m_s$. As long as the string fields enter the action quadratically
(if we do not resolve the algebraic equations for the auxiliary fields) the
one-loop approximation is exact.

Integration over the charged matter fields in (\ref{qu})
leads to
generation of  the Chern--Simons term \cite{Redlich,AlvgaumeWit,AHISS}
with the coefficient
 proportional to
\beq
\frac1{4\pi}\Big[{\rm sign}(a) +{\rm sign}(m-a)\Big]
\varepsilon_{nmk}A^{-}_n \pt_m A_{k}^{-}\,.
\label{cs}
\eeq
Another effect related to the one in  (\ref{cs}) by supersymmetry is
generation of a non-vanishing $D$-term,
\beq
\frac{D}{2\pi}
\Big[|m-a_{-}|-|a_{-}|\Big]
=\frac{D}{2\pi}(m-2a_{-})\,,
\label{Dterm}
\eeq
where $D$ is the $D$-component of the gauge supermultiplet.
As a result we get from (\ref{qu}) the following low-energy
effective action for the gauge multiplet:
\beqn
S_{2+1}
&=&
\int d^3 x \left\{ \frac{1}{2e^2 (a_{-})}
(\partial_n a_{-})^2+\frac{1}{4 e^2 (a_{-})} ( F_{mn}^{-} )^2
\right.
\nonumber\\[3mm]
&+&\left.
\frac1{2\pi}\varepsilon_{nmk}A^{-}_n \pt_m A_{k}^{-} +
\frac{e^2 (a_{-})}{8\pi^2}
\left(2a_{-}-m\right)^2\right\},
\label{eff}
\eeqn
where we also take into account  a finite renormalization of the bare
coupling constant $e^2$ \cite{IntrS,BHOO,BHO},
\beq
\frac1{e^2 (a_{-})}=\frac1{e^2} +\frac1{8\pi |a_{-}|} +
\frac1{8\pi |m-a_{-}|}\,.
\label{rencoup}
\eeq
This is a small effect since $1/e^2$ is the largest parameter
(see Fig.~\ref{fig:masses}).
Note that the coefficient in front of the Chern--Simons term is an integer
in Eq.~(\ref{eff}) (in the units of $1/(2\pi )$)  as required by gauge invariance.

The most dramatic effect in (\ref{eff}) is the generation of a potential for
the field $a_{-}$. Remember  $a_{-}$ is proportional to the separation $l$ between the walls.
The vacuum of (\ref{eff}) is located at
\beq
\bra a_-\ket=\frac{m}{2}\,,\qquad l=\frac{L}2\,.
\label{vac3}
\eeq
There are two extra solutions at $a_-=0$ and $a_-=m$,
but they lie outside the limits of applicability of our approach.

We see that the wall and antiwall are pulled apart; they want to be located
at the opposite sides of the
cylinder. Moreover, the potential is quadratically rising with the deviation
from the equilibrium point (\ref{vac3}). As was mentioned in the beginning
of this section, the wall and anti-wall interact with exponentially small potential
due to the overlap of their profiles. However, these interactions
are negligibly small at $l\gg R$ as compared to the interaction in
Eq.~(\ref{eff}). The interaction potential in (\ref{eff}) arises due to
virtual pairs of strings (``boojum loops")
 which pulls the walls apart.

\begin{figure}[h]
\centerline{\includegraphics[width=2.5in]{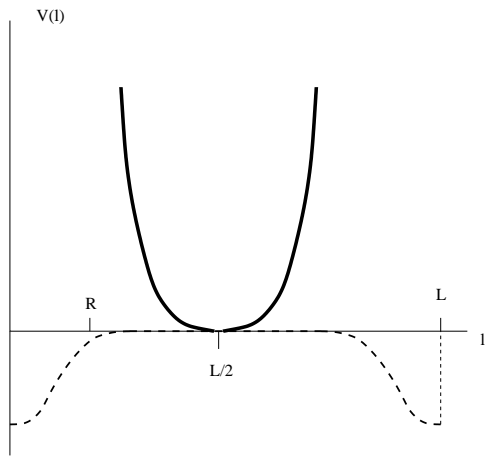}}
 \caption{\small  Classical and quantum wall-antiwall interaction
potential. The dashed line depicts the classical exponentially small potential
while the solid line the quantum potential presented  in Eq.~(\ref{eff}).
}
\label{fig:wint}
\end{figure}

Clearly, our description
of strings in the bulk theory was purely classical and we were unable
to see this quantum effect. The classical and quantum interaction potential
of the wall-antiwall system is schematically shown in Fig.~\ref{fig:wint}.
The quantum potential induced by virtual string loop is much larger
than the classical exponentially small $W\bar W$ attraction
at separations $l\sim L/2$. The quantum effect stabilizes the
classically unstable $W\bar W$ system at the equilibrium position (\ref{vac3}).

Note, that if the wall-antiwall interactions were mediated by particles
they would have exponential  fall-off at large separations
$l$ (there are no massless particles in the bulk).
 Quadratically rising potential would never be generated. In our case the
interactions are due to virtual pairs of extended objects -- strings.
Strings are produced as {\sl rigid} objects stretched between walls.
 The string excitations are not
taken into account  as they are too heavy. The fact that the strings come out
in our treatment as  rigid objects rather than  local particle-like
states propagating
between the walls is of a paramount importance. This is the reason why the
wall-antiwall potential does not fall off at large separations.
Note, that a similar effect,
power-law interactions between the domain
walls in \none QCD were recently obtained via a two-loop calculation
in the effective worldvolume theory \cite{ArmHol}.

The presence of the potential for the scalar field $a_{-}$ in
Eq.~(\ref{eff}) makes this
field massive, with mass
\beq
m_a=\frac{e^2}{\pi}\,.
\label{amass3}
\eeq
By supersymmetry, the photon is no longer massless too, it
acquires the same mass. This is associated with the Chern--Simons term in
(\ref{eff}). As it is clear from
Fig.~\ref{fig:masses}, $m_a\ll m_s\,.$ This shows that integrating out
massive string fields in (\ref{qu}) to get (\ref{eff}) makes sense.

Another effect seen in (\ref{eff}) is the renormalization of the
coupling constant which results in a non-flat metric. Of course, this
effect is very small in our range of parameters since $m_s\gg e^2$.
Still we see that the virtual string pairs induce additional power
interactions
between the walls through the nontrivial metric in (\ref{eff}).

\vspace{2cm}

\centerline{\includegraphics[width=1.3in]{extra3.eps}}

\newpage

\section{Conclusions}
\label{concl}

This concludes our travel diary in the land of supersymmetric solitons
in gauge theories. It is time to summarize lessons.

Advances
in supersymmetric solitons, especially in non-Abelian gauge theories,
that took place in the last decade or so,
are impressive. In the bulk of this review
we thoroughly discussed many aspects of the subject
at a technical level. Important and relevant technical details
presented above should not overshadow a big picture,
which is in the making since 1973. Sometimes people tend to
forget about this big
picture which is understandable: its development is painfully slow
and notoriously difficult.

Let us ask ourselves: what is the most remarkable feature of
quantum chromodynamics and QCD-like theories? The fact
that at the Lagrangian level one deals with quarks and gluons
while experimentalists detect pions, protons,
glueballs and other color singlet states --- never quarks and gluons ---
is the single most salient feature of non-Abelian gauge
theories at strong coupling. Color confinement makes
colored degrees of freedom inseparable.
In a bid to understand this phenomenon
Nambu, 't Hooft and Mandelstam suggested in the mid-1970s
(independently but practically simultaneously)
a ``non-Abelian dual Meissner effect." At that time their suggestion was more of a dream than a physical scenario.  According to their vision,
``non-Abelian monopoles" condense in the vacuum
resulting in formation of ``non-Abelian chromoelectric flux tubes"
between color charges, e.g. between a probe heavy quark and antiquark.
Attempts to separate these probe quarks would lead
to stretching of the flux tubes, so that the energy of the
system grows linearly with separation. That's how linear confinement was
visualized. However, at that time the
the notions of non-Abelian
flux tubes and  non-Abelian monopoles (let alone condensed monopoles
in  non-Abelian gauge
theories)   were nonexistent. Nambu, 't Hooft and Mandelstam
operated with nonexistent objects.

One may ask where did these theorists get their inspiration from?
There was one physical phenomenon known since long ago
and well understood theoretically which yielded a rather analogous picture.

In 1933 Meissner discovered that magnetic fields could not penetrate
inside superconducting media. The expulsion of the
magnetic fields by superconductors goes under the name of the
Meissner effect. Twenty years later Abrikosov posed the question:
``what happens if one immerses a magnetic charge and an anticharge
in type-II superconductors (which in fact he discovered)?" One can visualize
a magnetic charge as an endpoint of a very long and very thin solenoid.
Let us refer to the $N$ endpoint of such a solenoid as
a positive magnetic charge and the $S$ endpoint as a negative magnetic charge.

In the empty space the magnetic field will spread in the bulk,
while the energy of the magnetic charge-anticharge configuration will
obey the Coulomb $1/r$ law. The force between them will die off as $1/r^2$.

What changes if the magnetic charges are placed inside a large
 type-II superconductor?

 Inside the superconductor the Cooper pairs condense,
 all electric charges are screened, while the photon acquires a mass.
 According to modern terminology, the electromagnetic U(1)
 gauge symmetry is Higgsed. The magnetic field
 cannot be screened in this way; in fact, the magnetic flux is conserved.
 At the same time the superconducting medium does not tolerate
 the magnetic field.

 The clash of contradictory requirements
 is solved through a compromise. A thin tube is formed
 between the magnetic charge and anticharge immersed in the
superconducting medium. Inside this tube superconductivity is ruined ---
which allows the magnetic field to spread from the charge to
the anticharge through this tube. The tube transverse size
is proportional to the inverse photon mass while its tension is proportional to
the Cooper pair condensate. These tubes go under the name of Abrikosov vortices.
In fact, for arbitrary magnetic fields
he predicted lattices of such flux tubes.
A dramatic (and, sometimes, tragic) history
of this discovery is nicely described in Abrikosov's Nobel Lecture.

Returning to the magnetic charges immersed in the type-II superconductor
under consideration,
one can see that increasing the distance between these charges
(as long as they are inside the superconductor) does not lead to their
decoupling ---
the magnetic flux tubes become longer, leading to a linear growth
of the energy of the system.

The Abrikosov vortex lattices were experimentally
observed in the 1960s. This physical phenomenon inspired
Nambu, 't Hooft and Mandelstam's ideas on
non-Abelian confinement. Many people tried to quantify
these ideas. The first breakthrough, instrumental in all current developments,
came 20 years later,
in the form of the Seiberg--Witten solution of ${\mathcal N}=2$ super-Yang--Mills.
This theory has eight supercharges which makes dynamics quite``rigid"
and helps one to find the full analytic solution at low energies.
The theory bears a resemblance to quantum chromodynamics,
sharing common ``family treats." By and large, one can characterize it
as QCD's ``second cousin."

An important feature which distinguishes it from QCD is
the adjoint scalar field whose vacuum expectation value triggers
the spontaneous breaking of the gauge symmetry
SU(2)$\to$U(1). The 't Hooft--Polyakov monopoles ensue.
They are readily seen in the quasiclassical domain.
Extended supersymmetry and holomorphy in certain parameters which is associated
with it  allows one to analytically continue in the domain where
the monopoles become light --- eventually massless ---
and then condense after a certain small deformation breaking
${\mathcal N}=2$ down to ${\mathcal N}=1$ is introduced.
Correspondingly, at a much lower scale the (dual)
U(1) gauge symmetry breaks, so that the theory is fully Higgsed.
Electric flux tubes are formed.

This was the first ever demonstration of the dual Meissner effect
in non-Abelian theory, a celebrated analytic proof of linear
confinement, which caused much excitement and euphoria in the community.

It took people three years to realize that the flux tubes in the
Seiberg--Witten solution are not those we would like to have in QCD.
Hanany, Strassler and Zaffaroni who analyzed in 1997 the chromoelectric
flux tubes in the Seiberg--Witten solution
showed that these flux tubes are essentially Abelian (of the Abrikosov--Nielsen--Olesen
type) so that the hadrons they would create would have nothing to do with
those in QCD. The hadronic spectrum would be significantly reacher.
And, say, in the SU(3) case, three flux tubes
in the Seiberg--Witten solution would not annihilate into nothing, as they
should in QCD ...

Ever since
searches for genuinely non-Abelian
flux tubes and non-Abelian monopoles continued,
with a decisive breakthrough in 2003. By that time the program of
finding field-theory analogs of all basic constructions
of string/D-brane theory was in full swing.
BPS domain walls, analogs of D branes, had been identified in
supersymmetric Yang--Mills theory. It had been demonstrated that
such walls support gauge fields localized on them. BPS saturated
string-wall junctions had been constructed.
And yet, non-Abelian flux tubes, the basic element of the
non-Abelian Meissner effect, remained elusive.

They were first found in U(2) super-Yang--Mills theories with
extended supersymmetry, ${\mathcal N}=2$, and two matter hypermultiplets.
If one introduces a non-vanishing Fayet--Iliopoulos parameter $\xi$
the theory develops isolated quark vacua,
in which the gauge symmetry is fully Higgsed, and all elementary excitations are massive.
In the general case, two matter
mass terms allowed by  ${\mathcal N}=2$
are unequal, $m_1\neq m_2$.
There are free parameters whose interplay
determines dynamics of the theory:
the Fayet--Iliopoulos parameter $\xi$, the mass difference
$\Delta m$ and a dynamical scale parameter
$\Lambda$, an analog of the QCD scale $\Lambda_{\rm QCD}$.
Extended supersymmetry guarantees that some crucial dependences are holomorphic,
and there is no phase transition.

\begin{figure}[h]
\centerline{\includegraphics[width=4in]{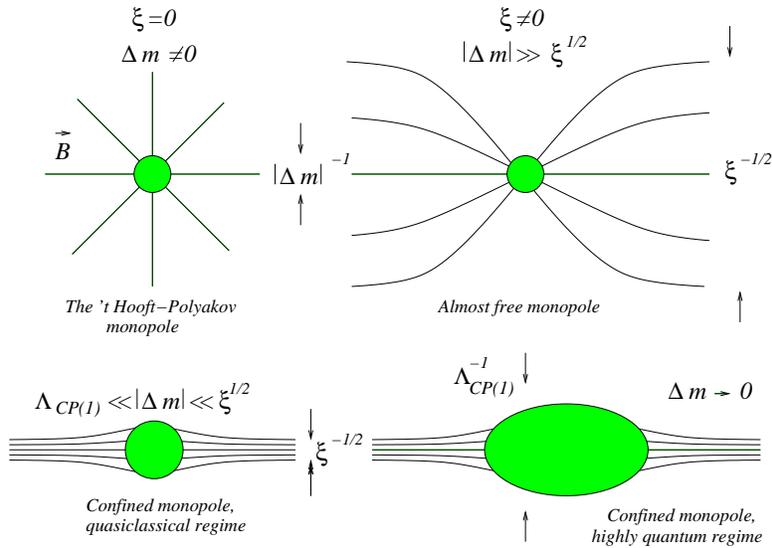}}
 \caption{\small  Various regimes for monopoles and strings
in the simplest case of two flavors.
}
\label{fig:evmon2}
\end{figure}

As various parameters vary, this theory evolves in a very
graphic way , see Fig.~\ref{fig:evmon2} which
is almost the same as Fig.~\ref{sixf} (the first stage of unconfined
't Hooft--Polyakov monopole is added in left upper corner ). At $\xi=0$ but
$\Delta m \neq 0$
(and $\Delta m \gg \Lambda$) it presents a very clear-cut example
of a model with the standard 't Hooft--Polyakov monopole.
The monopole is free to fly --- the flux tubes are not yet formed.

Switching on $\xi\neq 0$ traps the magnetic fields inside
the flux tubes, which are weak as long as $\xi\ll\Delta m$.
The flux tubes change the shape of the monopole far away from its core,
leaving the core essentially intact. Orientation of the
chromomagnetic field inside the flux tube is essentially fixed.
The flux tubes are Abelian.

With $|\Delta m|$ decreasing,
fluctuations in the orientation of the
chromomagnetic field inside the flux tubes grow.
Simultaneously, the monopole seen as the string junction,
looses resemblance with the 't Hooft--Polyakov monopole.
It acquires a life of its own.

Finally, in the limit $\Delta m\to 0$
the transformation is complete. A global SU(2) symmetry restores
in the bulk. Orientational moduli
develop on the string worldsheet making it truly non-Abelian.
The junctions of degenerate strings present what remains of the
monopoles in this highly quantum regime. It is remarkable that,
despite the fact we are deep inside the highly quantum regime,
holomorphy allows one to exactly calculate the mass of
these monopoles.

What remains to be done?
The most recent investigations zero in on ${\mathcal N}=1$
theories, which are much closer relatives of QCD than ${\mathcal N}=2$.
The $M$ model discussed in the bulk of this review
can be regarded as the first cousin of QCD since  the adjoint fields
typical of  ${\mathcal N}=2$
are eliminated in this theory. Even though supersymmetry is considerably weakened,
the overall qualitative picture survives. This is probably
one of the most important findings at the current stage.
Of course, to get a (semi) quantitative description of
color confinement in QCD we still have to dualize this picture.
This does not seem to be an insurmountable obstacle.

And then, ${\mathcal N}=0$ theories --- sister theories of QCD --- loom large ...

\vspace{2cm}

\centerline{\includegraphics[width=1.3in]{extra3.eps}}

\newpage

 \section*{Acknowledgments}
 \addcontentsline{toc}{section}{Acknowledgments}

We are grateful to Yaron Oz, Adam Ritz, David Tong,
and  Arkady Vainshtein for useful discussions.

The work of M.S. was
supported in part by DOE grant DE-FG02-94ER408.
The work of A.Y. was  supported
by  FTPI, University of Minnesota, by INTAS Grant No. 05-1000008-7865,
by RFBR Grant No. 06-02-16364a
and by Russian State Grant for
Scientific School RSGSS-11242003.2.

\newpage

\vspace{1mm}

\section*{Appendix A. Conventions and notation}
\addcontentsline{toc}{section}{Appendix A. Conventions and notation}

\renewcommand{\theequation}{A.\arabic{equation}}
\setcounter{equation}{0}

 \renewcommand{\thesubsection}{A.\arabic{subsection}}
\setcounter{subsection}{0}

 The conventions we use in Parts I and II of this review
 are slightly different. In Part I presenting mainly a conceptual introduction
 to the subject of supersymmetric solitons we choose the so-called
 Minkowski notation. Here our notation is very close (but not identical!)
to that
of Bagger and Wess \cite{BW}. The main distinction is the
conventional choice of the metric tensor $g_{\mu\nu}
=\mbox{diag}(+---)$ as opposed to the $\mbox{diag}(-+++)$
version of Bagger and Wess.
Both, the spinorial and vectorial indices will be denoted by the Greek letters.
To differentiate between them we will use the letters from the beginning of the
alphabet for the spinorial indices, e.g. $\alpha$, $\beta$ and so on, reserving
those from the end of the alphabet (e.g. $\mu$, $\nu$, {\em etc.}) for the vectorial
indices.

 Those readers who venture to delve in Part II will have to switch to the
 so-called Euclidean notation which is more convenient for technical studies.
 The distinctions between these two notations are summarized in Sect.~\ref{seca4}.

  \subsection{Two-dimensional gamma matrices}
  \label{seca1}

 In two dimensions we choose the gamma matrices as
follows
\beq
\gamma^{0}=\gamma^t=\sigma_2\,,\qquad \gamma^{1}=\gamma^z = i\sigma_1\,,\qquad \gamma^{5}
\equiv\gamma^0\gamma^1 = \sigma_3\,.
\label{sieeight}
\eeq
In three dimensions
\beq
\gamma^{t}=\sigma_2\,,\qquad \gamma^{x}= -i\sigma_3\,,\qquad \gamma^{z}
 = i \sigma_1\,.
\label{sieeightp}
\eeq

\subsection{Covariant derivatives}
  \label{seca2}

The covariant derivative in the Minkowski space is defined as
\beq
D_\mu =\partial_\mu - i A_\mu^a \, T^a\,.
\eeq
Then for the spatial derivatives we have
\beq
D_1 = \frac{\partial}{\partial x} + i A_x^a \, T^a\,,
\eeq
and similar for $D_{2,3}$.

 \subsection{Superspace and superfields}
\label{sec30}

The four-dimensional space $x^\mu$ can be promoted to superspace
by adding four Grassmann coordinates $\theta_\alpha$ and $\bar
\theta_{\dot\alpha}$, ($\alpha,\,\dot\alpha=1,2$). The coordinate transformations
\begin{equation}
\{x^\mu ,\theta_\alpha\,, \bar\theta_{\dot\alpha}\}:\qquad
\delta\theta_\alpha=\varepsilon_\alpha\,,\quad
\delta\bar\theta_{\dot\alpha}=\bar\varepsilon_{\dot\alpha}\,, \quad
\delta x_{\alpha\dot\alpha}= -2i\,\theta_{\alpha}\bar\varepsilon_{\dot\alpha}
-2i\,\bar\theta_{\dot\alpha}\varepsilon_{\alpha}
\label{susytr}
\end{equation}
add SUSY to the translational and Lorentz transformations.

Here  the Lorentz vectorial indices are transformed into  spinorial according to the
standard rule
\begin{equation}
A_{\beta\dot\beta} = A_\mu (\sigma^\mu )_{\beta\dot\beta}\,,\qquad
  A^\mu = \frac{1}{2}
A_{\alpha\dot\beta}(\bar\sigma^\mu )^{\dot\beta\alpha}\, ,
\end{equation}
where
\begin{equation}
 (\sigma^\mu )_{\alpha\dot\beta} = \{ 1,
\vec\tau\}_{\alpha\dot\beta}
\,,\qquad   (\bar\sigma^\mu
)_{\dot\beta\alpha}=(\sigma^\mu)_{\alpha\dot\beta}
  \, .
\end{equation}
We use the notation $\vec \tau$ for the Pauli matrices throughout the paper.
The lowering and raising of the indices is performed by virtue of the
$\epsilon^{\alpha\beta}$ symbol
($\varepsilon^{\alpha\beta}=i(\tau_2)_{\alpha\beta}$). For
instance,
\begin{equation}
(\bar\sigma^\mu
)^{\dot\beta\alpha}=\epsilon^{\dot\beta\dot\rho}\,\epsilon^{\alpha\gamma}\,
(\bar\sigma^\mu )_{\dot\rho\gamma} =\{1, -\vec\tau\}_{\dot\beta\alpha}
 \, .
\end{equation}
Note that
\beq
\varepsilon^{12} = - \varepsilon_{12} =1\,,
\eeq
and the same for the dotted indices.

Two invariant subspaces $\{x^\mu_L\,,\theta_\alpha\}$ and
$\{x^\mu_R\,,\bar\theta_{\dot\alpha}\}$ are spanned on 1/2 of the
Grassmann coordinates,
\begin{eqnarray}
\{x^\mu_L\,,\theta_\alpha\}:&\qquad&\delta\theta_\alpha =\varepsilon_\alpha\, ,
\quad \delta({x}_L)_{\alpha\dot\alpha}=-
4i\, \theta_{\alpha}\bar\varepsilon_{\dot\alpha}\,;\nonumber\\[0.2cm]
\{x^\mu_R\,,\bar\theta_{\dot\alpha}\}:&\qquad&\delta\bar\theta_{\dot\alpha}
=\bar\varepsilon_{\dot\alpha}\, ,
\quad \delta({x}_R)_{\alpha\dot\alpha}=-
4i\, \bar\theta_{\dot\alpha}\varepsilon_{\alpha}\,,
\label{sutra}
\end{eqnarray}
where
\begin{equation}
({x_{L,R}})_{\alpha\dot{\alpha}} = {x}_{\alpha\dot{\alpha}} \mp
2i\, \theta_{\alpha}\bar{\theta}_{\dot{\alpha}}\, .
\label{chcoor}
\end{equation}
The minimal supermultiplet of fields includes one complex scalar field $\phi (x)$
(two bosonic states) and one complex Weyl spinor $\psi^\alpha
(x)\, , \,\,\alpha = 1,2$ (two fermionic states).
Both fields are united in one {\em chiral superfield},
\begin{equation}
{\Phi ({x}_L,\theta )} = \phi ({x}_L) + \sqrt{2}\theta^\alpha
\psi_\alpha ({
x}_L) +  \theta^2 F({x}_L)\, ,
\label{chsup}
\end{equation}
where $F$
is an auxiliary component. The  field  $F$
appears in the Lagrangian without
the  kinetic term.

In the gauge theories one also uses a {\em vector superfield},
\begin{eqnarray}
V(x,\theta,\bar\theta)
&=&
 C+i\theta\chi -i \bar\theta \bar\chi
+\frac{i}{\sqrt{2}} \theta^2 M -\frac{i}{\sqrt{2}} \bar\theta^2 {\bar M}
\nonumber \\[2mm]
&-&
2\theta_\alpha \bar\theta_{\dot \alpha } v^{\dot\alpha\alpha} +\left\{
2i \theta^2 \bar\theta_{\dot \alpha} \left[\bar\lambda^{\dot \alpha} - \frac{i}{4}
\partial^{\alpha \dot \alpha} \chi\right] +\mbox{H.c.}\right\}
\nonumber \\[2mm]
&+&
 \theta^2
\bar\theta^2 \left[D - \frac 1 4 \partial^2 C\right]\,.
\label{vecsf}
\end{eqnarray}
The superfield $V$ is real, $V=V^\dagger$, implying that the bosonic fields
$C$, $D$ and $v^\mu=\sigma^\mu_{\alpha\dot\alpha}v^{\dot\alpha\alpha}/2$
are real. Other fields are complex, and the bar denotes the complex conjugation.
The field strength superfield has the form
\beq
W_\alpha = i \left(\lambda_\alpha + i\theta_\alpha \, D -\theta^\beta\, F_{\alpha\beta} - i\theta^2 D_{\alpha\dot \alpha}\bar\lambda^{\dot\alpha}
\right)\,.
\eeq
The gauge field strength tensor is denoted by
$F_{\mu\nu}^a$. Sometimes we use the abbreviation
$F^2$  for
\beq
F^2\equiv F_{\mu\nu}^a\, F^{\mu\nu\,\, a}\,,
\eeq
while
\beq
FF^{*}\equiv F_{\mu\nu}^a\, F^{*\mu\nu\,\, a}
\equiv \frac{1}{2} \epsilon^{\mu\nu\rho\sigma} F_{\mu\nu}^a F_{\rho\sigma}^a \,.
\eeq

The transformations~(\ref{sutra}) generate the SUSY transformations
of the fields which can be written as
\begin{equation}
\delta V= i \left(Q\varepsilon +  \bar Q\bar\varepsilon\right)\,V
\end{equation}
where $V$ is a generic superfield (which could be chiral as well). The
differential operators $Q$ and $\bar Q$ can be written as
\begin{equation}
Q_\alpha=-i \frac{\partial}{\partial \theta^\alpha} +\partial_{\alpha\dot\alpha}
\bar\theta^{\dot \alpha}\,,\quad
\bar Q_{\dot\alpha}=i\frac{\partial}{\partial \bar\theta^{\dot\alpha}} -
\theta^{\alpha}\partial_{\alpha\dot\alpha}\,, \quad \left\{Q_\alpha\,,\bar
Q_{\dot\alpha}\right\}=2i\partial_{\alpha\dot\alpha}\,.
\label{diffq}
\end{equation}
These differential operators give the explicit realization of the SUSY algebra,
\begin{equation}
\left\{Q_\alpha\,,\bar Q_{\dot\alpha}\right\}=2P_{\alpha\dot\alpha}\,,\quad
\left\{Q_\alpha\,, Q_{\beta}\right\}=0\,,\quad \left\{\bar Q_{\dot\alpha}\,,
\bar Q_{\dot\beta}\right\}=0\,,\quad
\left[Q_\alpha\,,P_{\beta\dot\beta}\right]=0\,,
\label{susyalgebra}
\end{equation}
where  $Q_\alpha$ and $\bar Q_{\dot\alpha}$ are the supercharges while
$P_{\alpha\dot\alpha}=i\partial_{\alpha\dot\alpha}$ is the energy-momentum
operator.  The {\em superderivatives} are defined as the differential operators
$\bar D_\alpha$, $D_{\dot\alpha}$
anticommuting with $Q_\alpha$ and $\bar
Q_{\dot\alpha}$,
\begin{equation}
D_\alpha=\frac{\partial}{\partial \theta^\alpha} -i \partial_{\alpha\dot\alpha}
\bar\theta^{\dot \alpha}\,,\quad
\bar D_{\dot\alpha}=-\frac{\partial}{\partial \bar\theta^{\dot\alpha}} +i
\theta^{\alpha}\partial_{\alpha\dot\alpha}\,, \quad \left\{D_\alpha\,,\bar
D_{\dot\alpha}\right\}=2i\partial_{\alpha\dot\alpha}\,.
\end{equation}

 \subsection{The Grassmann integration conventions}
\label{sec31}
 \beq
 \int d^2\theta \,\theta^2 =1\,,\qquad \int d^2\theta d^2\,\bar \theta \,\,\theta^2\,
\bar\theta^2=1\,.
 \eeq

 \subsection{$(1,0)$ and $(1,0)$ sigma matrices}
  \label{seca3dop}

  To convert the two-index spinorial symmetric representation
  in the vectorial representation we will need the following sigma
  matrices:
  \beqn
  \left(\vec \sigma\right)^{\alpha\beta} &=&
  \{\tau^3, \,\,   i, \,\, -\tau^1\}_{\alpha\beta}\,,\qquad
  \left(\vec \sigma\right)_{\alpha\beta} =
  \{-\tau^3, \,\,   i, \,\, \tau^1\}_{\alpha\beta}\,,
  \nonumber\\[3mm]
  \left(\vec \sigma\right)^{\dot\alpha\dot\beta} &=&
  \{\tau^3, \,\,   -i, \,\, -\tau^1\}_{\dot\alpha\dot\beta}\,,\qquad
  \left(\vec \sigma\right)_{\dot\alpha\dot\beta} =
  \{-\tau^3, \,\,  - i, \,\, \tau^1\}_{\dot\alpha\dot\beta}\,.
  \eeqn

\subsection{The Weyl and Dirac spinors}
  \label{seca3}

If we have two Weyl (right-handed) spinors
$\xi^\alpha$ and $\eta_\beta$, transforming in the
representations $R$ and $\bar R$ of the gauge group, respectively,
then the Dirac spinor $\Psi$ can be formed as
\beq
\Psi = \left(\begin{array}{l}
\psi^\alpha \\[2mm]
\bar{\tilde\psi}_{\dot\alpha}
\end{array}\right)\,.
\eeq
The Dirac spinor $\Psi$ has four components, while
$\xi^\alpha$ and $\eta_\beta$ have two components each.

\subsection{Euclidean notation}
  \label{seca4}

As was mentioned, in Part II
we switch to a formally Euclidean notations e.g.
\beq
F_{\mu\nu}^2 = 2F_{0i}^2 + F_{ij}^2\,,
\eeq
and
\beq
(\partial_\mu a)^2 = (\partial_0 a)^2 +(\partial_i a)^2\,,
\eeq
 etc.
This is appropriate, since we mostly consider
static (time-independent)
field configurations, and $A_0 =0$. Then the Euclidean action is
nothing but the energy functional.

Then, in the fermion sector  we have to define
 the Euclidean matrices
\beq
\sigma^{\alpha\dot{\alpha}}=(1,-i\vec{\tau})_{\alpha\dot{\alpha}}\,,
\eeq
and
\beq
\bar{\sigma}_{\dot{\alpha}\alpha}=(1,i\vec{\tau})_{\dot{\alpha}\alpha}
\,.
\eeq
 Lowing and raising
 of the spinor indices
is performed by
virtue of the antisymmetric tensor defined as
\beqn
\varepsilon_{12}=\varepsilon_{\dot{1}\dot{2}}=1\,,
\nonumber\\[2mm]
\varepsilon^{12}=\varepsilon^{\dot{1}\dot{2}}=-1\,.
 \eeqn
 The same raising and lowering convention applies to the flavor SU(2)$_{R}$
 indices $f$, $g$, etc.

When the contraction of the spinor indices is assumed inside the parentheses
we use the
following notation:
\beq
(\lambda\psi)\equiv \lambda_{\alpha}\psi^{\alpha},\qquad
(\bar{\lambda}\bar{\psi})\equiv \bar{\lambda}^{\dot{\alpha}}\bar{\psi}_{\dot{\alpha}}.
\eeq

\subsection{Group-theory coefficients}
  \label{seca5}

As was mentioned, the gauge group is assumed to be SU($N$). For a given
representation $R$ of SU($N$), the definitions of the
Casimir operators to be used below are
\beq
{\rm Tr} (T^a T^b )_R= T(R) \delta^{ab}\,,\qquad (T^aT^a)_R = C(R) \,I\,,
\label{casop}
\eeq
where  $I$ is the unit
matrix in this representation. It is quite obvious that
\beq
C(R) = T(R)\,\,\frac{{\rm dim} (G)}{{\rm dim} (R)}\,,
\label{gripp}
\eeq
where ${\rm dim} (G)$ is the dimension of the group
(= the  dimension of the adjoint representation).
Note that $T(R)$ is also known as (one half of) the Dynkin index,
or the dual Coxeter number.
For the adjoint representation, $T(R)$ is denoted by $T(G)$. Moreover,
 $T({\rm SU}(N)) = N$.

\subsection{Renormalization-group conventions}
  \label{seca6}

We use the following definition of the $\beta$ function and anomalous dimensions:
\beq
\mu\,\frac{\partial\alpha}{\partial \mu}\equiv \beta (\alpha)
=-\frac{\beta_0}{2\pi } \alpha^2 -\frac{\beta_1}{4\pi^2 } \alpha^3 +...
\eeq
while
\beq
\gamma  = -d \ln Z (\mu ) /d\ln\mu \,.
\label{defgin}
\eeq
In supersymmetric theories
\beq
\beta (\alpha) = -\frac{\alpha^2}{2\pi}\left[3\,T(G) -\sum_i T(R_i)(1-\gamma_i )
\right]\left(1-\frac{T(G)\,\alpha}{2\pi} \right)^{-1}
\, ,
\eeq
where the sum runs over all matter supermultiplets.
The anomalous dimension of the $i$-th  matter superfield is
\beq
\gamma_i =  -2 C(R_i)\,\frac{\alpha}{2\pi } + ...
\eeq
Sometimes, when one-loop anomalous dimensions
are discussed,  the coefficient in front of
$-\alpha/(2\pi )$ in (\ref{defgin})  is also
referred to as  an ``anomalous dimension."

\section*{Appendix B. Strings in \ntwo SQED}
\addcontentsline{toc}{section}{Appendix B. Strings in \ntwo SQED}

\renewcommand{\theequation}{B.\arabic{equation}}
\setcounter{equation}{0}

 \renewcommand{\thesubsection}{B.\arabic{subsection}}
\setcounter{subsection}{0}

In this Appendix we briefly review the Abelian Abrikosov--Nielsen--Olesen  strings
in \ntwo supersymmetric QED in four dimensions. The BPS strings in this theory were first considered in \cite{EFMG,VY}.

\subsection{\ntwo supersymmetric QED}
  \label{secb1}

\ntwo supersymmetric QED is discussed in Sect. \ref{n2qed}. Here we
summarize basic
features of this theory for convenience.
The field content of \ntwo supersymmetric QED consist of U(1) vector
\ntwo multiplet as well as $N_f$  matter hypermultiplets.
The mass terms are introduced via the superpotential
\beq
W =\sum_{A} \left( m_A Q^A \tilde{Q}_A + \frac{1}{\sqrt{2}}
{\mathcal A}  Q^A \tilde{Q}_A\right)\,.
\eeq
For the definition of the Fayet--Iliopoulos term
see Eq.~(\ref{sqed}). In this form it is the same in \none and \ntwo,
cf. Eq.~(\ref{fiterm}).
 The bosonic part of the action of this
theory is
\beqn
&&
S=\int d^4 x \left\{ \frac{1}{4 g^2} F_{\mu \nu}^2 + \frac{1}{g^2}
|\partial_\mu a|^2 +\bar{\nabla}_\mu \bar{q}_A \nabla_\mu q^A +
\bar{\nabla}_\mu \tilde{q}_A \nabla_\mu \bar{\tilde{q}}^A
\right.\nonumber\\[3mm]
 &&
 +
 \left.n_e^2\frac{g^2}{2}\left(|q^A|^2-|\tilde{q}_A|^2-\xi\right)^2+2n_e^2g^2
\left|\tilde{q}_A q^A\right|^2
\right.
 \nonumber\\ [3mm]
&&
\left.
+\frac{1}{2}(|q^A|^2+|\tilde{q}^A|^2)\left| a+\sqrt{2}m_A\right|^2 \right\},
\label{N2qed}
\eeqn
where
\beq
\nabla_\mu=\partial_\mu-in_e A_\mu\,,\qquad
\bar{\nabla}_\mu=\partial_\mu+in_eA_\mu\,.
\eeq
Here $\xi$ is the coefficient in front of the Fayet--Iliopoulos term,
we consider FI $D$-term here while
$g$ is the U(1)
gauge coupling and $n_e$ is the electric charge. It can be integer or half integer. The
index $A=1,..., N_f$ is the flavor index.
Below  we consider the case $N_f=1$. This is the simplest case
which admits BPS string solutions.

The FI term triggers the squark condensation.
The vacuum of this theory is given by
\beq
a=-\sqrt{2} m,\qquad q=\sqrt{\xi}, \qquad  \tilde{q}=0\,,
\label{qedvac}
\eeq
Hereafter in search for string solutions
we will stick to the {\em ansatz} $\tilde{q}=0$.

Now let us discuss the mass spectrum of light fields in this vacuum.
The spectrum  can be obtained by diagonalizing  the quadratic form
in  (\ref{N2qed}). This is done in Ref.~\cite{VY};  the result is as
follows:  one real component of field $q$ is eaten up by the Higgs
 mechanism to become the third components of the massive photon. Three
components of the massive photon, one remaining component of $q$ and
four real components of  the fields
$\tilde{q}$ and $a$ form one long \ntwo
multiplet (8 boson states + 8 fermion states), with mass
\beq
\label{mgammaqed}
m_{\gamma}^2=2n_e^2\, g^2\,\xi.
\eeq

\subsection{String solutions}
  \label{secb2}

As soon as fields $a$ and $\tilde{q}$ plays no role in string solutions we can
look for these solutions using reduced theory with these fields set to zero. The
bosonic action (\ref{N2qed}) reduces to
\beq
S = \int {\rm d}^4x\left\{
 \frac1{4g^2}F_{\mu\nu}^{2}
 + |\nabla_\mu q|^2+\frac{g^2}{2}\,n_e^2\,\left(
 | q|^2 - \xi \right)^2\right\}.
\label{redn2qed}
\eeq

Since  U(1) gauge group is  spontaneously broken,
the model supports
conventional ANO strings \cite{ANO}.
The topological stability of the ANO string is due to the fact that
$\pi_1({\rm U(1)}) = Z$.

Let us derive first order equations which determines string solution
  making use of the Bogomol'nyi representation \cite{B} of the model
(\ref{redn2qed}). We have for the string tension
\beqn
T
&=&
\int{d}^2 x   \left\{
\left[\frac1{\sqrt{2}g}F^{*}_{3} +
\frac{g }{\sqrt{2}}\,n_e
\left(|q|^2-\xi \right)\right]^2
\right.
\nonumber\\[3mm]
&+&
\left.
 \left|\nabla_1 \,q +
i\nabla_2\, q\right|^2
 +n_e\,\xi\,  F^{*}_3\right\},
\label{bogsqed}
\eeqn
where $F_{3}^{*}= F_{12}$
and we assume that fields depend only on coordinates $x_i$, $i=1,2$.

The Bogomol'nyi representation (\ref{bogsqed})
leads us to the following first-order equations:
\beqn
&& F^{*}_{3}+g n_e\left(|
q|^2-\xi\right)\,=0\, ,
\nonumber\\[3mm]
&& (\nabla_1+i\nabla_2)q=0\, .
\label{foesqed}
\eeqn
Once these equations are satisfied the energy of the BPS object is
given by the last surface term in (\ref{bogsqed}). Note that representation
(\ref{bogsqed}) can be written also with different sign in front of terms
proportional
to gauge fluxes. This would give first order equations for the anti-string
with negative values of gauge fluxes.

For topologically stable string solution the scalar field winds $n$ times in
U(1) gauge group when we move around the string along large circle in $(x,y)$ plane
(we assume that  the string stretches along $z$-axis),
$$
q\sim \sqrt{\xi}\,e^{in\alpha},
$$
\beq
A_i\sim \frac{n}{n_e}\,\pt_i \alpha, \qquad r\to\infty,
\label{stringassimpt}
\eeq
where $r$ and $\alpha$ are the polar coordinates in $(x,y)$ plane and $i=1,2$.
This ensures that the flux of the string is
\beq
\int d^2 x F^{*}_3=\frac{2\pi n}{n_e}
\label{stringflux}
\eeq
The tension of the string with winding number $n$ is determined by the surface term
in (\ref{bogsqed}),
\beq
T_n=2\pi n\,\xi.
\label{astrten}
\eeq

For the elementary $n=1$ string the solution can be found using the standard
{\em ansatz}  \cite{B}
\begin{equation}
q(x) = \phi(r)\,  e^{i\,\alpha}\;,\qquad
A_i(x) =\frac1{n_e}\,\pt_i \alpha\,[1- f(r)]\ ,
\label{astringprofile}
\end{equation}
where we introduced two profile functions $\phi$ and $f$ for scalar and gauge fields
respectively.

The {\em ansatz} (\ref{astringprofile})
goes through the set of equations
(\ref{foesqed}), and  we get the following two equations
on the profile functions:
\begin{equation}
-\frac1{r}\,\frac{ d f}{ dr} +n_e^2g^2\left(\phi^2-\xi\right) = 0\ ,\qquad
r\, \frac{  d\,\phi}{ dr}- f\,\phi= 0\ .
\label{astringfoe}
\end{equation}

The boundary conditions for the   profile functions
are the following. At large distances we have
\beq
\phi(\infty)=\sqrt{\xi},\qquad f(\infty)= 0\,.
\label{astrbcinf}
\eeq
At the origin
the smoothness of the field configuration at hand
 requires
\begin{equation}
\phi(0)=0,\qquad f(0)= 1\,.
\label{astrbczero}
\end{equation}
These boundary conditions are such that
  the scalar field reaches its vacuum value  at
infinity. The same first order equations arise for BPS vortex in
\none QED in (2+1) dimensions, see Sect. \ref{main}. The fermion zero modes for
BPS vortices in (3+1) and (2+1) dimensions are different, however.
Equations (\ref{astringfoe})    have  a
unique solution for the profile functions, which can be found numerically
\cite{dvsch}, see Fig.~\ref{figano}. The string transverse size is $\sim 1/m_{\gamma}$.

First order equations (\ref{astringfoe}) can be also obtained using
supersymmetry.
We start from  the supersymmetry transformations for
the fermion fields in  the  theory (\ref{N2qed}),
\begin{eqnarray}
\delta\lambda^{\alpha f}
&=&
\frac12(\sigma_\mu\bar{\sigma}_\nu\epsilon^f)^\alpha
F_{\mu\nu}+\epsilon^{\alpha p}F^m(\tau^m)^f_p\ +\dots,
\nonumber\\[3mm]
\delta\bar{\tilde\psi}_{\dot{\alpha}}^{A}
&=&
i\sqrt2\
\bar\nabla\hspace{-0.65em}/_{\dot{\alpha}\alpha}q_f^{A}\epsilon^{\alpha
f}\ +\cdots,
\nonumber\\[3mm]
\delta\bar\psi_{\dot{\alpha}A}
&=&
 i\sqrt2\
\bar\nabla\hspace{-0.65em}/_{\dot{\alpha}\alpha}\bar
q_{fA}\epsilon^{\alpha f}\ +\cdots.
\label{qedtransf}
\end{eqnarray}
Here $f=1,2$ is the SU(2)$_R$ index
so $\lambda^{\alpha f}$
are the fermions from the ${\mathcal N}=2$ vector supermultiplet,
 while $q^{Af}$ denotes SU(2)$_R$ doublet of squark fields
 $q^{A}$ and $\bar{\tilde{ q}}^{A}$ in the
quark hypermultiplets.
The parameters of SUSY transformations
in  are denoted as  $\epsilon^{\alpha f}$.
Furthermore, the $F$ terms in Eq.~(\ref{qedtransf}) are
\beq
 F^3=
-i\, n_e\,g^2\, \left({\rm Tr} \, |q|^2-\xi\right)\, ,\qquad F^1+ iF^2=0
\label{fterm}
\eeq
 The dots in (\ref{qedtransf})  stand for terms involving
the $a$ field which vanish on the string solution
 because the it is  given by its vacuum expectation value
(\ref{qedvac}).

In Ref.~\cite{VY} it was shown that the four (real) supercharges generated by
\beq
\epsilon^{12}, \qquad
\epsilon^{21}
\label{astrtrivQ}
\eeq
act trivially on the BPS string. Namely imposing conditions $\epsilon^{11}=\epsilon^{22}=0$
and  requiring that left hand sides of Eqs. (\ref{qedtransf}) are zero we
get first order equations (\ref{astringfoe}) upon substitution of the {\em ansatz}
(\ref{astringprofile}) \footnote{If we instead of (\ref{astrtrivQ}) impose
that different
combinations of SUSY transformation parameters act trivially
 we get the equations
for anti-string with opposite directions of gauge fluxes.}.

\subsection{The fermion zero modes}
  \label{secb3}

 The string is half-critical,
so 1/2 of supercharges (related to SUSY transformation parameters $\epsilon^{12}$
and $\epsilon^{21}$),  act trivially on the string solution.
The remaining four (real) supercharges
parametrized by $\epsilon^{11}$ and $\epsilon^{22}$
generate four (real)  supertranslational fermion zero modes.
They have the form \cite{VY}
\beqn
\bar{\psi}_{\dot{2}}
& = &
-2\sqrt{2}\,\frac{x_1+ix_2}{r^2}\,f\,\phi\,\alpha^{11}
\, ,
\nonumber\\[3mm]
\bar{\tilde{\psi}}_{\dot{1}}
& = &
2\sqrt{2}\,\frac{x_1-ix_2}{r^2}\,f\,\phi\,\alpha^{22} ,
\nonumber\\[3mm]
\bar{\psi}_{\dot{1}}
& = &
0\, , \qquad
\bar{\tilde{\psi}}_{\dot{2}}= 0\, ,
\nonumber\\[3mm]
\lambda^{11}
& = &
-in_e g^2\left(\phi^2-\xi\right)\alpha^{11}\, ,
\nonumber\\[3mm]
\lambda^{22}
& = &
in_e g^2\left(\phi^2-\xi\right)\alpha^{22}\, ,
\nonumber\\[3mm]
\lambda^{12}
& = & 0
\, ,\qquad  \lambda^{21}= 0\,,
\label{abzmodes}
\eeqn
where the modes proportional to complex Grassmann parameters
$\alpha^{11}$ and $\alpha^{11}$ are generated by $\epsilon^{11}$ and
$\epsilon^{22}$  transformations, respectively.

\end{document}